\documentclass[phd, 11pt, a4paper, inlinechaptertoc]{psuthesis}
\usepackage[left=3.3cm,top=4cm,right=2.1cm,bottom=3cm]{geometry} 
\usepackage{fancyhdr}
\usepackage{amsmath}
\usepackage{amssymb}
\usepackage{amsthm}
\usepackage{amsmath}
\usepackage{exscale}
\usepackage{dsfont}
\usepackage{fancybox}
\usepackage[mathscr]{eucal}
\usepackage{bm}
\usepackage{eqlist} 
\usepackage[final]{graphicx}
\usepackage[dvipsnames]{color}
\usepackage{bbding,hyperref}
\definecolor{maroon}         {cmyk}{0   , 0.93, 0.9   , 0.40}

\usepackage{braket}
\usepackage{tikz}

\usepackage{url}

\usepackage{soul}

\newtheorem{theorem}{Theorem}
\newtheorem{prop}{Proposition}

\DeclareGraphicsExtensions{.pdf, .jpg, .png}

\DeclareUnicodeCharacter{2009}{\,} 

\usepackage[Lenny]{fncychap}
\ChTitleVar{\Huge\sffamily\scshape}

\title{Quantum Groups and Asymptotic Symmetries
}

\author{Josua Unger}
\dept{Institute for Theoretical Physics}
\degreedate{November 2021}
\copyrightyear{2021}

\documenttype{Thesis}

\submittedto{Institute for Theoretical Physics}

\advisor[~~]
        {\textbf{Jerzy Kowalski-Glikman test}}
        {Professor of Institute for Theoretical Physics, University of Wroclaw}
\begin{document}

\frontmatter

%



\psutitlepage
\clearpage
\thispagestyle{empty}
\mbox{}
\clearpage




%
\pagestyle{plain}
\chapter{Abstract}
    \begin{singlespace}
~~

This thesis is devoted to the study of Lie bialgebra and Hopf algebra structures related to certain versions of non-commutative geometry constructed on infinite-dimensional Lie algebras that arise in the context of asymptotic symmetries of spacetime. We prove a number of theorems about cohomology groups that aid the classification of the Lie bialgebras and explicitly construct and analyze selected Hopf algebras. Particularly interesting behavior was found by studying the contraction limit of spacetimes with cosmological constant and the inclusion of central charges on the level of Lie bialgebras and Hopf algebras. Phenomenological consequences, like deformed in-vacuo dispersion relations, known from the study of $\kappa$-Poincar\'e quantum groups, are investigated. Furthermore, we examine how a new proposal in the context of the black hole information loss paradox and counterarguments against it are affected.
    \end{singlespace}
\newpage



%
\chapter{Acknowledgments}
\begin{singlespace}
\textit{I would like to thank my family, friends and everyone who supported me during my studies. In particular, I thank Remi, Thomas, Aleksander, Matze, Lennart, Giacomo, Andrzej and Jenni for proofreading parts of the thesis.}
\\[310pt]
\textit{I would like to thank my supervisor, Professor Jerzy Kowalski-Glikman, for his support and a fruitful collaboration during my studies}.\\[35pt]
\footnotesize
This PhD dissertation has been supported by the grants 2017/27/B/ST2/01902 and 2019/33/B/ST2/00050 from the National Science Center and the Grant M\l odych 2018/19 from the University of Wroc\l aw.

\newpage
\thispagestyle{empty}
~~
\newpage
\thispagestyle{empty}
~~\\[550pt]
\large
\begin{flushright}
    \textit{A well-chosen generalization supplies for a theory the effect of a frame on a painting, and enables it to be looked at `from the outside' as it were.}
\end{flushright}
\begin{flushright}
   \normalsize \textsc{Kirill C.H. Mackenzie} 
\end{flushright}
\newpage
\thispagestyle{empty}
\mbox{}

\end{singlespace}



\thesistableofcontents

\thesismainmatter

\allowdisplaybreaks{
%

\pagestyle{fancy}
\fancyhead{}  
\fancyfoot{}  
\fancyhead[LE,RO]{\thepage}

\chapter{Introduction}

After the dust had settled around the major scientific revolutions of the 20th century, general relativity and quantum mechanics, these two are standing as firm as ever. Predictions of general relativity, like the existence of black holes or gravitational waves, that seemed preposterous when they were first made, have been spectacularly confirmed. The standard model of particle physics, an incarnation of a quantum field theory, has passed (almost) all high-precision tests with flying colors. While attempts to reconcile these frameworks, a notoriously difficult quest known as quantum gravity, certainly led to new insights and theoretical progress, especially in String Theory and Loop Quantum Gravity, distinct predictions that can be tested in current experiments, remain elusive. A main reason for this is that the scale at which both general relativity and quantum mechanics are basically guaranteed to be relevant, the Planck scale (or Planck mass) $m_{\text{Pl}} \sim 10^{19}~GeV$, is orders of magnitude larger than what is currently obtainable in high energy experiments. This scale has its origin in a famous thought experiment; if we want to probe ever shorter distances e.g. with photons, their wavelength has to become shorter and thus the energy increases. When the Planck scale is reached, with which we could achieve resolutions of about the Planck length $l_{\text{Pl}} \sim 1/m_{\text{Pl}}$, the energy in the collision region is so high that a black hole is formed. On the one hand, this means that both general relativity and quantum mechanics are needed to describe the experiment, but also that shorter distances than $l_{\text{Pl}}$ are operationally meaningless. 

Some theories and frameworks bundled under the notion of non-commutative geometry can be seen as attempts to remedy parts of the problems mentioned above. In particular, the $\kappa$-Poincar\'e quantum group and related constructions feature an observer independent scale governing the breakdown of spacetime as a smooth manifold as heralded by the thought experiment from the previous paragraph. Its origin lays in the mathematical advances around Hopf algebras by the mathematical physicists Drinfeld, Fadeev, Jimbo and Woronowicz (\cite{Drinfeld:1985rx}, \cite{Jimbo:1985zk}, \cite{Faddeev:1987ih}, \cite{Woro87}) which in turn were used by Lukierski, Ruegg, Majid and collaborators in \cite{Lukierski:1991pn}, \cite{Lukierski:1993wx}, \cite{Majid:1994cy}. While we leave a more detailed introduction to chapter \ref{chap2}, it should be noted that instead of spacetime itself, its (deformed) symmetries play a fundamental role in $\kappa$-Poincar\'e. This feature is also motivated by the fact that an observer does not have direct access to spacetime coordinates and only measures the (angular) momenta of incoming particles associated with the spacetime symmetries via the Noether theorem. On its own, $\kappa$-Poincar\'e is not a complete theory of quantum gravity but rather aims to describe effective phenomena from such a (unspecified) theory. It is expected that in various ways these phenomena already contribute at some intermediate (``mesoscopic'') scale and thus provide some empirical clues to the nature of quantum gravity. An important example, especially in the age of multi-messenger astronomy \cite{Santander:2016bvv}, is the possibility of deformed in vacuo dispersion relations where effect sizes can be enhanced by cosmological length scales. 
As non-commutative geometry does not explicitly deal with dynamical aspects, a key part of its study is to consistently formulate general relativity and quantum field theory in terms of the non-commutative foundation. Considerable progress has been made in the understanding of QFT on so-called Moyal non-commutative spaces (\cite{Doplicher:1994zv}, \cite{Doplicher:1994tu}, \cite{Wulkenhaar:2006si} and references therein) but, despite several valiant efforts (\cite{Lukierski:1992dt},\cite{Kosinski:1998kw}, \cite{Freidel:2007hk}, \cite{Arzano:2007ef}, \cite{Daszkiewicz:2008bm}, \cite{Poulain:2018two}, \cite{Govindarajan:2009wt}, \cite{Mercati:2018hlc}, \cite{Arzano:2020jro}), there was no matching success for a QFT description in $\kappa$-Poincar\'e. In \cite{Aschieri:2005zs} a proposal for expressing the concepts underlying general relativity in a $\kappa$-Poincar\'e framework (or at least some variant of it) was brought forward.

Seemingly unrelated, in recent years there has been a renewed interest in studying the asymptotic symmetries of spacetime (cf. chapter \ref{chap3}), in part due to new theoretical understanding but also because of the prospect of testing phenomenological consequences like the gravitational memory effect. Also in the endeavor to reformulate physics in the ``bulk'' of spacetime in terms of a dual theory on the boundary, the so-called holographic principle with its most prominent example, the AdS/CFT correspondence \cite{Maldacena:1997re}, the asymptotic symmetries play a vital role. Furthermore, as reviewed in \cite{Strominger:2017zoo}, they are deeply connected to infrared phenomena like the soft theorems by Weinberg \cite{Wei65}.            

A hallmark of the corresponding symmetry algebras is that they are infinite dimensional but always contain the Poincar\'e algebra (or its analogon in spacetimes with constant curvature). Thus a natural question appears which will be at the heart of this thesis: How can we consistently formulate deformed quantum groups like the $\kappa$-Poincar\'e if they are to be embedded in the asymptotic symmetry algebras? And more generally, what kind of quantum groups are possible on the infinite dimensional algebras? Armed with the technical constraints coming from this analysis we can then investigate potential phenomenological consequences, some of which are known from the study of quantum groups, while others arise only in the setting of asymptotic symmetries. So far, the particular combination of non-commutative geometry and asymptotic symmetry algebras has not been studied from a physical point of view, however, several contributions to the relevant mathematical underpinnings have been made, e.g. by \cite{Ng:2000}. 

Even though our universe exhibits a four dimensional spacetime, it is interesting to study the three dimensional case as well beyond the obvious reason that many things are technically easier. On the gravity side this has been particularly fruitful as the three dimensional general relativity has no dynamical degrees of freedom and allows for quantization \cite{ch:I-2Witten:1988hc}. Similarly, the study of three dimensional spacetime revealed connections between quantum gravity and $\kappa$-Poincar\'e \cite{Cianfrani:2016ogm}. 

This thesis is structured as follows. In the second chapter, we start by motivating some of the mathematical ideas behind quantum groups (Hopf algebras) and develop its framework mainly following \cite{Chari:1994pz}, \cite{klimyk2011quantum}. We put special emphasis on deformations that can be obtained from a so-called twist. Then the mathematical structures are endowed with a physical interpretation focusing on the $\kappa$-Poincar\'e. 
In chapter \ref{chap3}, the derivation of the symmetry algebra of asymptotically flat four dimensional spacetime is reviewed and related to the algebra of the conserved soft charges. In a similar way the more recent constructions of symmetries of asymptotically (anti) de Sitter spacetimes are introduced in three and four dimensions and in the former case we highlight the possibility of viewing the asymptotically flat case as a contraction limit.  

After these introductory parts, the content of which is mostly well-known, chapter \ref{chap4} features mainly original research by the author based on the publications \cite{Borowiec:2018rbr}, \cite{Borowiec:2020ddg} and \cite{Borowiec:2021odp}. In particular, we classify Lie bialgebras on symmetry algebras of asymptotically flat and (A)dS spacetimes in three and four dimensions building on recent work by \cite{Borowiec:2017apk}, \cite{Borowiec:2015nlw}, \cite{Kowalski-Glikman:2019ttm} in the finite dimensional cases. This rather technical part is supplemented by theorems about certain cohomology groups for which we give elementary proofs in Appendix \ref{app-a}. Some of the theorems are already known in the mathematical literature \cite{Ng:2000}, \cite{Junbo:2010}. 
We then continue by explicitly constructing several Hopf algebras in all orders of the deformation parameter and analyze them. Thereby a special focus lies on the possibility of the so-called specialization, a requirement that turns out to be rather restricting in the infinite dimensional setting.
In chapter \ref{chap5}, we apply the previously obtained structures to non-commutative geometry and study potential phenomenological consequences. While the latter draws from results known in the $\kappa$-Poincar\'e literature, its application to asymptotic symmetries constitutes novel research. 
Furthermore, we review a potential solution to the black hole information loss paradox brought forward by Strominger, Hawking and Perry in \cite{Hawking:2016msc} and some critical responses to which we in turn reply with a new proposal that could remedy the original argument.

Finally, in chapter \ref{chap6} the results are summarized and put in the context of the main goals of the thesis. Some open questions and further research directions are highlighted.

\chapter{Quantum Groups and Non-commutative Geometry}\label{chap2}

In this chapter, the underlying ideas of quantum groups are introduced, following \cite{Chari:1994pz}, and the mathematical framework is developed \cite{Chari:1994pz},\cite{klimyk2011quantum}. We present the connection to non-commutative geometry, focussing on the so-called $\kappa$-Poincar\'e formulation and highlight conceptual differences in various approaches to non-commutativity.

\section{Quantum Groups}

In classical mechanics the state of a system at time $t$ is described as a point $m(t)$ on a Poisson manifold $M$. The evolution of functions $f$ of this state, representing an observable, is governed by the equation
\begin{align}\label{hamilton}
\frac{d f}{d t} (m(t)) = \{ \mathcal{H}, f\} (m(t))
\end{align}
where $\mathcal{H}$ is the classical Hamiltonian and $\{,\}$ denotes the Poisson bracket defined on smooth functions on $M$ (cf. next section for more details). For the coordinate functions, e.g. the canonical momentum $p$ and position $q$ of a particle in one dimension it is given by
\begin{align*}	
\{p, q\} = 1
\end{align*}
and \eqref{hamilton} become Hamilton's equation.
In contrast, the description of a state in quantum mechanics is given by a (complex) ray in a Hilbert space $V$ and observables of the system are represented by linear operators $A$ on $V$. Their time evolution is governed by the commutator
\begin{align*}
\frac{d A}{dt} = [\mathcal{H}, A]
\end{align*}
with the quantum Hamiltonian $\mathcal{H}$. 
Passing from the classical to the quantum mechanical description is known as the problem of quantization. Of course, there is no one to one correspondence between these descriptions but one possibility, due to Moyal \cite{Moyal:1949sk}, is to reformulate the problem in terms of finding a ``deformed'' product $\star_h$ on the algebra of functions of a manifold so that the commutative product is obtained in the limit of vanishing deformation parameter $h$ and the first order deformation satisfies 
\begin{align}
\underset{h \rightarrow 0}{\text{lim}} \frac{f_1 \star_h f_2 - f_2 \star_h f_1}{h} = \{ f_1, f_2\}.
\end{align}
Equipped with such a product, the algebra of functions is related to non-commutative geometry, as introduced by Connes \cite{Con94}, in the following way. A theorem by Gelfand and Naimark \cite{GelNeu43} states that there is a one to one correspondence of commutative $C^*$ algebras and compact Hausdorff spaces, i.e. there is a duality between the category of certain spaces and algebras of functions on that space. Extending this concept, one can view quantum spaces as a dual category of associative but not commutative algebras (see also \cite{Sitarz:2013yva} for a pedagogical introduction). These spaces are not necessarily manifolds and to describe the symmetries of them we need to generalize the notion of a group as well. Therefore it is helpfull to focus on functions on a group instead of the group itself, much like in the treatment of quantum spaces. As will be discussed in section \ref{sec2.1.4}, this introduces coproduct and antipode as pullback from the group multiplication and inverse as well as a counit and thus leads to the notion of a Hopf-algebra. 

It should be noted that different authors use varying more or less restrictive definitions of what a quantum group is. For example \cite{klimyk2011quantum} takes the stance to view every Hopf algebra with interesting applications as a quantum group. In particular the deformations of enveloping Hopf algebras of Lie algebras (see section \ref{sec2.1.6}) serve as a guiding example. Others have focussed on objects with extra structure that put more emphasis on the origin as quantized algebras of coordinate functions as e.g. the compact matrix pseudogroups of \cite{Woro87}. In this thesis we adopt the former point of view and loosely use the terms quantum group and Hopf algebra synonymously. 


\subsection{Lie Bialgebras and Poisson Lie Groups}

The concept of a Lie bialgebra can on the one hand be seen as an infinitesimal version of a Hopf algebra, on the other hand it arises also naturally in the context of Poisson Lie groups, as we will show in the following to motivate the definition \cite{Chari:1994pz}. 

A \textit{\textbf{Poisson Lie group}} is a Lie group that is also a Poisson manifold. A \textit{\textbf{Poisson manifold}} in turn is a manifold $M$ with a bilinear map defined for smooth real functions on $M$
$$
\{ , \} : C^{\infty}(M) \times C^{\infty} (M) \rightarrow C^{\infty} (M),
$$
which is antisymmetric, satisfies the Jacobi identity
$$
\{f_1, \{f_2, f_3\}\}+ \{f_3, \{f_1, f_2\}\} + \{f_2, \{f_3, f_1\}\} = 0
$$
and is a derivation
$$
\{f_1 f_2, f_3\} = f_1 \{f_2, f_3\} + \{f_1, f_3\} f_2.
$$
Additionally the group multiplication $\mu$ and the Poisson structure of the Poisson Lie group have to be compatible in the sense that $\mu$ is a Poisson map, i.e. 
\begin{align}\label{compat1}
\{ f_1, f_2 \} (x_0 y_0) = \{f_1(x y_0) , f_2(x y_0)  \} (x_0) + \{f_1(x_0 y) , f_2(x_0 y)  \} (y_0)
\end{align}
Restricted to the maximal ideal $I$ of functions vanishing at the unit $e$ of $M$, $I^2$, generated by functions of second order in local coordinates that also vanish at $e$, is a Lie-ideal, i.e. 
\begin{align}
\{ , \} : I \otimes I^2 \rightarrow I^2.
\end{align}
Thus the bracket can be defined on the coset $I/I^2$ consisting of equivalence classes $\{[f] = f + g, f\in I, g \in I^2\}$ and the result of taking a bracket does not depend on the representative of the equivalence class. One can show that $I/I^2$ is isomorphic to the cotangent space $T^*_e M$ \cite{reid1988undergraduate} which can be identified with the dual Lie algebra $\mathfrak{g}^*$ of $M$ where $\{ , \}$ is therefore defining a Lie bracket. Its dual $ \{ , \}^* \equiv \delta : \mathfrak{g} \rightarrow \mathfrak{g} \otimes \mathfrak{g}$ is called the cobracket which satisfies the Co-Jacobi identity
\begin{align}\label{cojac}
\text{Cycl}((\delta \otimes \text{id} ) \delta (x)) = 0
\end{align}
with $\text{Cycl}(a \otimes b \otimes c) = a \otimes b \otimes c + c \otimes a \otimes b + b \otimes c \otimes a$. As the Jacobi identity for a Lie algebra encodes the concept of associativity, the co-Jacoby identity is said to express the dual coassociativity.

Duality here refers to the dual vector space $V^*$ which can be seen as the vector space of linear functionals on $V$. The evaluation of these functionals on elements of $V$ defines a nondegenerate bilinear form $\braket{\cdot ,\cdot}$ called the dual pairing. 

Furthermore, \eqref{compat1} ensures a compatibility relation between the bracket and cobracket. To see this, we note that expressed in terms of the Poisson bivector $\Pi: M \rightarrow \bigwedge^2 T M$, defined by
\begin{align}
\{f, g\} = (df \otimes dg) \Pi,
\end{align}
\eqref{compat1} reads
\begin{align}
(d f \otimes dg) \Pi(x_0 y_0)  & =  (d (f \circ R_{y_0}) \otimes d(g \circ R_{y_0})) \Pi(x_0) + (d (f \circ L_{x_0}) \otimes d(g \circ L_{x_0})) \Pi(y_0) \nonumber \\
& = (df \otimes dg) ( d_x (R_{y_0}) \otimes  d_x (R_{y_0}) \Pi(x_0) + d_y (R_{x_0}) \otimes  d_y (R_{x_0}) \Pi(y_0)), \label{comp2}
\end{align}
where $R_x, L_x : M \rightarrow M, y \mapsto yx, xy$ are the left and right translation and the subscript on the exterior derivative $d_x$ means that the resulting map, e.g. $d_x R$, starts from the tangent space at $x$. In order to turn the bivector into a map to the exterior product of the Lie algebra $\bigwedge^2 \mathfrak{g} \simeq \bigwedge^2 T_e M$ one can use the right translation
\begin{align}
\tilde \Pi: M \rightarrow \bigwedge^2 \mathfrak{g}, \tilde \Pi (x) = d_e (R_{x^-1}) \Pi(x).  
\end{align}
Applying this to \eqref{comp2} yields
\begin{gather}
\tilde \Pi(x_0 y_0) = d_e (R_{y_0^{-1} x_0^{-1}}(d_x (R_{y_0}) \otimes  d_x (R_{y_0}) \Pi(x_0) + d_y (R_{x_0}) \otimes  d_y (R_{x_0}) \Pi(y_0)) \nonumber \\
= d_e (R_{x_0^-1}) \otimes d_e (R_{x_0^-1}) \Pi(x_0) + d_e (R_{x_0^{-1} }) d_e (L_{x_0}) d_{y_0} (R_{y_0^{-1}}) \otimes d_e (R_{x_0^{-1} }) d_e (L_{x_0}) d_{y_0} (R_{y_0^{-1}}) \Pi(y_0) \nonumber \\
= \tilde \Pi(x_0) + \text{Ad}_{x_0} \otimes \text{Ad}_{x_0} \tilde \Pi(y_0) \label{comp3}
\end{gather}
where we introduced the adjoint action\footnote{Cf. section \ref{sec2.1.5} for more details.} $\text{Ad}: \mathfrak{g} \otimes \mathfrak{g} \rightarrow \mathfrak{g}$ that is the derivative of the map $ \Phi_x :h \mapsto xhx^{-1} = L_x (R_{x^-1}(h))$. Since $M$ is also a Lie algebra we can write $x_0 = e^{t a}, y_0 = e^{tb}, a,b \in \mathfrak{g}$ and subtracting \eqref{comp3} from the same equation with $x_0 \leftrightarrow y_0$ gives in second order of $t$
\begin{align}\label{cocyclec}
d \tilde \Pi ([a,b]) = [a \otimes 1 + 1 \otimes a, d \tilde \Pi (b)] - [b \otimes 1 + 1 \otimes b, d \tilde \Pi (a)] 
\end{align}
and we identify the differential of the right-translated bivector $d \tilde \Pi$ with $\delta$. This compatibility relation between bracket and cobracket is called cocycle condition since it has a cohomological interpretation which will be explored in the next section.

To summarize, we showed that the Lie algebra $\mathfrak{g}$ of a Poisson Lie group has the structure of a \textit{\textbf{ Lie bialgebra}} (LBA), i.e. it is a Lie algebra with a cobracket $\delta: \mathfrak{g} \rightarrow \mathfrak{g} \otimes \mathfrak{g}$ satisfying \eqref{cojac} and \eqref{cocyclec}.


\subsection{Coboundary and Triangular Lie Bialgebras}\label{sec2.1.2}

In the following, we will show that all cobrackets of Lie bialgebras are cocycles of certain cohomology groups and introduce special coboundaries with properties that turn out to play an important role in the construction of quantum groups \cite{etingof2010lectures}.

Starting with a Lie algebra $\mathfrak{g}$ and a $\mathfrak{g}$-module $V$ (cf. \ref{sec2.1.5}) the vector spaces of \textbf{\textit{cochains}} are defined as $C^n = \text{Hom}( \Lambda^n \mathfrak{g}, V)$, i.e. homomorphisms from the $n$-th exterior product of $\mathfrak{g}$ into $V$. The coboundary operators $\partial_n : C^n \rightarrow C^{n+1}$ are given by
\begin{align}
\partial_n (f)(x_1 \wedge ... \wedge x_n+1) = & \sum_{i=1}^{n+1} (-1)^i x_i \triangleright f (x_1 \wedge ... \wedge \hat {x_i} ... \wedge x_{n+1}) \nonumber \\
&+ \sum_{i<j} (-1)^{i+j} f( [x_i, x_j] \wedge x_1 ... \wedge \hat{x_i} ... \wedge \hat{x_j} ... \wedge x_{n+1}), \label{coc-base}
\end{align}
where $\hat{x_i}$ means that the $i$-th tensor leg is dropped and $\triangleright$ denotes the left action on the module. One can check that $\partial_{n} \partial_{n-1} = 0$ so that the image $\text{Im } \partial_{n-1}$, the space of the $n$-\textbf{\textit{coboundaries}}, is contained in the Kernel $\text{Ker } \partial_{n}$ whose elements are called the $n$-\textbf{\textit{cocycles}}. The quotient $\text{Ker } \partial_n / \text{Im } \partial_{n-1}$ is the $n$-th \textit{\textbf{cohomology group}} of $\mathfrak{g}$ with values in $V$, denoted by $H^n(\mathfrak{g}, V)$. 

Considering the case $V = \bigwedge^2 \mathfrak{g}$, we see that the condition that $\delta : \mathfrak{g} \rightarrow \mathfrak{g} \wedge \mathfrak{g}$ is a 1-cocycle, i.e. $\partial_{1}(\delta)(x) = 0$ coincides with \eqref{cocyclec} justifying the name. 
The coboundaries can by definition be written in the form
\begin{align}
\delta_r (x) = \partial_0( r)(x) = [x \otimes 1 + 1 \otimes x, r]
\end{align}
for some $r \in \bigwedge^2 \mathfrak{g}$ and the cocycle condition is satisfied automatically because  $\partial_{n} \partial_{n-1} = 0$. We define a \textit{\textbf{coboundary Lie bialgebra}} to be the triple $( \mathfrak{g}, [ , ], r)$ such that $( \mathfrak{g}, [ , ], \delta_r)$ is a Lie bialgebra. By directly calculating that 
\begin{align}
\text{Cycl}((\delta_r \otimes \text{id} )) \delta_r (x) + [x, [[r, r]] ] = 0,
\end{align}
one can show that $r$ defines a coboundary Lie bialgebra if and only if
\begin{align}\label{mcybe}
[[r, r]] \equiv [r_{12}, r_{13}] + [r_{12}, r_{23}] + [r_{13}, r_{23}] = \Omega,
\end{align}
 i.e. if its Schouten bracket $[[ , ]] : \bigwedge^2 \mathfrak{g} \rightarrow \bigwedge^3 \mathfrak{g}$ is an ad-invariant (cf. section \ref{sec2.1.5}) element $\Omega$. This condition is the modified classical Yang Baxter equation (mCYBE) which can be seen as the infinitesimal version of the quantum YBE introduced in section \ref{sec2.1.7}. 
Coboundary Lie bialgebras with an r-matrix that satisfies the CYBE with a vanishing rhs are called \textit{\textbf{triangular}}, otherwise \textit{\textbf{quasitriangular}}.

Consider two coboundary Lie bialgebras over $\mathfrak{g}$ with r-matrices $r_1, r_2 = \tilde \phi (r_1))\equiv (\phi \otimes \phi)(r_1) $ that are related by an Lie algebra automorphism, i.e. a linear map $\phi: \mathfrak{g} \rightarrow \mathfrak{g}$ such that 
\begin{align}
[ \phi(x_i), \phi(x_j) ] = f^k_{ij} \phi (x_k) ,
\end{align}
 where $\{x_i\}$ is a basis of $\mathfrak{g}$ and the structure constants $f^k_{ij}$ are those of $( \mathfrak{g}, [, ])$. These two Lie bialgebras are then equivalent in the sense that the structure constants $g^{ij}_k$ of the cobracket $\delta_{r_1}$ (in other words $\delta_{r_1} (x_i) = g^{kj}_i x_k \wedge x_j$) are identical to those of $\delta_{r_2}$ because we have
 \begin{align}
\delta_{r_2} (\phi(x)) = [ \phi (x) \otimes 1 + 1 \otimes \phi(x), r_2] = \tilde \phi (\delta_{r_1}(x)).
\end{align}
 Therefore we are interested in equivalence classes of coboundary Lie bialgebras that are the orbits under Lie algebra automorphisms in the space of solutions to the mCYBE.

An important result about Lie bialgebras on semi-simple Lie algebras is the Whitehead Lemma \cite{jacobson1979}. A Lie algebra is \textit{\textbf{semi-simple}} if it is the direct sum of simple Lie algebras which have no proper non-trivial ideals, i.e. subalgebras $\mathfrak{I} \subset \mathfrak{g}$ such that $[ \mathfrak{g}, \mathfrak{I}] = \mathfrak{I}$. The Lemma states that all cohomology groups $H^n(\mathfrak{g}, V)$ of cocycles on a finite-dimensional semi-simple Lie algebra $\mathfrak{g}$ with values in any finite-dimensional module $V$ vanish. In particular, this means that all Lie bialgebras in a finite-dimensional semi-simple Lie algebra are coboundary.

\subsection{Real Forms of Complex Lie Algebras and $*$-Structures}\label{sec2.1.3}

In the context of symmetry algebras of space-times with cosmological constant we will encounter algebras that can be written as different real forms of a single complex Lie algebra which allows for a unified description. 

A complex Lie algebra is simply a Lie algebra defined over the field of complex numbers $\mathbb{C}$. Starting from a real algebra $W$ (defined over $\mathbb{R}$) one can construct a complex algebra by setting $W^{\mathbb{C}} = \mathbb{C} \otimes_{\mathbb{R}} W$ as vector spaces. $W^{\mathbb{C}}$ is called the complexification of $W$. We call any real algebra whose complexification is a complex algebra $A$ a \textit{\textbf{real form}} of $A$. There is a one to one correspondence between real forms and involutive antiautomorphisms, i.e. bijective maps $*: A \rightarrow A$ such that
\begin{align}
(a^*)^* = a, \quad (\alpha a)^* = \bar \alpha a^*, \quad \forall a \in A, \alpha \in \mathbb{C}.
\end{align}
This means that for a given $*$-map one can find a basis of $A$ such that the structure constants are real and $a^* = -a, \forall a \in A$. 

A $*$\textit{\textbf{-Lie bialgebra}} is a complex Lie bialgebra such that all the operations are \linebreak $*$-homomorphisms \cite{Chari:1994pz}. In particular for the bracket we have
\begin{align}
[a^*, b^*] = (ba)^*- (ab)^* = - [a,b]^*
\end{align}
and for the cobracket we demand
\begin{align}
\delta(a^*) = (\delta (a))^{* \otimes *}, \quad (a \otimes b)^{* \otimes *} \equiv a^* \otimes b^*.
\end{align}
For coboundary Lie bialgebras with r-matrix $r = r_1 \wedge r_2$ this condition implies
\begin{align}
[a^* \otimes 1 + 1 \otimes a^* , r] =& [a^*, r_1] \wedge r_2 + r_1 \wedge [a^*, r_2] \nonumber \\
\overset{!}{=}& [a \otimes 1 + 1 \otimes a , r]^* = - [a^*, r_1^*] \wedge r_2^* - r_1^* \wedge [a^*, r_2^*] \\
\Rightarrow r_1^* \wedge r_2^* =& - r_1 \wedge r_2
\end{align}
and by linearity all r-matrices have to satisfy
\begin{align}
r^{* \otimes *} = -r,
\end{align}
i.e. the reality condition constrains the consistent coboundary Lie bialgebras and different real forms of a complex Lie algebra might have different sets of allowed r-matrices.

\subsubsection*{Duality}

The notion of a finite dimensional Lie bialgebra is self-dual in the sense of the following proposition due to \cite{etingof2010lectures}.
\begin{prop}
 Let $(\mathfrak{g}, [, ], \delta)$ be a finite dimensional Lie bialgebra, and $[, ]^*: \mathfrak{g}^* \rightarrow \mathfrak{g}^*  \otimes \mathfrak{g}^*, \delta^*: \mathfrak{g}^* \otimes \mathfrak{g}^* \rightarrow \mathfrak{g}^*$ the respective dual maps of $[, ], \delta$. Then $( \mathfrak{g}^*, [, ]^*, \delta^*)$ is a Lie bialgebra.
\end{prop}
As we will see in concrete examples later on, the condition that $\mathfrak{g}$ is finite dimensional is indeed crucial.

To explicitly calculate the dual structures of a given Lie bialgebra one can make use of the dual pairing $\braket{ \cdot, \cdot}: \mathfrak{g}^* \otimes \mathfrak{g} \rightarrow \mathbb{C}$ satisfying
\begin{align}
\braket{\delta^*(f, g), x} & = \braket{f \wedge g, \delta(x)}, \\
\braket{[ , ]^*(f), x \wedge y} & = \braket{f, [x, y]}.
\end{align}

\subsection{Hopf Algebras}\label{sec2.1.4}

As motivated in the beginning of this chapter it is often useful to study the (algebra of) functions $\mathcal{F}(G)$ on a group $G$ instead of the group itself. The multiplication $\mu: G \times G \rightarrow G$ then induces a map $\mu^* \equiv \Delta: \mathcal{F}(G) \rightarrow \mathcal{F} (G \times G) \simeq \mathcal{F}(G) \otimes \mathcal{F}(G)$ by a pullback
\begin{align}
\Delta(f)(a, b) = f(\mu(a,b)) , \quad f \in \mathcal{F}(G); a,b \in G
\end{align}
and similarly from the inverse of the group $\iota : G \rightarrow G$ one obtains $\iota^*: \mathcal{F}(G) \rightarrow \mathcal{F}(G)$ by
\begin{align}
\iota^* (f)(a) = f(\iota(a)).
\end{align}
Evaluating a function at the unit $1_G$ of $G$ gives rise to the so-called counit $\varepsilon : \mathcal{F}(G) \rightarrow \mathbb{C}$
\begin{align}
\varepsilon(f) = f(1_G).
\end{align}
Thus, we motivated the following definition \cite{klimyk2011quantum}.
\newline
A \textit{\textbf{Hopf algebra}} $(H, \mu, \eta, \Delta, \varepsilon, S)$ over a field $\mathbb{K}$ is a unital algebra $H$ with associative multiplication $\mu: H \otimes H \rightarrow H$, i.e.
\begin{align}
\mu \circ (\text{id} \otimes \mu) = \mu \circ (\mu \otimes \text{id}),
\end{align}
unit $\eta: \mathbb{K} \rightarrow H$
\begin{align}
\eta( \lambda ) = \lambda 1_H, \quad \lambda \in \mathbb{K},
\end{align}
coassociative coproduct $\Delta: H \rightarrow H \otimes H$, i.e.
\begin{align}\label{coass}
(\Delta \otimes \text{id}) \circ \Delta = (\text{id} \otimes \Delta) \circ \Delta,
\end{align}
a counit $\varepsilon: H \rightarrow \mathbb{K}$ satisfying
\begin{align}
\varepsilon (1_H) = 1_{\mathbb{K}}
\end{align}
and an antipode $S: H \rightarrow H$ which has to fulfill
\begin{align}
\mu \circ (S \otimes \text{id}) \circ \Delta = \mu \circ (\text{id} \otimes S ) \circ \Delta = \eta \circ \varepsilon.
\end{align}
Thus, $H$ is an algebra with $\mu$ and $\eta$ as well as a coalgebra with $\Delta$ and $\varepsilon$.
Furthermore, the coproduct and the counit have to be algebra homomorphisms or, equivalently, the product and unit have to be coalgebra homomorphisms, i.e.
\begin{align}
\Delta \circ \mu= \mu_{H \otimes H} \circ (\Delta \otimes \Delta), \varepsilon \circ \mu = \mu_{\mathbb{K}}( \varepsilon \otimes \varepsilon),
\end{align}
where $\mu_{H \otimes H} = (\mu \otimes \mu) \circ (\text{id} \otimes \tau \otimes \text{id})$ with the flip $\tau: H \otimes H \rightarrow H\otimes H, a \otimes b \mapsto b \otimes a$.

When writing down relations for the coproduct the Sweedler notation will be useful. Instead of 
\begin{align}
\Delta (a) = \sum_i a_{1i} \otimes a_{2i}, \quad a, a_{1i}, a_{2i} \in H
\end{align}
we write
\begin{align}
\Delta (a) = a_{(1)} \otimes a_{(2)},
\end{align}
which implies the summation.

\subsection{Modules, Comodules and Module Algebras}\label{sec2.1.5}

A \textbf{\textit{representation}} of an algebra $A$ with multiplication $\mu$ on a vector space $V$ is an algebra homomorphism $\varphi$ mapping from  $A$ to the algebra of linear operators $\mathcal{L}(V)$ on $V$ \cite{klimyk2011quantum}. This entails linearity of $\varphi$ as well as 
\begin{align}
\varphi(a b) = \varphi(a) \varphi(b), \quad \varphi(1_A) = 1_{\mathcal{L}(V)}, \quad a, b \in A.
\end{align}
The same concept is formulated with a linear map $\varphi: A \otimes V \rightarrow V$,called a left \textbf{\textit{action}} and $V$, called a left $A$\textbf{\textit{-module}}, satisfying
\begin{align}
\varphi \circ(\text{id} \otimes \varphi) = \varphi (\mu \otimes \text{id}), \quad \varphi \circ (\eta \otimes \text{id}) = \text{id}.
\end{align}
Instead of $\varphi(a \otimes v)$ we may also write $a \triangleright v$. 
Similarly, a linear map $\varphi: V \otimes A \rightarrow V$ is a right action and $V$ a right $A$-module if 
\begin{align}
\varphi \circ(\text{id} \otimes \varphi) = \varphi (\text{id} \otimes \mu), \quad \varphi \circ (\text{id} \otimes \eta) = \text{id}
\end{align}
and a $A$-bimodule $V$ is a left as well as a right module that are compatible in the following sense
\begin{align}
(a \triangleright v) \triangleleft b = a \triangleright (v \triangleleft b).
\end{align}
Every associative algebra is a bimodule of itself with the algebra multiplication as the so-called regular action.  

A Hopf algebra $H$ with antipode $S$ canonically becomes a left $H$-module by endowing it with the left adjoint action $\text{ad}_L: H \rightarrow \mathcal{L}(H)$ defined as
\begin{align}
\text{ad}_L(a ) b = a_{(1)} b S (a_{(2)}), \quad a,b \in H.
\end{align}
In the case of a Lie algebra $\mathfrak{g}$ (which can be turned into a Hopf algebra in a canonical way as we will see in section \ref{sec2.1.6}) the adjoint action is just given by the Lie bracket
\begin{align}
\text{ad}_L(a) b = [a,b], \quad a,b \in \mathfrak{g}.
\end{align}
We denote the invariant (under the action) elements of a module by $$V_{A} \equiv \{ v \in V: a \triangleright v = 0, \forall a \in A \}.$$

The dual notion to the module of an algebra is defined for a coalgebra $B$ with coproduct $\Delta$ and counit $\varepsilon$. A left $B$\textbf{\textit{-comodule}} is a vector space $V$ such that there exists a linear map, the left coaction, $\varphi: V \rightarrow B \otimes V$ satisfying
\begin{align} \label{comod}
(\text{id} \otimes \varphi) \circ \varphi = (\Delta \otimes \text{id}) \circ \varphi, \qquad (\varepsilon \otimes \text{id}) \circ \varphi = \text{id}.
\end{align}
Using the coproduct, $B$ is itself a left $B$-module, i.e. $\Delta: B \rightarrow B \otimes B$ is a left coaction since \eqref{comod} is then just the coassociativity condition \eqref{coass}.
\newline

Importantly, if $A$ is a bialgebra one can construct the tensor product of representations, i.e. a representation on the tensor product of $A$-modules. Given two such (left) $A$-modules $V$ and $W$ with corresponding (left) actions $\varphi_V, \varphi_W$ $V \otimes W$ is a (left) $A$-module with action
\begin{align}\label{ten-rep}
\varphi_{V \otimes W} (a) = \varphi_V(a_{(1)}) \otimes \varphi_W(a_{(2)}), \quad a \in A.
\end{align}

For the dual description of non-commutative spacetime the following concept of module algebras is also of great importance. Consider a bialgebra $B$ with counit $\varepsilon$. A left $B$-module $W$ which is also an algebra with multiplication $\mu$ and unit $\eta$ is a $B$\textbf{\textit{-module algebra}} if $\mu$ and $\eta$ are $B$-module homomorphisms, i.e. if
\begin{align}
b \triangleright \mu(w,v) = \mu( b_{(1)} \triangleright w , b_{(2)} \triangleright v), \quad b \triangleright 1_W = \varepsilon (b) 1_W. 
\end{align}

\subsection{Deformations of Universal Enveloping Algebras}\label{sec2.1.6}

An important class of Hopf algebras, especially for our purposes, are deformations of universal enveloping algebras, which will be introduced in the following \cite{klimyk2011quantum}. 

First, we need the notion of a topological extension of a vector space in order to allow for formal power series in a formal parameter $h$, playing the role of a deformation parameter. Let $\mathbb{K}$ be a field with characteristic $0$. Then the set of 
\begin{align}
f = \sum_{n= 0}^{\infty} a_n h^n, \quad a_n \in \mathbb{K}
\end{align}
with the addition and multiplication defined by
\begin{align}\label{admu}
f + g = \sum_{n=0}^{\infty} (a_n + b_n) h^n, \quad fg = \sum_{n=0}^{\infty} \left( \sum_{r+s=n} a_r b_s \right) h^n
\end{align}
is a ring called $\mathbb{K}[[h]]$. For a $\mathbb{K}$ vector space $V$ we analogously define $V[[h]]$ as the set 
\begin{align}
f = \sum_{n= 0}^{\infty} a_n h^n, \quad a_n \in V
\end{align}
and multiplication and addition are given by \eqref{admu}, where the coefficients $a_n, b_n$ are elements of $V$. On a $\mathbb{K}[[h]]$ vector space $V$ one can introduce the so-called $h$-adic topology \cite{Bour61}, i.e. the topology in which the sets $\{ h^n V + v \vert n \in \mathbb{N}_0 \}$ are a basis of the neighborhood of $v \in V$. The completion of the tensor product $V \otimes_{\mathbb{K}[[h]]} W$ of two $\mathbb{K}[[h]]$ vector spaces $V, W$ in the $h$-adic topology is denoted by $V \hat \otimes W$. One can show that $V[[h]]$ is complete in the $h$-adic topology and that $V[[h]] \hat \otimes W[[h]] = (V \otimes W) [[h]]$.

Now we are ready to define the deformation of a Hopf algebra. For a Hopf algebra \linebreak $(H, \mu, \eta, \Delta, \varepsilon, S)$ a \textbf{\textit{deformation}} is a Hopf algebra $(H_h, \mu_h, \eta_h, \Delta_h, \varepsilon, S_h)$ such that
$H_h$ is isomorphic to $H[[h]]$ and 
\begin{align}
\mu_h = \mu \;\text{ mod } h, \quad \Delta_h = \Delta \; \text{ mod } h.
\end{align}
Two deformations $H_h$ and $H'_h$ of $H$ are said to be equivalent if there exists an isomorphism of Hopf algebras $f_h: H_h \rightarrow H'_h$ that is the identity $\text{mod } h$.

As stated above, of particular interest are deformations of the \textit{\textbf{universal enveloping algebra}} (so-called \textit{\textbf{quantum universal envelope}} QUE) $U (\mathfrak{g})$ of a Lie algebra $\mathfrak{g}$, i.e. the free algebra of finite powers of elements of $\mathfrak{g}$ with Lie bracket $[, ]$ subject only to the relation $ab - ba - [a,b]=0$. 
$U(\mathfrak{g})$ is naturally a Hopf algebra with the coproduct $\Delta$, antipode $S$ and counit $\varepsilon$ defined on the primitive elements by
\begin{align}
\Delta (x) = x \otimes 1 + 1 \otimes x, \quad S(x) = -x, \quad \varepsilon (x) = 0, \quad \forall x \in \mathfrak{g}.
\end{align}
We will call this Hopf algebra the undeformed Hopf algebra of $\mathfrak{g}$.
Note that $U(\mathfrak{g})$ also has a unit. 

The obstructions and uniqueness of constructing a full (all orders of $h$) deformation given a first order (i.e. $\text{mod }h$) deformation are quantified by a cohomology derived from the Hochschild cohomology of algebras and coalgebras (see \cite{Chari:1994pz} and references therein}). For deformations of a universal enveloping algebra of a semi-simple Lie algebra one finds e.g. that $U(\mathfrak{g})[[h]]$ is isomorphic to $U(\mathfrak{g})$ as an algebra, i.e. one can set $\mu_h = \mu$ for all deformations.

\subsection{Quasitriangular Hopf Algebras and Quantum $R$-Matrices}\label{sec2.1.7}

In quantum mechanics the existence of indistinguishable particles is encoded by using multi-particle states that are obtained from tensorproducts of one-particle states by \linebreak (anti)symmetrization. All permutations of a given tensorproduct of a vector space $V$(the one-particle states) are related by (multiple) flips $\tau: v \otimes w \mapsto w \otimes v$ and our goal is to find a quantum analog of this operation that will allow for covariant statistics on the multi-particle states (cf. \cite{Young:2007ag}). 

As described in section \ref{sec2.1.5}, given a representation $\varphi: H \otimes V \rightarrow V$ of a Hopf algebra $H$ on $V$, the map in \eqref{ten-rep} with $\varphi_V = \varphi_W \equiv \varphi$ is a representation $\tilde \varphi: H \otimes V \otimes V \rightarrow V \otimes V$. For an analog of $\tau$ on this tensorproduct to be covariant we mean that it is an intertwiner of this representation. In general, for two $H$ modules $V_1, V_2$ an \textbf{\textit{intertwiner}} is an isomorphism $\phi: V_1 \rightarrow V_2$ such that 
\begin{align}
\phi(\varphi_{V_1} (x, v)) = \varphi_{V_2}(x, \phi(v)).
\end{align}
With $V_1 = V_2 \equiv V \otimes V$ and $\phi = \tau$ this becomes 
\begin{align}
a_{(2)} \triangleright w \otimes a_{(1)} \triangleright v = a_{(1)} \triangleright w \otimes a_{(2)} \triangleright v \Leftrightarrow \Delta^{\text{op}} = \Delta.
\end{align}
It turns out that a general solution to this problem can be formulated by introducing \textit{\textbf{quasitriangular Hopf algebras}}, i.e. Hopf algebras $(H, \mu, \eta, \Delta, S, \varepsilon)$ with an invertible element $R \in H \otimes H$ such that
\begin{gather}
\Delta^{\text{op}}(x) = R \Delta(x) R^{-1}, \\
(\Delta \otimes \text{id}) R = R_{13} R_{23}, \quad (\text{id} \otimes \Delta) R = R_{13} R_{12}.
\end{gather}
$R$ is called a \textbf{\textit{universal quantum}} $R$\textbf{\textit{-matrix}} and $H$ is called triangular if $R_{21} = R^{-1}$.

Then, given a representation $T_V: H \rightarrow \mathcal{L}(V)$ the map $\tilde R \equiv \tau \circ (T_V \otimes T_V)(R)$ is an intertwiner of the action $\tilde T_V: H \rightarrow \mathcal{L}( V \otimes V)$ induced by the coproduct.
Analogously the problem of finding such an intertwiner can be formulated as finding representations of the Artin braid group, defined by the relations among the generators $\sigma_i, i \in \mathbb{N}$
\begin{gather}
\sigma_i \sigma_j = \sigma_j \sigma_i, \quad \text{for } |i-j| >1\\
\sigma_i  \sigma_{i+1} \sigma_{i} = \sigma_{i+1} \sigma_i \sigma_{i+1},
\end{gather}
since $\tilde R$ fulfills the cubic relations 
\begin{align}
\tilde R_{12} \tilde R_{23} \tilde R_{12} = \tilde R_{23} \tilde R_{12} \tilde R_{23}
\end{align}
if $R$ is a universal $R$-matrix and the quadratic relations are satisfied automatically. Furthermore, these cubic relations are equivalent to the so-called quantum Yang-Baxter equation (QYBE) \cite{klimyk2011quantum} for $R$
\begin{align}
R_{12} R_{13} R_{23}  = R_{23} R_{13} R_{12}.
\end{align}

The notation already hints at a connection with Lie bialgebras, the infinitesimal version or classical limit of QUE. Indeed, if a Hopf algebra $H$ is a QUE on a Lie algebra $\mathfrak{g}$, this Lie algebra is a Lie bialgebra with cobracket given by 
\begin{align}
\delta (x) = \frac{\Delta(x) - \Delta^{\text{op}}}{h} \vert_{h = 0}.
\end{align}
If $H$ is quasitriangular, so is the corresponding LBA as the linearized version of the QYBE for $R \sim 1 + r$ implies the CYBE for $r$. Note that $R$ is not necessarily anti-symmetric and a non anti-symmetric $r$ satisfying the CYBE does not imply that it defines a triangular LBA but one can obtain an equivalent one with $r'= -r'_{21}$ satisfying the mCYBE. Furthermore, the LBA is triangular iff $H$ is triangular \cite{Chari:1994pz}.

\subsection{Specialization and Classical Limit of the Standard Deformation}\label{sec2.1.8}

In section \ref{sec2.1.6} we encountered Hopf algebras with an $h$-adic topology with a formal parameter $h$. While this is a useful concept, it is ultimately problematic for the application to physical models where we want to interpret $h$ as an observable energy scale e.g. the Planck scale. The problem of finding a Hopf algebra (the so-called q-analog) with the same (co)algebra structure where $h$ can be specialized to a complex parameter $q$ is known as specialization \cite{Chari:1994pz}, \cite{klimyk2011quantum}, \cite{Borowiec:2014aqa}. Given a topological Hopf algebra $H [[h]]$ the q-analog is a subset $\tilde H \subset H[[h]]$ over the field $\mathbb{C}[q, q^{-1}]$ of finite monomials in $q, q^{-1}$.

An instructive example of such a q-analog can be constructed in the case of the standard deformation that we will analyze in the following \cite{klimyk2011quantum}. Consider the simple Lie algebra $sl(2, \mathbb{C})$ spanned by $\{H, E_+, E_-\}$ subject to the relations
\begin{align}
[H, E_{\pm}] = \pm 2 E_{\pm}, \quad [E_+, E_-] = H.
\end{align}
A deformation of the universal envelope of $sl(2, \mathbb{C})$, called $U_h(sl(2, \mathbb{C}))$ is defined on the set $P$ of polynomials in the generators endowed with the $h$-adic topology divided by the completion $\hat I$ (in the $h$-adic topology) of the ideal $I$ generated by the deformed algebra relations
\begin{align}\label{sl2def}
[H, E_{\pm}] = \pm 2 E_{\pm}, \quad [E_+, E_-] = \frac{e^{hH} -e^{-hH}}{e^h- e^{-h}},
\end{align}
i.e. $U_h(sl(2, \mathbb{C})) = P[[h]]/\hat I$. The coalgebra sector is given by
\begin{gather}\label{sl2co1}
\Delta_h (E_+) = E_+ \otimes e^{h H} + 1 \otimes E_+, \quad \Delta_h (E_-) = E_- \otimes 1 + e^{-hH} \otimes E_-, \\
\Delta_h (H) = H \otimes 1 + 1 \otimes H, \quad S_h(E_+) = - E_+ e^{-h H}, \\
S_h (E_-) = - e^{h H} E_-, \quad S_h(H) = -H, \quad \varepsilon_h(E_{\pm}) = \varepsilon_h(H) = 0. \label{sl2co2}
\end{gather}
This construction can be generalized to any finite dimensional semi-simple Lie algebra and is known as the standard or Drinfeld-Jimbo deformation \cite{Jim86}, \cite{Drin85}. It is the first important tool in the construction of quantum groups.

When trying to naively interpret $h$ as a complex number and use the formulas \eqref{sl2def}-\eqref{sl2co2} verbatim for the q-analog one encounters two problems. First, the coproducts and antipodes are infinite series in $H$ and also in $h$ and second that \eqref{sl2def} is not defined for $h =0$. Both of these can be remedied by passing to a slightly different formulation of $U_h(sl(2, \mathbb{C}))$ which we will call $\tilde U_q(sl(2, \mathbb{C}))$. Define $q = e^{h}, K = e^{hH}$ and 
\begin{align}
G = \frac{K - K^{-1}}{q-q^{-1}}.
\end{align}
The relations in \eqref{sl2def} (and an extra condition for $G$) for the new Hopf algebra with generators $E_{\pm}, K, K^{-1}, G$ then read
\begin{gather}
[G, E_{+}] =   E_{+} (qK + q^{-1} K^{-1}), \quad [G, E_-] = -(qK + q^{-1} K^{-1}) E_- , \\
[E_+, E_-] = G, \quad (q -q^{-1}) G = K - K^{-1},
\end{gather}
which contains finitely many terms but is not linear.
Similarly \eqref{sl2co1}-\eqref{sl2co2} can be rewritten and we have the additional relations
\begin{gather}
\Delta(G) = G \otimes K + K^{-1} \otimes G, \quad \Delta(K) = K \otimes K, \quad \Delta (K^{-1}) = K^{-1} \otimes K^{-1}, \\
S(G) = -G, \quad S(K) = K^{-1} , \quad S(K^{-1}) = K, \\
 \varepsilon(G) = 0, \quad \varepsilon(K) = \varepsilon(K^{-1}) = 1, 
\end{gather}
which are also finite polynomials in the new generators and defined for all values of $q$. With these relation $\tilde U_q(sl(2, \mathbb{C}))$ is a suitable q-analog of $U_h(sl(2, \mathbb{C}))$ and one can now try to find a classical limit. In the algebra sector for $q=1$ one finds that $K$ is in the center and $K^2 = 1$. Indeed, $\tilde U_1(sl(2, \mathbb{C}))$ is just the central extension of the universal envelope of $sl(2, \mathbb{C})$ by a central $K$ with $K^2 =1$.

For the coalgebra sector the classical limit means the corresponding LBA (cf. previous section) and one easily obtains that it has a cobracket defined with classical r-matrix
\begin{align}
r = E_+ \wedge E_-.
\end{align}
As the Schouten bracket yields the ad-invariant element
\begin{align}
[[r,r ]] = 4 E_+ \wedge E_- \wedge H,
\end{align}
the LBA is quasitriangular.

\subsection{Drinfeld Twist Deformation}\label{sec2.1.9}

In section \ref{sec2.1.7} we saw that one can easily obtain a LBA from a quantum group and in the previous section the first method of constructing a quantum group given a LBA was presented. That method, however, only works for semi-simple Lie algebras. Another method of obtaining new Hopf algebras that does not have this restriction is the so-called Drinfeld twist presented in the following \cite{Drin90}.

Let $(H, \mu, \eta, \Delta, \varepsilon, S)$ be a Hopf algebra and $\mathcal{F} = f_{\alpha} \otimes f^{\alpha}$ an invertible element in $H \otimes H$ satisfying the 2-cocycle condition
\begin{align}\label{coc-cond}
\mathcal{F}_{12} ( \Delta_0 \otimes 1) (\mathcal{F}) = \mathcal{F}_{23} (1 \otimes \Delta_0)(\mathcal{F})
\end{align}
 and the normalization 
 \begin{align}
\varepsilon (f_{\alpha}) f^{\alpha} = 1.
\end{align}
Then $(H, \mu, \eta, \Delta_{\mathcal{F}}, \varepsilon, S_{\mathcal{F}})$ with 
\begin{align}
\Delta_{\mathcal{F}} (x) = \mathcal{F} \Delta(x) \mathcal{F}^{-1}, \quad S_{\mathcal{F}}(x) = v S(x) v^{-1}, \nonumber \\
v = \mu \circ ( \text{id} \otimes S ) (\mathcal{F}), \quad v^{-1} = \mu \circ (S \otimes \text{id}) (\mathcal{F}^{-1}) , \quad \forall x \in H
\end{align}
is a Hopf algebra as well. In particular, $\Delta_{\mathcal{F}}$ is coassociative 
\begin{align}
(\Delta_{\mathcal{F}} \otimes \text{id}) \circ \Delta_{\mathcal{F}} = & \left(\mathcal{F}_{12} ( \Delta_0 \otimes \text{id}) \mathcal{F}^{-1}_{12} \right) \circ  \left( \mathcal{F} \Delta_0 \mathcal{F}^{-1} \right) \nonumber \\
= & \mathcal{F}_{12} \left[  ( \Delta_0 \otimes \text{id})( \mathcal{F}) ( \Delta_0 \otimes \text{id}) \circ \Delta_0 ( \Delta_0 \otimes \text{id}) \left( \mathcal{F}^{-1} \right) \right] \mathcal{F}^{-1}_{12} \nonumber \\
= & \mathcal{F}_{23} \left[  (\text{id}  \otimes \Delta_0)( \mathcal{F})  (\text{id}  \otimes \Delta_0) \circ \Delta_0  (\text{id}  \otimes \Delta_0)\left( \mathcal{F}^{-1} \right) \right] \mathcal{F}^{-1}_{23} \nonumber \\
= &  (\text{id}  \otimes \Delta_{\mathcal{F}}) \circ \Delta_{\mathcal{F}},
\end{align}
where equation \eqref{coc-cond}, the analog relation for the inverse twist element and the coassociativity of $\Delta_0$ was used in the third line. 
If the original Hopf algebra was quasitriangular with universal R-matrix $R$ so is the twisted one with $R_{\mathcal{F}} = \mathcal{F}_{21} R \mathcal{F}^{-1}$. 

Starting from the undeformed Hopf algebra of the universal envelope of some Lie algebra, this construction can be used to construct deformed quasitriangular QUE with $R = \mathcal{F}_{21} \mathcal{F}^{-1}$. We already know that the linearized and anti-symmetrized $R$ yields a classical r-matrix $r$ and if it is given by a twist $r$ is even triangular. Conversely, one can show that given a triangular r-matrix $r$ there is a twist such that the associated R-matrix has $r$ as the classical limit \cite{Chari:1994pz}.
A related, even more general result from \cite{Etingof1996}, \cite{Etingof98} states that all first order deformations encoded in a LBA structure can be extended to a full deformation. In particular, it is stated that the category of Lie bialgebra  over a field with h-adic topology and the category of QUE are equivalent. 

\subsection{Star Product}\label{sec2.1.10}

We saw in \eqref{sec2.1.5} that the action of a Hopf algebra $H$ on a product in a module algebra $A$ with product $\mu$ involves the coproduct. If the coproduct is deformed one can ask how the product in the module has to change so that the deformed action is compatible with it. In the case of a twist deformation with twist $\mathcal{F}$ the answer can be explicitly given in the form of the so-called star product $\star : A_{\star} \otimes A_{\star} \rightarrow A_{\star}$ where $A =A_{\star}$ as a vector space. It is defined by \cite{Aschieri:2005zs}
\begin{align}
a \star b = \mu \left( \mathcal{F}^{-1} \triangleright ( a \otimes b) \right) 
\end{align}
and the compatibility with the action
\begin{align}\label{compac}
h \triangleright (a \star b) = \left( h_{(1)^{\mathcal{F}}}  \triangleright a \right) \star \left( h_{(2)^{\mathcal{F}}}  \triangleright b \right) ,
\end{align}
where the twisted coproduct is written in the Sweedler notation $\Delta_{\mathcal{F}}(h) =  h_{(1)^{\mathcal{F}}} \otimes  h_{(2)^{\mathcal{F}}}$, follows from the compatibility of the undeformed action with the product $\mu$
\begin{align}
 h \triangleright \left( \mu \left( \mathcal{F}^{-1} \triangleright ( a \otimes b) \right) \right) & = \mu \left( \Delta(h) \triangleright \left(\mathcal{F}^{-1} \triangleright ( a \otimes b) \right)\right) \nonumber \\
 & = \mu \left(\mathcal{F}^{-1}\triangleright \left( \Delta^{\mathcal{F}} (h) \triangleright (a \otimes b) \right) \right) = \left( h_{(1)^{\mathcal{F}}}  \triangleright a \right) \star \left( h_{(2)^{\mathcal{F}}}  \triangleright b \right).
\end{align}
Furthermore, the star product is associative.

Two examples of star products on modules will be especially important in this work. The first is the Hopf algebra itself via the adjoint action which we discuss in the following. 
Starting from a universal envelope $U \mathfrak{g}$ of a Lie algebra $\mathfrak{g}$ by a twist deformation one obtains a QUE $U g^{\mathcal{F}}$. Using the adjoint action a star product
\begin{align}
\star: U \mathfrak{g} \otimes U \mathfrak{g} \rightarrow U \mathfrak{g}, \quad h \star g = (\bar f^{\alpha} \triangleright h)( \bar f_{\alpha} \triangleright g )
\end{align}
 can be defined on $U \mathfrak{g}_{\star} \equiv ( U \mathfrak{g}, \star)$. On the generators of $\mathfrak{g}$ the star product can be used to define a braided commutator
\begin{align}
[u, v]_{\star} = u \star v - \left(\bar R^{\alpha}\triangleright v \right) \star \left( \bar R_{\alpha} \triangleright u \right),
\end{align}
where $R^{-1} = \mathcal{F} \mathcal{F}_{21}^{-1} \equiv \bar R^{\alpha} \otimes \bar R_{\alpha}$ is the inverse R-matrix. The braided commutator closes in $\mathfrak{g}$ and satisfies braided antisymmetry and the braided Jacobi identity
\begin{align}
[u, v]_{\star} & = -[\bar R^{\alpha} (v), \bar R_{\alpha}(u)]_{\star}, \\
[u, [v, t]_{\star}]_{\star} & = [[u, v]_{\star}, t]_{\star} + [\bar R^{\alpha} (v), [\bar R_{\alpha} (u), t]_{\star} ]_{\star},
\end{align}
turning $\mathfrak{g}_{\star} \equiv (\mathfrak{g}, [, ]_{\star})$ into a deformed Lie algebra. Furthermore, one can show that $\mathfrak{g}_{\star}$, sometimes referred to as a quantum Lie algebra, generates $U \mathfrak{g}$ .

On $U \mathfrak{g}_{\star}$ one can naturally define a Hopf algebra structure by
\begin{align}\label{copstar}
\Delta_{\star} (u) = u \otimes 1 + \bar f^{\alpha} \bar R^{\beta} S^{-1}(\bar f_{\alpha}) \otimes \bar R_{\beta} (u), \quad S^{-1}_{\star}(u) = - (\bar R^{\beta} \triangleright u ) \star ( \bar f^{\alpha} \bar R_{\beta} S^{-1}(\bar f_{\alpha})) ,
\end{align}
which is isomorphic to $(U\mathfrak{g}^{\mathcal{F}}, \mu, \Delta_{\mathcal{F}}, S_{\mathcal{F}}, \varepsilon)$ with the isomorphism
\begin{gather}
D: U \mathfrak{g}_{\star} \rightarrow U \mathfrak{g}^{\mathcal{F}}, \quad D( u \star v) = D(u) D(v) , \\
D^{-1} : a \mapsto D^{-1} (a) = (\bar f^{\alpha} \triangleright a ) \bar f_{\alpha} \equiv a^{\mathcal{F}}.
\end{gather}
It will be useful to express the Hopf algebra structures of $U \mathfrak{g}_{\star}$ in terms of $a^{\mathcal{F}}$, i.e.
\begin{align}
 [a^{\mathcal{F}}, b^{\mathcal{F}}]_{\star} & = a^{\mathcal{F}} \star b^{\mathcal{F}} - (\bar R^{\alpha} \triangleright b )^{\mathcal{F}} \star ( \bar R_{\alpha} \triangleright a)^{\mathcal{F}}, \\
 \Delta_{\mathcal{F}} (a^{\mathcal{F}}) & = a^{\mathcal{F}} \otimes 1 + \bar R^{\alpha} \otimes ( \bar R_{\alpha} \triangleright a)^{\mathcal{F}}. \label{itsthebase}
\end{align}

The deformed bracket can also be expressed as \cite{Aschieri:2017ost}
\begin{align}
[a^{\mathcal{F}}, b^{\mathcal{F}}]_{\star} = a^{\mathcal{F}}_{(1^{\mathcal{F}})} b^{\mathcal{F}} S_{\mathcal{F}} ( a^{\mathcal{F}}_{(2^{\mathcal{F}})}).
\end{align}
An important property of the deformed coproduct \eqref{copstar} is 
\begin{align}\label{dcprop}
\Delta_{\star} (\mathfrak{g}_{\star}) \subset \mathfrak{g}_{\star} \otimes 1 + U \mathfrak{g}_{\star} \otimes \mathfrak{g}_{\star},
\end{align}
which does not hold in general for e.g. $\Delta_{\mathcal{F}}$. This property allows for the construction of a deformed differential calculus in the sense of Woronowicz \cite{Woro89} (see also \cite{Sitarz:1994rh} for an explicit construction for $\kappa$-Minkowski)which in \cite{Aschieri:2005zs} was used to define a theory of gravity in a non-commutative setting. The deformed infinitesimal diffeomorphisms under which the theory is invariant are described by $U \mathfrak{g}_{\star}$ giving an argument that this form is preferred over other, isomorphic forms like $U \mathfrak{g}_{\mathcal{F}}$.
In particular, it is possible to define a deformed quadratic wave operator on the algebra of the infinitesimal  diffeomorphisms (vector fields) that can be expressed as \cite{Aschieri:2017ost}
\begin{align}
\Box^{\mathcal{F}} \equiv D (\Box),
\end{align}
where $\Box$ is the undeformed wave operator.

The second example for making use of the star product is directly related to non-commutative geometry and will be discussed in the next section.

\section{$\kappa$-Poincar\'e and Non-commutative Geometry}\label{sec2.2}

In this section, the $\kappa$-Poincar\'e quantum group is introduced and used to illustrate the relation of non-trivial coalgebra structure and non-commutative geometry (see e.g. \cite{Kowalski-Glikman:2017ifs}).

In the seminal paper by Lukierski et.al. \cite{Lukierski:1991pn} the $\kappa$-Poincar\'e quantum group was first introduced. They employed the Drinfeld-Jimbo deformation discussed in section \ref{sec2.1.8} for the symmetry algebra of anti de Sitter (AdS) spacetime (cf. chapter \ref{chap3}). That symmetry algebra is the semi-simple $so(3,2)$ which is a real form of $\mathfrak{o}(5, \mathbb{C})$. When taking the contraction limit (which will also be discussed in greater detail in chapter \ref{chap3} and \ref{chap4}) of $so(3,2)$, the Poincar\'e algebra in four dimensions
\begin{gather}
[M_{\mu \nu}, M_{\rho \lambda}] = i (\eta_{\mu \lambda} M_{\nu \rho} - \eta_{\nu \lambda} M_{\mu \rho} + \eta{\nu \rho} M_{\mu \lambda} - \eta_{\mu \rho} M_{\nu \lambda}), \\
[M_{\mu \nu}, P_{\rho}] = i (\eta_{\nu \rho} P_{\mu}- \eta_{\mu \rho} P_{\nu}), \quad [P_{\mu}, P_{\nu} ] = 0,
\end{gather}
with the Minkowski metric $\eta_{\mu \nu} = \text{diag}(1, -1,-1,-1)$ is obtained on the algebraic level. 

In the bicrossproduct formulation (\cite{Majid:1994cy}) the algebra and coalgebra relations of the $\kappa$-Poincar\'e are both deformed, i.e. one has
\begin{align}
[M_{0i} , k_j ] = i\delta_{ij}\left( \frac{\kappa}{2}\left( 1- e^{-2 k_0/\kappa} \right) + \frac{{\bf{k}}^2}{2 \kappa} \right) - i \frac{1}{\kappa} k_i k_j,
\end{align}
for the algebra sector while the rest of the relations are undeformed. The coproducts take the form
\begin{align}\label{kpco}
\Delta(k_i) = k_1 \otimes 1 + e^{k_0/\kappa} \otimes k_i, \quad \Delta (k_0) = k_0 \otimes 1 + 1 \otimes k_0.
\end{align}
However, one can choose a basis (the so-called classical basis, also discussed in \cite{Kosinski:1994br}, \cite{Kresic-Juric:2007vgu}, \cite{Borowiec:2009vb})
\begin{align}\label{pk}
P_{0} (k) = & \kappa \sinh \left( \frac{k_0}{\kappa} \right) + \frac{{\bf{k}}^2}{2 \kappa} e^{\frac{k_0}{\kappa}}, \\
P_i (k) = & k_i e^{\frac{k_0}{\kappa}},
\end{align}
so that the algebra relations are undeformed and the coalgebra sector can be expressed as \cite{Borowiec:2014aqa}
\begin{align}
\Delta_{\tau} (P_{\mu}) =& P_{\mu} \otimes \Pi_{\tau} + 1 \otimes P_{\mu} - \frac{\tau_{\mu}}{\kappa} P^{\rho} \Pi_{\tau}^{-1} \otimes P_{\rho} - \frac{ \tau_{\mu}}{2 \kappa^2} C_{\tau} \Pi_{\tau}^{-1}, \\
\Delta_{\tau}(M_{\mu \nu}) =& M_{\mu \nu} \otimes 1+ 1 \otimes M_{\mu \nu} + \frac{1}{\kappa} P^{\rho} \Pi_{\tau}^{-1} \otimes ( \tau_{\nu} M_{\rho \mu} - \tau_{\mu} M_{\rho \nu}) \nonumber \\
& - \frac{1}{2\kappa^2}C_{\tau} \Pi_{\tau}^{-1} \otimes (\tau_{\mu} \tau^{\rho}M_{\rho \nu} - \tau_{\nu} \tau^{\rho}M_{\rho \mu}),
\end{align}
where  the symbols
\begin{align}
\Pi_{\tau} & = \frac{1}{\kappa} \tau^{\mu} P_{\mu} + \sqrt{1 + \frac{\tau^2}{\kappa^2} C}, \\
C_{\tau} & = \frac{2\kappa^2}{\tau^2} \left(\sqrt{1 + \frac{\tau^2}{\kappa^2} C}-1 \right), \quad C = P^{\mu} P_{\mu},
\end{align}
have been used\footnote{The quadratic element $C_{\tau}$ is argued to play the role of a deformed Casimir operator in \cite{Borowiec:2009ty} because it satisfies $[C_{\tau}, x_{\mu}] = 2 P_{\mu}$ when acting on the dual non-commutative coordinates (cf. below). In chapter \ref{chap4} we will employ a different construction of deformed Casimir operators based on the generators discussed in section \ref{sec2.1.10}.} and $C_{\tau} = C$ for $\tau =0$. The original version of Lukierski et.al. corresponds to the choice $\tau_{\mu} = (1,0,0,0) \rightarrow \tau^2 =1$ which is also called the time-like $\kappa$-Poincar\'e but other choices are possible, i.e. the spacelike $\tau^2 = -1$ and light-like (or light-cone) $\tau^2 = 0$ versions. The light-like form is special since it is triangular and can thus be obtained from a twist. 

\subsection{Non-commutative Geometry}\label{sec2.2.1}

From ordinary quantum mechanics we know that the operators for position and momenta are dual to each other in the sense that e.g. the momentum operator can be represented as $P_{\mu} \equiv -i \partial_{\mu}$ and thus acts on the coordinates via
\begin{align}\label{canac}
\braket{P_{\mu}, x_{\nu}} = P_{\mu} \triangleright x_{\nu} = -i \eta_{\mu \nu},
\end{align}
leading to the Heisenberg relation
\begin{align}
[P_{\mu}, x_{\nu}] \equiv \braket{P_{\mu}, x_{\nu}}  = -i \eta_{\mu \nu}.
\end{align}
This canonical action on the dual Minkowski algebra can be understood as dual to the regular action, i.e. the first equality in \eqref{canac} follows from
\begin{align}
\braket{ P_{\nu} \otimes P_{\rho},\Delta(x_{\mu})} & =\braket{ P_{\nu} P_{\rho},x_{\mu}} = \braket{P_{\nu}  , P_{\rho}\triangleright^*x_{\mu}} , \\
P_{\rho}\triangleright^*x_{\mu} & = \braket{x_{\mu (1)}, P_{\nu}} x_{\mu (2)},
\end{align}
if we endow the dual algebra with a primitive coproduct $\Delta (x) = x \otimes 1 + 1 \otimes x$.

Now, as we saw in section \ref{sec2.1.5}, if the algebra of momenta is a Hopf algebra one can look for a representation on the algebra of the form
\begin{align}
P_{\mu} \triangleright (x^{\nu} \cdot x^{\rho}) = (P_{\mu (1)} \triangleright x^{\nu}) \cdot (P_{\mu (2)} \triangleright x^{\rho}). 
\end{align}
When comparing to \eqref{compac} it becomes apparent that the product $\cdot$ that satisfies this module algebra relation for a deformed coproduct is the star product which allows us to calculate what we will interpret as deformed commutation relations for Minkowski space. For example, with the coproduct from \eqref{kpco} one finds e.g.
\begin{align}
k_j \triangleright [x^0, x^i] = (k_{j (1)} \triangleright x^0) (k_{j (2)} \triangleright x^i) - (k_{j (1)} \triangleright x^i) (k_{j (2)} \triangleright x^0) = \frac{1}{\kappa} \delta{ij},
\end{align}
indicating that 
\begin{align}\label{kmink}
[x^0, x^i] = \frac{i}{\kappa}x^i.
\end{align}
This relation (with the rest of the commutators vanishing) constitutes the time-like $\kappa$-Minkowski algebra and the other variations can be included in the description again with the help of $\tau_{\mu}$ 
\begin{align}\label{kmink2}
[x^{\mu}, x^{\nu}] = \frac{i}{\kappa}\left(\tau^{\mu} x^{\nu} - \tau^{\nu} x^{\mu}\right).
\end{align}

Besides commutation relations of Lie algebra type that we just encountered there is also the so-called canonical or Moyal-Weyl non-commutative spacetime defined by \cite{Moyal:1949sk}
\begin{align}
[x^{\mu}, x^{\nu}] = \Theta^{\mu \nu},
\end{align}
which will be discussed in the next section. Note that in this situation the direct formulation in terms of the star product (cf. below) is more suitable because the $k_j$ acting on a constant would just be zero. 

Another point of view to describe the relation between non-commutativity and quantum groups is to start from the commutation relations \eqref{kmink} which form what is known in the mathematical literature as the $an(3)$ algebra. A five dimensional representation of $an(3)$ is given by the matrices \cite{Kowalski-Glikman:2017ifs}
\begin{align}
x^0 = - \frac{i}{\kappa} 
  \begin{pmatrix}
    0 & {\bf{0}} & 1\\
    {\bf{0}} & {\bf{0}} & {\bf{0}}\\
    1 & {\bf{0}} & 0
  \end{pmatrix}
, \quad x^i = \frac{i}{\kappa} 
\begin{pmatrix}
    0 & {\bf{\epsilon}}_i^T & 0\\
    {\bf{\epsilon}}_i& {\bf{0}} & {\bf{\epsilon}}_i\\
    0 & -{\bf{\epsilon}}_i^T & 0
  \end{pmatrix},
\end{align}
where ${\bf{\epsilon}}_i$ is the unit vector with a single entry at the $i$'th position.
As usual, the associated Lie group ($AN(3)$) is obtained by exponentiation and thus a generic group element can be written as
\begin{align}
\hat e_k = e^{i k_i x^i} e^{ik_0 x^0},
\end{align}
parametrized by the functions $k_{\mu}$ on the group manifold, which becomes
\begin{align}
\hat e_k = \frac{1}{\kappa} 
  \begin{pmatrix}
    \bar P_4 & -{\bf{P}} e^{-k_0/\kappa} & P_0\\
    -{\bf{P}} & \kappa 1 & -{\bf{P}}\\
    \bar P_0 & {\bf{P}} e^{-k_0/\kappa} & P_4
  \end{pmatrix}
\end{align}
in terms of the matrix representation. The momenta $P_{\mu} (k)$ are given by \eqref{pk} and 
\begin{align}
\bar P_0 (k) = & \kappa \sinh \left( \frac{k_0}{\kappa} \right) - \frac{{\bf{k}}^2}{2 \kappa} e^{\frac{k_0}{\kappa}}, \quad \bar P_4 (k) = \kappa \cosh  \left( \frac{k_0}{\kappa} \right) + \frac{{\bf{k}}^2}{2 \kappa} e^{\frac{k_0}{\kappa}}, \\
P_4(k) = & \kappa \cosh  \left( \frac{k_0}{\kappa} \right) - \frac{{\bf{k}}^2}{2 \kappa} e^{\frac{k_0}{\kappa}}.
\end{align}
The vector $(0, {\bf{0}}, \kappa) \in \mathbb{R}^5$, the origin of the momentum space, has norm $\kappa^2$ and acting on it with $\hat e_k$ gives
\begin{align}
\hat e_k   
\begin{pmatrix}
    0\\
    {\bf{0}} \\
    \kappa
  \end{pmatrix} = 
  \begin{pmatrix}
    P_0\\
    -{\bf{P}} \\
    P_4
  \end{pmatrix}
\end{align}
so that the invariance of the norm implies
\begin{align}
P_4^2 - P_0^2 + {\bf{P}}^2 = \kappa^2, \quad P_0 + P_4 = e^{k_0/\kappa} > 0.
\end{align}
But this just means that the momentum space is a submanifold of the four dimensional de Sitter space with curvature radius $1/\kappa$.

From the group perspective the deformed coproduct acquires the interpretation of a non-standard addition ($ \hat +$) of momenta in the group multiplication, i.e.
\begin{align}
\hat e_{k \hat + l} \equiv \hat e_{k} \hat e_{l} = e^{i (k_i + e^{-k_0/\kappa} l_i)x^i} e^{i(k_0 + l_0)x^0},
\end{align}
with the help of the Baker Campbell Hausdorff formula.

A connection between the non-commutative algebra as in \eqref{kmink2} and a star product between commutative coordinate functions $\text{x}^{\mu}$ is given by a Weyl map (see e.g. \cite{Agostini:2002de}) $\Omega$
\begin{align}
\Omega  \left(e^{k_{\mu} \text{x}^{\mu}}\right) & \equiv \hat e_k =  e^{i k_i x^i} e^{ik_0 x^0}, \\
(f \star g) & = \Omega^{-1} ( \Omega (f) \Omega(g)),
\end{align}
for coordinate functions $f,g$ of $\text{x}^{\mu}$. Then the deformed commutator is written as
\begin{align}\label{starnc}
[\text{x}^{\mu}, \text{x}^{\nu}] \equiv \text{x}^{\mu} \star \text{x}^{\nu} - \text{x}^{\nu} \star \text{x}^{\mu}.
\end{align}
Note that while for the star product we also have the unique prescription from the previous section, the Weyl map is in general ambiguous, i.e. it depends on ordering ambiguities. For a further star product construction related to $\kappa$-Minkowski see also \cite{Durhuus:2011ci}.

\subsection{Broken and Deformed Symmetries}\label{sec2.2.2}

Now that we saw how the formalism of Hopf algebras relates to spacetime non-commutativity let us explore two different types of physical motivation of this phenomenon \cite{Amelino-Camelia:2008aez}. 

An instructive analogy for this distinction is the treatment of the speed of light in Galilean relativity, where the laws of motion (Newton's law) hold in all frames related by a galilean transformation. Such transformations relate frames that differ by constant relative motion and velocities simply add as vectors. When Maxwell introduced his equations for electromagnetism the speed of light $c$ was introduced and two different ways of incorporating it into Galilean relativity are conceivable. First, one could treat $c$ just as a ordinary velocity that transforms like a vector. Such a view \textit{breaks} the relativistic symmetry as the Maxwell equation would have a different form in distinct inertial frames, singling out a frame where $c$ has the value that one measures in our universe (ether). Alternatively, Einstein postulated that $c$ has the same value in all inertial frames. This \textit{deforms} the Galilean into special relativity, e.g. the addition of velocities is no longer vectorial but no frame is singled out.

When spacetime non-commutativity of the form \cite{Wess:2003da}
\begin{align}
[x^{\mu}, x^{\nu}] = \frac{1}{\kappa^2} \Theta^{\mu \nu} ( \kappa x)
\end{align}
is introduced, where the rhs is expanded as \cite{Lukierski:2005fc}
\begin{align}\label{expth}
\Theta^{\mu \nu} ( \kappa x) = \Theta^{\mu \nu}_{(0)} + \kappa \Theta^{\mu \nu}_{(1) \rho} x^{\rho} + \kappa^2 \Theta^{\mu \nu}_{(2) \rho \tau} x^{\rho} x^{\tau} + ..., 
\end{align}
one similarly has the choice how to treat the energy scale $\kappa$ or the coefficients $\Theta$. One possibility is to assume that $\Theta^{\mu \nu}$ transforms as an ordinary tensor. This choice can be motivated by a simple system of quantum mechanical particles in a strong magnetic field as follows \cite{Szabo:2001kg}. Consider the Lagrangian of a particle in the $x,y$ plane with mass $m$, charge $q$ and  velocity $v$ in the magnetic field $B$
\begin{align}
L = \frac{m v^2}{2} + q B x v_y.
\end{align}
The canonical momenta are simply given by
\begin{align}
p_x = \frac{\partial L}{\partial \dot x} = m v_x, \quad p_y = m v_y + q B x
\end{align}
and the canonical commutation relations $[p_y, y] = i$ yield for large $B$ 
\begin{align}
\frac{i}{q B} = \frac{1}{qB} [p_y, y] = \frac{1}{qB} [x, m v_y + qB x] \simeq [y, x].
\end{align}
The physical interpretation of this emerging effective non-commutativity is that even though the model allows for states in which both $x$ and $y$ can be measured sharply simultaneously these states correspond to high energy if $B$ is very large and thus a low-energy observer can not excite them. 
An analogous mechanism was discovered in string theory where the Kalb-Ramond field $B^{ij}$ (which generalizes the  electromagnetic potential) plays the role of the magnetic field \cite{Kalb:1974yc} \cite{Szabo:2001kg}. This low-energy effective non-commutativity was mostly studied in the case of constant $B^{ij}$ corresponding to the zeroth order in \eqref{expth} and is also known as canonical or Moyal non-commutativity \cite{Moyal:1949sk}.

Another possibility consists in postulating that the scale $\kappa$, often identified with the Planck mass $m_{\text{Pl}}$, is observer independent. Such a scenario is called doubly special relativity \cite{Amelino-Camelia:2000stu} because of the status as a fundamental constant of nature both of the speed of light and $\kappa$ and it has been mostly studied in the context of $\kappa$-Poincar\'e \cite{Kowalski-Glikman:2017ifs},\cite{Bruno:2001mw} introduced above (the first order in \eqref{expth}) but is not limited to it \cite{Amelino-Camelia:2010lsq}. Usually the interpretation is that the momentum space described by a Hopf algebra is more fundamental in the sense that it is immediately accessible by experiments whereas the spacetime (module algebra) is emergent and properties like locality are generally relative for different observers. Thus, the classical notion of a local continuous spacetime is recovered only in a limit.

In this thesis we will adopt the latter point of view and focus mainly on the Lie algebra type non-commutativity but also the zeroth order (which we will also call canonical non-commutativity and second order (called quadratic non-commutativity, cf. \cite{Lukierski:2005fc}) terms in \eqref{expth} can be consistently described in this framework.

Let us close with a general remark on the motivation for deformation. If a new physical theory is discovered the old ones are not fully abandoned, they usually remain valid as an approximation, i.e. they can be seen as limits of the new theory. The parameter(s) controlling the limit are often new constants of nature. Deformations, as we met them in this chapter, are in some sense the attempt to reverse this procedure, to look for new physics in the ``vicinity'' of established theories. Also special relativity and quantum mechanics are, in hindsight, motivating examples for this strategy.

\chapter{Asymptotic Symmetries of Spacetime}\label{chap3}

In the following chapter, we review the symmetries of spacetimes with certain asymptotic behavior in the framework of general relativity. We mainly focus on the algebraic features and investigate four and three dimensional spacetimes with different values for the cosmological constant. Also the corresponding conserved charges and their respective algebras are studied.

\section{4D Gravity}

In the three spatial and one time dimensions of spacetime the phenomenon of gravity is described by Einsteins theory of general relativity (GR). This theory is formulated in terms of a differentiable manifold representing spacetime endowed with a semi-Riemannian metric $g_{\mu \nu}$ that is affected by the energy-momentum tensor $T_{\mu \nu}$ via the Einstein field equations
\begin{align}\label{efe}
R_{\mu \nu} - \frac{1}{2} R g_{\mu \nu} + \Lambda g_{\mu \nu} = 8 \pi G T_{\mu \nu}
\end{align}
where $R_{\mu \nu}$ and $R$ are the Ricci curvature and scalar determined by the metric. The theory is, as the name suggests, relativistic in the sense of special relativity, i.e. physical laws hold equally in all inertial frames and do not single out a preferred one. This implies that in the absence of sources and a cosmological constant the symmetry group of the vacuum is the Poincar\'e group. GR generalizes the relativity principle even further by postulating that it holds for non-rotating, freely falling frames of reference which locally can not be distinguished from a uniformly moving frame in the absence of a gravitational field. Therefore the metric can be expanded locally around the Minkwski metric of flat space and it transforms covariant under general coordinate transformations.
In \eqref{efe} $\Lambda$ denotes the cosmological constant and can be interpreted as the energy density of spacetime. Its origin is still not very well understood and observations suggest that it has a small but positive value in our universe. On shorter scales the effect of $\Lambda$ is negligable but it has important implications for the global structure of spacetime as we will explore later.

\subsection{Asymptotic Flatness}\label{sec3.1.1}
As already mentioned above, \eqref{efe} also suggests that in the absence of sources and vanishing $\Lambda$ the curvature vanishes and it is well-known that the resulting flat spacetime is invariant under Lorentz transformations and translations which form the Poincar\'e group whose Lie algebra we encountered in the previous chapter. When Bondi, Metzner, Sachs and van der Burgh \cite{Bondi:1962px}, \cite{Sachs:1962wk}, \cite{Sachs:1962zza} studied the symmetries of a spacetime that is arbitrary in a finite region but approaches flat space in the limit of infinite radial coordinate $r$, they probably also expected to find the Poincar\'e group. However, it turned out that the symmetry group of such an asymptotically flat spacetime is enhanced to an infinite dimensional group. 

In the following we will shortly review this finding. 
It is useful to first introduce Penrose diagrams which allow for a visualization of the infinite distances of spacetime while highlighting the causal structure (see e.g. \cite{Strominger:2017zoo}). 

\begin{figure}
\centering
\includegraphics[width=10cm]{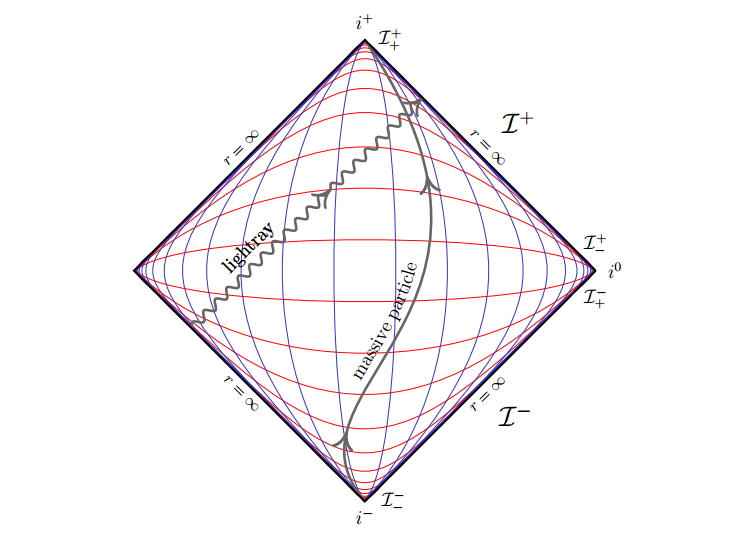}
\caption{A Penrose diagram of Minkowski space. Red lines indicate surfaces of constant time $t$ and blue lines surfaces of constant radius $r$ \cite{Strominger:2017zoo}.}\label{fig1}
\end{figure}

As figure \ref{fig1} shows in the example of an eternal Minkowski spacetime the infinite region is squished into a finite one by a conformal transformation in such a way that lightrays move in a $45^{\circ}$ angle. Every pair of points with the same radial coordinate $r$ and time $t$ is actually a two-sphere and the lightrays start from the lower boundary called past light-like infinity $\mathcal{I}^-$, where $r=\infty =t$ but the advanced time $v = t+r$ is finite, and move to the future light-like infinity $\mathcal{I}^+$ parametrized by the retarded time $u= t-r$. Massive particles that start from the time-like infinity $i^-$ on the other hand will always be overtaken on their way to $r=\infty$ by all lightrays and end up in the future time-like infinity $i^+$. The sphere $i^0$ labels the so-called spatial infinity. 

\subsubsection*{Bondi Metric}

Because of the (gauge) invariance of GR under general coordinate transformations one has to make a gauge choice fixing some of the metric components. The choice
\begin{align}
g_{rr} = 0, \quad g_{r A} = 0, \partial_r \text{det}(g_{AB}/r^2) = 0,
\end{align}
defining the Bondi gauge, implies that $u$ is labeling null hypersurfaces as the normal $n^{\mu} = g^{\mu \nu} \partial_{\nu} u$ is null. Furthermore, the angular coordinates $x^A = (z, \bar z)$ are defined such that their directional derivative along $n^{\mu}$ vanishes. Here $z$ and its complex conjugate $\bar z$ are coordinates of a stereographic projection of the two-sphere into the complex plane. A general metric in the Bondi gauge
\begin{align}\label{bondimet}
ds^2 = -U du^2 - 2e^{\beta} du dr + g_{AB} \left( d x^A + \frac{1}{2} U^A du \right) \left( d x^B + \frac{1}{2} U^B du \right)
\end{align}
is parametrized by functions $U, U^A, \beta$ which can depend on all coordinates. In order to obtain a metric that describes asymptotically flat spacetime one thus has to impose fall-off conditions on these functions that restricts their behavior for large $r$. Even though there is no general consensus as to what exactly these conditions should be, they are ultimately tied to observation and should allow for phenoma that are known to exist but exclude the unphysical ones. We will first adapt some of the loosest restrictions and discuss possible objections later on (see also \cite{Ruzziconi:2020cjt})\footnote{We will sometimes use the term BMS algebra to loosely refer to asymptotic symmetry algebras and name specific algebras with calligraphic letters.}. After imposing the fall-off conditions the general metric reads
\begin{align}
ds^2 =& -du^2 -2 du dr + 2r^2 \gamma_{z\bar z} dz d \bar z\nonumber \\
&  + \frac{2 m_B}{r} du^2  + r C_{zz} dz^2 + r C_{\bar z\bar z} d\bar z^2 + D^z C_{zz} du dz +  D^{\bar z} C_{\bar z \bar z} du d\bar z \nonumber \\
& + \frac{1}{r} \left( \frac{4}{3} (N_z  + u \partial_z m_B) - \frac{1}{4} \partial_z (C_{zz} C^{zz}) \right) du dz + c.c. + \mathcal{O}(1/r^2),
\end{align}
with the Bondi mass aspect $m_B$ describing the mass of a system as measured at null infinity, the angular momentum aspect $N_A$ and the traceless and symmetric tensor $C_{AB}$. $C_{AB}$ is related to the Bondi news tensor  by $N_{AB} = \partial_u C_{zz}$ whose square is proportional to the energy flux of gravitational waves (propagating disturbances in the curvature) across $\mathcal{I}^+$. On the sphere one has the metric $\gamma_{ z \bar z} = \frac{2}{(1+z \bar z)^2}$ and $D_A$ denotes the covariant derivative with respect to $\gamma_{A B}$.

\subsubsection*{Supertranslations and Superrotations}

Gauge transformations of \eqref{bondimet} that preserve the Bondi gauge are infinitesimal diffeomorphisms $\xi$ satisfying \cite{Strominger:2017zoo}
\begin{align}\label{xireq}
\mathcal{L}_{\xi} g_{rr} =0, \quad \mathcal{L}_{\xi} g_{rA} = 0, \quad \mathcal{L}_{\xi} \partial_r \text{det} \left( g_{AB}/r^2\right) = 0,
\end{align}
where $\mathcal{L}_{\xi}$ denotes the Lie derivative with respect to $\xi$ and the asymptotic requirements
\begin{align}\label{asreq}
\mathcal{L}_{\xi} g_{uu} = \mathcal{O} \left( r^{-1} \right), \quad \mathcal{L}_{\xi} g_{ur} = \mathcal{O} \left( r^{-2} \right), \quad \mathcal{L}_{\xi} g_{uA} = \mathcal{O} \left( r^{0} \right), \quad \mathcal{L}_{\xi} g_{AB} = \mathcal{O} \left( r^{1} \right).
\end{align}
 We have
\begin{align}
\xi(f,R^A) = & \left( f + \frac{u}{2} D_A R^A + \mathcal{O}(r^0)\right) \partial_u + \left( R^A - \frac{1}{r} D^A f + \mathcal{O}(r^{-1}) \right)\partial_A  \nonumber \\
& + \left( - \frac{r+u}{2} D^A R^A + \frac{1}{2} D_A D^A f + \mathcal{O}(r^0) \right) \partial_r 
\end{align}
and we call these solutions of \eqref{xireq} and \eqref{asreq} asymptotic Killing vectors as they leave the form of \eqref{bondimet} invariant. 
The asymptotic Killing vectors are parametrized by functions $f$ and $R^A$ on the sphere. While $f$ is unconstrained, the last equation in \eqref{asreq} enforces the conformal Killing equation on the two-sphere for $R^{A}$
\begin{align}
D_A R_B + D_B R_A = \gamma_{AB} D_C R^C
\end{align}
which in turn implies that $R^z$ (respective $R^{\bar z}$) is holomorphic (antiholomorphic),i.e $\partial_z R^{\bar z}=0 (\partial_{\bar z} R^{z} =0)$. The Lie derivative of vector fields naturally defines a Lie bracket via $\mathcal{L}_{\xi} (\xi') \equiv [\xi, \xi']$ that reads in first order
\begin{gather}\label{bms1}
[\xi(f, R^A), \xi(f', {R'}^A) ] = \xi (\hat f, \hat{R^A}),  \\
\hat f = R^A D_A f' + \frac{1}{2} f D_A {R'}^A - {R'}^A D_A f -  \frac{1}{2} f' D_A R^A,  \\
\hat{R^A} = R^B D_B {R'}^A - {R'}^B D_B R^A
\end{gather}
for the asymptotic Killing vectors. The vector fields $\xi(f, 0)$ form an ideal of the Lie algebra and are called supertranslations. Acting on the Minkowski vacuum defined by $m_B = N_{ AB} = C_{AB} =0$ with a supertranslation does not change the mass or angular momentum aspect but it transforms
\begin{align}
\mathcal{L}_f C_{z z} = -2 D^2_z f.
\end{align}
The vanishing of the curvature requires that 
\begin{align}
C_{z z} = -2 D^2_z C
\end{align}
for some function $C(z, \bar z)$ and because one has $\mathcal{L}_f C = f$, supertranslations preserve this condition and lead to a degeneracy of the gravitational vacuum already at the classical level. Only the modes that are annihilated by $D^2_z$ are not spontaneously broken in this sense and are just the usual spacetime translations. 

Expanding into monomials in $z, \bar z$ according to $f_{mn} \equiv \frac{z^m \bar z^n}{1+z \bar z} \equiv T_{mn}$ and similar for the superrotations $R^z_n \equiv z^{n+1} \equiv l_n, R^{\bar z}_n \equiv \bar z^{n+1}\equiv \bar l_n$ \cite{Barnich:2009se}, one can write \eqref{bms1} as
\begin{align}\label{bms4d}
[l_m, l_n] & = (m-n) l_{m+n}, & [l_m, T_{pq} ] & = \left( \frac{m+1}{2}-p \right) T_{m+p, q}, \\
[\bar l_m , T_{pq}] & = \left( \frac{m+1}{2}-q \right) T_{p, q+m}, & [T_{mn}, T_{pq}] & = [l_m, \bar l_n] =0,\label{bms4d2}
\end{align}
the infinite dimensional BMS algebra in four dimensions denoted by $\mathfrak{B}_4$ in the following.
It is the semi-direct product of two copies of the algebra of vector fields on the sphere, the Witt algebra (denoted by $ \mathfrak{W}$) defined by the first equation in \eqref{bms4d} and the abelian ideal $V_4$ generated by the $T_{pq}$.
Restricting to $m, n, p, q \in \{0, \pm 1\}$ the algebra \eqref{bms4d} is indeed just the ordinary Poincar\'e algebra which is therefore a subalgebra of $\mathfrak{B}_4$. However, there exist infinitely many subalgebras generated by $$\{l_0, \bar l_0, l_{\pm(1-2m)}, \bar l_{\pm(1-2m)}, T_{mm}, T_{1-m, m}, T_{m, 1-m}, T_{1-m, 1-m}\vert m \in \mathbb{Z} \}$$ that are isomorphic to the Poincar\'e after the rescaling
\begin{align}
l_i \rightarrow \frac{l_i}{n}, \quad \bar l_i \rightarrow \frac{\bar l_i}{n},
\end{align}
where $n= 1-2m$. These so-called \textbf{\textit{embeddings}} correspond to classical vacua which are left invariant by a unique subgroup of the BMS group \cite{Strominger:2017zoo}. 
An observer describes the symmetry of the vacuum he/she is in with the Poincar\'e algebra ($n=1$) and transitions to another vacuum by supertranslations/-rotations. 

A useful basis for $\mathfrak{B}_4$ is obtained by the transformation
\begin{align}
k_n & = l_n + \bar l_n, & \bar k_n & = -i ( l_n - \bar l_n), \\
S_{mn} & = \frac{1}{2} (T_{mn} + T_{nm}) , & A_{mn} & = - \frac{i}{2} (T_{mn}-T_{nm}),
\end{align}
and the algebra relations then read
\begin{align} \label{4dembed1}
[k_n, k_m] & = (n-m) k_{n+m}, \quad [\bar k_n, \bar k_m] = -(n-m) k_{n+m}, \\
[k_n, \bar k_m] & = (n-m) \bar k_{n+m},\\
[k_n, S_{pq}] & = \left( \frac{n+1}{2} -p \right) S_{p+n, q} + \left( \frac{n+1}{2} -q \right) S_{p, q+n}, \\
[\bar k_n, S_{pq}] & = \left( \frac{n+1}{2} -p \right) A_{p+n, q} - \left( \frac{n+1}{2} -q \right) A_{p, q+n}, \\
[k_n, A_{pq}] & = \left( \frac{n+1}{2} -p \right) A_{p+n, q} + \left( \frac{n+1}{2} -q \right) A_{p, q+n}, \\
[\bar k_n, A_{pq}] & = -\left( \frac{n+1}{2} -p \right) S_{p+n, q} + \left( \frac{n+1}{2} -q \right) S_{p, q+n}.\label{4dembed2}
\end{align}
In this basis the Poincar\'e algebra in light-cone coordinates
\begin{align}
P_{\pm} = \frac{P_0 \pm P_3 }{\sqrt{2}}, \quad M_{\pm \, 1/2} = \frac{M_{0\, 1/2} \pm M_{3\, 1/2}}{\sqrt{2}}, \quad M_{ + -} = M_{0 3}
\end{align}
 is embedded according to 
\begin{align}
k_0 & = \frac{i}{n}  M_{+-}, \quad \bar k_0 = - \frac{i}{n} M_{12}, \quad k_n = \frac{i}{n} \sqrt{2} M_{+1}, \\
 k_{-n} & = \frac{i}{n} \sqrt{2} M_{-2}, \quad \bar k_n = \frac{i}{n} \sqrt{2} M_{+2}, \quad \bar k_{-n} = \frac{i}{n} \sqrt{2} M_{-1}, \\
S_{mm} & = \frac{i}{n} \sqrt{2} P_-, \quad S_{1-m, 1-m} = - \frac{i}{n} \sqrt{2} P_+, \\
S_{m, 1-m} & = - \frac{i}{n}  P_1, \quad A_{m,1-m} = - \frac{i}{n} P_2.
\end{align}

\subsubsection*{Phenomenology}

One might ask now whether the enlargement of the Poncar\'e algebra has any observable consequences. This is indeed the case and the easiest way to confirm it is by considering the degenerate vacua related by supertranslations \cite{Strominger:2014pwa}. An experiment might consist of two inertial detectors near $\mathcal{I}^+$ such that the Bondi news vanishes at early and late times but in the intermediate time the passing of a gravitational wave results in non-zero Bondi news. From the geodesic equation one finds that their relative separations $(s^z, s^{\bar z})$ are altered by the gravitational wave and after it has passed the spacetime is flat but characterized by a different $C_{zz}$, i.e. it was supertranslated. The residual change in the separations is proportional to 
\begin{align}\label{displm}
\Delta s^{\bar z} = \frac{\gamma^{z \bar z}}{2r} \Delta C_{zz} s^z
\end{align}
and yields a finite result. This effect is called the gravitational displacement memory effect (cf. also chapter \ref{chap5}).

In the case of the superrotations the situation is a bit more subtle. With the general parametrization where $m \in \mathbb{Z}$ there are singularities at $z = 0, \infty$, i.e. the functions are only meromorphic. Only the ordinary Lorentz rotations and boosts with $m= 0, 1, 2$ are holomorphic on the entire sphere and thus globally well defined as the vector fields $R_m \equiv z^m \partial_z, m<0$ and, after redefining $\omega = z^{-1}$,$R_m = \omega^{2-m} \partial_{\omega}, m>2$, have a singularity at the origin \cite{Compere:2018aar}. 
A general superrotation will thus map a flat spacetime to one that is flat everywhere except at the singularities which is then called asymptotically locally flat (ALF) \cite{Strominger:2016wns}. While some authors argued that superrotations have to be excluded because of the singularities, others suggested that they might have a physical interpretation related to cosmic strings. Cosmic strings are one-dimensional topological defects which were linked to the following construction due to Penrose \cite{Pen72}, \cite{Strominger:2016wns}. Consider a Minkowski spacetime that is cut along the light-cone $u=0$. After performing a diffeomorphism on the patch with $u>0$ it is glued together such that the metric is continuous across the cut. This procedure introduces singularities, e.g. at $z=0, \infty$ which describe a cosmic string. In our universe such strings might exist if at some early time the universe was not simply connected yet (cf. also section \ref{sec5.4.2}). Vacua with strings would then be ALF spacetimes and the snapping of a string can therefore be seen as a transition from an ALF spacetime to an inequivalent asymptotically flat one which is indeed just a superrotation.
In analogy to the supertranslations a further consequence of the existence of superrotations was suggested to be the (sub-leading) spin-memory effect \cite{Pasterski:2015tva}.


\subsection{Surface Charges}\label{sec3.1.2}

A cornerstone of theoretical physics is the principle that continuous symmetries lead to conserved quantities as expressed by the Noether theorem. In the case of a ``rigid'' or global symmetry of a field theory on the $D$-dimensional manifold $\mathcal{M}$ with a parameter $\epsilon$ \cite{Oblak:2016eij}
\begin{align}\label{gtrafo}
x \mapsto x + \delta_{\epsilon} x, \quad \phi \mapsto \phi + \delta_{\epsilon} \phi,
\end{align}
one can obtain the associated Noether current $j^{\mu}$ by replacing $\epsilon$ with an arbitrary function $\epsilon (x)$ and reading it off from the variation of the action
\begin{align}\label{ds}
\delta S = - \int_{\mathcal{M}} d^D x j^{\mu} \partial_{\mu} \epsilon.
\end{align}
If the transformation \eqref{gtrafo} is indeed a symmetry of the theory the variation vanishes on-shell, i.e. if the fields satisfy the equations of motion, which in turn implies that $j^{\mu}$ is conserved. For a space-like slice $\Sigma$ of $\mathcal{M}$ the associated conserved Noether charge then reads
\begin{align}
Q = \int_{\Sigma}  d^{D-1} x n_{\mu} j^{\mu},
\end{align}
where $n_{\mu}$ is the time-like normal vector field of $\Sigma$. If, however, we are dealing with a gauge symmetry then $\epsilon$ will be depending on the spacetime coordinates from the start and thus the rhs of \eqref{ds} vanishes identically leaving us with no possibility to determine $j^{\mu}$. The situation can be rescued by noting that we can modify the current $j^{\mu} \rightarrow j^{\mu} + \partial_{\nu} k^{\mu \nu}$ and \eqref{ds} is unaffected if $k^{\mu \nu}$ is antisymmetric. The Noether charge then receives a contribution 
\begin{align}\label{qs}
Q_{\text{surface}} = \int_{ \Sigma} d^{D-1} x n_{\mu} \partial_{\nu} k^{\mu \nu} = \int_{ \partial \Sigma } d^{D-2} n_{\mu} n'_{\nu} k^{\mu \nu},
\end{align}
where $n'_{\nu}$ is normal to $\partial \Sigma$. We can now associate a two-form $k^{\mu \nu}$ to the symmetry by using \eqref{ds} with $\partial_{\nu} k^{\mu \nu}$ and take \eqref{qs} as a definition for the charge, it is conserved if $\nabla_{\mu} k^{\mu \nu} = 0$ on-shell, where $\nabla_{\mu}$ is the covariant derivative. 

Conserved charges $Q$ of a symmetry described by a group $G$ generate infinitesimal symmetry transformations in the following way \cite{Oblak:2016eij}. Let $\mathcal{M}'$ be a Poisson manifold with Poisson bracket $\{, \}$, the classical phase space of a theory and $F$ an observable, i.e. a function on $\mathcal{M}'$. One can identify the infinitesimal generator $\xi_{X}$ of the action of $G$ on $\mathcal{M}'$, which is a vector field on $\mathcal{M}'$ related to an element $X$ of the Lie algebra $\mathfrak{g}$ associated to $G$ by
\begin{align}\label{xicom}
[\xi_X, \xi_Y] = - \xi_{[X, Y]},
\end{align}
with the charge via
\begin{align}
- \{ Q[\xi_X], \} \equiv \xi_{X}, \rightarrow \{ Q[\xi_X], F \} = \xi_X F \equiv \delta_X F.
\end{align}
One says that $Q$\footnote{Or rather the map $\tilde Q: \mathcal{M}' \rightarrow \mathfrak{g}^*$, $ \braket{\tilde Q( ), X} \equiv Q[\xi_X]$.} is a momentum map of the action of $G$ on $\mathcal{M}'$. 

For the symmetry transformation on the fields $\phi$ we can write
\begin{align}
\{ Q[\xi], \phi \} = - \delta_{\xi} \phi
\end{align}
and the Jacobi identity for the Poisson bracket implies (using \eqref{xicom})
\begin{align}
\{ \phi, \{Q[\xi], Q[\zeta]\}\} = - \{ Q[\zeta], \{ Q[\xi], \phi\}\} - \{ Q[\xi] , \{ \phi, Q[\zeta]\}\} = [\delta_{\xi}, \delta_{\zeta}] \phi = \{ \phi, Q[ [\xi, \zeta]] \}.
\end{align}
As this has to hold for arbitrary fields, one can conclude that the terms in the bracket have to coincide up to a constant term
\begin{align}
\{ Q[\xi], Q[\zeta] \} = Q [[ \xi, \zeta]] + c (\xi, \zeta)
\end{align}
such that $\{c, \phi\} = 0$.
The Jacobi identity then enforces
\begin{align}\label{2coc}
c( \{ Q[\xi_1], Q [\xi_2]\} , Q[\xi_3]) + \text{cycl.} = 0,
\end{align}
which is the condition that $c$ is a 2-cocycle of the Chevalley-Eilenberg cohomology with values in the field $\mathbb{C}$ \footnote{The first line in \eqref{coc-base} vanishes because the action of the field is trivial and thus we obtain \eqref{2coc}}. Also, if $c$ were a coboundary derived from a 1-cocycle $c'$ we could redefine 
\begin{gather}
Q'[\cdot] = Q[ \cdot]  + c'(\cdot) \\
\rightarrow \{Q'[\xi_1] , Q'[\xi_2] \} = Q[[\xi_1, \xi_2]] + c(\xi_1, \xi_2) = Q[[\xi_1, \xi_2]] + c'([\xi_1, \xi_2]) = Q'[[\xi_1, \xi_2]]
\end{gather}
so we find that the algebra of the surface charges is in general a central extension of the algebra of the symmetry generators classified by the cohomology group $H^2(\mathfrak{g}, \mathbb{C})$.

In the flat $(\Lambda = 0)$ case one finds for the surface charges associated to the supertranslations \cite{Strominger:2017zoo}
\begin{align}
Q[f] = \frac{1}{4 \pi G} \int d^2 z \gamma_{z \bar z} f m_B,
\end{align}
where the integration is over the boundary of a space-like slice, e.g. a sphere on $\mathcal{I}^{\pm}$ which is a point on the boundary of the Penrose diagram. Defining $Q^+[f]$ and $Q^-[f]$ as the charges from integration over $\mathcal{I}^+_-$ and $\mathcal{I}^-_+$, one has a matching condition implying the conservation of the charges, i.e.
\begin{align}
Q^+[f] = Q^-[f].
\end{align}
Similarly, for the superrotations we have
\begin{align}\label{4dcharge}
Q^+[R^A] = \frac{1}{8 \pi G} \int_{\mathcal{I}^+_-} d^2 z(R_{\bar z} N_z + R_z N_{\bar z} ) =  \frac{1}{8 \pi G} \int_{\mathcal{I}^-_+} d^2 z(R_{\bar z} N_z + R_z N_{\bar z} ) = Q^-[R^A],
\end{align}
where the conservation follows from a matching condition for the angular momentum aspect $N_A$.

For the $\mathfrak{B}_4$(eq. \eqref{bms4d}-\eqref{bms4d2}) it was shown in \cite{Safari:2019zmc} that the only consistent central extensions are those of the Witt subalgebras. There, the second cohomology group with values in the trivial module is one-dimensional and given by the Gelfand-Fuks cocycle leading to the Virasoro algebra ($\mathfrak{Vir}$), i.e. the centrally extended Witt algebra
\begin{align}\label{ce1}
[l_m, l_n] = (m-n) l_{m+n} + \frac{c_l}{12} (m^3 -m) \delta_{m+n, 0}
\end{align}
with central charge $c_l$. We denote this surface charge algebra by $\mathfrak{B}_{4,c} \equiv V_4 \rtimes (\mathfrak{Vir}\oplus \mathfrak{Vir}) $. In four dimensions the calculation of these charges $(c_l, c_{\bar l})$ in general is quite involved as they are non-integrable in the presence of gravitational waves \cite{Wald:1999wa}.

\subsection{Asymptotically (Anti) de Sitter Spacetimes}\label{sec3.1.3}

As the solutions to the Einstein equations for a globally source-free configuration are, besides the Minkowski spacetime, also (Anti) de Sitter ((A)dS) spacetimes depending on the cosmological constant it is natural to investigate the concept of asymptotic symmetries for these solutions. 
Recently, this has been done in \cite{Compere:2019bua}, where the symmetry group of a spacetime that is asymptotically Anti de Sitter was investigated. 
Even though the de Sitter case is algebraically very similar there are several unresolved conceptual difficulties in the gravitational description of its asymptotics as described for example in \cite{Ashtekar:2014zfa}. The main issue appears to be that the null infinities $\mathcal{I}^{\pm}$ are space-like for $\Lambda >0$ which makes it difficult to even formulate boundary conditions. 

In \cite{Compere:2019bua}, just as in the asymptotically flat case, a generic spacetime metric is constrained by fall-off conditions in the radial coordinate $r$ such that in the limit $r \rightarrow \infty$ the AdS metric is recovered. The Killing vectors  $\xi_{f, R} = f \partial_u + R^A \partial_A$ of this asymptotically AdS metric satisfy in first order of $1/r$
\begin{gather}
[\xi_1, \xi_2] = \hat \xi = \hat f \partial_u + \hat R^A \partial_A, \\
\hat f = R_1^A \partial_A f_2 + \frac{1}{2} f_1 \partial_A R^A_2 - (1 \leftrightarrow 2), \quad \hat R^A = R^B_1 \partial_B R_2^A + \frac{\Lambda}{3} f_1 \gamma^{AB} \partial_B f_2 - (1 \leftrightarrow 2).\label{mkv}
\end{gather}
Parametrizing the Killing vector fields by whole numbers
\begin{align}
\xi_{f_{pq}, 0} \equiv T_{pq} \equiv \frac{z^p {\bar z}^q}{1+ z \bar z} \partial_u, \quad \xi_{0, R^z_n} \equiv l_n \equiv -z^{n+1} \partial_z , \quad \xi_{0, R^{\bar z}_n} \equiv \bar l_n \equiv - \bar {z}^{n+1} \partial_{\bar z} ,
\end{align}
yields the following structure
\begin{align}
[T_{p,q}, T_{m,n}] & =  \Lambda \frac{z^p \bar z^q}{1+ z \bar z} \frac{(1+ z\bar z)^2}{2} \left( \partial_{\bar z} \frac{z^m \bar z^n}{1+ z \bar z} \partial_z + \partial_{ z} \frac{z^m \bar z^n}{1+ z \bar z} \partial_{\bar z} \right) - (mn \leftrightarrow pq) \nonumber \\
& =  \Lambda ((n-q) z^{p+m} \bar z^{q+n-1} \partial_z + (m-p) z^{p+m-1} \bar z^{q+n} \partial_{\bar z}) \label{algebroid} \\
[l_m, l_n] & = (m-n) l_{m+n}, \quad [\bar l_m, \bar l_n] = (m-n) \bar l_{m+n} \\
[l_m, T_{p,q}] & = \left(\frac{m+1}{2} -p \right) T_{p+m, q}, \quad [\bar l_m, T_{p,q}]  = \left(\frac{m+1}{2} -q \right) T_{p, q + m}. 
\end{align}
Thus, the result \eqref{algebroid} can not be written as the sum of basis elements $\xi_{f_{mn}, R^A_p}$ but is a general vector field on the sphere, i.e. the four dimensional $\Lambda$-BMS is a Lie algebroid. Technically this is a consequence of the appearance of the coordinate dependent metric on the sphere in \eqref{mkv}. However, it is still true that the ordinary symmetry algebra of Anti de Sitter spacetime (denoted by $AdS_4$)
\begin{align}\label{4dads}
[K_+, K_-] & =  -i\Lambda M_{+-}, \quad [K_+, K_a] =  -i\Lambda M_{+a},  \\
 [K_-, K_a] & =  -i \Lambda M_{-a},  \quad [K_a, K_b] = -i \Lambda M_{ab},  \\
 [M_{\mu \nu}, M_{\rho \sigma}] & = i( \eta_{\mu \sigma} M_{\nu \rho} - \eta_{\nu \sigma} M_{\mu \rho} + \eta_{\nu \rho} M_{\mu \sigma} - \eta_{\mu \rho} M_{\nu \sigma}), \\
 [M_{\mu \nu}, K_{\rho}] &= i ( \eta_{\nu \rho} K_{\mu} - \eta_{\mu \rho} K_{\nu}),
\end{align}
is embedded into \eqref{algebroid} which can be seen by considering the subalgebras spanned by 
\begin{align*}
l_0, l_{1-2m}, l_{-1+2m},  \bar l_0, \bar l_{1-2m}, \bar l_{-1+2m}, T_{mm}, T_{1-m, m}, T_{m, 1-m}, T_{1-m,1-m},
\end{align*}
with the rescaling 
\begin{align}
l_n \rightarrow \frac{l_n}{1-2m}, \quad \bar l_n \rightarrow \frac{\bar l_n}{1-2m}, \quad \Lambda \rightarrow (1-2m)^2 \Lambda.
\end{align}
We have e.g. 
\begin{align}
[K_-, K_+] &= [T_{mm}, T_{1-m, 1-m}] = \Lambda (1-2m) (z \partial_z + \bar z \partial_{\bar z})  \nonumber \\
& = (1-2m)^2 \Lambda (\xi_{l_0} + \xi_{\bar l_0}) = i(1-2m)^2 \Lambda M_{+-},
\end{align}
and similar for the rest of the $AdS_4$. 

Although there are attempts to define generalizations of bialgebras on some versions of algebroids, e.g. in \cite{Brzezinski:2002}, \cite{Lu96}, there is no established notion of quantum groups on a Lie algebroid and therefore deformations of asymptotically (A)dS spacetime symmetries in four dimensions will not be studied directly in this work. Nonetheless, one can hope to infer valuable insights from the three dimensional theory of gravity.

\section{3D Gravity}

As we will see, Einsteins theory of GR in two spatial and one time dimension is a useful toy model for studying asymptotic symmetries as well as other problems that are harder to address in four dimensional gravity. On the first glance, the theory seems to be rather trivial and completely different from what one would associate with gravity because it does not have a Newtonian limit and massive particles do not attract each other \cite{ch:I-1Staruszkiewicz:1963zza}, \cite{Deser:1983tn}, \cite{Deser:1983nh}. In fact, there are no local gravitational degrees of freedom at all in three dimensions! This means in particular that no gravitational waves exist and gravitational energy can not be radiated away. Formally this is a consequence of the fact that the three dimensional Riemann tensor, which can be expressed in terms of the Ricci and the Weyl tensor, has the same number of degrees of freedom as the Ricci tensor. Thus, the Weyl tensor vanishes leading to the peculiar features mentioned above. 

Nonetheless, because of topological and asymptotic properties, there are still interesting physical phenomena. For example particles can be introduced in the form of topological defects with the geometry of a cone \cite{Trzesniewski:2017sqa}. The fact that topological but no local gravitational degrees of freedom exist is also exemplified by the equivalence to Chern-Simons theory which is a topological field theory \cite{Witten:1988hf}. Because Chern-Simons theory can be quantized, it is hoped that the reformulation can shed light on the problem of quantum gravity \cite{ch:I-2Witten:1988hc}.
It was also shown that black hole solutions with an event horizon exist \cite{Banados:1992wn}. These so-called BTZ black holes admit (if they are non-extremal) a finite temperature implying that they decay like their four dimensional counterparts.

\subsection{Asymptotic Symmetries}

Depending on the value of the cosmological constant, the three dimensional spacetime is locally isometric to either Minkowski, de Sitter or Anti de Sitter space. We will start the discussion of the asymptotic symmetries in the case of curved spacetime (following \cite{Oblak:2016eij}) and then view the asymptotically flat spacetime as a contraction limit $\Lambda \rightarrow 0$. 

Analogously to the four dimensional case one can start from a general metric 
\begin{align}
ds^2 = - \left( 1 - r^2 \Lambda \right) dt^2 +  \left( 1 - r^2 \Lambda \right)^{-1} dr^2 + r^2 d \phi^2
\end{align}
 and impose fall-off conditions so that it asymptotically approaches (A)dS. The Killing vectors are $\xi_{f, R} = f \partial_u + R \partial_z$ where $R = R(z), f = T(z) + u \partial_z R$ and we have
\begin{gather}
[\xi_1, \xi_2] = \hat \xi \equiv \hat f \partial_u + \hat R \partial_z \\
\hat f  = R_1 \partial_z f_2 + f_1 \partial_z R_2 - (1 \leftrightarrow 2) , \quad \hat R = R_1 \partial_z R_2 - \Lambda f_1 \partial_z f_2  - (1 \leftrightarrow 2).
\end{gather}
Here, $u = t -r$ is again the retarded time and $z = e^{i \phi}$ with the angular coordinate $\phi$. 
Parametrizing $f_m \equiv T_m = z^{m+1}$ and $R_m \equiv l_m = z^{m+1}$, we find
\begin{align}\label{3dbms}
[l_m, l_n] & = (m-n) l_{m+n}, \quad [l_m, T_n] =  (m-n)T_{m+n}, \\
[T_m, T_n] & = -\Lambda(m-n) l_{m+n}.
\end{align}
As in the four dimensional case the symmetry algebra of the globally AdS spacetime ($AdS_3$)
\begin{gather}\label{ads3}
[\tilde P_+, \tilde P_- ] = - i \eta_{+-} \Lambda M_{+-}, \quad [\tilde P_{\pm}, \tilde P_2] = - i \eta_{+-} \Lambda M_{\pm 2}, \\
[M_{+2}, M_{-2}] = i(-\eta_{22})  M_{+-}, \quad [M_{+-}, M_{\pm 2}] = \pm i \eta_{+-} M_{\pm 2}, \quad [M_{+-}, \tilde P_{\pm}] = \pm i \eta_{+-} \tilde P_{\pm}, \\
[M_{\pm 2}, \tilde P_2] = i\eta_{22} \tilde P_{\pm}, \quad [M_{\pm 2}, \tilde P_{\mp}] = - i \eta_{+-}  P_2, \label{ads3l}
\end{gather}
where we choose the metric to be $\eta_{+-} = 1, \eta_{22} =-1$,
is a subalgebra of \eqref{3dbms} and there are infinitely many embeddings of the form
\begin{align}\label{pt}
\tilde P_2 = -i \frac{T_0}{n}, \quad \tilde P_{\pm} = i \frac{T_{\pm n}}{\sqrt{2}n}, \quad M_{+-} = -i \frac{l_0}{n}, \quad M_{\pm 2} = \pm \frac{i}{\sqrt{2}n} l_{\pm n}.
\end{align}
The embeddings are indeed isomorphic to \eqref{ads3} with the rescaling $\Lambda \rightarrow n^2 \Lambda$. Alternatively, one could set $n=1$ in \eqref{pt} and rescale the ambient metric $\eta_{\mu \nu} \rightarrow \frac{\eta_{\mu \nu}}{n^2}$ instead of $\Lambda$.

\eqref{ads3} can also be obtained from the four dimensional Lorentz algebra
\begin{align}
[M_{\mu \nu} , M_{\rho \lambda}] = i (\eta_{\mu \lambda} M_{\nu \rho} -  \eta_{\nu \lambda} M_{\mu \rho} + \eta_{\nu \rho} M_{\mu \lambda} - \eta_{\mu \rho} M_{\nu \lambda})
\end{align}
if $M_{\pm 3}, M_{2 3}$ is identified with $K_{\pm }, K_2$. Here the metric is given by $\eta_{+-} = 1, \eta_{22} =-1, \eta_{33} = -1$. $K_i$ has to be rescaled according to
\begin{align}
\tilde P_i = R^{-1} K_i,
\end{align}
where $R^{-1} = \sqrt{|\Lambda|}$ is the inverse AdS radius. Thus, the generators $ \tilde P_i$ are actually boosts in a higher dimensional spacetime and acquire the interpretation as momenta only in the contraction limit $\Lambda \rightarrow 0$ when they commute among themselves and form an ideal. 

\subsubsection*{$(A)dS_3$ as real forms}

On the algebraic level one can also obtain the de Sitter case by simply setting $\Lambda >0$. Depending on the choice of $\Lambda$, \eqref{3dbms} is a different real algebra with one reality condition\footnote{We warn the reader that keeping it real might also go wrong, as exemplified in \cite{Chap20}.} 
\begin{align}\label{real1}
l^{\dagger}_m = - l_m, \quad T_m^{\dagger} = - T_m.
\end{align}
These two algebras can also be expressed as two different real forms of the complex Lie algebra $\mathfrak{W}\oplus \mathfrak{W}(\mathbb{C})$ \footnote{In the following, if no real form is specified, we often refer to the complex algebra without making it explicit in the notation.} defined by
\begin{align}
[L_m, L_n] = (m-n) L_{m+n}, \quad [\bar L_m, \bar L_n] = (m-n) \bar L_{m+n}, \quad [L_m, \bar L_n] = 0. \label{cwalg}
\end{align}
This is established by the isomorphism
\begin{align}\label{isoL}
L_m = \frac{1}{2}\left( l_m+ \frac{1}{\sqrt{-\Lambda}} T_m\right), \quad \bar L_m = \frac{1}{2}\left( l_m- \frac{1}{\sqrt{-\Lambda}} T_m\right) .
\end{align}
The embeddings now take the form
\begin{align} \label{embed1}
L_0, \bar L_{\pm n}, \bar L_0, \bar L_{\pm n}, \\
L_m \rightarrow \frac{L_m}{n}, \quad \bar L_m \rightarrow \frac{\bar L_m}{n}, \label{embed2}
\end{align} 
they are isomorphic to the algebra $\mathfrak{o}(4, \mathbb{C})$
\begin{align}\label{o4}
[L_0, L_{\pm 1}] = \mp L_{\pm 1}, \qquad [L_{+1}, L_{-1}] = 2 L_0, \\
[\bar L_0, \bar L_{\pm 1}] = \mp \bar L_{\pm 1}, \qquad [\bar L_{+1}, \bar L_{-1}] = 2 \bar L_0.
\end{align}
For negative $\Lambda$, i.e. the AdS case, we have from \eqref{isoL} and \eqref{real1} that 
\begin{align}\label{invoc1}
L_m^{\dagger} = - L_m, \quad \bar L_m^{\dagger} = -\bar L_m,
\end{align}
whereas for the de Sitter case one finds
\begin{align}\label{invoc2}
L_m^{\ddagger} = -\bar L_m, \quad \bar L_m^{\ddagger} = -L_m.
\end{align}
Restricted to \eqref{o4}, which is related to the standard Cartan-Weyl form
\begin{align}\label{o4cw}
[H, E_{\pm }] = \pm E_{\pm }, \qquad [E_{+}, E_{-}] = 2 H, \\
[\bar H, \bar E_{\pm }] = \pm \bar E_{\pm }, \qquad [\bar E_{+}, \bar E_{-}] = 2 \bar H, 
\end{align}
 via
\begin{align}
H= -L_0, \quad E_{\pm} = i L_{\pm 1}, \quad \bar H = - \bar L_0, \quad \bar E_{\pm} = i \bar L_{\pm 1},
\end{align}
the real forms correspond to two copies of $\mathfrak{sl}(2, \mathbb{R}) \simeq \mathfrak{o}(2,1)$ for AdS and  the Kleinian algebra $\mathfrak{o}(2,2) \simeq \mathfrak{o}(2,1) \oplus \bar {\mathfrak{o}}(2,1)$ for dS expressed by
\begin{align}
 H^{\dagger} & = - H, \quad E_{\pm}^{\dagger} = E_{\pm}, \quad \bar H^{\dagger} = - \bar H, \quad \bar E_{\pm}^{\dagger} = \bar E_{\pm}, \\
 H^{\ddagger} & = - \bar H, \quad E^{\ddagger}_{\pm} = \bar E_{\pm}.
\end{align}
This can be identified with the real structures listed in \cite{Borowiec:2017apk} with the automorphism $E_{\pm} \rightarrow - E_{\pm}, \bar E_{\pm} \rightarrow - \bar E_{\pm}$.  Note that this automorphism of the $\mathfrak{sl}(2)$ 
\begin{align}\label{auto}
\Phi (E_{\pm}) = - E_{\pm}, \quad \Phi(H) = H, \rightarrow \Phi (L_{\pm 1}) = - L_{\pm 1}, \quad \Phi(L_0) = L_0,
\end{align}
can be extended uniquely to an automorphism of the Witt algebra. On the algebraic level we thus can describe both dS and AdS spacetime symmetries simultaneously by switching between the two real forms.

\subsection{Algebra of $AdS_3$ Surface Charges} 

Following \cite{Oblak:2016eij}, one can also introduce light-cone coordinates $x^{\pm} = \frac{t}{\sqrt{- \Lambda}} \pm \phi$ leading to the metric (in Fefferman-Graham gauge)
\begin{align}
ds^2 = -\frac{1}{r^2 \Lambda} dr^2 - \left(r dx^+ - \frac{4G}{\sqrt{-\Lambda} r} \bar p(x^-) dx^- \right) \left(r dx^- - \frac{4G}{\sqrt{-\Lambda} r} p(x^+) dx^+ \right) 
\end{align}
and instead of $f$ and $R$ the Killing vectors are then parametrized by $X(x^+)$ and $\bar X(x^-)$. In analogy to \eqref{4dcharge} one has
\begin{align}
Q[X, \bar X] = \frac{1}{2 \pi} \int_0^{2 \pi} d \phi \left( X (x^+) p(x^+) + \bar X(x^-) \bar p(x^-)\right).
\end{align}
The algebra of these charges is a central extension of the algebra of the Killing vectors, i.e. $\mathfrak{W} \oplus \mathfrak{W}$. As noted at the end of section \ref{sec3.1.2}, the Witt algebras only have one possible central extension and so the general form of the central charge algebra is
\begin{align}\label{surf1}
[L_m, L_n] = &(m-n) L_{m+n} + \frac{c_L}{12} (m^3 -m) \delta_{m+n, 0},\\
 [\bar L_m,\bar L_n] = & (m-n) \bar L_{m+n} + \frac{c_{\bar L}}{12} (m^3 -m) \delta_{m+n, 0}, \\
[L_m, \bar L_n] = & 0.\label{surf2}
\end{align}
Brown and Henneaux calculated explicitly that \cite{Brown:1986nw}
\begin{align}\label{bhcc}
c_L = c_{\bar L} = \frac{3}{2 G \sqrt{-\Lambda}}.
\end{align}

\subsection{Asymptotic Flatness}\label{sec3.2.3}

One possibility of obtaining the symmetry algebra of asymptotically flat spacetime in three dimensions would be to consider a general metric, impose fall-off conditions and analyze the algebra of the Killing vectors just as it was done for the four dimensional case. However, since we already know the symmetry algebra of asymptotically (A)dS spacetime and it has the form of a Lie algebra, we can simply calculate the contraction limit\footnote{On the algebraic level such a procedure is also called Inonu-Wigner contraction  \cite{Inonu53}.} $\Lambda \rightarrow 0$ \cite{Compere:2018aar}. Starting from the algebra of the surface charges \eqref{surf1}-\eqref{surf2} and using \eqref{isoL} with the automorphism $\bar L_m \rightarrow - \bar L_{-m}$, i.e.
\begin{align}
T_m = \sqrt{-\Lambda} (L_m + \bar L_{-m}), \quad l_m = L_m - \bar L_{-m},
\end{align}
we obtain
\begin{align}
[T_m, T_n] =&  -\Lambda ( [L_m, L_n] + [\bar L_{-m}, \bar L_{-n}]) = -\Lambda (m-n) l_{m+n} \nonumber\\
& + \delta_{m+n, 0} (m^3-m) \frac{c_L}{12} + \delta_{-m-n, 0} ((-m)^3+m) \frac{\bar c_L}{12}, \\
[l_m, l_n] = & (m-n) l_{m+n} + \delta_{m+n, 0} (m^3-m) \frac{c_L}{12} + \delta_{-m-n, 0} ((-m)^3+m) \frac{\bar c_L}{12}, \\
[l_m , T_n] = & (m-n) T_{m+n} + \sqrt{-\Lambda}\left( \delta_{m+n, 0} (m^3-m) \frac{c_L}{12} - \delta_{-m-n, 0} ((-m)^3+m) \frac{\bar c_L}{12}\right) .
\end{align}
Now the contraction limit and \eqref{bhcc} yields the algebra of the surface charges in three dimensions
\begin{align}\label{bms3c1}
[T_m, T_n] & = 0, \qquad [l_m, l_n] = (m-n) l_{m+n}, \\
[l_m, T_n] & = (m-n) T_{m+n} + \delta_{m+n, 0} (m^3-m) c'_L, \label{bms3c2}
\end{align}
with $c'_L = \frac{1}{4 G}$, which we denote $\mathfrak{B}_{3, c}$ in the following. Without the central extension terms, \eqref{bms3c1}-\eqref{bms3c2} defines the three dimensional BMS algebra $\mathfrak{B}_3$.

The three dimensional Poincar\'e algebra ($\mathfrak{P}_3$)
\begin{align}\label{pc1}
[ M_{+ 2}, M_{-2} ] & = i\eta_{1 1} M_{+-}\,, \quad [M_{+-}, M_{\pm 2}] = \mp i \eta_{+ -}M_{\pm 1} \,, \\
[M_{+ -}, P_{\pm} ] & = \mp i \eta_{+ -} P_{\pm}\,, \quad [M_{\pm 2}, P_2] = -i \eta_{2 2} P_{\pm}\,, \\
[M_{\pm 2}, P_{\mp} ] & = i \eta_{+ -} P_2\,, \label{pc2}
\end{align}
is infinitely often embedded in $\mathfrak{B}_3$ with
\begin{align}
M_{+-}  = -i l_0, \quad M_{\pm 2} = \pm \frac{i}{\sqrt{2}} l_{\pm n}, \quad P_2 = -i T_0, \quad P_{\pm} = \frac{i}{\sqrt{2}} T_{\pm n},
\end{align}
if the metric is rescaled as $\eta_{+-} = \eta_{-+} = - \eta_{22} = n$, otherwise the Lorentz generators have to be rescaled by $1/n$.

Note that while the symmetry algebras without central extension in three dimensions are subalgebras of their four dimensional counterparts, this is not true for the centrally extended algebras, i.e. $\mathfrak{B}_{3,c} \not\subseteq \mathfrak{B}_{4,c}$. Thus, arguments that results from the three dimensional theory can be transferred to four dimensions as one is a subset of the other, have to be particularly wary of the influence of contributions from the central extension.

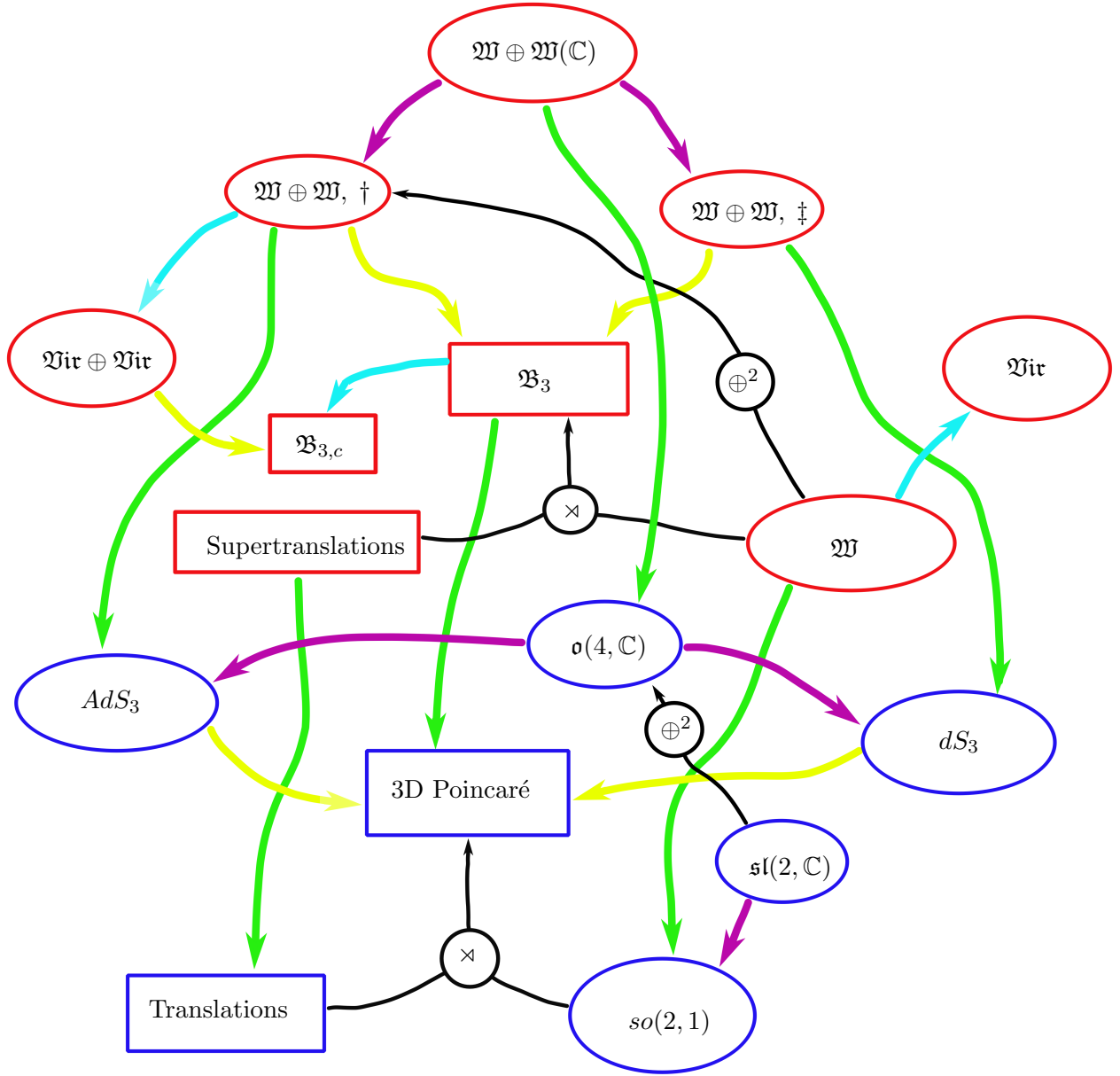
\begin{figure}
\centering
\begin{tikzpicture}
    \draw (0, 0) node[inner sep=0] {\includegraphics[width=18cm]{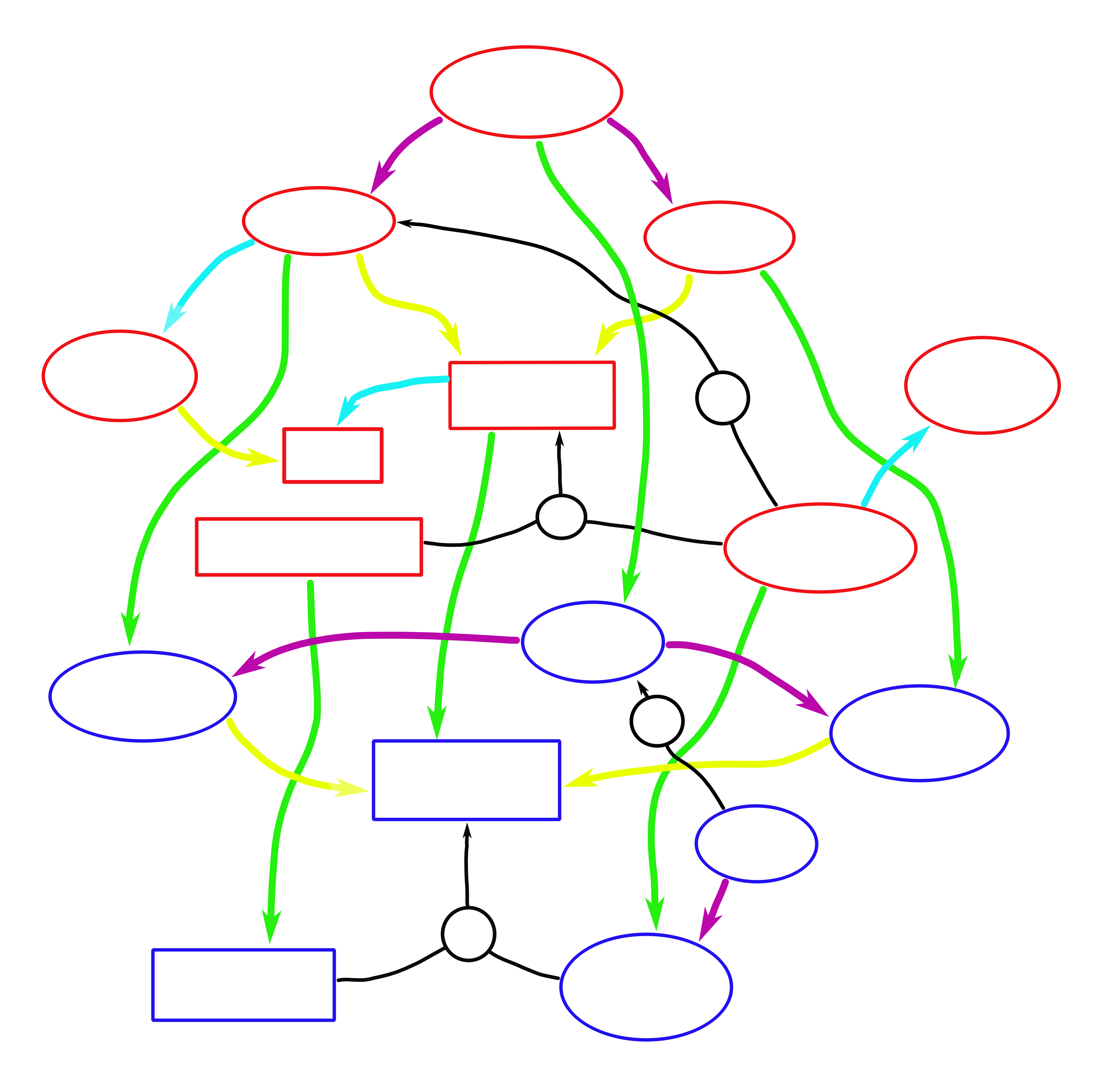}};
    \draw (-0.5, 7.3) node {$\mathfrak{W}\oplus \mathfrak{W}(\mathbb{C})$};
	\draw (-3.8, 5.2) node {$\mathfrak{W} \oplus \mathfrak{W},\; \dagger$};
	\draw (2.7, 4.9) node {$\mathfrak{W} \oplus \mathfrak{W},\; \ddagger$};
	\draw (-0.5, 2.4) node {$\mathfrak{B}_3$};
	\draw (-3.7, 1.4) node {$\mathfrak{B}_{3, c}$};
	\draw (0.05, 0.45) node {$\rtimes$};
	\draw (2.6, 2.4) node {$\oplus^2$};
	\draw (1.6, -2.8) node {$\oplus^2$};
	\draw (-7.0, 2.7) node {$\mathfrak{Vir}\oplus \mathfrak{Vir}$};
	\draw (6.8, 2.6) node {$\mathfrak{Vir}$};
	\draw (-3.9, -0.1) node {Supertranslations};
	\draw (4.1, -0.1) node {$\mathfrak{W}$};
	\draw (0.6, -1.6) node {$\mathfrak{o}(4, \mathbb{C})$};
	\draw (3.3, -4.9) node {$\mathfrak{sl}(2, \mathbb{C})$};
	\draw (-6.8, -2.4) node {$AdS_3$};
	\draw (5.8, -3.0) node {$dS_3$};
	\draw (-1.6, -3.7) node {3D Poincar\'e};
	\draw (-1.5, -6.2) node {$\rtimes$};
	\draw (-5.2, -7.0) node {Translations};
	\draw (1.5, -7.2) node {$so(2,1)$};
\end{tikzpicture}
\caption{Schematic overview of algebras relevant to asymptotic symmetries in three dimensions. The following properties/relations are encoded in the boundaries of the vertices: round shape $\hat =$ semi-simple, red $\hat =$ infinite dimensional, blue $\hat =$ finite dimensional, and the arrows: violet $\hat =$ real forms, green $\hat =$ finite dimensional embeddings, yellow $\hat =$ contraction limit, light-blue $\hat =$ central extension.} \label{alg-overview}
\end{figure}

An overview over all the relevant algebras related to the three dimensional symmetry algebras is given in figure \ref{alg-overview}.

\chapter{Quantum Groups on Asymptotic Symmetry Algebras}\label{chap4}

In this chapter, the consistent quantum groups on the symmetry algebras of three dimensional asymptotically (anti) de Sitter spacetimes are analyzed. To this end, we first classify Lie bialgebra structures and for several examples we will explicitly construct the associated Hopf algebras and investigate their properties and also restrictions they might impose on the symmetries themselves based on both physical and mathematical constraints. Subsequently the asymptotically flat limit and the four dimensional case are considered and compared.

\section{Deformations of Asymptotically (Anti) de Sitter Symmetries in 3D Gravity}

\subsection{Lie Bialgebra Structures on $\mathfrak{W} \oplus \mathfrak{W}$}\label{sec4.1.1}

Recall from section \ref{sec2.1.2} that all Lie bialgebra structures on an algebra $\mathfrak{g}$ are either coboundary or have a cobracket from the cohomology group $H^1( \mathfrak{g}, \bigwedge^2 \mathfrak{g})$. In Appendix \ref{app-a}, the following theorems are proven\footnote{In \cite{Junbo:2010} the authors independently prove a slightly more general result than our first theorem at the cost of a longer proof. The proofs presented here conceptually follow the proof of $H^1( \mathfrak{W}, \mathfrak{W}) = \{0\}$ in \cite{Ecker:2019thw}.}

\begin{theorem}\label{Wth}
The first cohomology group $H^1\left( \mathfrak{W}, \bigwedge^2 \mathfrak{W} \right)$ vanishes.
\end{theorem}

\begin{theorem}\label{W+Wth}
The first cohomology group $H^1\left( \mathfrak{W}\oplus \mathfrak{W}, \bigwedge^2 \mathfrak{W} \oplus \mathfrak{W} \right)$ vanishes.
\end{theorem}

Therefore all LBA can be obtained from a classical r-matrix satisfying the mCYBE \eqref{mcybe}. This is also true for the LBA on the symmetry algebra of globally (A)dS space, the real forms of $\mathfrak{o}(4, \mathbb{C})$, as a consequence of the Whitehead lemma. What is different, though, is that in the infinite dimensional algebras $\mathfrak{W}$ and $\mathfrak{W} \oplus \mathfrak{W}$ there is no ad-invariant element $\Omega \in \left( \bigwedge^3 \mathfrak{W} \right)_{\mathfrak{W}}$, resp. $\left( \bigwedge^3 (\mathfrak{W} \oplus \mathfrak{W}) \right)_{\mathfrak{W}\oplus \mathfrak{W}}$ anymore. This follows easily from
\begin{align}
0 \overset{!}{=}& L_m \triangleright \sum_i \alpha_i L_{p_i} \wedge L_{q_i} \wedge L_{s_i} \nonumber \\
=& \sum_i \alpha_i ((m-p_i)  L_{p_i +m} \wedge L_{q_i} \wedge L_{s_i} + (m-q_i)  L_{p_i} \wedge L_{q_i+m} \wedge L_{s_i} \nonumber \\
&+ (m-s_i) L_{p_i} \wedge L_{q_i} \wedge L_{s_i+m})  
\end{align}
if we choose $m>0$ and focus on the term with $s_{i'} > s_i, p_i, q_i$ without loss of generality for the Witt algebra. Similarly, one shows that 
\begin{align}
\left( \bigwedge^3 (\mathfrak{W} \oplus \mathfrak{W}) \right)_{\mathfrak{W}\oplus \mathfrak{W}} = \{0\}, \qquad  \left( \bigwedge^2 \mathfrak{W} \right)_{\mathfrak{W}} = \left( \bigwedge^2 (\mathfrak{W} \oplus \mathfrak{W}) \right)_{\mathfrak{W}\oplus \mathfrak{W}}= \{0\}.
\end{align} 
The latter implies that the map from an r-matrix to the cobracket derived from it is injective. Otherwise, i.e. if there was an ad-invariant element $a \wedge b \in \bigwedge^2 \mathfrak{W}$, the cobrackets from $r$ and $r+a \wedge b$ would be identical.
Furthermore, as noted in section \ref{sec2.1.2}, only r-matrices belonging to different orbits under automorphisms of the algebra lead to inequivalent LBA.

Thus, one can conclude that all LBA on $ \mathfrak{W}\oplus \mathfrak{W}$ are uniquely given by triangular r-matrices that are classified up to automorphism. The simple observation that, in contrast to the situation for the embedded $\mathfrak{o}(4, \mathbb{C})$, only triangular LBA are coassociative will exclude a wide range of deformations.

\subsection{Classification of Triangular r-matrices}\label{sec4.1.2}

Since the Witt algebra and its central extension, the Virasoro algebra, play a vital role in various parts of math and physics, including string theory \cite{Witten:1987ty}, there were attempts to classify all LBA structures on them, most notably by \cite{Ng:2000}. These results remained incomplete though except for some subalgebras, including the one-sided Witt algebra $\mathfrak{W_+}$ (generated by $\{L_m, m \geq -1\}$), where all r-matrices were shown to be r-matrices of one of the embeddings. For the two copies of the Witt algebra that are studied here it is also easy to see that e.g. any element of the form 
\begin{align}\label{genr}
\left(\sum_i \alpha_i L_i \right) \wedge \left(\sum_j \beta_j \bar L_j \right) 
\end{align}
is a triangular r-matrix. 

However, one has to take into account that the asymptotic symmetry is spontaneously broken in the bulk in the sense that the vacua related by supertranslations and superrotations are physically distinguishable (cf. section \ref{sec3.1.1} and \cite{Strominger:2017zoo}). 
There is a correspondence between these vacua and the embeddings of Poincar\'e or (A)dS subalgebras which leave the associated vacuum invariant. Therefore, we require that the restriction of the Hopf algebra deformed by the twist associated with a given r-matrix to the (A)dS algebra (which is the $n=1$ embedding for an observer) is a sub-Hopf algebra and we are interested in r-matrices from (real forms of) $\mathfrak{o}(4, \mathbb{C})$, where $\{H, E_{\pm}, \bar H, \bar E_{\pm}\}$ is replaced with the embedding\footnote{In the following, we often leave the embedding characterized by the index $n$ unspecified. Also note that from a mathematical point of view there is no reason the generators that appear in the r-matrix have to be part of a specific embedding and we will comment on this possibility later on.}.
Note that while in the case of positive $\Lambda$ the involution mixes left and right-handed elements, this is not happening for negative $\Lambda$, leaving the potential possibility to use different embeddings for them.


The classification of r-matrices of (real forms of) $\mathfrak{o}(4, \mathbb{C})$ was done in \cite{Borowiec:2015nlw} and has been discussed as well in \cite{Kowalski-Glikman:2019ttm} and \cite{Borowiec:2017apk}. Filtering out the quasi-triangular r-matrices we are left with
\begin{align}\label{c1}
r_{I} & = \chi (E_+ - \bar E_+ ) \wedge (H + \bar H), \\
r_{II} & = \chi E_+ \wedge H + \bar \chi \bar E_+ \wedge \bar H + \zeta E_+ \wedge \bar E_+, \label{lcfc} \\
r_{III} & = \eta H \wedge \bar H. \\
r_{V} & = \bar \chi \bar E_+ \wedge \bar H + \rho H \wedge \bar E_+. \label{c2}
\end{align}
An important caveat remains; the classification \eqref{c1}-\eqref{c2} is up to automorphisms of the $\mathfrak{o}(4, \mathbb{C})$ but not all of these automorphisms are also automorphisms of  $ \mathfrak{W}\oplus \mathfrak{W}$, only the inverse statement holds in general\footnote{Here we understand the automorphisms of $\mathfrak{W}\oplus \mathfrak{W}$ to be restricted to the subalgebra $\mathfrak{o}(4)$ and write $\text{Aut}'( \mathfrak{W}\oplus \mathfrak{W})$ for this (and similar) restriction(s) so that quotients are well defined.}.  
Therefore, in Appendix \ref{app-b} the classification of triangular r-matrices on $\mathfrak{o}(4)$ is revised  along the lines of \cite{Borowiec:2015nlw} but using only the $\mathfrak{W} \oplus \mathfrak{W}$ 
automorphisms 
\begin{gather}\label{au1}
\varphi_{(\gamma,\bar\gamma,\epsilon,\bar\epsilon)}(L_m) = \gamma^m \epsilon L_{\epsilon m}, \quad \varphi_{(\gamma,\bar\gamma,\epsilon,\bar\epsilon)}(\bar L_m) = \bar \gamma^m \bar \epsilon  \bar L_{\bar \epsilon m}, \\
\varphi'(L_m) = \bar L_m , \quad \varphi'(\bar L_m) = L_m, \label{au2}
\end{gather}
where $0\neq \gamma,\bar\gamma\in \mathbb{C}, \epsilon,\bar\epsilon=\pm 1$.
As a result we obtain the following classes of r-matrices
\begin{align}\label{classa}
r_{1'} & = \beta (L_1 + L_{-1} + 2 L_0) \wedge ( \bar L_1 + \bar L_{-1} + 2 \bar L_0) + a_1 + \bar a_1, \\
r_{2'} & = \beta L_{1} \wedge \bar L_0 + a_2 , \\ 
r_{3'} & = \beta (L_1 + \epsilon L_0 + \epsilon' L_{-1} ) \wedge (\bar L_1  + \bar L_{-1} + 2 \bar L_0) + \bar a_1 + (1- \epsilon) (1- \epsilon') a_2, \\
r_{4'} & = \beta_1  L_1 \wedge \bar L_1 + \beta_2  (L_1+ \bar L_1) \wedge (L_0 +  \bar L_0 ) , \\
r_{5'} & = \beta (L_1 + L_{-1}) \wedge ( \bar L_1 + \bar L_0) + a_1, \\
r_{6'} & = (\beta L_1 + \beta_0 L_0  + \epsilon \beta L_{-1}) \wedge ( \bar \beta \bar L_1 + \bar \beta_0 \bar L_0 + \bar \epsilon \bar \beta \bar L_{-1}), \\
r_{7'} & = L_1 \wedge ( \bar \beta \bar L_1 + \bar \beta \bar L_0 + \epsilon \bar \beta \bar L_{-1}) + a_2, \\
r_{8'} & = \beta L_1 \wedge \bar L_{1} + a_2 + \bar a_2, \label{classb}
\end{align}
where $\epsilon, \epsilon', \bar \epsilon \in \{0,1\}$ and
\begin{align}
a_1 & = \alpha ( L_1 \wedge L_0 +  L_1 \wedge L_{-1} -L_{-1} \wedge L_0), \quad 
a_2  = \alpha L_1 \wedge L_0, \\
\bar a_1 & = \bar \alpha (\bar L_1 \wedge \bar L_0 + \bar L_1 \wedge \bar L_{-1} - \bar L_{-1} \wedge \bar L_0), \quad 
\bar a_2  =\bar \alpha \bar L_1 \wedge \bar L_0.
\end{align}

In the complex $\mathfrak{W} \oplus \mathfrak{W}$ all the parameters in \ref{classa}-\ref{classb} can take values in $\mathbb{C}$ but for the real forms associated with the involutions $\dagger, \ddagger$ \eqref{invoc1}, \eqref{invoc2} constrains the choice of parameters. For $\dagger$ in all classes except $r_{6'}$ this enforces $\beta, \beta_1, \beta_2, \alpha, \bar \alpha \in i \mathbb{R}$ and in $r_{6'}$ one can restrict $\beta, \beta_0,\bar \beta, \bar \beta_0 \in \sqrt{i} \mathbb{R}$ without loss of generality.
In the case of positive $\Lambda$, i.e. the involution $\ddagger$, the reality condition is more restrictive. In particular,
\begin{align*}
r_{1'}:& \quad \alpha = \bar \alpha \in i \mathbb{R}, \beta \in \mathbb{R}, \qquad r_{2'}: \quad \beta = 0, \alpha \in i \mathbb{R}, \\ 
r_{3'}:& \quad \bar \alpha =0, \epsilon = \epsilon' = 1, \beta \in \mathbb{R}, \qquad r_{4'}: \quad \beta_1 \in \mathbb{R}, \beta_2 \in i \mathbb{R}, \\
r_{5'}:& \quad \text{excluded}, \qquad r_{6'}: \quad \epsilon = \bar \epsilon, \beta_0 \bar \beta = - \frac{\beta^*}{\beta_0^*} \text{ or } (\beta_0 = \bar \beta_0 =0, \beta, \bar \beta \in \mathbb{R}), \\
r_{7'}:& \quad \text{excluded}, \qquad r_{8'}: \quad \beta \in \mathbb{R}, \alpha = \bar \alpha \in i \mathbb{R}.
\end{align*}

As the r-matrices from \eqref{classa}-\eqref{classb} that are not included in \eqref{c1}-\eqref{c2} are at least in  $\mathfrak{o}(4)$ automorphism orbits containing them, one can always use the inverse of the automorphisms to obtain the full twists. For example, $r= (L_1 - L_{-1}) \wedge (\bar L_1  - \bar L_{-1})$ (being automorphic to $r_{6'}$ with $\beta_0 = \bar \beta_0 =0, \epsilon = \bar \epsilon =1$ ) is automorphic to $L_0 \wedge \bar L_0 \hat{=} r_{III}$ by 
\begin{align}
\varphi(L_1) = -\frac{1}{2}(L_1 + L_{-1}) + L_0, \quad \varphi(L_{-1}) =  -\frac{1}{2}(L_{1} + L_{-1}) -L_0, \quad \varphi(L_0) =  \frac{1}{2} (L_1 - L_{-1}). 
\end{align}
From the abelian twist for $r_{III}$ 
\begin{align}
\mathcal{F}_{III} = \exp( \eta L_0 \wedge \bar L_0)
\end{align}
one then gets the twist
\begin{align}
\mathcal{F} = \left(\varphi^{-1} \otimes \varphi^{-1}\right) \mathcal{F}_{III} = \exp (\eta (L_1 - L_{-1}) \wedge (\bar L_1 - \bar L_{-1})).
\end{align}

\subsection{Full Deformations via Twist}\label{sec4.1.3}

From section \ref{sec2.1.9}, it is known that all triangular LBA can be extended to a deformation in all orders by a twist satisfying the 2-cocycle condition \eqref{coc-cond}. This does not necessarily entail an explicit method of constructing the twists but for (the real forms of) $\mathfrak{o}(4, \mathbb{C})$ all of them are given in \cite{Borowiec:2017apk}. By the final remark in the previous section it follows that the twists corresponding to the r-matrices \eqref{classa}-\eqref{classb} can be obtained.

\subsubsection*{Abelian Twist}

The first example we construct in detail is the abelian twist (see \cite{Resh90}) corresponding to the r-matrix $r_{III}$ from \eqref{c2}. Its name is motivated by the fact that the two elements used for its construction are commuting. While there are several other r-matrices with this feature (like \eqref{genr}), $r_{III}$ is special in the sense that it is the only r-matrix from elements that are contained in all embeddings. Thus, the abelian twist deformation does not single out any particular embedding. 

The full abelian twist can be expressed as
\begin{align}\label{abtwist1}
\mathcal{F}_A = \exp \left(\eta \bar L_0 \wedge L_0 \right) \exp \left(\eta(L_0 \otimes L_0 - \bar L_0 \otimes \bar L_0 )\right),
\end{align}
which factorizes into the twist $\mathcal{F}_{3''}$ from \cite{Borowiec:2017apk} and a factor that only contributes symmetric deformations of the coproduct. This particular construction is chosen to make contact with the deformations in the flat limit $(\Lambda \rightarrow 0)$. Explicitly, by making use of the Hadamard formula 
\begin{align}
e^A B e^{-A} = \sum_{n=0}^{\infty} \frac{1}{n!} \underset{\text{n times}}{[\underbrace{A, [A, ...[A, }} B]...]
\end{align}
for general elements of an algebra $A$ and $B$ one finds the deformed coproducts
\begin{align}
\Delta_{\mathcal{F}_{3''}}(L_m) = e^{-m \eta \bar L_0} \otimes L_m + L_m \otimes e^{m \eta \bar L_0},  \\
\Delta_{\mathcal{F}_{3''}}(\bar L_m) = e^{m \eta   L_0} \otimes \bar L_m + \bar L_m \otimes e^{-m \eta L_0} 
\end{align}
and 
\begin{gather}\label{copab1}
\Delta_{\mathcal{F}_A}(L_m) = e^{-m \eta (\bar L_0 + L_0)} \otimes L_m + L_m \otimes e^{m \eta (\bar L_0 - L_0)} \\
\Delta_{\mathcal{F}_A}(\bar L_m) = e^{m \eta (\bar L_0 + L_0)} \otimes \bar L_m + \bar L_m \otimes e^{-m \eta ( L_0 -  \bar L_0)} \\
\Delta_{\mathcal{F}_A}(L_0) = L_0 \otimes 1 + 1 \otimes L_0, \quad \Delta_{\mathcal{F}_A}(\bar L_0) = \bar L_0 \otimes 1 + 1 \otimes \bar L_0. \label{copab2}
\end{gather}
The antipodes can also be inferred easily from 
\begin{align}\label{antip}
m \circ (S \otimes \text{id}) \circ \Delta = 1 \circ \epsilon 
\end{align}
and turn out to be
\begin{align}
S_{\mathcal{F}_{3''}}(L_m) =& -L_m, & S_{\mathcal{F}_{3''}}(\bar L_m) & = - \bar L_m \\ 
S_{\mathcal{F}_A}(L_m) =& -L_m  e^{-2m \eta   L_0}, &  S_{\mathcal{F}_A}(\bar L_m) & = -\bar L_m  e^{2m \eta  \bar L_0} .
\end{align}
The abelian twist thus induces non-cocommutative coproducts, and consequently the antipodes are non-trivial, except on $L_0$ and $\bar L_0$. Before discussing further properties let us consider another example of a twist. 

\subsubsection*{Jordanian Twist}

One of the first known examples of twists besides of the abelian twist was the Jordanian twist. In a two dimensional Lie algebra with elements $X, Y$ and (the only non-abelian) bracket $[X,Y] = \alpha Y$ one has 
\begin{align}
\mathcal{F} = \exp \left( X \otimes \log \left(1 + \alpha Y \right) \right).
\end{align}
Because the Lorentz sector ( containing (super)rotations and boosts) of all the relevant asymptotic symmetry algebras is semi-simple, there are always Cartan elements, e.g. $L_0, \bar L_0$ for $\mathfrak{W} \oplus \mathfrak{W}$, that diagonalize the adjoint action and thus have the structure of the two dimensional example. As a result, many of the consistent twists are of this prototypical Jordanian form or have it as a building block. 

For the r-matrix $r_I$ one has
\begin{align}\label{jtwi}
\mathcal{F}_{J, n} = \exp \left( - \frac{1}{n} (L_0 + \bar L_0) \otimes \log \left(1 + \tilde a (L_n - \bar L_n) \right) \right) ,
\end{align}
so that 
\begin{align}
\Delta_{\mathcal{F}_{J, n}}(L_0) & = L_0 \otimes 1 +1 \otimes L_0  - \frac{1}{n}(L_0 + \bar L_0) \otimes \frac{d \sigma_n}{d(L_n - \bar L_n)} n L_n \nonumber \\
& = L_0 \otimes 1 + 1 \otimes L_0 - \tilde a (L_0 + \bar L_0) \otimes L_n \Pi_{+n}^{-1},  \\
\Delta_{\mathcal{F}_{J, n}}(\bar L_0) & = \bar L_0 \otimes 1 + 1 \otimes \bar L_0 + \tilde a (L_0 + \bar L_0) \otimes \bar L_n \Pi_{+n}^{-1}, 
\end{align}
where 
\begin{align}
\sigma_n \equiv \log \left(1 + \tilde a (L_n - \bar L_n)\right), \quad \Pi_{+n} = e^{\sigma_n}. \label{defpi}
\end{align}
Using \eqref{antip} and the previous equations we find
\begin{align}
S_{\mathcal{F}_{J, n}}(L_0 ) =& - (L_0 + \tilde a (\bar L_0 L_n + L_0 L_n))\frac{\Pi_{+n}^{-1}}{1-  \Pi_{+n}^{-1} (L_n - \bar L_n)} \nonumber \\
=& - (L_0 + \tilde a (\bar L_0 L_n + L_0  L_n)),  \label{SL0} \\
S_{\mathcal{F}_{J, n}}(\bar L_0) =&  - ( \bar L_0 -  \tilde a (\bar L_0 \bar L_n + L_0 \bar L_n)).
\end{align}

For general generators we find 
\begin{align}
\Delta_{\mathcal{F}_{J, n}}(L_m) & = \mathcal{F}_{J, n} (L_m \otimes 1 + 1 \otimes L_m) \mathcal{F}^{-1}_{J, n} \nonumber \\
& = L_m \otimes \Pi_{+ n}^{\frac{m}{n}} +  \sum_{l=0}^{\infty} \frac{1}{l !} \left(\frac{L_0 + \bar L_0}{n } \right)^{l}  \otimes \big[ \sigma_n, \big[..., \big[\sigma_n, L_m\big]...\big], \label{delgen}
\end{align}
with
\begin{align}\label{comm}
[\sigma_n, L_m] =& \bigg[ - \sum_{j = 0}^{\infty} \frac{(- \tilde a)^j}{j} (L_n - \bar L_n)^j, L_m \bigg] \\
\big[\left(L_n- \bar L_n \right)^{j}, L_m \big] =& \sum_{s=0}^{j-1} (L_n - \bar L_n)^s (n-m) L_{m+n} (L_n - \bar L_n)^{j-s} \nonumber \\
 =&\sum_{k=1}^{j} \bigg( \sum_{s_1 = k-1}^{j-1} ... \sum_{s_k = 0}^{s_{k-1}-1} L_{m+ kn} (L_n - \bar L_n)^{j-k} \prod_{p=0}^{k-1} (n-(m+pn)) \bigg),
\end{align}
where in the last line we iteratively commuted the $s$ terms in front of $L_{m+kn}$ to the right. 
From this formula, it can be seen that in general there is an infinite number of terms in the coproduct involving a tower of infinitely many different generators. In the h-adic topology this is well defined but in the light of the comments in section \ref{sec2.1.8} one could ask in which cases finite expressions are obtained. We find that this can be achieved when restricting to (two copies of) the one-sided Witt algebra ($\mathfrak{W}_-$) spanned by $\{ L_m, \bar L_m; m \leq 1 \}$. In that case there are only two possible embeddings with $n= \pm 1$ and by choosing $n=1$ the sum over $k$ in \eqref{comm} terminates after \textit{min}$\{1-m, j\}$ terms, instead of when $k=j$ which is not finite as we sum $j$ to infinity. As a consequence also the sum over $l$ terminates after $(1-m)$ terms, only a finite number of generators appear and as we will see also only a finite number of terms.

In that case we proceed from \eqref{delgen} with the identity
\begin{align}\label{ide}
 \sum_{s_1 = k-1}^{j-1} ... \sum_{s_k = 0}^{s_{k-1}-1}= \frac{j!}{(j-k)! k!}  = {j \choose k} .
\end{align}
It is straightforward to see that it holds for $k=1$. Now assume that it holds for $k=k', 1 \leq k' < j$ and it follows that the lhs for $k=k'+1$ is
\begin{align}
\sum_{s'_1= k'}^{j-1} \sum_{s'_2 = k'-1}^{s'_1-1} ... \sum_{s'_{k'} = 0}^{s'_{k'-1} -1} = \sum_{s'_1= k'}^{j-1} {s'_1 \choose k'} = {j \choose k'+1} ,
\end{align}
where in the last step the hockey stick identity was used. Thus, we proved \eqref{ide} by induction.
Therefore,
\begin{align}
[\sigma_1, L_m]  =&  - \sum_{j = 0}^{\infty} \frac{(- \tilde a)^j}{j} \big[\left(L_n- \bar L_n \right)^{j}, L_m \big] \nonumber \\
=& L_m - \sum_{j = 1}^{\infty} \sum_{k=1}^{\text{min} \{j, 1-m\}}   \frac{(- \tilde a)^j}{j} {j \choose k} L_{m+k} (L_1 - \bar L_1)^{j-k} \prod_{p=0}^{k-1} (1-(m+p)) \label{sigcom}
\end{align}
and comparing with 
\begin{align}
\frac{d^k \sigma_1}{d L_1^k} = -\sum_{j = k}^{\infty} \frac{(- \tilde a)^j}{j} \frac{j!}{(j-k)!} (L_1 - \bar L_1)^{j-k},
\end{align}
one finds that the summand in \eqref{sigcom} and
\begin{align}\label{sigcom2}
L_m + \sum_{k=1}^{1-m} L_{m+k} \frac{d^k \sigma_1}{d L_1^k} \frac{1}{k!} \prod_{p=0}^{k-1} (1-(m+p))
\end{align}
are identical. Splitting the sums in \eqref{sigcom} and \eqref{sigcom2} according to
\begin{align*}
\left(\sum_{j=1}^{1-m} \sum_{k=1}^j + \sum_{j=2-m}^{\infty}\sum_{k=1}^{1-m} \right) ... \\
\left( \sum_{k=1}^{1-m}\sum_{j=k+1}^{1-m} + \sum_{k=1}^{1-m}\sum_{j=2-m}^{\infty} \right) ...
\end{align*}
and using
\begin{align}
\sum_{j =1}^{1-m} \sum_{k = 1}^j ... = \sum_{k=1}^{1-m} \sum_{j=k}^{1-m} ...,
\end{align}
it follows that \eqref{comm} is indeed given by \eqref{sigcom2} if the one-sided Witt algebra with $n=1$ is considered.
Plugging the result into \eqref{delgen} yields
\begin{align}
\Delta_{\mathcal{F}_{J, 1}}(L_m) =& L_m \otimes \Pi_{+}^{m} + \sum_{l=1}^{1-m} \frac{1}{l!}\bigg(\sum_{k_1 = 0}^{1-m} \frac{1}{k!} \prod_{p_1=0}^{k_1-1} (1-(m+p_1)) \nonumber \\
&\times  \bigg( \sum_{k_2 = 0}^{1-(m+k_1)} \frac{1}{k_2!} \prod_{p_2=0}^{k_2-1} (1-(m+k_1 +p_2)) \nonumber \\
& \times \bigg(... \bigg( \sum_{k_l =0}^{1-(m+ k_1 + ... k_{l-1})} \frac{1}{k_l !} \prod_{p_l=0}^{k_l-1} (1-(m+k_1 +... + k_{l-1} +p_l)) A\otimes B \bigg)... \bigg), \label{delres}
\end{align}
where
\begin{align}
A \otimes B =& \left(L_0 + \bar L_0 \right)^l \otimes L_{m+k_1 + ...+k_l} \frac{d^{k_1} \sigma_1}{d L_1^{k_1}}...  \frac{d^{k_l}\sigma_1}{d L_1^{k_l}}, \\
 \frac{d^{k}\sigma_1}{d L_1^{k}} =& - (- \tilde a)^k \Pi_+^{-k} k!
\end{align}
and all sums are finite for $m \leq 1$. Similarly for $\bar L_m$ one finds
\begin{align}
\Delta_{\mathcal{F}_{J, 1}}(\bar L_m) =& \bar L_m \otimes \Pi_{+}^{m} + \sum_{l=1}^{1-m} \frac{1}{l!}\bigg(\sum_{k_1 = 0}^{1-m} \frac{1}{k!} \prod_{p_1=0}^{k_1-1} (1-(m+p_1)) \nonumber \\
&\times  \bigg( \sum_{k_2 = 0}^{1-(m+k_1)} \frac{1}{k_2!} \prod_{p_2=0}^{k_2-1} (1-(m+k_1 +p_2)) \nonumber \\
& \times \bigg(... \bigg( \sum_{k_l =0}^{1-(m+ k_1 + ... k_{l-1})} \frac{1}{k_l !} \prod_{p_l=0}^{k_l-1} (1-(m+k_1 +... + k_{l-1} +p_l)) \bar A\otimes \bar B \bigg)... \bigg),
\end{align}
where
\begin{align}
\bar A \otimes \bar B = \left(L_0 + \bar L_0  \right)^l \otimes \bar L_{m+k_1 + ...+k_l} \frac{d^{k_1} \sigma_1}{d \bar L_1^{k_1}}...  \frac{d^{k_l}\sigma_1}{d \bar L_1^{k_l}}.
\end{align}

Then the antipodes follow from \eqref{antip} and read
\begin{align}
S_{\mathcal{F}_{J,1}} (L_m) =& -  \sum_{l=1}^{1-m} \frac{1}{l!}\bigg(\sum_{k_1 = 0}^{1-m} \frac{1}{k!} \prod_{p_1=0}^{k_1-1} (1-(m+p_1)) \nonumber \\
&\times  \bigg( \sum_{k_2 = 0}^{1-(m+k_1)} \frac{1}{k_2!} \prod_{p_2=0}^{k_2-1} (1-(m+k_1 +p_2)) \nonumber \\
& \times \bigg(... \bigg( \sum_{k_l =0}^{1-(m+ k_1 + ... k_{l-1})} \frac{1}{k_l !} \prod_{p_l=0}^{k_l-1} (1-(m+k_1 +... + k_{l-1} +p_l)) S(A)\otimes  B \bigg)... \bigg) \Pi_+^{-m},
\end{align}
and $S(A)$ can be inferred from \eqref{SL0}
\begin{align}
S(A) = S ((L_0+\bar L_0)^l) = (-(L_0  + \bar L_0) \Pi_+)^l.
\end{align}

We summarize that for the full $\mathfrak{W} \oplus \mathfrak{W}$ as well as the one-sided $\mathfrak{W}_- \oplus \mathfrak{W}_-$ the coproducts deformed by the Jordanian twist are non-cocommutative for all the generators. In the latter case the algebra and coalgebra relations contain only a finite number of generators.

\subsection{Specialization and One-sided Witt Algebra}\label{sec4.1.4}

In section \ref{sec2.1.8} it was argued that physical applications of a Hopf algebra $H$, entailing a numeric value for the deformation parameter, require the so-called specialization of $H$.

Let us study the specialization (or q-analog) on the examples of the abelian and the Jordanian twist respectively. For the abelian twist the formulas \eqref{copab1}-\eqref{copab2} show that only a finite number of generators appear in the coproducts but there are infinite power series in $a \equiv \frac{i}{\kappa }$. Thus, the full $\mathfrak{W} \oplus \mathfrak{W}$ can be turned into the q-analog by adding the elements 
\begin{align}
e^{a L_0} \equiv K, \quad e^{-aL_0}  \equiv K^{-1}, \quad e^{a \bar L_0} \equiv \bar K, \quad e^{-a \bar L_0}  \equiv \bar K^{-1},
\end{align}
to the algebra. Furthermore, define $q = e^{a}$ and the extra commutation relations become
\begin{align}
e^{a L_0 } L_m e^{-a L_0 } =& \sum_{j=0}^{\infty} \frac{a^j}{j!} [L_0 , [...,[ L_0 , L_m]...] = e^{-am} L_m = q^{-m} L_m \\
\Rightarrow [K, L_m] = & q^{-m} L_m e^{aL_0 } - L_m e^{a L_0 } = (q^{-m} -1) L_m K
\end{align}
and similarly
\begin{align}
[K^{-1}, L_m] =& (q^m -1)L_m K^{-1}, \qquad [K, \bar L_m] = 0, \\
[K^{-1}, \bar L_m] = & 0, \qquad [\bar K, \bar L_m] = (q^{-m} -1) \bar L_m \bar K\\
[\bar K^{-1}, L_m] = & 0, \qquad [\bar K^{-1}, \bar L_m] = (q^m-1) \bar L_m \bar K^{-1}, \\
[K, K^{-1}] = & [K, \bar K] = [K, \bar K^{-1}] =0.  
\end{align}
It is also easy to compute
\begin{align}
\Delta_{\mathcal{F}_A}(K) = & K \otimes K, \qquad \Delta_{\mathcal{F}_A}(K^{-1}) = K^{-1} \otimes K^{-1}, \\
S_{\mathcal{F}_A}(K) = & -K , \qquad S_{\mathcal{F}_A}(K^{-1}) = - K^{-1}
\end{align}
and reexpressing \eqref{copab1}-\eqref{copab2} gives
\begin{align}
\Delta_{\mathcal{F}_A}(L_m) = & K^m \bar K^m \otimes L_m + L_m \otimes K^{-m} \bar K^m , \\
\Delta_{\mathcal{F}_A}(\bar L_m) = & K^{-m} \bar K^{-m} \otimes \bar L_m + \bar L_m \otimes K^{m} \bar K^{-m} .
\end{align}
Endowed with this algebra and coalgebra structures the set of polynomials in the generators $\{ L_m, \bar L_m, K, K^{-1}, \bar K, \bar K^{-1}\}$ does indeed form a q-analog of the twisted Hopf algebra and it can be defined for any $q \in \mathbb{C}$. In particular, the classical limit $\kappa \rightarrow \infty \leftrightarrow q \rightarrow 1$ gives simply the Lie algebra $\mathfrak{W} \oplus \mathfrak{W}$, but extended by the central elements $K, \bar K$.

In the case of the Jordanian twist the situation is different. We discovered in \eqref{delgen} that for $L_m$ with $m \in \mathbb{Z}$ the coproduct contains infinitely many different generators and thus it would be impossible to define a q-analog. However, by restricting to two copies of the one-sided Witt algebra $\mathfrak{W}_-$ containing $L_m$ with  $m \leq 1$ it was shown that all coproducts contain finitely many terms. Similarly, one could use the embedding corresponding to $n=-1$ and restrict to $\mathfrak{W}_+$ containing $L_m, m \geq -1$. In order to express all algebra and coalgebra relations involving only finite powers of $1/\kappa$ the elements $\Pi_+$, defined in \eqref{defpi}, and its inverse $\Pi_+^{-1}$ are used. The additional commutation relations then read
\begin{align}
[\Pi_+, L_m] =& \tilde a (1-m) L_{m+1}, \\
[\Pi_+^{-1}, L_m] =& \sum_{j = 0}^{\infty} \frac{(-1)^{\underline{j}}}{j!}  \tilde a^j [(L_1 - \bar L_1)^j, L_m] \nonumber \\
=& \sum_{j=0}^{\infty} \sum_{k=1}^{\text{min}\{1-m, j\}} \frac{(-1)^{\underline{j}}}{j!} {j \choose k} L_{m+k} \tilde a^j (L_1 - \bar L_1)^{j-k} \left( \prod_{r = 0}^{k-1} (1-m-r) \right) \nonumber \\
=& \sum_{k=1}^{1-m} L_{m+k} \frac{d^k e^{-\sigma_1}}{d L_1^k} \left( \prod_{r = 0}^{k-1} (1-m-r) \right) \nonumber \\
= & \sum_{k=1}^{1-m} \frac{(-1)^{\underline{k}}}{k!} \tilde a^k L_{m+k} \Pi_+^{k-1} \left( \prod_{r = 0}^{k-1} (1-m-r) \right) 
\end{align}
and similarly
\begin{align}
[\Pi_+, \bar L_m] =& -\tilde a (1-m) \bar L_{m+1}, \\
[\Pi_+^{-1}, \bar L_m] = &  \sum_{k=1}^{1-m} \frac{(-1)^{\underline{k}}}{k!} (-\tilde a)^k \bar L_{m+k} \Pi_+^{k-1} \left( \prod_{r = 0}^{k-1} (1-m-r) \right) .
\end{align}
From \eqref{delres} one has in particular
\begin{align}
\Delta_{\mathcal{F}_J} (L_1) = L_1 \otimes \Pi_+ + \otimes L_1, \quad \Delta_{\mathcal{F}_J} (L_1) = \bar L_1 \otimes \Pi_+ + \otimes \bar L_1,
\end{align}
leading to
\begin{align}
\Delta_{\mathcal{F}_J} (\Pi_+) & = \Pi_+ \otimes \Pi_+, \qquad \Delta_{\mathcal{F}_J} (\Pi_+^{-1}) = \Pi_+^{-1} \otimes \Pi_+^{-1} , \\
S_{\mathcal{F}_J} (\Pi_+) & = -\Pi_+, \qquad S_{\mathcal{F}_J} (\Pi_+^{-1}) = - \Pi_+^{-1}. 
\end{align}
All these formulas are well defined for $\tilde a \in \mathbb{C}$ and for $\kappa \rightarrow \infty$ the elements $\Pi_+, \Pi_+^{-1}$ become central. Thus, similar to the abelian twist, the classical limit is the centrally extended Lie algebra $\mathfrak{W}_- \oplus \mathfrak{W}_-$. 

It turns out that all twist deformations except for the abelian twist do not have a q-analog on the full Witt algebra. However, these q-analogs can always be constructed on the one-sided Witt algebras similar to the Jordanian twist if and only if the corresponding twist (or r-matrix) does not contain both $L_1$ and $L_{-1}$ or $\bar L_1$ and $\bar L_{-1}$ simultaneously. Removing the r-matrices that do not admit a q-analog when quantized yields a set that is still bigger than \eqref{c1}-\eqref{c2}. E.g. the r-matrix $r_{4'}$ can be obtained from $r_I$ by an $\mathfrak{o}(4)$ automorphism of the form 
\begin{align}
\varphi(L_0) = \beta L_1 + \beta' L_0, \quad \varphi (L_1) = \beta'' L_1, \quad \varphi (\bar L_m) = (-1)^m \bar L_m 
\end{align}
but this is not an automorphism of $\mathfrak{W} \oplus \mathfrak{W}$, and also not of $\mathfrak{W}_- \oplus \mathfrak{W}_-$, as can be seen from the impossibility of solving
\begin{align}
[\varphi(L_0), \varphi (L_{-2})] = 2 \varphi(L_{-2}).
\end{align}

It is not hard to prove (cf. Appendix \ref{app-a}), although it does not directly follow from theorem \ref{W+Wth}, that also all LBA on $\mathfrak{W}_+ \oplus \mathfrak{W}_+$ are coboundary, i.e. we have\footnote{The same result obviously also holds for $\mathfrak{W}_- \oplus \mathfrak{W}_-$.}
\begin{theorem}\label{W++W+th}
The first cohomology group $H^1\left( \mathfrak{W}_+ \oplus \mathfrak{W}_+, \bigwedge^2 \left( \mathfrak{W}_+ \oplus \mathfrak{W}_+ \right) \right)$ vanishes.
\end{theorem}

Some physical consequences of considering one-sided subalgebras will be discussed in section \ref{sec5.4.2}.

\subsection{Deformations of the Surface Charge Algebra}\label{sec4.1.5}

From a physical point of view we are interested in the conserved charges and their algebra is the central extension of $\mathfrak{W} \oplus \mathfrak{W}$ given in section \ref{sec3.1.2}. A natural question is whether the central charges have any impact on the allowed deformations or on their precise form which will be investigated in the following.

We show in Appendix \ref{app-a} that the theorems stated at the beginning of the chapter generalize to the Virasoro algebra, i.e. one has\footnote{For one copy of the Virasoro algebra this was already shown in \cite{Ng:2000} independently.}
\begin{theorem}\label{Virth}
The first cohomology group $H^1\left( \mathfrak{Vir} \oplus \mathfrak{Vir}, \bigwedge^2 \left( \mathfrak{Vir} \oplus \mathfrak{Vir} \right) \right)$ vanishes.
\end{theorem}

To see how the classification of the r-matrices is affected let us first note that on the entire $\mathfrak{Vir} \oplus \mathfrak{Vir}$ there is no automorphism that mixes central elements and other generators. Furthermore, it is easy to see that any
\begin{align}
r_c = c_L \wedge X,
\end{align}
where $X$ is any linear combination in $\mathfrak{W} \oplus \mathfrak{W}$, is triangular. Then we only have to calculate the Schouten brackets of combinations of $r_c$ with \eqref{classa}-\eqref{classb} and find triangular r-matrices 
\begin{align}\label{crm1}
    r = c_L \wedge L_p + L_p \wedge \left(\sum_q \bar L_q \right), \quad    r = c_L \wedge L_p + \bar r, \quad  r = c_L \wedge L_p + \alpha L_p \wedge L_0, 
\end{align}
where $\bar r$ is a triangular r-matrix generated by $\bar L_q$. The only exception is that $a_1$ with an embedding ($L_{\pm1} \rightarrow L_{\pm n}, n \neq 1$) (and thus r-matrices containing it) is no longer triangular in $\mathfrak{Vir} \oplus \mathfrak{Vir}$, i.e. one has
\begin{align}
    [[a_1, a_1]] = \frac{(n^3-n)}{6} (L_n \wedge c_L \wedge L_{-n} + L_n\wedge c_L \wedge L_0 + L_{-n} \wedge c_L \wedge L_0).
\end{align}

The central extension thus can impact the LBAs. For example with $r = c_L \wedge L_p$ we obtain
\begin{align}
    \delta_r (L_m) = c_L \wedge (m-p) L_{m+p}, \quad \delta_r(\bar L_m) = 0.
\end{align}
But also Lie bialgebras from r-matrices that contain no central element can contribute extra terms, e.g. $r = \chi L_0 \wedge L_n$ yields
\begin{align}
    \delta_r(L_{-n}) = -n \chi L_{-n} \wedge L_n - \chi \frac{(n^3-n)}{12} L_0 \wedge c_L.
\end{align}
For all r-matrices from \eqref{crm1} we can write down the twist. This is easy to see in the first two cases as everything is abelian. In the last case note that we can obtain the twist from a Jordanian twist by simply redefining $L_0 \rightarrow L_0 - \alpha c_L$ which leaves $[L_0, L_p]$ invariant.

From the twists we can directly compute the coalgebra structures. In general these will contain infinite expressions which, as we did before, could be tried to remedy by considering one-sided algebras. Even though this works in the case of the Witt algebras, with a central extension the one-sided algebra is pointless to consider as the central elements do not appear in the algebra sector. Thus, out of \eqref{crm1} only $r =  c_L \wedge (\chi L_0 + \bar \chi \bar L_0)$ leads to a finite coalgebra sector. In particular the twist $\mathcal{F} = \exp( c_L  \otimes (\chi L_0 + \bar \chi \bar L_0))$ yields
\begin{align}
    \Delta (L_m) = L_m \otimes 1 + \exp ( -m \chi c_L) \otimes L_m, \quad \Delta (\bar L_m) = \bar L_m \otimes 1 + \exp (-m \bar \chi c_L) \otimes \bar L_m
\end{align}
and the specialization is straightforward with adding $\Pi \equiv \exp (\chi c_L), \bar \Pi \equiv \exp (\bar \chi c_L)$ and the respective inverse elements.

\section{Deformations of Asymptotically Flat Symmetries in 3D Gravity}

Following the analysis of deformations on asymptotically (anti) de Sitter spacetimes, we will now focus on the flat limit. From a physical intuition one might expect that there is a smooth limit $\Lambda \rightarrow 0$ implying that this case is qualitatively similar to $\Lambda \neq 0$. However, as addressed in the beginning of section \ref{sec3.1.3}, there are indeed conceptual differences in the asymptotic analysis of different values for $\Lambda$. Also, from a mathematical perspective the fact that the BMS algebra in three dimensions ($\mathfrak{B}_3$) is not semi-simple complicates some of the results from the previous sections.

\subsection{Lie Bialgebra Structures on $\mathfrak{B}_3$} 

As stated above the $\mathfrak{B}_3$ is not semi-simple but it has the form of a semi-direct product of the semi-simple Witt algebra and the ideal of supertranslations. 
Using this fact in combination with theorem \ref{Wth} of section \ref{sec4.1.1} in Appendix \ref{app-a}, we prove the following
\begin{theorem}\label{B3th}
The first cohomology group $H^1( \mathfrak{B}_3, \bigwedge^2 \mathfrak{B}_3)$ vanishes.
\end{theorem}

Additionally, all the results about ad-invariant elements carry over, i.e. one has
\begin{align}
\left( \bigwedge^3 \mathfrak{B}_3 \right)_{\mathfrak{B}_3} = \left( \bigwedge^2 \mathfrak{B}_3 \right)_{\mathfrak{B}_3} = \{0\}.
\end{align}
We again conclude that the LBA structures on $\mathfrak{B}_3$ are uniquely defined by triangular r-matrices classified up to automorphism.

\subsection{Contraction Limit and Classification of Triangular r-matrices}\label{sec4.2.2}

In order to obtain r-matrices in $\mathfrak{B}_3$ one can perform a contraction limit $\Lambda \rightarrow 0$ on the r-matrices of $\mathfrak{W} \oplus \mathfrak{W}$. 
For the finite dimensional embeddings this has been done in \cite{Kowalski-Glikman:2019ttm} and the results were compared to the classification of r-matrices in the three dimensional Poincar\'e algebra given by \cite{Stachura_1998}.

It was claimed that the latter r-matrices can all be obtained from a contraction of $\mathfrak{o}(4, \mathbb{C})$ r-matrices.  It should be noted that the contraction is ambiguous, i.e. the contraction of a class of r-matrices can be performed in different non-equivalent ways (cf. below) and not injective, i.e. there are r-matrices in the 3D Poincar\'e that can be obtained as a contraction from distinct $\mathfrak{o}(4)$ r-matrices. Whether the contraction is actually surjective is an interesting question because an affirmative answer could grant us a constructive recipe for quantization as we will see below.

Let us examine the previous points by explicitly contracting the r-matrices from \eqref{classa}-\eqref{classb} starting from $r_{1'}$. To this end, $r_{1'}$ is expressed in terms of $l_m, T_m$ with the help of \eqref{isoL} and subsequently expanded in powers of $1/\sqrt{-\Lambda}$
\begin{align}
r_{1'} =& \frac{\alpha + \bar \alpha}{-\Lambda} (T_1 \wedge T_0 - T_{1}\wedge T_{-1} - T_{-1} \wedge T_0)  \nonumber \\
& + \frac{\beta}{\sqrt{-\Lambda}}  (l_1 +l_{-1} + 2l_0)\wedge (T_1 +T_{-1} + 2T_0) + \frac{\alpha}{\sqrt{-\Lambda}} (...) + \beta(...)
\end{align}
In order to obtain a finite result we have to rescale $(\alpha + \bar \alpha) \rightarrow (\hat \alpha + \hat {\bar \alpha})(-\Lambda)$ and $\beta \rightarrow \hat \beta \sqrt{-\Lambda}$. Then taking the limit $\Lambda \rightarrow 0$ yields
\begin{align}
\hat r_{1',a} = (\hat \alpha + \hat{\bar \alpha}) (T_1 \wedge T_0 + T_{1} \wedge T_{-1}  - T_{-1} \wedge T_0) + \hat \beta (l_1 +l_{-1} + 2l_0)\wedge (T_1 +T_{-1} + 2T_0).
\end{align}
This is not the only possibility to obtain a finite limit, in the case $\alpha =-\bar \alpha$ one can also rescale $\alpha \rightarrow \hat \alpha \sqrt{-\Lambda}$ to get
\begin{align}
\hat r_{1',b} =&\hat \beta (l_1 \wedge T_1 + l_{-1} \wedge T_{-1} + 4 l_0 \wedge T_0) + (2 \hat \beta + \hat \alpha)(l_1 \wedge T_0 + l_0 \wedge T_{-1}) \nonumber \\
&+ (2 \hat \beta  - \hat \alpha) (l_0 \wedge T_1 + l_{-1} \wedge T_0) + ( \hat \beta + \hat \alpha) l_1 \wedge T_{-1} + (\hat \beta - \hat \alpha) l_{-1} \wedge T_1.
\end{align}
Similarly, one obtains the contraction limits for the rest of the r-matrices, i.e. 
\begin{align}
\hat r_{2',a} & = ( \hat\alpha  - \hat\beta) T_1 \wedge T_0, \\ 
\hat r_{2',b} & =-2 \hat\alpha l_0 \wedge T_1,
\end{align}
with $\alpha = \beta $ for $\hat r_{2',b}$,
\begin{align}
\hat r_{3',a} & = (\hat{\bar \alpha} + \hat\alpha (1-\epsilon )(1-\epsilon') -2 \hat\beta) T_1 \wedge T_0 + (\hat{\bar \alpha} - \hat\beta (1- \epsilon')) T_1 \wedge T_{-1} \nonumber \\
& \phantom{=} + (- \hat{\bar \alpha} + \hat\beta (\epsilon -2 \epsilon')) T_{-1} \wedge T_0, \\
\hat r_{3',b} & = \hat r_{1',b}(\hat\alpha =0),
\end{align}
with $\bar \alpha =0$ for $\hat r_{3',b} $ (if $\epsilon = \epsilon'=1 \Rightarrow \hat r_{3', a} \subset \hat r_{1',a}$), 
\begin{align}
\hat r_{4', a} & = \hat\beta_1 l_1 \wedge T_1 + \hat\beta_2 l_1 \wedge l_0 , \\
\hat r_{5', a} & = (\hat\alpha - \hat\beta) T_1 \wedge T_0 + (\hat\alpha + \hat\beta) T_1 \wedge T_{-1} -(\hat\alpha + \hat\beta) T_{-1} \wedge T_0, \\
\hat r_{6',a} & = (\hat\beta T_1 + \hat\beta_0 T_0 + \epsilon \hat\beta T_{-1}) \wedge (\hat{\bar \beta} T_1 + \hat{\bar \beta_0} T_0 + \bar \epsilon \hat{\bar \beta} T_{-1}), \\
\hat r_{6',b} & = ( \hat\beta l_1 + \hat\beta l_{-1} + \hat\beta_0 l_0) \wedge (\hat\beta T_1 + \hat\beta T_{-1} + \hat\beta_0 T_0),
\end{align}
with $\beta_0 = \bar \beta_0, \epsilon = \bar \epsilon , \beta = \bar \beta$ for $\hat r_{6',b}$,
\begin{align}
\hat r_{7',a} & = (\hat\beta + \hat\alpha) T_1 \wedge T_0 + \epsilon \hat\beta T_1 \wedge T_{-1}, \\
\hat r_{7',b} & =  \hat\alpha T_1 \wedge (l_1 + l_0),
\end{align}
with $\epsilon = 0, \bar \beta =  \alpha$ for $\hat r_{7',b}$,
\begin{align}
\hat r_{8',a} & = (\hat\alpha + \hat{\bar \alpha}) T_1 \wedge T_0 -2 \hat\beta l_1 \wedge T_1, \\
\hat r_{8',b}  & =  - \hat\beta l_1 \wedge T_1  + \hat\alpha (l_1 \wedge T_0 - l_0 \wedge T_1), 
\end{align}
with $ \alpha = - \bar \alpha$ for $\hat r_{8',b}$.
One easily checks that the reality condition is satisfied under the involution 
\begin{align}
l_m \mapsto -l_m, \quad T_m \mapsto -T_m,
\end{align}
if the parameter restrictions from $\dagger$ are adopted but not necessarily for those from $\ddagger$. In particular, the r-matrices corresponding to the Jordanian and extended Jordanian twist ($\hat r_{2', b}, \hat r_{8', b}(\hat \beta =0)$) are incompatible with the restrictions listed in section \ref{sec4.1.2} for $\ddagger$. 

The contraction results can be compared to the classification of r-matrices in $\mathfrak{P}_3$ up to $\mathfrak{P}_3$ automorphisms from \cite{Stachura_1998}. Considering only the triangular parts these read\footnote{After removing the quasitriangular part, the r-matrices $r_2, r_3$ are $\mathfrak{P}_3$-automorphic but this does not extend to $\mathfrak{B}_3$.},
\begin{align}\label{classp1}
r_1 & = l_0 \wedge (l_1 - l_{-1}) - l_1 \wedge l_{-1}, \\
r_2 & = l_0 \wedge T_0 + T_0 \wedge (\Theta_1 T_1 + \Theta_{-1} T_{-1}), \label{r2} \\
r_3 & = (l_1 - l_{-1}) \wedge (T_1 - T_{-1}) + T_0 \wedge (\Theta_1 T_1 + \Theta_{-1} T_{-1}), \\
r_4 & = \frac{\chi}{2}(l_1 \wedge T_0 - l_0 \wedge T_1) - \frac{\zeta}{2} l_1 \wedge T_1 + T_1\wedge (\Theta_1 T_0 + \Theta_{0} T_{-1}) + \delta_{\zeta , \frac{\chi}{2}} \Theta_{-1} T_0 \wedge T_{-1} , \\
r_5 & = l_0 \wedge T_1 +  T_1\wedge (\Theta_1 T_0 + \Theta_{0} T_{-1}).\label{classp2}
\end{align}
Following the argumentation from section \ref{sec4.1.2} we are interested in r-matrices up to $\text{Aut}(\mathfrak{B}_3)$ and we see that e.g. the r-matrix $\hat r_{1',b}$ is in fact not $\mathfrak{B}_3$-automorphic to any r-matrix from \eqref{classp1}-\eqref{classp2} as e.g. the term $l_0 \wedge T_0$ is not contained there. This term could only be obtained from $r_3$ by a $\mathfrak{P}_3$ automorphism.
As we will see in section \ref{sec5.1.2}, terms in the r-matrices of the form $ T \wedge T$, $l \wedge T$ and $l \wedge l$ correspond to canonical, Lie algebra type and quadratic non-commutativity, respectively.

\begin{figure}
\centering
\begin{tikzpicture}
    \draw (0, 0) node[inner sep=0] {\includegraphics[width=16cm]{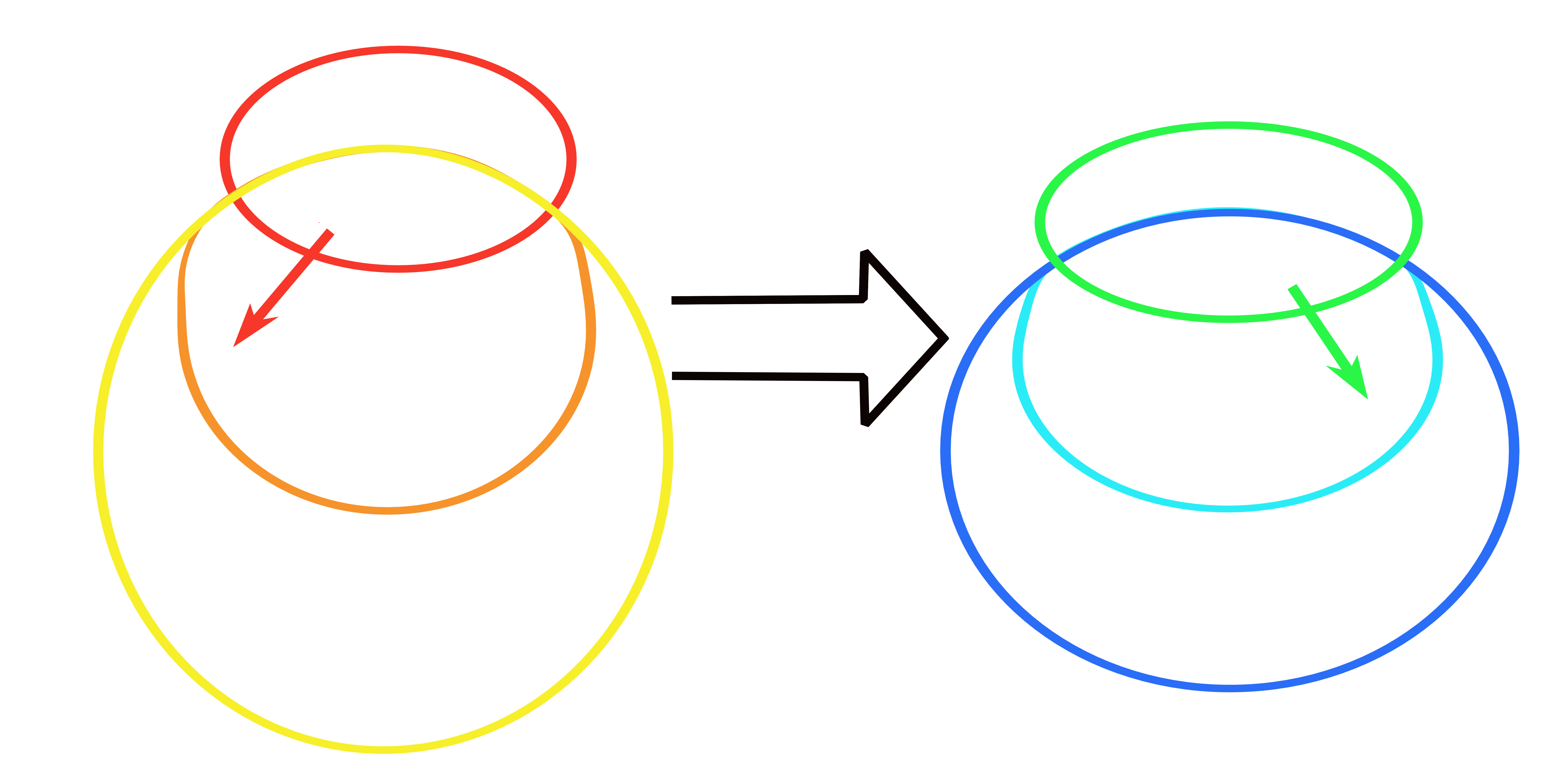}};
    \draw (-3.8, 2.0) node {$r \in \bigwedge^2 \mathfrak{o}(4)$};
	\draw (-3.8, 2.8) node {quasitriangular};
	\draw (4.5, 2.1) node {quasitriangular};
	\draw (0, 0.5) node {contraction};
	\draw (4.3, 1.4) node {$r \in \bigwedge^2 \mathfrak{P}_3$};
    \draw (-4.0, 0.0) node {\begin{tabular}{l}
    $r \in \bigwedge^2 \mathfrak{o}(4)$  \\
	up to $\text{Aut}'(\mathfrak{W}\oplus \mathfrak{W})$
    \end{tabular}};
    \draw (-3.8, -2.0) node {\begin{tabular}{l}
     $r \in \bigwedge^2 \left(\mathfrak{W}\oplus \mathfrak{W}\right)$ \\
     triangular 
     \end{tabular}};
	\draw (4.5, -1.8) node {\begin{tabular}{l}
	$r \in \bigwedge^2 \mathfrak{B}_3$ \\
    triangular 
     \end{tabular}};	
	\draw (4.5, -0.2) node {\begin{tabular}{l}
    $r \in \bigwedge^2 \mathfrak{P}_3$  \\
	up to $\text{Aut}'(\mathfrak{B}_3)$
    \end{tabular}};
\end{tikzpicture}
\caption{Schematic depiction of r-matrices in three dimensional asymptotic symmetry algebras. The red arrow represents the quotient $\text{Aut}(\mathfrak{o}(4))/\text{Aut}'(\mathfrak{W} \oplus \mathfrak{W})$ and the green arrow $\text{Aut}(\mathfrak{P}_3)/\text{Aut}'(\mathfrak{B}_3)$. All the r-matrices inside the yellow and blue circle are triangular.} \label{fig4.1}
\end{figure}
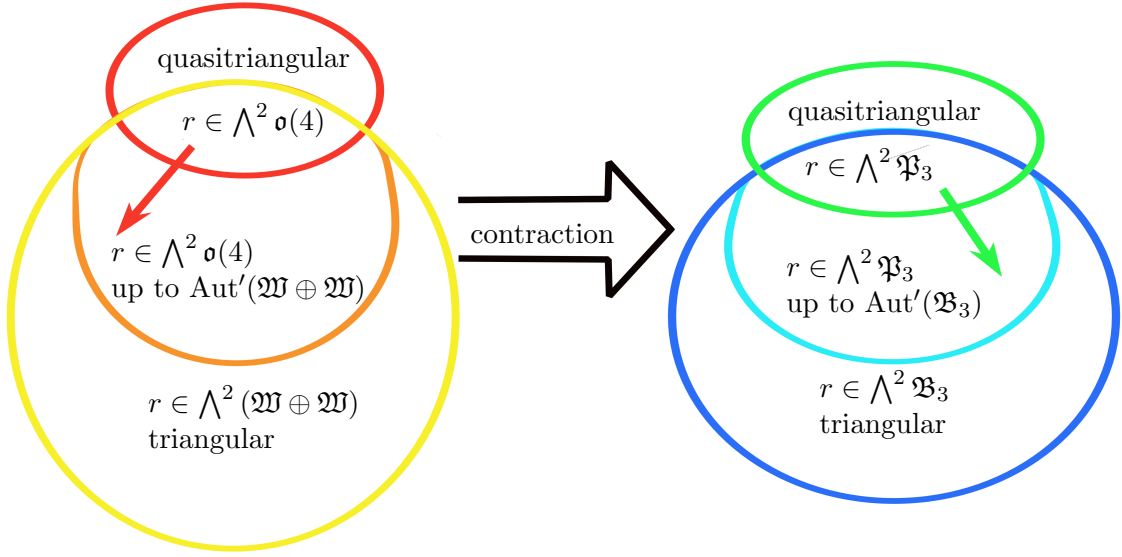

\subsubsection*{Surjectivity of the Contraction Limit}
To address the question of surjectivity, i.e. whether all triangular $\mathfrak{B}_3$ r-matrices can be obtained from a contraction limit of a triangular $\mathfrak{o}(4)$ r-matrix, there are two more aspects that have to be taken into account. First, there might be quasitriangular r-matrices in $\mathfrak{o}(4)$ that contract to triangular $\mathfrak{B}_3$ r-matrices. As shown in \cite{Kowalski-Glikman:2019ttm} this is indeed happening but in this case there is always also a triangular $\mathfrak{o}(4)$ r-matrix with the same limit.
Secondly, it is possible to perform the contraction limit along different axis as explained in the following. 
The (anti) de Sitter algebra \eqref{ads3}-\eqref{ads3l} is isomorphic to 
\begin{align}
[M_{AB}, M_{CD}] = \delta_{AC} M_{BD} - \delta_{BC} M_{AD} + \delta_{BD} M_{AC} - \delta_{AD} M_{BC},
\end{align}
where the indices range from $1$ to $4$ and 
\begin{gather}
M_{+-} = M_{13}, \quad M_{+2} = M_{12} + M_{32}, \quad M_{-2} = M_{12}- M_{32}, \\
K_{\pm}  = M_{14} \pm M_{34}, \quad K_2 = M_{24},
\end{gather}
i.e. the fourth axis is chosen for contraction. With the isomorphism 
\begin{align}
H = & -\frac{i}{2} (M_{12} + M_{34}), \quad \bar H = \frac{i}{2} (M_{12} - M_{34}), \\
E_{\pm} = & -\frac{i}{2} (M_{23} + M_{14}) \mp \frac{1}{2}(M_{24} - M_{13}),\\
\bar E_{\pm} = & \frac{i}{2} (M_{23} - M_{14}) \mp \frac{1}{2}(-M_{24} - M_{13}),
\end{align}
one can express the r-matrices in terms of $M_{AB}$. Depending on which axis is chosen the result of the contraction differs, e.g. when choosing the second instead of the fourth axis
\begin{align}
K_{i} = M_{i2}, \quad J_{i} = \epsilon_{ijk} M_{jk}
\end{align}
one finds for 
\begin{align}
r = (E_+ + \bar E_+) \wedge (H + \bar H),
\end{align}
which is automorphic to $r_I$, 
\begin{align}
r = \frac{i}{2} (i J_3 + J_4) \wedge J_1,
\end{align}
where the contraction limit can be taken without rescaling. The $J_i$ satisfy 
\begin{align}
[J_1, J_3] = J_4, \quad [J_1, J_3] = J_4, \quad [J_3, J_4] = J_1
\end{align}
and can thus be related to the three dimensional Lorentz sector via
\begin{align}
J_3 = -l_0, \quad J_1 = \frac{l_1+l_{-1}}{2},\quad J_4 = \frac{l_1-l_{-1}}{2},
\end{align}
leading to 
\begin{align}
\hat r = l_1 \wedge l_{-1} + l_0 \wedge(l_1 + l_{-1}) .
\end{align}
Note that this r-matrix of the $\mathfrak{P}_3$ can not be obtained from a contraction of \eqref{classa}-\eqref{classb} which are associated with the fourth axis, so the consideration of different axis actually makes a difference. 
But even after taking into account the possibility to contract along different axis it turns out that there are triangular r-matrices in $\mathfrak{P}_3$, e.g.
\begin{align}
r = l_0 \wedge T_1 + \Theta_1 T_1 \wedge T_0 + \Theta_2 T_1 \wedge T_{-1},
\end{align}
that can {\textit{not}} be obtained in its general form from a contraction of a triangular r-matrix. It can, however, be obtained from the sum of two different contraction limits, a possibility also mentioned in \cite{Kowalski-Glikman:2019ttm}.

To summarize, in this weak sense (including sums of contractions and contractions along different axis) all r-matrices in $\bigwedge^2 \mathfrak{P}_3$ up to $\text{Aut}(\mathfrak{P}_3$) (the intersection of the green and dark-blue circle in figure \ref{fig4.1}) can be obtained from a contraction of r-matrices from $\bigwedge^2 \mathfrak{o}(4)$ up to $\text{Aut}'(\mathfrak{W} \oplus \mathfrak{W})$ (the orange circle in figure{fig4.1}). In fact these contractions form a larger set as we saw with explicit examples. The full set represented by the light-blue circle in figure \ref{fig4.1} could in principle be derived using the quotient of automorphisms $\text{Aut}( \mathfrak{P}_3)/ \text{Aut}'(\mathfrak{B}_3)$ or alternatively by redoing the classification using only $\text{Aut}(\mathfrak{B}_3$. Since we did not explicitly carry out these procedures, it remains an open question whether the contraction of $r \in \bigwedge^2 \mathfrak{o}(4)$ up to $\text{Aut}' (\mathfrak{W} \oplus \mathfrak{W})$ is surjective.

\subsubsection*{Contraction of Twists}
As stated above, the surjectivity might be relevant for obtaining the associated Hopf algebras. Namely there is the possibility in performing the contraction limit on the level of the full twist \cite{Borowiec:2006fc}. As an example let us consider the light-cone twist (the name will be justified later)
\begin{align}\label{lctw}
\mathcal{F}_{\text{LC}} = e^{ L_0 \otimes \log \left( 1+ a L_n \right)} e^{ \bar L_0 \otimes \log \left( 1 - a \bar L_n \right)} 
\end{align}
corresponding  to \eqref{lcfc} with $a = \chi = - \chi, \zeta =0$. We express the $L_m, \bar{L}_m$ in terms of $l_m, T_m$
 \begin{align}
 \mathcal{F}_{\text{LC}} = & \exp \left( \frac{1}{2} \left(l_0 + \frac{T_0}{\sqrt{-\Lambda}}\right) \otimes \log  \left(1 + \frac{a}{2}\left(l_n + \frac{T_n}{\sqrt{-\Lambda}} \right) \right) \right)  \nonumber \\
 & \times \exp \left( \frac{1}{2} \left(l_0 - \frac{T_0}{\sqrt{-\Lambda}}\right) \otimes \log  \left(1 - \frac{a}{2}\left(l_n - \frac{T_n}{\sqrt{-\Lambda}} \right) \right) \right) . \label{lctwist}
 \end{align}
In order to obtain a finite contraction limit we have to rescale $a \rightarrow a' = a/\sqrt{-\Lambda}$. Expanding the first exponent of the twist in powers of $\sqrt{-\Lambda}$ and taking the limit then results in
\begin{align}
\frac{1}{2} \left( l_0 + \frac{T_0}{\sqrt{-\Lambda}} \right) & \otimes \sum_{j=1} \frac{- (-a')^j}{j} \left( \sqrt{-\Lambda}l_n + T_n\right)^j \nonumber \\
= & \frac{1}{2} l_0 \otimes \log ( 1 + a' T_n) + \frac{1}{2} T_0 \otimes \sum_{j=1} \frac{-(-a')^j}{j} j l_n T_n^{j-1} + \mathcal{O}( \sqrt{-\Lambda}) \nonumber \\
= & \frac{1}{2} l_0 \otimes \log ( 1 + a' T_n) + \frac{1}{2} T_0 \otimes l_n a'\left( 1 + a' T_n\right)^{-1}.
\end{align}
After repeating this procedure for the second exponent the final twist is given by
\begin{align}\label{lc2}
\mathcal{F}_{\text{LC}} = \exp \left( l_0 \otimes \log (1 + a' T_n) + T_0 \otimes  a' l_n \left( 1+ a' T_n \right) \right).
\end{align}
It automatically satisfies the 2-cocycle condition in $\mathfrak{B}_3$ since the twist \eqref{lctwist} satisfies it in $\mathfrak{W}\oplus \mathfrak{W}$. The associated classical r-matrix turns out to be
\begin{align}\label{lcrm}
r_{\text{LC}} = a' (l_0 \wedge T_n - l_n \wedge T_0),
\end{align}
which corresponds indeed to the three dimensional version of the light-cone $\kappa$-Poincar\'e (for some embedding).

But such a simple procedure is not always possible. As an example, we could try to use the same rescaling for the twist of $r_{2'}$ as for the r-matrix $\hat r_{2',b}$. The $\mathfrak{o}(4)$ twist for $r_{2'}$ reads \cite{Borowiec:2017apk}
\begin{align}
\mathcal{F} = \exp \left( - \frac{\beta}{\alpha} \bar L_0 \wedge \log ( 1 + \alpha L_1) \right) \exp \left( - L_0 \otimes \log (1 + \alpha L_1) \right)
\end{align}
and after rescaling we obtain in first order in the deformation parameters
\begin{align}
 \mathcal{F} = 1\otimes 1 + \frac{1}{\sqrt{- \Lambda}} (\hat \beta T_0 \wedge T_1 - \hat \alpha T_0 \otimes T_1) + \mathcal{O}(1),
\end{align}
i.e. the result diverges because the relevant terms only cancel in the antisymmetrized r-matrix but not in the twist. Even though we know the twist for $\hat r_{2',b}$ (being just a Jordanian twist), this illustrates the limitations of the contraction on the level of the twist. 
In particular, we notice that for r-matrices of the form $l \wedge T + T \wedge T$ this means, together with the fact that some r-matrices of this form can only be expressed as a sum of different contractions\footnote{Note that there is also no $\mathfrak{B}_3$ automorphism that mixes superrotations and -translations after the contraction.}, that there is no simple recipe to obtain all the twists for the LBA's on $\mathfrak{P}_3$ up to $\mathfrak{B}_3$ automorphisms. However, it is possible to explicitly write down the twists for r-matrices of the form $l \wedge T$. In the classification \eqref{classp1}-\eqref{classp2} this remains to be shown only for $r_4(\Theta_{1/0/-1} =0) \hat = r_{8',b}$, other twists for r-matrices that are not equivalent in $\mathfrak{B}_3$ can always be obtained from a $\mathfrak{P}_3$ automorphism. For $r_{8',b}$ one can contract the twist of $r_{8'} \hat = r_{II}$, which is given by \cite{Borowiec:2017apk}
\begin{align}
\mathcal{F} = e^{\frac{\beta}{-\alpha^2} \log  \left( 1+ \alpha L_1 \right) \wedge \log  \left( 1- \alpha \bar L_1 \right) } e^{ L_0 \otimes \log \left( 1+ \alpha L_1 \right)} e^{ \bar L_0 \otimes \log \left( 1 - \alpha \bar L_1 \right)} .
\end{align}
Apart from the first exponent, this is just our example \eqref{lctw} and the contraction is done analogously. The first exponent becomes simply $\exp \left( 2 \hat \beta l_1 \wedge T_1 \right)$ and thus 
\begin{align}
\mathcal{F}_{r_{8',b}} = \exp \left( 2 \hat \beta l_1 \wedge T_1 \right) \exp  \left( l_0 \otimes \log (1 + \hat \alpha T_1) + T_0 \otimes  \hat \alpha l_1 \left( 1+ \hat \alpha T_1 \right) \right).
\end{align}

\subsection{Uniqueness of Twist Deformation}

In the previous subsection a twist for the light-cone r-matrix \eqref{lcrm} was found via contraction. From \cite{Borowiec:2020ddg} (cf. also \cite{Borowiec:2014aqa}, \cite{Kulish99}) it is known that the extended Jordanian twist of the form 
\begin{align}
 \mathcal{F}_{\text{eJ}} = \exp \left( \frac{a''}{2} l_n \otimes T_0 \right) \exp \left( - \frac{l_0}{n} \otimes \log \left(1 + a'' n T_n \right) \right)
\end{align}
exists for the same r-matrix.  However, the inequivalence is only superficial as we have to take into account automorphisms on the universal envelope. We find there exist invertible elements $\omega \in U \mathfrak{B}_3 [[1/\kappa]]$ inducing the automorphisms  
\begin{align}
f(l_m) = \omega^{-1} l_m \omega, \quad f (T_m) = \omega^{-1} T_m \omega
 \end{align}
 by a similarity transformation (see \cite{Bor19} and references therein for a detailed discussion). In general for every twist deformed Hopf algebra with $\mathcal{F}$ one can obtain a ``gauge equivalent'' twist via $\mathcal{F}_{\omega} = \omega^{-1} \otimes \omega^{-1} \mathcal{F} \Delta (\omega)$. The new twist then satisifies the 2-cocycle condition because
 \begin{align}
 {\mathcal{F}_{\omega}}_{12} ( \Delta \otimes 1) ( \mathcal{F}_{\omega}) &= (\omega^{-1} \otimes \omega^{-1}) F_{12} (\Delta(\omega) \otimes \omega) \Delta( \omega^{-1} ) \otimes \omega^{-1}) (\Delta \otimes 1) \mathcal{F} (\Delta(\omega) \otimes \omega) \nonumber \\
 &=  (\omega^{-1} \otimes \omega^{-1}) \mathcal{F} _{23} ( 1 \otimes  \Delta) \mathcal{F}  (\Delta (\omega) \otimes \omega) \nonumber \\
 &= {\mathcal{F}_{\omega}}_{23} (1 \otimes \Delta) \mathcal{F}_{\omega} 
 \end{align}
 and $f(x) \equiv \omega x \omega^{-1}$ establishes the isomorphism between the twisted Hopf algebras
 \begin{align}
 \Delta_{\mathcal{F} } \circ f = (f \otimes f) \circ \Delta_{\mathcal{F}_{\omega}}.
 \end{align}
 If the untwisted Hopf algebra admits a $*$-structure, the twist has to satisfy
 \begin{align}
 \mathcal{F}^{* \otimes *} = \mathcal{F}^{-1},
 \end{align}
 i.e. be unitary in order to preserve the $*$-structure. On the invertible element $\omega$ this enforces {the unitarity condition}
 \begin{align}
 \mathcal{F}_{\omega}^{* \otimes *}  &= \Delta (\omega^*) \mathcal{F}^{-1} ( {\omega^*}^{-1} \otimes  {\omega^*}^{-1}) \overset{!}{=}  \mathcal{F}_{\omega}^{-1} = \Delta (\omega^{-1}) \mathcal{F}^{-1} (\omega \otimes \omega) \\
 \Leftrightarrow \omega^* &= \omega^{-1}.
 \end{align}
 
 In our particular example we find that in first order of $1/\kappa$ the isomorphism induced from the element 
 \begin{align}
 \omega = e^{- \frac{a'}{4} (l_n  T_0 + T_0 l_n)}
 \end{align}
 relates the extended Jordanian and the contraction limit of the light-cone twist for $a'' = - a'$. It is easy to see that it is hermitian with respect to the reality condition $l_m^* = -l_m, T_m^* = -T_m$ and $a' \in i \mathbb{R}$.
 
Such a resolution as a ``gauge equivalence'' of the apparent ambiguity is in line with the statement that the full deformation is already determined by the first order quoted at the end of section \ref{sec2.1.9}.

\subsection{Full Deformations via Twist}\label{sec4.2.4}

Even though we did not find a general recipe for all twists, let us construct and examine the full quantum group structures in some example cases where it is possible. 
A special focus will again be put on the possibility of finding a q-analog. We also noted in section \ref{sec2.2} that the fate of the light-like $\kappa$-Poincar\'e is of interest as it has been studied in the finite dimensional Poincar\'e algebra.

\subsubsection*{Abelian Twist}

There is again a unique abelian r-matrix, i.e. $r_2(\Theta =0) $ from eq.\eqref{r2}, that does not single out any embedding as its constituents are present in each of the embeddings. It is associated with the twist
\begin{align}\label{abtw}
\mathcal{F}_A = \exp \left( \eta l_0 \otimes T_0 \right),
\end{align}
which does not come from a contraction of a $\mathfrak{o}(4)$ twist but the construction is straightforward since it is abelian. The twisting element \eqref{abtw} induces the coalgebra relations
\begin{align}
\Delta_{\mathcal{F}_A}(l_m) & = 1 \otimes l_m - m \eta l_0 \otimes T_m + l_m \otimes \exp \left( -m \eta T_0 \right), \\
\Delta_{\mathcal{F}_A}(T_m) & = 1 \otimes T_m + T_m \otimes \exp \left( -m \eta T_0\right) .
\end{align}
Because the coproduct is cocommutative (only for) $l_0, T_0$ we infer
\begin{align}
S_{\mathcal{F}_A} (l_0) =  -l_0 , \quad S_{\mathcal{F}_A} (T_0) = -T_0  
\end{align}
and then
\begin{align}
S_{\mathcal{F}_A} (l_m) = - (l_m + m \eta l_0 T_m) e^{ m \eta T_0}, \quad S_{\mathcal{F}_A} (T_m) = - T_m e^{m \eta T_0}.
\end{align}
One can see that the qualitative behavior is very similar to the abelian twist in the $\mathfrak{W} \oplus \mathfrak{W}$ algebra (eq.\eqref{abtwist1} and following).

\subsubsection*{Jordanian and Extended Jordanian Twist}

In the following, we will first construct a Jordanian twist in $\mathfrak{B}_3$ since a lot of the consistent twists are either of this form or have it as a building block, which we already noted in section \ref{sec4.1.3}. Furthermore, the extended Jordanian twist will be investigated as it will turn out to exhibit qualitatively different behavior and also because of its relevance in the context of $\kappa$-Poincar\'e (cf. section \ref{sec2.2} and \ref{sec5.1}).

For the r-matrix $r_5$ from \eqref{classp2} with $\Theta_{1/0} =0$ one obtains the Jordanian twist
\begin{align}\label{jtwx}
\mathcal{F}_J = \exp \left( -\frac{l_0}{n} \otimes \log \left( 1 + \chi T_n \right) \right),
\end{align}
leading to the coproducts
\begin{align}\label{jordco1}
\Delta_{\mathcal{F}_J} (l_m) & = 1 \otimes l_m  - \frac{n-m}{n} \chi l_0 \otimes T_{m+n} \Pi_{+n}^{-1} + l_m \otimes \Pi_{+n}^{\frac{m}{n}}, \\
\Delta_{\mathcal{F}_J} (T_m) & = 1 \otimes T_m + T_m \otimes \Pi_{+n}^{\frac{m}{n}},\label{jordco2}
\end{align}
where 
\begin{align}
\Pi_{+n} = \left( 1 + \chi T_n \right).
\end{align}
From 
\begin{align}
\Delta_{\mathcal{F}_J} (l_0) = 1 \otimes l_0 + l_0 \otimes (1- \chi T_n) \Pi_{+n}^{-1} , \quad \rightarrow S_{\mathcal{F}_J} (l_0) = -l_0 (1- \chi T_n) \Pi_{+n}^{-1},
\end{align}
we infer
\begin{align}
S_{\mathcal{F}_J} (l_m) & = -(l_m - \frac{n-m}{n} \chi  (1- \chi T_n) T_{m+n} \Pi_{+n}^{-2} ) \Pi_{+n}^{-\frac{m}{n}}, \\
S_{\mathcal{F}_J} (T_m) & = - T_m \Pi_{+n}^{-\frac{m}{n}}.
\end{align}
We can already note that, in contrast to the asymptotically (A)dS symmetry, the coalgebra structures of the Jordanian deformation only involve a finite number of generators for the entire algebra. As a consequence there is no necessity to restrict to a subset of the algebra and thereby to a specific embedding.

For the light-cone deformation described by the r-matrix \eqref{lcrm} it was already noted that there are two possibilities for a twist. To perform the calculations we choose the extended Jordanian twist 
\begin{align}\label{ejtwx}
\mathcal{F}_{eJ} = \exp \left( \frac{\chi}{n} l_n \otimes T_0 \right) \mathcal{F}_J
\end{align}
which, as the name suggests, also contains the Jordanian twist. Using again the Hadamard formula we calculate
\begin{align}
\Delta_{\mathcal{F}_{eJ}} (l_m) = & \exp \left( \frac{\chi}{n} l_n \otimes T_0 \right) \Delta_{\mathcal{F}_J} (l_m) \exp \left( -\frac{\chi}{n} l_n \otimes T_0 \right), \nonumber \\
= & \Delta_{\mathcal{F}_{J}} (l_m) - \frac{m}{n} \chi l_n \otimes T_m + \frac{n-m}{n} \chi^2 l_n \otimes T_0 T_{m+n} \Pi_{+n}^{-1} \nonumber \\
& + \sum_{k=1}^{\infty} \frac{\chi^k }{n^k k !} \prod_{p=0}^{k-1} (n-(m+pn))  l_{m+kn} \otimes T_0^k \Pi_{+n}^{\frac{m}{n}} ,\label{copej} \\
\Delta_{\mathcal{F}_{eJ}} (T_m)  = &\Delta_{\mathcal{F}_{J}} (T_m) + \sum_{k=1}^{\infty} \frac{\chi^k }{n^k k !} \prod_{p=0}^{k-1} (n-(m+pn))  T_{m+kn} \otimes T_0^k \Pi_{+n}^{\frac{m}{n}}. \label{copej2}
\end{align}
If there is no $k \in \mathbb{N}$ such that $n = m + kn$ there appear infinite power series that also contain infinitely many different generators. Excluding either $m >1$ and choosing $n=1$ or $m<-1$ with $n=-1$ solves the latter problem as the sum over $k$ in \eqref{copej}-\eqref{copej2} terminates. Choosing the former then results in finite coproducts, i.e. the sums in \eqref{copej} and \eqref{copej2} terminate at $k = 1-m$ (for $m = 1$ the sum is set to zero) which also allows to iteratively compute the antipodes from
\begin{align}
S_{eJ} (l_m) = & - \left(l_m + (1-m) \chi l_0 T_{m+1} \Pi_{+}^{-1} + m \chi l_1 T_m -(1-m) \chi^2 l_1 T_0 T_{m+1} \Pi_+^{-1} \right) \Pi_+^{-m} \nonumber \\
& - \sum_{k=1}^{1-m} \frac{\chi^k }{ k !} \prod_{p=0}^{k-1} (1-(m+p))  S(l_{m+k})  T_0^k , \\
S_{eJ} (T_m) = & -T_m \Pi_+^{-m} - \sum_{k=1}^{1-m} \frac{\chi^k }{ k !} \prod_{p=0}^{k-1} (1-(m+p))  S(T_{m+k})  T_0^k.
\end{align}

Because the $T_i$ form an abelian ideal all the formulas for the Jordanian twist remain finite and all embeddings are possible. The extended Jordanian twist on the other hand behaves more like the Jordanian twist in $\mathfrak{W} \oplus \mathfrak{W}$ as the one-sided algebra has to be chosen to obtain finite coalgebra structures.

As in section \ref{sec4.1.4}, the theorem \ref{B3th} has a one-sided analogon proven in Appendix \ref{app-a}.
\begin{theorem} \label{B3+th}
The first cohomology group $H^1 \left( \mathfrak{B}_{3 +}, \bigwedge^2 \mathfrak{B}_{3 +} \right)$ vanishes.
\end{theorem}

\subsubsection*{Specialization}

The specialization of the Hopf algebra deformed by the abelian twist can be performed analogously to section \ref{sec4.1.4}. This means we add the elements
\begin{align}
K \equiv e^{\eta T_0}, \quad K^{-1} \equiv e^{-\eta T_0}
\end{align}
to the algebra and then the coalgebra structures can be reformulated as
\begin{align}
\Delta_{\mathcal{F}_A}(l_m) & = 1 \otimes l_m - m \eta l_0 \otimes T_m  + l_m \otimes K^{-m}, \\
\Delta_{\mathcal{F}_A} (T_m) & = 1 \otimes T_m + T_m \otimes K^{-m}, \\
S_{\mathcal{F}_A} (l_m) & = - (l_m + m \eta l_0 T_m) K^m , \quad S_{\mathcal{F}_A} (T_m) = -T_m K^m
\end{align}
and 
\begin{align}
\Delta_{\mathcal{F}_A}(K) = K \otimes K , \quad \Delta_{\mathcal{F}_A}(K^{-1}) = K^{-1} \otimes K^{-1}.
\end{align}
In the algebraic sector the extra relations read
\begin{gather}
[l_m, K] =  -m \eta T_m K, \quad [l_m, K^{-1}] = m \eta T_m K^{-1}, \\
[T_m, K] = [T_m, K^{-1}] = [K, K^{-1}] = 0
\end{gather}
and everything is well defined for all complex $\eta$ so we can set $\eta = \frac{i}{\kappa }$. In the classical limit $\kappa \rightarrow \infty \leftrightarrow \eta \rightarrow 0$ the elements $K, K^{-1}$ become central.

When deforming the primitive Hopf algebra with a Jordanian twist only a finite number of generators appear. However, the coproducts \eqref{jordco1}-\eqref{jordco2} involve fractional powers in $\Pi_{+ n}$. Thus, after singling out an embedding for the twist we add the elements
\begin{align}\label{addpi}
\Pi \equiv \Pi_{+n}^{\frac{1}{n}}, \qquad \Pi^{-1} \equiv  \Pi_{+n}^{- \frac{1}{n}}
\end{align}
to the algebra and the coalgebra sector can be expressed using only finite powers
\begin{align}
\Delta_{\mathcal{F}_J} (l_m) & = 1 \otimes l_m - \frac{n-m}{n} \chi l_0 \otimes T_{m+n} \Pi^{-n} + l_m \otimes \Pi^m, \\
\Delta_{\mathcal{F}_J} (T_m) & = 1 \otimes T_m + T_m \otimes \Pi^m.
\end{align}
Furthermore we have
\begin{align}
\Delta_{\mathcal{F}_J} ( \Pi) =  \Pi \otimes \Pi, \qquad \Delta_{\mathcal{F}_J} ( \Pi^{-1}) =  \Pi^{-1} \otimes \Pi^{-1}.
\end{align}
Adding the fractional powers of $\Pi_{+ n}$ comes at the cost of slightly more complicated algebraic relations
\begin{align}
[\Pi^{\frac{1}{n}}_{+n}, l_m] =  & \sum_{j=0}^{\infty} \frac{(1/n)^{\underline j}}{j!} \chi^j [ T_n^j, l_m] = (n-m) T_{m+n} \frac{d \Pi_{+n}^{\frac{1}{n}}}{d T_n} = \chi \frac{(n-m)}{n} T_{m+n} \Pi^{\frac{1}{n}-1}_{+n}, \label{specalj}\\
[\Pi_{+n}^{-\frac{1}{n}} , l_m] =  & -\chi \frac{(n-m)}{n} T_{n+m} \Pi_{+n}^{-\frac{1}{n}-1}, \\
[\Pi^{\frac{1}{n}}_{+n}, T_m] = & [\Pi_{+n}^{-\frac{1}{n}}, T_m] = 0, \label{specalj2}
\end{align}
but all formulas are well defined for general complex $\chi$. Again, the classical limit $\chi \rightarrow 0$ results in central $\Pi, \Pi^{-1}$.

We thus find that the q-analog can be constructed on the full algebra for this particular twist deformation which contains elements with non-zero index in contrast to the situation in $\mathfrak{W} \oplus \mathfrak{W}$ where this was true only for the deformation defined by the abelian $r \sim L_0 \wedge \bar L_0$. In section \ref{sec4.2.2} we saw that while the r-matrix corresponding to $\mathcal{F}_J$ can be obtained as a contraction limit from the triangular $r_{2'} \in \bigwedge^2 \left(\mathfrak{W} \oplus \mathfrak{W}\right)$, the same limit does not work on the level of the twist. Such a result is compatible with the different behavior in the specialization of the Jordanian twist and the twist for $r_{2'}$. Again, we conclude that the flat case is more than just the contraction of (A)dS but in this example the difference appeared only at the level of the full quantum group.

For the extended Jordanian twist , implementing the light-like $\kappa$-Poincar\'e, it is obvious that one has to restrict to the one-sided BMS algebra ($\mathfrak{B}_{3, -}$), generated by $\{l_m, T_p | m, p \leq 1\}$ (or $\mathfrak{B}_{3, +}$), to avoid infinite series in the deformed coproducts. Adding the elements from \eqref{addpi} to the algebra one can reexpress \eqref{copej}-\eqref{copej2} so that it contains finite powers in the deformation parameter. As we added the same elements as for the Jordanian twist, the extra relation for the algebra sector are again \eqref{specalj}-\eqref{specalj2}.

\subsection{Deformations of the Surface Charge Algebra}\label{sec4.2.5}

When considering the algebra of the surface charge generators in asymptotically flat three dimensional spacetime $\mathfrak{B}_{3,c}$ we, somewhat surprisingly, find (proved in Appendix \ref{app-a})
\begin{theorem}\label{B3cth}
The first cohomology group $H^1( \mathfrak{B}_{3,c}, \bigwedge^2 \mathfrak{B}_{3,c}) $ contains only the cocycle $\delta_{\alpha}$.
\end{theorem}
Here, 
\begin{align}\label{dalp}
\delta_{\alpha} : l_m \mapsto 0, \; T_m \mapsto \alpha c'_L \wedge T_m,
\end{align}
and this cocycle clearly defines a Lie bialgebra as the co-Jacobi identity is trivially satisfied. This result pinpoints a clear situation where the flat case can not be seen as just a limit from the (A)dS case.

Besides this single possibility of a non-coboundary LBA let us investigate how the classification of r-matrices is affected by the central element. First, we calculate the Schouten brackets of \eqref{classp1}-\eqref{classp2} in $\mathfrak{B}_{3c}$ and find that all of them vanish with the exception that if we consider a different embedding ($l_1, T_1 \rightarrow l_n, T_n$ ) then we have to demand $\Theta_1 = - \Theta_{-1}$ for $r_3$ and $\Theta_{-1} = 2 \Theta_0$ for $r_4$. As the $\mathfrak{P}_3$ automorphisms do not necessarily extend to $\text{Aut}( \mathfrak{B}_{3c})$ it is not surprising that there exist r-matrices classified up to  $\text{Aut}( \mathfrak{B}_{3})$ that are not triangular in $\mathfrak{B}_{3c}$. This is the case for $\hat r_{1',b}$, which is a contraction of $r_{1'}$ and we already noted in section \ref{sec4.1.5} that the $a_1$ part of $r_{1'}$ is not triangular in $\mathfrak{Vir}\oplus \mathfrak{Vir}$. However, the other possible contraction $\hat r_{1',a}$ is triangular in $\mathfrak{B}_{3c}$.

Furthermore, the addition of the central element makes it possible to define triangular r-matrices which are any linear combination of 
\begin{align}\label{clt}
r = c'_L \wedge l_m, \quad r = c'_L \wedge T_m.
\end{align}
A tedious but straightforward computation reveals that also the following combinations with the r-matrices we discussed in \ref{sec4.2.2} define coboundary LBA
\begin{align}\label{crm21}
 r_{1''} & = c'_L \wedge (\alpha T_0 + \beta l_0 ) + l_0 \wedge T_0, \quad r_{2''} = c'_L \wedge T_0 + \alpha l_n  \wedge T_n + \beta T_0 \wedge T_n, \\ 
r_{3''} & = c'_L \wedge (\alpha T_n + \beta T_0 + \gamma T_{-n}) + T_0 \wedge ( \Theta_1 T_n + \Theta_{-1} T_{-n}) + \Theta_0 T_n \wedge T_{-n}, \\
r_{4''} & = c'_L \wedge (\alpha T_n + \beta l_n) + \chi (l_n \wedge T_0 - l_0 \wedge T_n) - \zeta l_n \wedge T_n \nonumber \\
& \phantom{=} + \Theta_1 T_n \wedge T_0 + \delta_{\beta, 0} \delta_{\zeta, \chi/2} \Theta_{-1} T_0 \wedge T_{-n}, \\
r_{5''} & = c'_L \wedge (\alpha T_0 + \beta T_n + \gamma l_n) + \chi l_0 \wedge T_{n} + \Theta_1 T_0 \wedge T_n + \delta_{\gamma, 0} \Theta_0 T_n \wedge T_{-n}, \\
 r_{6''} & = c'_L \wedge (l_n + \alpha l_0) + \chi_1 l_n \wedge T_n + \delta_{\alpha, 0} \chi_2 l_n \wedge l_0, \\
 r_{7''} & = c'_L \wedge (\alpha T_n + \beta T_0) + T_n \wedge (\chi_1 l_n + \chi_2 l_0) .\label{crm22}
\end{align}

As in the asymptotically (A)dS setting also LBA from r-matrices that do not contain central elements can receive extra contributions, e.g. with $r = \chi l_0 \wedge T_n$ we get
\begin{align}
\delta_r(l_{-n}) = \chi  \left((-n) l_{-n} \wedge T_n -2n l_0 \wedge T_0 - l_0 \wedge (n^3-n) c'_L \right).
\end{align}

\subsubsection*{Quantization}

Here we want to qualitatively analyze how the Lie bialgebras discussed in the previous section can be quantized to quantum groups in all orders of the deformation parameter(s) and highlight this with several examples.

We already established that there is one LBA, defined by the cobracket \eqref{dalp} that is not coboundary and consequently the associated Hopf algebra can not be obtained by a twist. However, we simply make an educated guess, namely
\begin{gather}\label{dac1}
\Delta_{\alpha} (l_m) = l_m \otimes 1 + 1 \otimes l_m, \quad \Delta_{\alpha} (c'_L) = c'_L \otimes 1 + 1 \otimes c'_L, \\
\Delta_{\alpha} (T_m) = T_m \otimes 1 + e^{\alpha c'_L} \otimes T_m, \\
S_{\alpha} (l_m) = -l_m, \quad S_{\alpha} (c'_L) = -c'_L, \\
S_{\alpha} (T_m) = - e^{- \alpha c'_L} T_m.\label{dac2}
\end{gather}
This indeed defines a Hopf algebra with a coassociative coalgebra and the correct limit. When trying to find quantum groups for LBA with a cobracket $\delta = \delta_r + \delta_{\alpha}$ it is in general not clear that there exists a constructive method even if we know the twist associated with $\delta_r$. For $r = -\frac{\chi}{n} l_0 \wedge T_n + \frac{\alpha}{n} c_L \wedge l_0$, containing our standard example, we are in luck though; the situation exactly resembles method 5 from \cite{Lya08} if we identify $-\frac{l_0}{n} = H, e^{\alpha c'_L}= C, T_n = D$. Then the coalgebra \eqref{dac1}-\eqref{dac2} can be deformed with the twist 
\begin{align}\label{twc3}
\mathcal{F} = \exp \left( - \frac{l_0}{n} \otimes \log \left( e^{\alpha c'_L} + \chi T_n\right) \right).
\end{align}
For the supertranslations this then yields
\begin{align}
\Delta_{\mathcal{F}, \alpha} (T_m) = \mathcal{F} \Delta_{\alpha}(T_m) \mathcal{F}^{-1} = T_m \otimes \left( e^{\alpha c'_L} + \chi T_n \right)^{\frac{m}{n}} + e^{\alpha c'_L} \otimes T_m,
\end{align}
which has indeed $\delta = \delta_r + \delta_{\alpha}$ as a classical limit.

While the twist implementing the deformation defined by (linear combinations of) \eqref{clt} is just an abelian twist, it is not easy to find explicit twists for the combined r-matrices in \eqref{crm21}-\eqref{crm22} in general. In some cases we can use automorphisms on the subalgebra spanned by the elements appearing in the r-matrix. E.g. by redefining $l_0 \rightarrow l'_0 =l_0 + \frac{\beta}{\chi} c'_L$ we can bring $r_{5''} ( \gamma =0 = \Theta_1 = \Theta_0)$ in the form $ r = \chi  l'_0 \wedge T_n + \alpha c'_L \wedge T_0$ consisting of a Jordanian part and a second term commuting with everything else. Thus, the twist is given by
\begin{align}\label{twc32}
\mathcal{F} = \exp \left( \left(l_0 +  \frac{\beta}{\chi} c'_L \right) \otimes \log \left( 1 + \chi T_n \right) \right) \exp ( \alpha c'_L \otimes T_0).
\end{align}

$r$-matrices that do not contain $c'_L$ explicitly admit the same twist as in the $\mathcal{B}_3$ algebra, however, there can be extra contributions to the coalgebra sector. We illustrate this fact with the standard example \eqref{jtwx} leading to \eqref{jordco1}-\eqref{jordco2} except for $l_{-n}$ where we obtain
\begin{align}
\Delta_{\mathcal{F}_J} (l_{-n}) = 1 \otimes l_{-n} - 2 \chi l_0 \otimes \Pi_{+ n}^{-1} \left( T_0 + c'_L (n^2-1) \right) + l_{-n} \otimes \Pi_{+n}^{-1}.
\end{align}

\subsubsection*{Specialization}

The process of specialization is not much different in the centrally extended BMS algebra. For example, adding the simple redefinition $\Pi_c = e^{\alpha c'_L}$ to the Hopf algebra with coalgebra sector \eqref{dac1}-\eqref{dac2} allows the expression in finite powers.

Still, the general comment applies that any Hopf algebra derived from an r-matrix containing $l_m, m \neq 0$ will not admit a q-analog on the whole algebra. It is noteworthy though that the restriction to the one-sided algebra does not make much sense as then no central charge appears in the algebra sector.

\section{Deformations of Asymptotically Flat Symmetries in 4D Gravity}

After the investigation of quantum groups in the three dimensional toy model we will now try to carry the insights over to the realistic four dimensional spacetime. Again, first the ``infinitesimal deformations'', i.e. the Lie bialgebras, are studied. Then, full quantum groups will be constructed and analyzed, focusing on the comparison between three and four dimensions.

\subsection{Lie Bialgebras and Classification of Triangular r-matrices}\label{sec4.3.1}

As expected, it is fairly straightforward to prove the following theorem similar to theorem \ref{B3th} for the three dimensional case.
\begin{theorem}\label{B4th}
The first cohomology group $H^1( \mathfrak{B}_4, \bigwedge^2 \mathfrak{B}_4)$ vanishes.
\end{theorem}

Furthermore, all Lie bialgebras have to be triangular and are uniquely determined by equivalence classes of r-matrices since 
\begin{align}
\left( \bigwedge^3 \mathfrak{B}_4 \right)_{\mathfrak{B}_4} = \left( \bigwedge^2 \mathfrak{B}_4 \right)_{\mathfrak{B}_4} = \{0\}.
\end{align}

For the four dimensional Poincar\'e algebra a classification of the r-matrices was carried out in \cite{Zak97} (see Appendix \ref{app-b}). Even though it is not complete, the only missing parts are some quasi-triangular solutions. By calculating the Schouten brackets of the r-matrices we can filter out their triangular parts that are listed in Appendix \ref{app-b} (in the basis \eqref{4dembed1}-\eqref{4dembed2}).
We will mostly focus on the $l \wedge T$ part for the sake of simplicity and because it corresponds to the Lie algebra type non-commutativity. Then the distinct r-matrices of this form are 
\begin{align}\label{4dclass1}
r_2 & = S_{11} \wedge \bar k_0, & r_7 & =   S_{11} \wedge k_0 + S_{01} \wedge k_1 + A_{01} \wedge \bar k_1  + \beta S_{11} \wedge k_1, \\
r_{9} & = S_{01} \wedge \bar k_1 + S_{11} \wedge k_1, & r_{10} & = A_{01} \wedge k_1 , \\
r_{13} & =  (S_{11} - S_{00}) \wedge \bar k_0, & r_{14} & = S_{11} \wedge \bar k_0, \\
r_{17} & = S_{11} \wedge (k_0 + \beta \bar k_0).\label{4dclass2}
\end{align}
In their general form all of these are not part of the $\mathfrak{B}_3$ subalgebra. It is easy to find a twist for $r_2, r_{10}, r_{12}, r_{13}, r_{14}$ since they are abelian and $r_{17}$ corresponds to a Jordanian twist. Furthermore, $r_{7}$ is the four dimensional light-cone r-matrix and thus the deformation can be performed with an extended Jordanian twist. Only for $r_9$ no twist is explicitly known\footnote{One can show that finding a twist for $r_9$ is equivalent to finding one for a matrix of th form $l \wedge T + T \wedge T$ in three dimensions.}.

Another difference to the three dimensional setting is that the only deformation that does not single out a particular embedding is given by the abelian r-matrix $k_0 \wedge \bar k_0$ which does not lead to a non-commutativity of Lie algebra type. The simple reason for this is that in four dimensions the embeddings do not ``overlap'' in the supertranslation sector.  

\subsection{Full Deformations via Twist}\label{sec4.3.2}

Quantum groups in all orders of the deformation parameter are constructed again by using the twist deformation of the coalgebra sector. The abelian twist and the Jordanian twist are also twists in $\mathfrak{B}_4$ and the corresponding deformations from section \ref{sec4.2.4} define Hopf subalgebras of four dimensional Hopf algebras. For example, the Jordanian twist
\begin{align}\label{4jtwx}
\mathcal{F}_{J} = \exp \left( - \frac{k_0}{n} \otimes \log \left( 1 + \chi S_{1-m, 1-m} \right) \right)
\end{align}
leads to 
\begin{gather}
\Delta_{\mathcal{F}_J} (S_{pq}) = S_{pq} \otimes \Pi_{+, 1-m}^{-\frac{1-p-q}{n}} + 1 \otimes S_{pq}, \\
\Delta_{\mathcal{F}_J} (A_{pq}) = A_{pq} \otimes \Pi_{+, 1-m}^{-\frac{1-p-q}{n}} + 1 \otimes A_{pq}, \\
\Delta_{\mathcal{F}_J} (k_p) = k_{p} \otimes \Pi_{+, 1-m}^{\frac{p}{n}} + 1 \otimes k_{p} - \chi \frac{k_0}{n} \otimes (n -p) S_{1-m+p, 1-m} \Pi_{+, 1-m}^{-1}, \\
\Delta_{\mathcal{F}_J} (\bar k_p) = \bar k_{p} \otimes \Pi_{+, 1-m}^{\frac{p}{n}} + 1 \otimes \bar k_{p} - \chi \frac{k_0}{n} \otimes (n -p) A_{1-m+p, 1-m} \Pi_{+, 1-m}^{-1}, 
\end{gather}
with
\begin{align}\label{pij4}
\Pi_{+,n} \equiv  (1 + \chi S_{n n}), \quad 1-2m = n.
\end{align}
The antipodes are given by
\begin{align}
S_{\mathcal{F}_J} (S_{pq}) & = - S_{pq} \Pi_{+,1-m}^{\frac{1-p-q}{n}}, \quad S_{\mathcal{F}_J} (A_{pq}) = - A_{pq} \Pi_{+, 1-m}^{\frac{1-p-q}{n}}, \\
S_{\mathcal{F}_J} (k_p) & = - \left( k_p  + \frac{\chi}{n} (n -p) S_{\mathcal{F}_J}(k_0) S_{1-m -p, 1-m} \Pi_{+, 1-m}^{-1} \right) \Pi_{+, 1-m}^{- \frac{p}{n}}, \\
S_{\mathcal{F}_J} (\bar k_p) & = - \left(\bar k_p  + \frac{\chi}{n} (n -p) S_{\mathcal{F}_J}(k_0) A_{1-m -p, 1-m} \Pi_{+, 1-m}^{-1} \right) \Pi_{+, 1-m}^{- \frac{p}{n}},
\end{align}
where
\begin{align}
S_{\mathcal{F}_J}(k_0) = - k_0 (1- \chi S_{1-m, 1-m}) \Pi_{+, 1-m}^{-1}.
\end{align}

The extended Jordanian twist on the other hand includes the momentum generators from the extra spatial dimension but the general form is analogous to that of section \ref{sec4.2.4}
\begin{align}\label{4ejtwx}
\mathcal{F}_{eJ} = \exp \left( \frac{\chi}{n} ( k_{n} \otimes S_{1-m, m}- \bar k_{n} \otimes A_{1-m, m} \right) \mathcal{F}_{J}.
\end{align}
This complicates the calculations of the deformed coproducts and we find 
\begin{align}
\Delta_{\mathcal{F}_{eJ}} (S_{pq}) = &  \exp \left( \frac{\chi}{n} ( k_{n} \otimes S_{1-m, m}- \bar k_{n} \otimes A_{1.m, m} \right) \left(\Delta_{\mathcal{F}_J} (S_{pq}) \right) \nonumber \\
& \times \exp \left(- \frac{\chi}{n} ( k_{n} \otimes S_{1-m, m}- \bar k_{n} \otimes A_{1.m, m} \right) \\
= & 1  \otimes S_{pq} + \sum_{l =0}^{\infty} \frac{1}{l!} \left(\frac{\chi}{n}\right)^l  \sum_{r+s = l} {r \choose l} \Bigg(  A_{p+rn, q+sn} \otimes g^{(pq)}_{rs}  \Pi_{+, 1-m}^{\frac{p+q-1}{n}}  \nonumber \\
&+  S_{p+rn, q+sn} \otimes f^{(pq)}_{r s}  \Pi_{+, 1-m}^{\frac{p+q-1}{n}} \Bigg), \label{ej41}\\
\Delta_{\mathcal{F}_{eJ}} (A_{pq})  = & 1  \otimes A_{pq} + \sum_{l =0}^{\infty} \frac{1}{l!} \left(\frac{\chi}{n}\right)^l  \sum_{r+s = l} {r \choose l} \Bigg(A_{p+rn, q+sn} \otimes h^{(pq)}_{rs}  \Pi_{+, 1-m}^{\frac{p+q-1}{n}}  \nonumber \\
&+ S_{p+rn, q+sn} \otimes j^{(pq)}_{r s}  \Pi_{+, 1-m}^{\frac{p+q-1}{n}} \Bigg),\label{ej42}
\end{align}
with the recursively defined coefficients
\begin{gather}
g^{(pq)}_{00} = 0, \quad f^{(pq)}_{00} = 1, \\
g^{(pq)}_{10} = -(1-m -pn) A_{1-m, m} , \quad g^{(pq)}_{01} = (1-m-qn) A_{1-m, m}, \\
f^{(pq)}_{10} = (1-m-pn) S_{1-m, m}, \quad f^{(pq)}_{01} = (1-m-qn) S_{1-m, m}, \\
g^{(pq)}_{r+1, s} = (1-m -(p+r)n) S_{1-m, m} g^{(pq)}_{rs} - (1-m-(p+r)n) A_{1-m,m} f^{(pq)}_{rs} , \\
g^{(pq)}_{r, s+1} = (1-m -(q+s)n) S_{1-m, m} g^{(pq)}_{rs} + (1-m-(q+s)n) A_{1-m,m} f^{(pq)}_{rs} , \\
f^{(pq)}_{r+1, s} = (1-m -(p+r)n) S_{1-m, m} f^{(pq)}_{rs} + (1-m-(p+r)n) A_{1-m,m} g^{(pq)}_{rs} , \\
f^{(pq)}_{r, s+1} = (1-m -(q+s)n) S_{1-m, m} f^{(pq)}_{rs} - (1-m-(q+s)n) A_{1-m,m} g^{(pq)}_{rs} ,
\end{gather}

\begin{gather}
h^{(pq)}_{00} = 1, \quad j^{(pq)}_{00} = 0, \\
h^{(pq)}_{10} = (1-m -pn) S_{1-m, m} , \quad h^{(pq)}_{01} = (1-m-qn) S_{1-m, m}, \\
j^{(pq)}_{10} = -(1-m-pn) A_{1-m, m}, \quad j^{(pq)}_{01} = (1-m-qn) A_{1-m, m}, \\
h^{(pq)}_{r+1, s} = (1-m -(p+r)n) S_{1-m, m} h^{(pq)}_{rs} - (1-m-(p+r)n) A_{1-m,m} j^{(pq)}_{rs} , \\
h^{(pq)}_{r, s+1} = (1-m -(q+s)n) S_{1-m, m} h^{(pq)}_{rs} + (1-m-(q+s)n) A_{1-m,m} j^{(pq)}_{rs} , \\
j^{(pq)}_{r+1, s} = (1-m -(p+r)n) S_{1-m, m} j^{(pq)}_{rs} + (1-m-(p+r)n) A_{1-m,m} h^{(pq)}_{rs} , \\
j^{(pq)}_{r, s+1} = (1-m -(q+s)n) S_{1-m, m} j^{(pq)}_{rs} - (1-m-(q+s)n) A_{1-m,m} h^{(pq)}_{rs} ,
\end{gather}
and similarly
\begin{align}
\Delta_{\mathcal{F}_{eJ}} (k_p) = & \Delta_{\mathcal{F}_{J}} (k_p) -\frac{\chi}{n} k_n \otimes  \left( \left(\frac{p+1}{2} - (1-m)\right) S_{1-m+p, m} + \left(\frac{p+1}{2} -m)\right) S_{1-m, m+p} \right)  \nonumber \\
& - \frac{\chi}{n} \bar k_n \otimes  \left( -\left(\frac{p+1}{2} - (1-m)\right) A_{1-m+p, m} -
 \left(\frac{p+1}{2} -m)\right) A_{1-m, m+p} \right) \nonumber \\
& - \chi (n-p)  \left(k_n \otimes S_{1-m, m} - \bar k_n \otimes A_{1-m, m} \right)  \left( 1 \otimes S_{1-m+p, 1-m} \Pi_{+, 1-m}^{-1} \right) \nonumber \\
& + \sum_{l = 1}^{\infty} \frac{1}{l!} \left( \frac{\chi}{n} \right)^l \sum_{r+s=l}{r \choose l } \Bigg(   k_{p+rn} \otimes \tilde g^{(p)}_r \Pi_{+, 1-m}^{\frac{p}{n}} + \bar k_{p+sn} \otimes \tilde f^{(p)}_s \Pi_{+, 1-m}^{\frac{p}{n}} \Bigg), \label{ej43}\\
\Delta_{\mathcal{F}_{eJ}} (\bar k_p) = & \Delta_{\mathcal{F}_{J}} (\bar k_p) -\frac{\chi}{n} k_n \otimes  \left( \left(\frac{p+1}{2} - (1-m)\right) A_{1-m+p, m} - \left(\frac{p+1}{2} -m)\right) A_{1-m, m+p} \right)  \nonumber \\
& - \frac{\chi}{n} \bar k_n \otimes  \left( \left(\frac{p+1}{2} - (1-m)\right) S_{1-m+p, m} -
 \left(\frac{p+1}{2} -m)\right) S_{1-m, m+p} \right) \nonumber \\
& - \chi (n-p)  \left(k_n \otimes S_{1-m, m} - \bar k_n \otimes A_{1-m, m} \right)  \left( 1 \otimes A_{1-m+p, 1-m} \Pi_{+, 1-m}^{-1} \right) \nonumber \\
& + \sum_{l = 1}^{\infty} \frac{1}{l!} \left( \frac{\chi}{n} \right)^l \sum_{r+s=l}{r \choose l } \Bigg(   k_{p+rn} \otimes \tilde h^{(p)}_r \Pi_{+, 1-m}^{\frac{p}{n}} + \bar k_{p+sn} \otimes \tilde j^{(p)}_s \Pi_{+, 1-m}^{\frac{p}{n}} \Bigg),\label{ej44}
\end{align}
with
\begin{gather}
\tilde g^{(p)}_1 = (n-p) S_{1-m, m}, \quad \tilde f^{(p)}_1 = -(n-p) A_{1-m,m}, \\
\tilde g^{(p)}_{r+1} = (n-(p+r)n) S_{1-m, m} \tilde g^{(p)}_r - ( n - (p+r)n) A_{1-m, m} \tilde f^{(p)}_r, \\
\tilde f^{(p)}_{s+1} = (n-(p+s)n) S_{1-m, m} \tilde f^{(p)}_r + ( n - (p+s)n) A_{1-m, m} \tilde g^{(p)}_r, \\
\tilde h^{(p)}_1 = (n-p) A_{1-m, m}, \quad \tilde j^{(p)}_1 = (n-p) S_{1-m,m}, \\
\tilde h^{(p)}_{r+1} = (n-(p+r)n) S_{1-m, m} \tilde h^{(p)}_r + ( n - (p+r)n) A_{1-m, m} \tilde j^{(p)}_r, \\
\tilde j^{(p)}_{s+1} = (n-(p+s)n) S_{1-m, m} \tilde j^{(p)}_r - ( n - (p+s)n) A_{1-m, m} \tilde h^{(p)}_r.
\end{gather}
Despite the more involved relations the extended Jordanian twist on $\mathfrak{B}_3$ and $\mathfrak{B}_4$ share most conceptual properties. E.g. also for \eqref{ej41},\eqref{ej42},\eqref{ej43}, \eqref{ej44} the expressions admit only a finite number of generators if the one-sided $\mathfrak{B}_{4 -}$ (spanned by $\{ k_p, \bar k_p, S_{pq}, A_{pq}| p,q \leq 1\}$) is considered, implying the embedding with $m = 0 \hat = n= 1$. In that case one can also calculate the antipodes recursively from
\begin{align}
0 = & S (S_{pq}) + \sum_{l =0}^{\infty} \frac{1}{l!} \left(\chi\right)^l \sum_{r+s = l} {r \choose l} \Bigg(  A_{p+r, q+s} S\left( g^{(pq)}_{rs}  \Pi_{+}^{p+q-1n} \right) \nonumber \\
&+S_{p+r, q+s} S \left( f^{(pq)}_{r s}  \Pi_{+}^{p+q-1} \right) \Bigg) ,
\end{align}
starting from 
\begin{gather}
\Delta_{\mathcal{F}_{eJ}} (S_{11}) = \Delta_{\mathcal{F}_{J}} (S_{11}) = S_{11} \otimes \Pi_+ + 1 \otimes S_{11}, \\
\Delta_{\mathcal{F}_{eJ}} (k_{1}) = \Delta_{\mathcal{F}_{J}} (k_{1}) = k_1 \otimes \Pi_+ + 1\otimes k_1, \quad \Delta_{\mathcal{F}_{eJ}} (\bar k_{1})  = \bar k_1 \otimes \Pi_+ + 1 \otimes \bar k_1  \\
\rightarrow S_{\mathcal{F}_{eJ}}(S_{11}) = - S_{11} \Pi_+^{-1}, \quad S_{\mathcal{F}_{eJ}}(k_1) = - k_1 \Pi_+^{-1}, \quad S_{\mathcal{F}_{eJ}}(\bar k_1) = - \bar k_1 \Pi_+^{-1}.
\end{gather}
Thus we were able to implement the unique deformed Hopf algebra that contains the light-like $\kappa$-Poincar\'e as a sub Hopf algebra in the symmetry algebra of a three and four dimensional asymptotically flat spacetime. In both cases the restriction to the respective one-sided algebras is necessary to assign a numerical value to the deformation parameter $1/\kappa$.

\subsection{Comparison with 3D Deformations}\label{sec4.3.3}

Since, as exemplified by the extended Jordanian twist discussed above, most aspects of the deformation are similar in three and four dimensions even though the formulas tend to be more bloated, we will mostly discuss key differences and similarities in this section.

A first distinction, that there is no deformation leading to Lie algebra type non-{\linebreak}commutativity which does not single out an embedding, was already noted in section \ref{sec4.3.1}. This implies, in contrast to the three dimensional case, that any deformation qualitatively changes the picture of the asymptotic symmetries as presented in chapter \ref{chap3}. Namely, no choice of vacuum is preferred a priori but the deformation would violate this ambiguity.

\subsubsection*{Specialization}

We already saw that the formulas for the coproducts deformed by an extended Jordanian twist contain a finite amount of generators under the same conditions as in three dimensions and it is straightforward to show that these also allow for a specialization. Similarly one finds that the Jordanian twist deformation admits a specialization on the entire $\mathfrak{B}_4$ by extending it with the elements (cf. \eqref{pij4})
\begin{align}
\Pi_{+,1-m}^{\frac{1}{n}}, \quad \Pi_{+, 1-m}^{- \frac{1}{n}} .
\end{align}

Even though we excluded r-matrices that combine different embeddings for physical reasons (cf. section \ref{sec4.1.2}) we can make some general statements about all mathematically possible deformations. In the three dimensional case every r-matrix containing $l_p, p \neq 0$ will not have a quantization with q-analog.
 Thus, the only allowed (in the sense of the specialization) terms are either of the form $T \wedge T$ or the Jordanian $l_0 \wedge T_n$. Focusing only on the latter we find that these r-matrices can not contain multiple embeddings since e.g. $l_0 \wedge (T_n +T_{n'})$ is not triangular for $n \neq n'$. 
In $\mathfrak{B}_4$, however, there are triangular r-matrices of the form
\begin{align}\label{mixem}
r = \bar k_0 \wedge ( S_{p p} + S_{p' p'})
\end{align}
which admit a q-analog for arbitrary $p,p'$ (cf. section \ref{sec5.3.3}). The right tensor leg in \eqref{mixem} can also contain more generators $S_{p'' p''}$ but otherwise this is the only possibility as can be checked by directly computing the Schouten brackets of general r-matrices containing $k_0, \bar k_0$.

\subsubsection*{Surface Charge Algebra}

Despite the different algebraic structure of the algebra of surface charges in four dimensional asymptotically flat spacetime $\mathfrak{B}_{4,c}$ we obtain a similar result in Appendix \ref{app-b}
\begin{theorem}\label{B4cth}
The first cohomology group $H^1( \mathfrak{B}_{4,c}, \bigwedge^2 \mathfrak{B}_{4,c})$ contains only   $\delta_{\alpha, \bar \alpha}$.
\end{theorem}
Here the cocycle
\begin{align}
\delta_{\alpha, \bar \alpha}: l_m \mapsto 0, \quad \bar l_m \mapsto 0, \quad T_{pq} \mapsto (\alpha c_l + \bar \alpha c_{\bar l} ) \wedge T_{pq}
\end{align}
defines a LBA since the co-Jacobi identity is also satisfied. 

The rest of the LBA are derived from classical r-matrices and in contrast to the three dimensional BMS all of the r-matrices in $\mathfrak{B}_{4}$ of the form $l \wedge T$ are triangular in $\mathfrak{B}_{4c}$ as well due to the different form of the central extension. 
Furthermore, any linear combination of 
\begin{align}
r = (\alpha c_l + \bar \alpha c_{\bar l}) \wedge T_{pq}, \quad r = (\alpha c_l + \bar \alpha c_{\bar l}) \wedge l_p, \quad r = (\alpha c_l + \bar \alpha c_{\bar l}) \wedge \bar l_p
\end{align}
defines a LBA. We refrain from classifying all possible combinations with \eqref{4dclass1}-\eqref{4dclass2} except for our standard example, the Jordanian $r = \chi k_0 \wedge S_{1-m,1-m}$, where we find
\begin{align}
r =&   \chi k_0 \wedge S_{1-m,1-m} +  (\alpha c_l + \bar \alpha c_{\bar l}) \wedge ( \beta_1 S_{m, 1-m} + \beta_2 A_{m, 1-m} \nonumber \\
& + \beta_3 S_{1-m, 1-m} + \beta_4 \bar k_n + \beta_5 \bar k_{-n} + \beta_6 \bar k_0). 
\end{align}

While it is still true that some LBA defined by r-matrices in $\bigwedge^2 \mathfrak{B}_4$ receive extra terms with central charges in the cobrackets, these do not extend to Hopf algebras which admit a q-analog on the entire algebra because they necessarily contain $k_n$ or $\bar k_n$ with $n \neq 0$.

The rest of the discussion of the coalgebr sector of full quantum groups in section \ref{sec4.2.5} carries over to the four dimensional case similarly. In particular, the Hopf algebra associated with $\delta_{\alpha, \bar \alpha}$ only has the non-primitive coproducts
\begin{align}
\Delta_{\alpha, \bar \alpha} (T_{pq}) = T_{pq} \otimes 1 + e^{\alpha c_l + \bar \alpha c_{\bar l}} \otimes T_{pq}
\end{align}
and the example with the twist \eqref{twc3} can be defined in four dimensions if we replace $l_0 \rightarrow k_0, T_n \rightarrow S_{1-m, 1-m}, \alpha c'_L \rightarrow \alpha c_l + \bar \alpha c_{\bar l}$. The same redefinition also works to define the twist \eqref{twc32} in $\mathfrak{B}_{4c}$.

%

\chapter{Applications and Phenomenology}\label{chap5}

In the last chapter all Lie bialgebra structures and (almost all) Hopf algebras that are compatible with the asymptotic symmetry algebras in three and four dimensions were constructed. Now, let us examine applications and consequences for these deformations some of which are motivated from the finite dimensional symmetry algebras and the $\kappa$-Poincar\'e in particular (cf. also chapter \ref{chap2}). Of special interest is the interplay of constraints on the deformations and the gravitational aspects of the asymptotic symmetries. Furthermore, novel applications that require the existence of the infinite dimensional symmetry algebra are investigated in the context of the black hole information loss paradox.
Even though the three dimensional spacetime does not fully capture the universe we live in it will again be a useful toy model that allows for many explicit calculations.

\section{Dual Lie Bialgebras and Non-commutative Spacetime}\label{sec5.1}

In section \ref{sec2.2.1} it was shown that the coordinates dual to a non-cocommutative Hopf algebra of momenta satisfy non-vanishing commutation relations. A direct interpretation of the dual of an infinite dimensional ``momentum algebra'' is of course hard to justify and the symmetry only holds asymptotically so we demand that a local observer that picks a vacuum (embedding) should find just the (deformed) spacetime dual to the finite embedding. This requirement will turn out to be non-trivial which is not surprising from a mathematical point of view, i.e. in general the set which is dual to a subalgebra is not necessarily a closed subalgebra itself.

Recall from section \ref{sec2.1.3} that for a finite dimensional Lie bialgeba $(\mathfrak{g}, [,], \delta)$ a dual can be constructed on the dual vector space by simply dualizing the bracket and cobracket. The dual Lie bialgebra is not necessarily coboundary also if $(\mathfrak{g}, \delta)$ is.
In the infinite dimensional case we will encounter two types of problems with performing this dualization. First, a general problem is that, given an infinite dimensional vector space $V$, the tensor product of its dual is not necessarily the dual of the tensorproduct, i.e. $V^* \otimes V^* \overset{?}{=} ( V \otimes V)^*$. Thus the cobracket $\delta^{\circ} \equiv [, ]^*: \mathfrak{g}^* \rightarrow ( \mathfrak{g} \otimes \mathfrak{g})^*$ might not be closed. To fix this, one can define a subspace $\mathfrak{g}^{\circ} \subseteq \mathfrak{g}^*$ as a sum of all so-called good subspaces, i.e. subspaces $\mathcal{S} \subset \mathfrak{g}^*$ such that $\delta^{\circ}(\mathcal{S}) \subseteq \mathcal{S} \otimes \mathcal{S}$ (see e.g. \cite{Taft1, Taft2, Griffing}). 
The second problem will be discussed in a concrete setting in the next section.

Looking at the examples in section \ref{sec2.2.1}, one can see that the first order deformation of the coproduct is enough to determine the dual brackets for the Lie algebra type non-commutativity and so we will be content with studying only the LBA and its duals in order to spare ourselves the burden of constructing full Hopf algebras for the dual spaces.

\subsection{Three Dimensional Case}

Consider a Lie bialgebra consisting of the BMS algebra in three dimension and a cobracket defined by a classical r-matrix from \eqref{classp1}-\eqref{classp2}. The generators of $\mathfrak{B}_3$ can be expressed in terms of monomials in a formal commuting parameter $a$ and its formal inverse (cf. e.g. \cite{Song:2019}), i.e. as a Laurent polynomials algebra from $\mathbb{C}[a^{\pm 1}]$
\begin{align}
l_m \equiv ( a^{m+1}, 0), \quad T_m \equiv ( 0, a^{m+1}).
\end{align}
For the complex, associative and commutative algebra $A \equiv \mathbb{C}[a^{\pm 1}]$ with basis $\{x^j \vert j \in \mathbb{Z} \}$ the algebraic dual $A^*$ admits a pseudo basis $\{\varepsilon^j \vert j \in \mathbb{Z} \}$ \cite{Taft2, Zhifeng}. This means that while each element $ f \in A$ can be decomposed into $f = \sum_n f_n a^n$ with only a finite number of non-zero coefficients, the decomposition $ A^* \ni h = \sum_n h_n \varepsilon^n$ has no restrictions on the coefficients. 
From \cite{Taft93} we have the following theorem.
\begin{theorem}
For the dual $A^*$ of $A \equiv \mathbb{C}[a^{\pm 1}]$ the good subspace $A^{\circ}$ is the subspace of all sequences $(h_n)_{n \in \mathbb{Z}}$ satisfying linearly recursive relations of the form $h_n = \sum_{j=1}^r c_j h_{n-j}$ for some $1 \leq r \in \mathbb{N}$ and some coefficients $c_1, ..., c_r \in \mathbb{C}, c_r \neq 0$. 
\end{theorem}

Since for a direct sum $B = B_1 \oplus B_2$ the good subspace is given by $B^{\circ} = B^{\circ}_1 \oplus B^{\circ}_2$ and $\mathfrak{B}_3 \sim A \oplus A$ we have $\mathfrak{B}^{\circ}_3 \sim A^{\circ} \oplus A^{\circ}$. Dual to the pseudo basis of $\mathfrak{B}_3$ we set
\begin{align}
\zeta^m \equiv ( \varepsilon^{m+1}, 0), \quad \chi^m \equiv (0, \varepsilon^{m+1}),
\end{align}
which is technically only valid in the one-sided BMS (or Witt) algebra. This is due to the second problem, namely that the dual to the Witt algebra bracket is given by
\begin{align}
\delta^{\circ}( \varepsilon^m ) = \sum_{i+j =m} (j-i) \varepsilon^i \otimes \varepsilon^j,
\end{align}
which is only an expression with finitely many terms if the indices are restricted to e.g. $i,j \leq 1$. Upon closer inspection we find that it is only necessary to consider the one-sided part of the superrotation sector and the supertranslations and their duals could have indices from $\mathbb{Z}$. This is in contrast with the requirements from the specialization for certain deformations but it still would single out the $n=1$ embedding.
 In the following we shall only focus on the supertranslations and calculate the dual algebra for specific deformations based on the Lie bialgebras discussed in chapter \ref{chap4}.

\subsubsection*{Abelian Twist}

The cobracket associated with the abelian twist (defined by eq.\eqref{abtw} with $\eta = \frac{i}{n \kappa}$)
\begin{align}
\delta_A(l_m) = m \frac{i}{n \kappa} (l_m \wedge T_0 + l_0 \wedge T_m), \quad \delta_A (T_m) = m\frac{i}{n \kappa} T_m \wedge T_0, 
\end{align}
leads to
\begin{align}
\braket{ \chi^k \otimes \chi^j, \delta_A(T_m)} & = \braket{[\chi^k, \chi^j]_A, T_m}, \\
\rightarrow [\chi^k, \chi^j]_A & = \frac{i}{\kappa} ( k \delta^j_0 \chi^k - j \delta^k_0 \chi^j). \label{duab}
\end{align}
Considering an embedding with the dual coordinates
\begin{align}\label{embco}
x^{\pm} \equiv - i \sqrt{2} \chi^{\pm n}, \quad x^2 \equiv i \chi^0,
\end{align}
results in
\begin{align}
[x^+, x^2]_A = \frac{1}{\kappa} x^+, \quad [x^-, x^2] = - \frac{1}{\kappa} x^-,
\end{align}
with all other commutators vanishing. Thus, the three dimensional (deformed) Minkowski subalgebra closes as required.

\subsubsection*{Jordanian Twist}

Similarly for the Jordanian twist (defined by eq. \eqref{jtwx} with $\chi = \frac{i}{\sqrt{2}\kappa}$)
\begin{align}
\delta_J(l_m) = \frac{i}{n\sqrt{2}\kappa} ( m l_m \wedge T_n + (m-n) l_0 \wedge T_{n+m}, \quad \delta_J (T_m) = \frac{i}{n\sqrt{2}\kappa} m T_m \wedge T_n, 
\end{align}
is dual to 
\begin{align}
[\chi^k, \chi^j]_J = \frac{i}{n\sqrt{2}\kappa} (k \delta^j_n \chi^k - j \delta^k_n \chi^j)
\end{align}
and the deformed Minkowski algebra closes again as
\begin{align}
[x^+, x^-]_J = - \frac{1}{\kappa} x^-.
\end{align}

\subsubsection*{Extended Jordanian Twist}

In the case of the extended Jordanian (or light-cone) twist (defined by eq.\eqref{ejtwx} with $\chi = \frac{i}{\sqrt{2}\kappa}$) one has
\begin{align}
\delta_{eJ} (l_m) & = \delta_J(l_m) - \frac{i}{n \sqrt{2}\kappa} ( (m-n) l_{m+n} \wedge T_0 + m l_n \wedge T_m) , \\
\delta_{eJ} (T_m) & = \delta_J(T_m) - \frac{i}{n \sqrt{2}\kappa} (m-n) T_{m+n} \wedge T_0, \\
[\chi^k, \chi^j]_{eJ}&  =  \frac{i}{n \sqrt{2}\kappa} ( k \delta^j_n \chi^k - j \delta^k_n \chi^j - (k-2n) \delta^j_0 \chi^{k-n} + (j-2n) \delta^k_0 \chi^{j-n},
\end{align}
yielding
\begin{align}\label{kmink31}
[x^+, x^-]_{eJ} & = - \frac{1}{\kappa} x^-, \quad [x^+, x^2]_{eJ} = -\frac{1}{\kappa} x^2, \\
[x^-, x^2]_{eJ} & = - \frac{3i}{\kappa} \chi^{-2n}.\label{kmink32}
\end{align}
This subalgebra does not close and even the restriction to the one-sided $\mathfrak{B}_3$ can not remedy this fact. Note that the other one-sided subalgebra with indices $\geq -1$ would close the algebra but then the construction of the q-analog were not possible.

When we compare \eqref{kmink31}-\eqref{kmink32} with the three dimensional light-cone $\kappa$ Minkowski algebra from section \ref{sec2.2.1} (i.e. with $\tau^2 =0$), it can be seen that they are identical up to the extra term $\sim \chi^{-2n}$ that leads to the algebra being not closed. Thus, one of our main motivations, to generalize the $\kappa$-Poincar\'e to the infinite dimensional asymptotic symmetry algebras can not be realized in such a way that it remains a sub (Hopf) algebra in momentum space and its dual.

Furthermore, it is noteworthy that all other deformations (from r-matrices that correspond to Lie algebra type non-commutativity (NC)) which do not admit a q-analog on the full algebra also do not close the dual Minkowski subalgebra.

\subsubsection*{Automorphisms on the Dual Algebra}

A potential caveat could arrise by using duals of automorphisms. In general one has
\begin{align}
\braket{ \delta_{\tilde \varphi(r)}\varphi(T), \chi \otimes  \chi' } & =  \braket{\tilde \varphi (\delta_r (T)), \chi \otimes \chi'}  \nonumber \\
& = \braket{T, [\varphi^{\dagger} (\chi), \varphi^{\dagger} (\chi')]_r}, 
\end{align}
where $\tilde \varphi = \varphi \otimes \varphi$, $\varphi^{\dagger}$ is the adjoint of $\varphi$ and the lhs is also equal to 
\begin{align}
\braket{ \delta_{\tilde \varphi(r)}\varphi(T), \chi \otimes  \chi' }  = \braket{ T, \varphi^{\dagger}( [\chi, \chi']_{\tilde \varphi (r)})},
\end{align}
so that
\begin{align}\label{duau}
 [\chi, \chi']_{\tilde \varphi (r)} =\left(\varphi^{\dagger} \right)^{-1}[\varphi^{\dagger} (\chi), \varphi^{\dagger} (\chi')]_r.
\end{align}

For the r-matrix $r = (l_1 - l_{-1}) \wedge (T_1 - T_{-1}) \subset r_3$ with
\begin{align}
\delta_r (T_m) & = (m-1) T_{1+m} \wedge (T_1 - T_{-1}) - (m+1) T_{m-1}  \wedge (T_1 - T_{-1}) , \\
[\chi^k, \chi^j]_r & = (k-2) \chi^{k-1} (\delta^j_1 - \delta^j_{-1}) - (k+2) \chi^{k+1}  (\delta^j_1 - \delta^j_{-1}) - (k \leftrightarrow j) \label{dur3}
\end{align}
one could apply the $\mathfrak{P}_3$ automorphism
\begin{align}
\varphi (l_{\pm 1} ) & = - \frac{1}{2} (l_1 + l_{-1}) \pm l_0, \quad \varphi (l_0) = \frac{1}{2} (l_1 - l_{-1}), \\
\varphi (T_{\pm 1} ) & = - \frac{1}{2} (T_1 + T_{-1}) \pm T_0, \quad \varphi (T_0) = \frac{1}{2} (T_1 - T_{-1}),
\end{align}
so that $\varphi (r) = l_0 \wedge T_0$. From the dual automorphism 
\begin{align}
\varphi^{\dagger} (\chi^{\pm 1}) = - \frac{1}{2} ( \chi^1 + \chi^{-1}) \pm \chi^0
\end{align}
and \eqref{duau}, \eqref{duab}
\begin{align}\label{ducl}
[\chi^1, \chi^{-1}]_r & = 2 \chi^0, \quad [\chi^1, \chi^0]_r = 2 ( \chi^1 + \chi^{-1}), \\
[\chi^{-1}, \chi^{0}]_r & =   -2 ( \chi^1 + \chi^{-1}) 
\end{align}
follows. Thus, the deformed Minkowski algebra closes in contrast to \eqref{dur3} leading e.g. to
\begin{align}
[\chi^1, \chi^{-1}]_r  = 2 \chi^0 + 3( \chi^{2} + \chi^{-2}).
\end{align}
However, $\varphi$ does not extend to an automorphism of $\mathfrak{B}_3$ and consequently the closing subalgebra \eqref{ducl} is not dual to the whole three dimensional BMS.

\subsection{Four Dimensional Case}\label{sec5.1.2}

In a completely analogous way one can construct the dual algebra to the supertranslation sector of $\mathfrak{B}_4$. We first define the dual pseudo basis as
\begin{gather}
\chi^{pq}_S, \quad \chi^{pq}_A, \\
\braket{S_{kj}, \chi^{pq}_S} = \delta^p_k \delta^q_j, \quad \braket{A_{kj}, \chi^{pq}_A} = \delta^p_k \delta^q_j,
\end{gather}
with $p \leq q$. 

A deformation with the Jordanian twist (defined in eq.\eqref{4jtwx} with $\chi = \frac{-i\sqrt{2}}{\kappa}$) leads to the cobrackets
\begin{align}
\delta_J(k_p) & =  -ip k_p \wedge S_{1-m,1-m} - i(p-n) S_{1-m+p, 1-m}, \\
\delta_J(\bar k_p) & = -i p \bar k_p \wedge S_{1-m,1-m} - i(p-n) A_{1-m+p, 1-m}, \\
\delta_J( S_{pq}) & = i(1-p-q)S_{pq} \wedge S_{1-m, 1-m}, \quad \delta_J (A_{pq}) =0,
\end{align}
which in turn imply
\begin{align}
[\chi^{kj}_S, \chi^{k' j'}_S]_J = -i(1-k-j) \chi^{kj}_S \delta^{k'}_{1-m} \delta^{j'}_{1-m} - ( (kj) \leftrightarrow (k' j')).
\end{align}
With 
\begin{align}
x^+ \equiv \frac{i}{\sqrt{2}} \chi^{1-m, 1-m}_S, \quad x^- \equiv - \frac{i}{\sqrt{2}} \chi^{m m}_S, \quad x^1 = i \chi^{m, 1-m}_S, \quad x^2 = i \chi^{m, 1-m}_A
\end{align}
we finally obtain the closed deformed Minkowski algebra
\begin{align}
[x^+, x^-]_J = - \frac{1}{\kappa} x^-.
\end{align}

The cobracekts for the extended Jordanian twist (defined by eq.\eqref{4ejtwx} with $\chi = \frac{-i \sqrt{2}}{\kappa}$) reads
\begin{align}
\delta_{eJ} (S_{pq}) = & \delta_J (S_{pq}) -i\left( \frac{n+1}{2} - p \right) S_{p+n, q} \wedge S_{1-m, m} -i \left( \frac{n+1}{2} - q \right) S_{p, q+n} \wedge S_{1-m, m} \nonumber \\
& +i \left( \frac{n+1}{2} - p \right) A_{p+n, q} \wedge A_{1-m, m} -i   \left( \frac{n+1}{2} - q \right) A_{p, q+n} \wedge A_{1-m, m}, \\
\delta_{eJ} (A_{pq}) = &  -i \left( \frac{n+1}{2} - p \right) A_{p+n, q} \wedge S_{1-m, m} -i \left( \frac{n+1}{2} - q \right) A_{p, q+n} \wedge S_{1-m, m} \nonumber \\
& -i \left( \frac{n+1}{2} - p \right) S_{p+n, q} \wedge A_{1-m, m} -i   \left( \frac{n+1}{2} - q \right) S_{p, q+n} \wedge A_{1-m, m}.
\end{align}
This is dual to 
\begin{align}
[\chi^{kj}_S, \chi^{k' j'}_S]_{eJ} = & [\chi^{kj}_S, \chi^{k' j'}_S]_{J} + i\bigg( \left( \frac{n+1}{2}-(k-n) \right) \chi^{k-n, j}_S \delta^{k'}_{1-m} \delta^{j'}_m  \nonumber \\
& + \left( \frac{n+1}{2}-(j-n) \right) \chi^{k, j-n}_S \delta^{k'}_{1-m} \delta^{j'}_m  - ((kj) \leftrightarrow (k'j')) \bigg) , \\
[\chi^{kj}_A, \chi^{k' j'}_A]_{eJ} = & i \bigg(- \left( \frac{n+1}{2}-(k-n) \right) \chi^{k-n, j}_S \delta^{k'}_{1-m} \delta^{j'}_m  \nonumber \\
& + \left( \frac{n+1}{2}-(j-n) \right) \chi^{k, j-n}_S \delta^{k'}_{1-m} \delta^{j'}_m  - ((kj) \leftrightarrow (k'j')) \bigg) , \\
[\chi^{kj}_A, \chi^{k' j'}_S]_{eJ} = & i \bigg( \left( \frac{n+1}{2}-(k-n) \right) \chi^{k-n, j}_A \delta^{k'}_{1-m} \delta^{j'}_m  \nonumber \\
&+ \left( \frac{n+1}{2}-(j-n) \right) \chi^{k, j-n}_A \delta^{k'}_{1-m} \delta^{j'}_m  - ((kj) \leftrightarrow (k'j')) \bigg) ,
\end{align}
and thus we obtain
\begin{gather}
[x^+, x^-]_{eJ} = \frac{1}{\kappa} x^-, \quad [x^+, x^1]_{eJ} = \frac{1}{\kappa} x^1, \\
[x^+, x^2]_{eJ} = \frac{1}{\kappa} x^2, \quad [x^-, x^1]_{eJ} = \frac{2i}{\kappa} \chi_S^{3m-1, m}, \\
[x^-, x^2]_{eJ} = \frac{2i}{\kappa} \chi_A^{3m-1, m},
\end{gather}
so the Minkowski algebra does not close as in the three dimensional case.

In the previous section we remarked that deformations that do not admit a specialization on the full algebra do not close the dual Minkowski subalgebra which is also true in four dimensions. Furthermore, the converse is only true if we insist on the requirement that only one embedding is present in the r-matrix. Otherwise there is the counterexample \eqref{mixem} which admits a q-analog on $\mathfrak{B}_4$ but does not close the Minkowski sector for $p \neq p'$.

Thus, we can conclude that imposing the constraint that the sector dual to the Poincar\'e subalgebra has to be a closed subalgebra as well would, in combination with the specialization requirement, rule out all (but in four dimensions not only) deformations which can be specialized only on the one-sided BMS algebra.

\subsubsection*{Other Types of Non-commutativity}

Recall from section \ref{sec2.2.2} that also other types of non-commutativity are possible. In particular, r-matrices of the form $T \wedge T$ lead to canonical NC. For example from $r = \Theta T_0 \wedge T_n$ with 
\begin{align}
\mathcal{F}^{-1}_r \equiv \bar f^{\alpha} \otimes \bar f_{\alpha} = \exp \left( -\frac{\Theta}{2} T_0 \wedge T_n \right)
\end{align}
one obtains
\begin{align}
[\chi^p, \chi^q] & = \left( \bar f^{\alpha} \triangleright \chi^p \right) \left( \bar f_{\alpha} \triangleright \chi^q \right) - (p \leftrightarrow q)  =- \Theta (\delta^p_0 \delta^q_n - \delta^p_n \delta^q_0), \\
 \rightarrow [x^2, x^+]  & = - \sqrt{2} \Theta , \quad [x^+, x^-] = [x^2, x^-] =0.
\end{align}
Here we have used the definition \eqref{starnc} and \eqref{embco}. In this specific example the deformed Minkowski algebra closes which is not necessarily true for arbitrary r-matrices. Since the supertranslations commute all r-matrices are mathematically allowed. Only in combination with terms of the form $l \wedge T$ there are restrictions for the canonical NC (zeroth order in \eqref{expth}). Unfortunately, already in the three dimensional case we do not know the explicit form of the twist for those r-matrices. 

Similarly, the classifications \eqref{classp1}-\eqref{classp2} and \eqref{appb1}-\eqref{appb2} allow us to infer constraints on mixed Lie algebra type NC and quadratic NC\footnote{Cf. \cite{Lukierski:2005fc} for detailed construction in the Poincar\'e algbra.}. From the former it is easy to see that in the three dimensional case there are no r-matrices that would result in such mixed NC, whereas in the four dimensional asymptotic symmetry algebras they are possible but much more restricted than each of the NC individually. The same is also true for combinations of canonical and quadratic NC and all three types of NC are not allowed simultaneously in three and four dimensions.
Note that some of these constraints are direct consequences of the triangularity requirements. For example, in the finite dimensional $\mathfrak{P}_3$ where also quasitriangular r-matrices exist combinations of quadratic and Lie algebra NC are allowed (cf. the classification of \cite{Stachura_1998}). Similarly, it is known that combinations of all three types of NC are possible in $\mathfrak{P}_4$ with quasitriangular r-matrices (cf. \cite{Zak97}).

Concerning the centrally extended algebra of the surface charges, it is easy to see that contributions to the cobrackets containing central charges do not alter the dual brackets as we have $\braket{c_l, \chi^p} = 0$. 
  
In principle, non-commutative spacetime coordinates could have directly observable consequences like uncertainty relations (implying a minimal unvertainty) (see e.g. \cite{Sasakura:1999xp} or \cite{Bolonek:2002cc}) but these would be out of reach for experiments in the near future. From a conceptional point of view the idea also adopted in this thesis is that the entire concept of spacetime breaks down at the smallest scales and is only valid as an emergent phenomenon. The dual momentum quantum group is the more fundamental and directly accessible object and potentially also allows to test deformations in current experiments as we explore in the following sections. 

\section{Quantum Lie Algebra and Deformed Dispersion Relations}

Recall from section \ref{sec2.1.10} that the property \eqref{dcprop} of the deformed coproduct in the so-called quantum Lie algebra basis is crucial to define a deformed differential calculus and a deformed wave operator expressed in the formulation admitting said property. As all the deformations we studied in chapter \ref{chap4} are given by a twist, we can readily construct the isomorphism $D$ and calculate the wave operator of any given Poincar\'e subalgebra, i.e.
\begin{align}
\Box^{\mathcal{F}} = D (\Box) = P_{\mu}^{\mathcal{F}} P^{\mu \, \mathcal{F}}.
\end{align}
A scalar field with mass $m$ in the non-commutative spacetime described by a wave function $\phi (p)$  in momentum space then satisfies
 \begin{align}
\left( \Box^{\mathcal{F}} + m^2 \right) \phi (p) = 0,
\end{align}
leading to a potentially deformed vacuum dispersion relation. If the non-commutativity is seen as an effective phenomenon from QG one could interpret these effects on the propagation in vacuum as arising from interactions with the non-trivial structure (at the QG scale) of spacetime itself.

For all generators of the Poincar\'e algebra we have 
\begin{align}\label{qc1}
[P_{\mu}^{\mathcal{F}}, \Box^{\mathcal{F}}]_{\star} & \equiv P_{\mu \, (1^{\mathcal{F}})}^{\mathcal{F}} \Box^{\mathcal{F}} S_{\mathcal{F}} (P_{\mu \, (2^{\mathcal{F}})}) = 0, \\
[M_{\mu \nu}^{\mathcal{F}}, \Box^{\mathcal{F}}]_{\star} & =0,\label{qc2}
\end{align}
i.e. the wave operator is the quadratic Casimir of the quantum Lie algebra.
Since \eqref{qc1}-\eqref{qc2} obviously also holds for the undeformed wave operator and the undeformed adjoint action, it follows that $\Box^{\mathcal{F}}$ is always undeformed if the $n=1$ embedding is used for the twist. 
Only if we allow for deformations that include different embeddings there are potentially differences in the wave operator involving generators of soft charges and thus the wave equation can be influenced by soft charges.


\subsection*{Three Dimensional Case}

The most trivial example is the abelian twist deformation in three dimensions (cf. section \ref{sec4.2.4}). Since it does not single out an embedding the quadratic Casimir has to be undeformed by the previous remarks. Indeed, expressing it in terms of the algebra generators\footnote{Summation convention applies to ``relativistic indices'' (greek letters) but not to ``BMS indices'' (latin letters).}
\begin{align}
P_{\mu } P^{\mu} = T_n T_{-n} - T_0^2 
\end{align}
and calculating
\begin{align}
D_A(T_m) = T_m e^{m \eta T_0},
\end{align}
we find
\begin{align}
D_A (\Box) = T_n T_{-n} - T_0^2 = \Box.
\end{align}

For the Jordanian twist deformation we know that one could consistently use different embeddings but from
\begin{align}
D_J (T_m) = T_m \Pi_{+, n'}^{\frac{m}{n'}}
\end{align}
where $\Pi_{+, n'} = (1 + \chi T_{n'})$, we find that
\begin{align}
D_J ( \Box) = \Box
\end{align}
again, i.e. the wave operator is undeformed independently from the embedding.

Our third standard example, the light-cone $\kappa$-Poincar\'e defined on the BMS with an extended Jordanian twist, requires the restriction to the one-sided $\mathfrak{B}_{3 -}$ for the specialization of the deformation parameter (cf. section \Ref{sec4.2.4}) which only admits one Poincar\'e subalgebra resulting in an undeformed $\Box^{\mathcal{F}}$. In general the quantum Lie algebra generators can be calculated from
\begin{align}
D_{eJ} (T_m) & = \sum_{k,k' = 0}^{\infty} \frac{1}{k ! k'!} ( - \chi)^{k'} (l_0^k l_{n'}^{k'}) \triangleright (T_m)  (\log (1 + \chi T_{n'}) )^k T_0^{k'}, \\
l_{n'}^{k'} \triangleright T_m & = \prod_{r=1}^{k'-1} (n' - (m +r n')) T_{m + k'},
\end{align}
which also contains infinitely many generators for unrestricted $m$. Ignoring the technical constraints for a moment and treating the deformation as an effective theory, one has in first order of the deformation parameter
\begin{align}
D_{eJ} (T_m) = T_m - \chi (n'-m) T_{n'+m} T_0 - \chi m T_m T_{n'} + \mathcal{O} (\chi^2)
\end{align}
resulting in 
\begin{align}\label{defbox}
D_{eJ} (\Box) = \Box - \chi \left((n'-n) T_{n+n'} T_0 T_{-n} +(n'+n) T_{-n + n'} T_0 T_n  +2 n' T_{n'} T_0^2 \right) + \mathcal{O} (\chi^2).
\end{align}
For $n' =n =1$ this is just the undeformed $\Box$ as it should be but otherwise the wave operator is deformed. Unlike the scenario in \cite{Aschieri:2017ost}, where a dilation operator is included in the algebra and used for deformation,  $\Box^{\mathcal{F}}$ is not proportional to $\Box$ and thus the wave equation would be also altered for massless fields.

Including the central charges does not change the conclusions considerably. In that case only the $n=1$ embedding has a quadratic Casimir operator and if we only permit deformations which allow for q-analog on the full $\mathfrak{B}_{3c}$ algebra $\Box^{\mathfrak{F}}$ is undeformed. For the non-coboundary LBA and the corresponding Hopf algebra it is not even clear how to define a deformed Casimir operator.

\subsection*{Four Dimensional Case}

In the four dimensional spacetime there is no abelian twist that does not single out an embedding and corresponds to non-commutative geometry of Lie algebra type (see section \ref{sec4.3.3}). For the Jordanian and extended Jordanian twist the results are qualitatively similar to the three dimensional scenario. In particular, for the Jordanian twist from \eqref{4jtwx} we find
\begin{align}
D_J (S_{pq}) = S_{pq} \Pi_{+, 1-m}^{\frac{p+q-1}{n'}}
\end{align}
and the quadratic Casimir for a given embedding
\begin{align}
\Box \equiv P_{\mu} P^{\mu} = S_{1-m,1-m} S_{m m} - S_{1-m, m}^2 - A_{1-m, m}^2
\end{align}
is undeformed, i.e.
\begin{align}
D_J (\Box ) = \Box.
\end{align}

The light-cone $\kappa$-Poincar\'e in four dimensions from the twist \eqref{4ejtwx} is not well defined in all orders for general  elements $S_{pq}, A_{pq}$ but in a first order approximation one obtains 
\begin{align}
D_{eJ} (S_{pq}) = & D_J (S_{pq}) + \chi \left( \left( \frac{n'+1}{2} -p \right) S_{p+n', q} + \left( \frac{`n+1}{2} -q \right) S_{p, ,q+n'} \right) S_{1-m', m'} \nonumber \\
&- \chi \left( \left( \frac{n'+1}{2} -p \right) A_{p+n', q} - \left( \frac{n'+1}{2} -q \right) A_{p, ,q+n'} \right) A_{1-m',m'}, \\
D_{eJ} (A_{pq}) = & D_J (A_{pq}) + \chi \left( \left( \frac{n'+1}{2} -p \right) A_{p+n', q} + \left( \frac{n'+1}{2} -q \right) A_{p, ,q+n'} \right) S_{1-m',m'}\nonumber \\
& - \chi \left(- \left( \frac{n'+1}{2} -p \right) S_{p+n', q} + \left( \frac{n'+1}{2} -q \right) S_{p, ,q+n'} \right) A_{1-m',m'}, 
\end{align}
leading to the deformed wave operator
\begin{align}
D_{eJ} ( \Box) = & \Box + \chi ( (n'-n) S_{1-m+n', 1-m} S_{1-m', m'} - (n'-n) A_{1-m+n',1-m} A_{1-m',m'} )S_{m, m} \nonumber \\
& + \chi ( (n'+n) S_{m+n', m} S_{1-m',m'} - (n'+n) A_{m+n',m} A_{1-m',m'}) S_{1-m, 1-m}  \nonumber \\
&- \chi (( (n'-n) S_{1-m+n', m} + (n'+n) S_{1-m, m+n'})S_{1-m',m'}  \nonumber \\
& - ((n'-n)A_{1-m+n',m} -(n'+n) A_{1-m,m+n'})A_{1-m',m'} ) S_{1-m, m}  \nonumber \\
&-  \chi (( (n'-n) A_{1-m+n', m} + (n'+n) A_{1-m, m+n'})S_{1-m',m'}  \nonumber \\
&- ( -(n'-n)S_{1-m+n',m} +(n'+n) S_{1-m,m+n'})A_{1-m',m'} ) A_{1-m, m}.\label{defdis5}
\end{align}
For $n'=n$ this expression vanishes and otherwise the deformation is not purely multiplicative as in three dimensions.

Without explicitly presenting the calculations\footnote{In four dimensions only twists from r-matrices of the form \eqref{mixem} have to be checked additionally.} we note that for all deformations associated to Lie algebra type non-commutativity the wave operator is either undeformed for all embeddings or the deformation can not be consistently performed in all orders if we insist on the specialization.

\subsection{Phenomenology of Deformed Dispersion Relations}

Deformed dispersion relations are one of the most studied consequences of various versions of non-commutativity as a QG effect because of the possibility to offset the suppression of the effect size by the smallness of the QG length scale $l_{\text{{QG}}} = 1/\kappa$ with large distances over which e.g. cosmic rays can travel as we will see below. In the context of $\kappa$-Poincar\'e deformed dispersion relations can arise for example when a basis is chosen such that the deformation affects also the algebra (cf. section \ref{sec2.2}) and consequently the Casimir operator of that algebra is deformed. Note that this is different from the choices made in this thesis where the algebra of symmetries itself is undeformed.

Also other frameworks like Loop Quantum Gravity or theories involving Lorentz invariance violation appear to predict deformed dispersion relations and usually (but not always, as we shall see) this results in an energy dependent speed of light $v(E) \equiv d E/ dp$ \cite{Amelino-Camelia:2008aez}. A model independent ansatz (in this limited sense) \cite{Amelino-Camelia:1997ieq}
\begin{align}\label{defdisan}
{\it{p}}^2 = E^2 \left( 1 + \xi (l_{\text{{QG}}} E)^{\alpha} + \mathcal{O} (l_{\text{{QG}}} E)^{\alpha +1} \right) , \\
\Rightarrow v(E) \approx 1 + \xi \frac{1-\alpha}{2} (l_{\text{{QG}}} E)^{\alpha}, 
\end{align}
with the parameters $\alpha \in \mathbb{N}$ and $\xi = \pm 1$ leads to the time delay at the detector of photons of different energy emitted simultaneously from a source at distance $L$
\begin{align}
\vert \Delta t \vert = L  \left \vert \frac{1}{v(E)} - \frac{1}{v(E + \Delta E)} \right \vert \approx L \left \vert \frac{(1- \alpha)\alpha l_{\text{{QG}}}^{\alpha} E^{\alpha-1}}{ 2} \right \vert \Delta E.
\end{align}
From this formula, one can see that a larger effect size can be achieved either by observing very high energy photons or having a large distance to the source. The former is very hard to obtain as it is currently not possible to produce them terrestrially. Even though high energy photons from astronomical sources are rare, they have the advantage that also the distance they traveled can be gigantic.

Some data has been collected for example on high-energy Gamma-ray bursts \cite{Fermi-LAT:2009owx}, \cite{Fermi-LAT09} and analyzed for signatures of deformed dispersion relations \cite{Abdo09}. While the raw accuracy of the experiments should already be high enough to detect such signatures, a big conceptual problem remains, namely that the emission process in astronomical sources is not understood very well (see \cite{Amelino-Camelia:2016ohi} and references therein). This makes it currently impossible to estimate whether photons were emitted sufficiently simultaneously but an improved understanding or larger effect sizes in the near future could make it possible to test deformed dispersion relations of the form \eqref{defdisan}.

In the previous section it was already noted that the twist performed in \cite{Aschieri:2017ost} causes a multiplicative deformation of $\Box$ and thus massless fields have to satisfy an undeformed equation of motion. Only massive fields would obey a deformed dispersion relation in such a scenario.

The only scenarios from all deformations we discussed in chapter \ref{chap4} that predict an effect in the dispersion relation (which is only possible if the specialization requirement is relaxed) would not just cause a multiplicative change in $\Box$. However, the deformed wave operator (e.g. \eqref{defbox}, \eqref{defdis5}) contains generators of supertranslations and in analogy to the calculations in \eqref{defdisan} we might compute $v(E, T)$. For example from the dispersion of a massless field with \eqref{defbox} which moves in the $P_2$ direction $(P_1 =0)$ it follows that
\begin{align}
v_2 \equiv \frac{d E}{d P_2} \approx \pm 1  \pm \chi  4 n' T_{n'}  + \frac{\chi}{2} [(n'+n) T_{n'-n} + (n'+n) T_{n'+n}] + \mathcal{O}\left(\chi^2\right) .
\end{align}
Thus, any deviations in the time of flight would not be depending on the energy of the quanta but on soft charges which makes this effect practically undetectable.

\section{Deformed Leibniz Rule and Black Hole Information}

An important application of a (deformed) coproduct on momentum space is that it induces a recipe for the addition of the total momentum of multi-particle states. This aspect will be explored in the following by applying it not only to the classical ``hard'' momenta but also to supertranslation and superrotation charges introduced in section \ref{sec3.1.1}. We then investigate a potential connection to the proposal by Hawking, Perry and Strominger \cite{Hawking:2016msc} that superrotation charges can (at least partially) solve the black hole information loss paradox.

\subsection{Addition of Momenta, Supertranslation/-rotation Charges}

With the additional structure of a Hopf algebra with primitive coproduct on the Poincar\'e algebra one can write the usual simple addition of momenta as the consequence of eq.\eqref{ten-rep}. To be more explicit, let us denote a single-particle state  $\ket{p_{\mu}}$ by its momentum eigenvalue (it may also have other discrete quantum numbers which we ignore for the moment) and in the momentum space representation the momentum operator $P_{\mu}$ acts via
\begin{align}
P_{\mu} \triangleright\ket{ p_{\nu}} = p_{\mu}\ket{p_{\nu}}
\end{align}
and
\begin{align}\label{defleib}
P_{\mu} \triangleright \left( \ket{p_{1, \nu}} \otimes \ket{p_{2, \nu}} \right) = P_{\mu (1)} \triangleright \ket{p_{1, \nu}} \otimes P_{\mu (2)} \triangleright \ket{p_{2, \nu}} = (p_{1, \mu} + p_{2 , \mu} ) \left( \ket{p_{1, \nu}} \otimes \ket{p_{2, \nu}} \right) 
\end{align}
for two-particles states, where the primitivity of the coproduct was used in the last equality. In other words, the total momentum of a multi-particle state is just the sum of the single-particle states. However, the addition might involve other, non-linear terms if the coproduct is deformed. This feature, sometimes referred to as deformed Leibniz rule, was discussed for example in \cite{Lukierski:2002wf}, \cite{Bruno:2001mw} and more recently also in \cite{Aschieri:2017ost}. In some scenarios the addition rules lead to a saturation at the QG scale (a maximal momentum) as discussed e.g. in \cite{Bruno:2001mw} and ideas to use this feature for automatic UV regularization were early motivations to study deformed symmetries \cite{Snyder47}.

Another peculiarity about the deformed Leibniz rule is the apparent lack of symmetry under the exchange of particles (via the flip of the tensorproduct) resulting from a non-symmetric coproduct. But recall from section \ref{sec2.1.7} that in the deformed setting the representation of the braid group is not given by the flip but instead dictated by a quantum R-matrix \cite{Aschieri:2017ost}
\begin{align}
\ket{p_1} \otimes \ket{p_2} \mapsto \tilde R^{-1} (\ket{p_1} \otimes \ket{p_2}) = \bar R^{\alpha} \triangleright  \ket{p_2} \otimes   \bar R_{\alpha} \triangleright  \ket{p_1},
\end{align}
restoring some form of symmetry.
Here $R^{-1} \equiv \bar R^{\alpha} \otimes \bar R_{\alpha} = \mathcal{F} \mathcal{F}^{-1}_{21}$ is known from the twist of the deformation.

When using the standard Leibniz rule to add up momenta, one does not need to care whether the individual momenta already belong to composite systems or elementary particles. Such a prescription applied to the deformed Leibniz rule would immediately lead to a contradiction with experiment because the momenta of macroscopic objects (as e.g. a soccer ball) can easily be bigger than the mass scale $\kappa \sim m_{\text{Pl}}$ that might appear for a $\kappa$-deformed coproduct in \eqref{defleib} leading to noticeable differences. For elementary particles on the other hand momenta of this scale are unobtainable in current experiments by orders of magnitude and the deviations for the resulting total momentum would be minuscule. The necessity to distinguish between elementary and composite states is, however, a big conceptual issue discussed in the literature under the term soccer-ball problem (see \cite{Kowalski-Glikman:2004fsz}, \cite{Amelino-Camelia:2013zja}, \cite{Amelino-Camelia:2011dwc} and also \cite{Hossenfelder:2012vk}).

All the deformations discussed in this thesis imply a deformed Leibniz rule (see section \ref{sec5.3.3} for two examples) which (depending on the embedding for the deformation) persists also if only the Poincar\'e sector is considered.

\subsection{Black Hole Information Loss Paradox and Soft Hair Solution}\label{sec5.3.2}

Black holes, astronomical objects so dense that not even light rays can escape their gravitational pull beyond the event horizon, are a prime example of a phenomenon where the necessity of a theory combining gravitational as well as quantum mechanical features becomes apparent. For once, this is due to the singularity of physical quantities like the curvature at the center of a black hole as predicted by GR. Another, more subtle point is that Hawking showed in 1974 \cite{Haw74} that black holes, contrary to what was previously believed, lose mass in the form of thermal radiation called Hawking radiation. Hawking used a semi-classical approach, meaning that the background geometry of the black hole is described by classical GR and the matter/radiation fields in the vicinity of the horizon by a quantum field theory. This simplification seems to be justified as the gravitational tidal forces are weak at the horizon for large black holes. In the original calculation the spectrum of the Hawking radiation is completely thermal, i.e. it is described by a mixed state. Since we can form a black hole out of matter from which we have perfect information, i.e. a pure state, the decay into thermal radiation and a unique vacuum is incompatible with the unitary (therefore reversible) evolution in quantum mechanics of the matter/black hole/radiation that would evolve pure states to pure states and this tension is called the black hole information loss paradox. A central argument for the paradox to persist is the so-called no-hair theorem stating that a black hole is characterized solely by the Poincar\'e momenta and gauge charges. In particular, multipole moments of infalling matter settle into a completely spherical black hole configuration long before the process of Hawking radiation starts if the black hole is large enough because the power radiated away is inversely proportional to the cube of the mass \cite{Bek73}, \cite{Haw75}. 

Several proposed solutions of the black hole information loss paradox have been put forward as e.g. recently in \cite{Almheiri:2019qdq} but the debate remains, so far, inconclusive. As stated above, here we will only consider the proposal by Strominger et. al.

\subsubsection*{Poincar\'e Hair}

The existence of the ten Poincar\'e charges does introduce correlations between early and late radiation but it is by far not enough to purify the state completely. These correlations are not included in Hawkings calculation and lead to a slight departure from a completely thermal spectrum. As a warm-up for the soft charges, we explicitly describe the hard Poincar\'e charges as follows \cite{Mirbabayi:2016axw}. We start with a black hole with mass $M$ described by $\ket{M}$ from which some early quanta with total mass $m$ were emitted (or alternatively in which an object of mass $m$ was dropped) and that decays into the configuration $\ket{X}$, i.e.
\begin{align}
\mathcal{S} \ket{M} = \ket{X} = \sum_{p} \mathcal{S}_{M \rightarrow p} \ket{p},
\end{align}
where $\mathcal{S}$ is a unitary S-matrix and $\ket{p} = \prod_{i} a^{\dagger}_{p_i} \ket{0}$ are radiation modes with measurable momentum $p_i$ created by $a^{\dagger}_{p_i}$ acting on the vacuum. Boosting the state $\ket{M}$ with a Lorentz transformation $U( \Lambda)$ such that the energy differs by $m$ then leads to 
\begin{align}
\mathcal{S} U (\Lambda) \ket{M} = U(\Lambda) \ket{X} = \sum_{p} \mathcal{S}_{M \rightarrow p} \ket{\Lambda p}
\end{align}
as Lorentz invariance implies the commutation of the S-matrix with $U$. The radiation gets boosted and the vacuum remains invariant, i.e. $U(\Lambda) \ket{0} = \ket{0}$ and thus an observer measuring the late radiation can distinguish between the black hole states with and without the early radiation (or the infalling object) corresponding to $\ket{M}$ and $U(\Lambda) \ket{M}$ respectively. In other words, the existence of the ten Poincar\'e charges (``hair'') is measurable by the observer.

\subsubsection*{Soft Hair Proposal}

The question is now whether charges related to the supertranslations and -rotations allow to retrieve some information in the same way.
In \cite{Hawking:2016msc} (cf. also \cite{Strominger:2017zoo}) Strominger et.al. argue that this is indeed the case. The crucial caveat in the argumentation leading to the paradox, they argue, is that while the no-hair theorem (correctly) implies that all stationary black hole solutions are diffeomorphic to the Kerr spacetime (a black hole described by angular momentum, mass and charge), the diffeomorphisms do not necessarily act trivial. We saw in section \ref{sec3.1.1} that the onset of a supertranslation changes the Bondi news $N_{AB}$ and superrotations similarly lead to physically inequivalent states. 
Corresponding to these transformation there exist conserved surface charges (cf. section \ref{sec3.1.2} and \cite{Strominger:2017zoo}) of the form
\begin{align}\label{chargesplit}
Q^+[f] = \frac{1}{4 \pi G} \int_{\mathcal{I}^+_-} d^2 z \gamma_{z \bar z} f m_B = \frac{1}{4 \pi G} \int_{\mathcal{I}^+} du d^2 z \gamma_{z \bar z} f \left[ T_{uu} - \frac{1}{4} \left( D^2_z N^{zz} + D^2_{\bar z} N^{ \bar z \bar z} \right) \right],
\end{align}
where $T_{uu}= 1/4 N_{AB} N^{AB} + 4 \pi G \underset{r \rightarrow \infty}{\text{lim}} \left[r^2 T_{uu}^M \right]$ is the $uu$ component of the stress energy tensor containing contributions from the matter stress energy tensor $T^M$. From \eqref{chargesplit} one observes that the charges are split into a  ``hard'' and a ``soft'' part (the first and second term on the rhs). The soft part in particular is unchanged by Poincar\'e ($ \hat= n=1$ embedding) transformations but can receive contributions from supertranslations and -rotations by the same arguments that were used in section \ref{sec3.1.1}. In analogy to the electromagnetic case discussed in section 2.5 of \cite{Strominger:2017zoo} the soft charge here promoted to an operator creates a zero energy graviton mode. 

So far the surface charges have been defined on future/past null infinity $\mathcal{I}^{\pm}$. However, in the presence of e.g. a Schwarzschild black hole $\mathcal{I}^+$ is no longer a Cauchy surface (a submanifold of the spacetime at a specific time for which the evolution equations of GR can be solved uniquely) and one has to include the future event horizon of the black hole $\mathcal{H}^+$ on which surface charges are defined analogously.
Strominger et. al. proceed to show that dropping matter into a black hole by e.g. sending in a shockwave induces a supertranslation which does change the superrotation horizon charges, which means soft (zero energy) gravitons are created at the horizon. While these soft modes  are basically by definition not directly observable (or the detection would take longer and longer time as the energy goes to zero), they are related to the supertransformations which  (as mentioned in section \ref{sec3.1.1}) can be observed in principal via memory effects in finite time \cite{Strominger:2017zoo}. Thus, after the entire black hole is evaporated information might be stored in the degenerate vacuum.

\subsection{Decoupling of Soft and Hard Modes}\label{sec5.3.3}

The soft hair proposal has been met with a series of papers (\cite{Mirbabayi:2016axw}, \cite{Bousso:2017dny},\cite{Javadinezhad:2018urv}) refuting that the soft hair have any bearing on the information loss paradox. A central argument, which will be detailed below, is that the soft and hard modes decouple completely and thus the change of the soft charges altered e.g. by infalling matter have no effect on late, measurable (i.e. hard) radiation.

We start our review of this critique (mostly following \cite{Mirbabayi:2016axw}) by noting that the splitting of the charges in a hard and a soft part also implies the factorization of the in- and outgoing Hilbert space $\mathcal{H}^{\pm}$ according to 
\begin{align}
\mathcal{H}^{\pm} = \mathcal{H}^{\pm}_h \otimes \mathcal{H}^{\pm}_s,
\end{align}
where $\mathcal{H}^{\pm}_{h/s}$ contain hard and soft modes respectively. An ingoing state can then be written as $\ket{\varphi_{in}} = \ket{a} \ket{\alpha}$ and a generic process is described by
\begin{align}
\bra{\beta} \bra{b} \mathcal{S} \ket{a} \ket{\alpha} \equiv \mathcal{S}_{b, \beta ; a, \alpha}.
\end{align}
Weinbergs soft theorem\footnote{This theorem is a gravitational generalization of soft theorems originally developed to tame infrared divergences in quantum electrodynamics by \cite{Bloch37}.} \cite{Wei65} in the form
\begin{align}
\mathcal{S}_{b, \beta ; a, \alpha} = \frac{\bra{\beta} \Omega (b) \Omega^{\dagger}(a)\ket{\alpha}}{\braket{0| \Omega(b) \Omega^{\dagger}(a) |0}} \mathcal{S}_{b, 0 ; a, 0}
\end{align}
implies
\begin{align}\label{splitscat}
\bra{\beta} \bra{b} \mathcal{S} \ket{a} \ket{\alpha}  = \braket{b | \hat{ \mathcal{S}} | a} \braket{\beta| \Omega(a) |\alpha},
\end{align}
with the soft graviton operators $\Omega$ and $\hat{\mathcal{S}}_{b,a} \equiv \frac{\mathcal{S}_{b, 0 ; a, 0}}{\braket{0| \Omega(b) \Omega^{\dagger}(a) |0}}$. The Weinberg theorem and the independence of $\hat {\mathcal{S}}$ in particular of the soft operators is a direct consequence of the conservation of the surface charges $Q$. For an extensive discussion of the relation between asymptotic symmetries and soft theorems also in the electromagnetic case see \cite{Strominger:2017zoo}.
Using the ``dressing'' \cite{Mirbabayi:2016axw}
\begin{align}
\ket{\ket{a, \alpha}} \equiv \Omega(a) \ket{a} \ket{\alpha},
\end{align}
the decoupling of soft and hard part of the scattering amplitude
\begin{align}
\bra{\bra{b, \beta}} \mathcal{S} \ket{\ket{a, \alpha}} = \braket{b|\hat{\mathcal{S}}| a} \braket{\beta |\alpha}
\end{align}
follows directly from \eqref{splitscat} using the unitarity of $\Omega$. 

In analogy to the case of the Poincar\'e hair discussed above let us consider a black hole of mass $M$ from which some early soft radiation has been emitted which is described by the action of a generator of supertranslations $\ket{\tilde M} = Q \ket{M}$ that creates a soft graviton with the associated charge $\beta$. But, using the dressed states with decoupled dynamics of soft and hard modes derived above, infalling hard matter could not be responsible for this change. Suppose now $\ket{M}$ decays into $\ket{X}$, i.e.
\begin{align}\label{softh}
\mathcal{S} \ket{\tilde M} =Q \ket{X} = \sum_{p} \hat{\mathcal{S}}_{M \rightarrow p} Q \ket{\ket{p, 0}} = \sum_{p} \hat{\mathcal{S}}_{M \rightarrow p} \ket{\ket{p, \beta}},
\end{align}
where the commutation of $\hat{\mathcal{S}}$ with $Q$ is a consequence of the BMS invariance. While the state $\ket{\beta}$ is in principle distinguishable for different $\beta$, it can only be used to retrieve information on the early soft modes which evolve trivially but not on the early hard infalling matter (or outgoing radiation). Thus, the authors of \cite{Mirbabayi:2016axw} conclude that soft hair are not contributing to solving the black hole information loss paradox.

\subsubsection*{Deformed Leibniz Rule for Surface Charges}

A natural question we investigate in the following is whether the deformed Leibniz rule for the surface charges can change the conclusion that soft hair have no influence on the hard scattering. 

This time we start directly in four spacetime dimensions with the Jordanian twist (cf. section \ref{sec4.3.2}). Its R-matrix is given by 
\begin{align}
R^{-1} & =  \mathcal{F} \mathcal{F}^{-1}_{21} = \exp \left( - \frac{k_0}{n'}\otimes \log \left( 1 + \chi S_{1-m', 1-m'} \right) \right) \nonumber \\
& = \sum_{s, t =0}^{\infty} \frac{1}{s! t!} \left( -\frac{k_0}{n'} \right)^s \left( \log \left( 1 + \chi S_{1-m', 1-m'} \right) \right)^t  \otimes \left(\log \left( 1 + \chi S_{1-m', 1-m'} \right) \right)^s \left( \frac{k_0}{n'} \right)^t
\end{align}
and then eq.\eqref{itsthebase} yields 
\begin{align}
\Delta_{\star} (S^{\mathcal{F}}_{pq}) & = S^{\mathcal{F}}_{pq} \otimes 1 + \sum_{t =0}^{\infty} \frac{1}{t!} \left( \log \left( 1 + \chi S_{1-m', 1-m'} \right) \right)^t \otimes \left( \frac{1-p-q}{n'} \right)^t S^{\mathcal{F}}_{pq} \nonumber \\
& = S^{\mathcal{F}}_{pq} \otimes 1 +\Pi_{+, n'}^{\frac{1-p-q}{n'}} \otimes S^{\mathcal{F}}_{pq}. \label{costarj}
\end{align}
To express also $\Pi_{+, n'}$ in terms of the deformed generators the general formula
\begin{align}
S^{\mathcal{F}}_{pq} = S_{pq} \Pi_{+, n'}^{\frac{1-p-q}{n'}}
\end{align}
is used so that
\begin{align}
S_{1-m',1-m'} = \frac{S^{\mathcal{F}}_{1-m',1-m'}}{1 - \chi S^{\mathcal{F}}_{1-m',1-m'}}, \\
\Pi_{+, n'} = \frac{1}{1- \chi S^{\mathcal{F}}_{1-m',1-m'}}
\end{align}
can be plugged into \eqref{costarj}
\begin{align}\label{costar21}
\Delta_{\star} (S^{\mathcal{F}}_{pq}) =  S^{\mathcal{F}}_{pq} \otimes 1 + \frac{1}{1- \chi S^{\mathcal{F}}_{1-m',1-m'}} \otimes S_{pq}^{\mathcal{F}}, \\
\Delta_{\star} (A^{\mathcal{F}}_{pq}) =  A^{\mathcal{F}}_{pq} \otimes 1 + \frac{1}{1- \chi S^{\mathcal{F}}_{1-m',1-m'}} \otimes A_{pq}^{\mathcal{F}}.\label{costar22}
\end{align}
Choosing an embedding $n' = n$ (\eqref{4dembed1}-\eqref{4dembed2}) and $\chi = \frac{i\sqrt{2}}{n \kappa}$ the addition of Poincar\'e momenta is dictated by the coproduct
\begin{align}\label{costarp}
\Delta_{\star} (P^{\mathcal{F}}_{\mu}) = P^{\mathcal{F}}_{\mu} \otimes 1 +  \frac{1}{1+P^{\mathcal{F}}_{+}/\kappa} \otimes P^{\mathcal{F}}_{\mu}.
\end{align}
Since the Poincar\'e sector forms a sub Hopf algebra (as demanded), the addition of hard momenta is not influenced by soft charges. On the other hand, the coproduct of generators of soft charges involves Poincar\'e generators and thus their addition rule depends on the hard charges as well. But the soft factor $\Omega$ from the soft theorems already depends on the hard momenta in the undeformed case and even though the altered Leibniz rule changes this dependence it is not enough to invalidate the decoupling.  

The only way for the deformation to have any bearing on the information loss paradox would be if the soft charges influence the hard scattering. In \cite{Bousso:2017dny} the authors acknowledge this possibility but immediately dismiss it because in that case ``experiments at the LHC could not be analyzed without detailed knowledge of the cosmic microwave background''. In our formalism these concerns can be rebutted because of the suppression of the effect size by the small scale $1/\kappa$. 
However, the deformation would have to include generators belonging to an embedding that does not correspond to the choice of the observer/detector. This could be realized by setting $n'$ (the index of the deformation) to a different value than $n =1$ (vacuum choice of the observer) in \eqref{costar21}-\eqref{costar22}. Then instead of \eqref{costarp} one finds
\begin{align}
\Delta_{\star} (P^{\mathcal{F}}_{\mu}) = P^{\mathcal{F}}_{\mu} \otimes 1 + \frac{1}{1-\chi S^{\mathcal{F}}_{1-m',1-m'}} \otimes P^{\mathcal{F}}_{\mu}.
\end{align}

Another possibility would be to pick a deformation that encompasses multiple embeddings. As we saw in section \ref{sec4.3.3}, this is compatible with specialization only in four dimensional spacetime e.g. by the r-matrix \eqref{mixem} with $p = 1-m, p'= 1-m'$. 
Then the inverse R-matrix is given by 
\begin{align}
R^{-1} = \sum_{s,t =0}^{\infty} \frac{1}{s! t!} (-\bar k_0)^s \left( \log \left( \Pi_+ \right) \right)^t \otimes \left( \log \left( \Pi_+ \right) \right) (\bar k_0)^t, \quad \Pi_+ \equiv (1+ \chi S_{1-m,1-m}  + \chi' S_{1-m',1-m'}),
\end{align}
resulting in the coproducts
\begin{align}
\Delta_{\star} (S^{\mathcal{F}}_{pq})  = S^{\mathcal{F}}_{pq} \otimes 1  + \cos \left( (p-q) \Pi_+ \right) \otimes S^{\mathcal{F}}_{pq} - \sin \left( (p-q) \Pi_+ \right) \otimes A^{\mathcal{F}}_{pq}, \\
\Delta_{\star} (A^{\mathcal{F}}_{pq})  = A^{\mathcal{F}}_{pq} \otimes 1  + \cos \left( (p-q) \Pi_+ \right) \otimes A^{\mathcal{F}}_{pq} + \sin \left( (p-q) \Pi_+ \right) \otimes S^{\mathcal{F}}_{pq},
\end{align}
with $\Pi_+ = \Pi_+^{\mathcal{F}}$ because
\begin{align}
S^{\mathcal{F}}_{pq} = S_{pq} \cos \left( (p-q) \Pi_+ \right) - A_{pq} \sin \left( (p-q) \Pi_+ \right), \\
S_{1-m,1-m} = S^{\mathcal{F}}_{1-m,1-m}.
\end{align}
The hard momenta of the embedding $n$ then add according to the coproducts
\begin{gather}
\Delta_{\star} (P^{\mathcal{F}}_{\pm}) = P^{\mathcal{F}}_{\pm} \otimes 1 + 1 \otimes   P^{\mathcal{F}}_{\pm}, \\
\Delta_{\star} (P^{\mathcal{F}}_{1/2}) =  P^{\mathcal{F}}_{1/2} \otimes 1 + \cos \left( n \Pi_+ \right) \otimes P^{\mathcal{F}}_{1/2} \mp \sin \left( n \Pi_+ \right) \otimes P^{\mathcal{F}}_{2/1}.
\end{gather}

Note, however, that both of these examples which break the decoupling of the soft and hard modes do not form a sub Hopf algebra when restricted to the Poincar\'e algebra ($n=1$) in the bulk. This fact is not changed by including the central extension of the algebra of surface charges but there are additional deformations that only alter the Leibniz rules for the rotation/boost sector. For example from $r = \chi T_{pq} \wedge c_l $ with $\mathcal{F} = \exp (\chi T_{pq} \otimes c_l)$ we get
\begin{align}
\Delta_{\star} ( l^{\mathcal{F}}_n) & = l^{\mathcal{F}}_n \otimes 1 + 1 \otimes l^{\mathcal{F}}_n - \chi c_l \otimes \left( \frac{n+1}{2}-p \right) T^{\mathcal{F}}_{p+n, q}, \\
\Delta_{\star} ( T^{\mathcal{F}}_{rs}) & =  \Delta_0(  T^{\mathcal{F}}_{rs}).
\end{align}

Furthermore, the constraint of being able to define a q-analog ensures that the Poincar\'e momenta only depend on supertranslations but not on superrotations. But as we reviewed at the end of section \ref{sec5.3.2}, the superrotation charges act as soft hair and hence on the full $\mathfrak{B}_4$ one can not avert the decoupling of soft and hard modes in a way that rescues the soft hair proposal.

\section{BMS Phenomenology}

In this section we will briefly discuss possible experiments which can give us information about the nature of the asymptotic symmetries of spacetime. Since there is no widely agreed-upon theoretical criterion (cf. section \ref{sec3.1.1}) one has to be open-minded about different possibilities and we are particularly interested in the implications for the deformation of the corresponding symmetry algebras.

\subsection{Gravitational Waves and Memory Effect}

The discovery of gravitational waves in 2015 \cite{LIGOScientific:2016aoc} marked the beginning of a new era in cosmology that enables us to investigate aspects of GR out of reach of previous methods. As already mentioned in section \ref{sec3.1.1}, one of the most interesting phenomena for our purposes is the gravitational (displacement) memory effect. In principle this effect is observable by the same means as the gravitational wave itself since it leads to the residual displacement of the detectors (according to \eqref{displm}) and the wave is detected by a temporary displacement. However, the effect size is much smaller for the memory effect and therefore the stacking of multiple waves in the same direction would be necessary for the current generation of experiments (LIGO/Virgo) to detect  it \cite{Hubner:2019sly}. With future setups like LISA (cf. \cite{LISA}, \cite{Hubner:2019sly}) also direct detection might be possible. 

Such a discovery would directly confirm the existence of physically distinct vacua and thus the transformations relating them, i.e. the supertranslations, have to be asymptotic symmetries of our spacetime. 

By the same argument the detection of the spin memory effect would prove the existence of superrotations as asymptotic symmetries. This effect would be observable in the relative delay between counter orbiting circular light beams \cite{Pasterski:2015tva} but since it is subleading (in the $1/r$ expansion) the detection is even harder. 

While the details of the asymptotic symmetries (inclusion of superrotations, supertranslations with poles, cf. also next section) are not universally agreed upon, the most restrictive requirement imposed on the deformations, the triangularity, is already a consequence of the existence of supertranslations. In other words, all subalgebras of the various asymptotic symmetry algebras we discussed that contain at least the supertranslations ($T_p$ or $T_{pq}$ with index $\geq -1$) and the Poincar\'e algebra have no ad-invariant element in the triple exterior product. Without a central extension we can also conclude from our findings that all consistent LBA are coboundary.

\subsection{Decay of Cosmic Strings}\label{sec5.4.2}

As we will detail in the following, cosmic strings also provide an opportunity to probe the nature of the asymptotic symmetries and potentially also deformations.

Recall from section \ref{sec3.1.1} that cosmic strings are one-dimensional topological defects of cosmological size. They might have formed after a stage of the universe when it was not simply connected yet or could be inflated fundamental strings from string theory.  

Even though there has been to date no conclusive evidence for the existence of cosmic strings, there have been attempts to identify gravitational lensing effects in which light-sources appear to be doubled due to cosmic strings \cite{Schild:2004uv}. Further experimental signatures could be found in gravitational waves from the following mechanism. In the collision of cosmic strings kink structures could emerge, leading to the breaking off of loops. These loops would then form cusps and kinks as the string velocity reaches the speed of light momentarily, subsequently emit gravitational waves and decay. Strominger and Zhiboedov \cite{Strominger:2016wns} also describe the possibility that the string snaps and decays via black hole pair creation where the black holes are at the endpoints of the snapped string and are dragged towards null infinity. 

As mentioned in section \ref{sec3.1.1}, the decay of a cosmic string would be a transition between inequivalent asymptotically locally flat and globally asymptotically flat vacua and therefore show that the asymptotic symmetries of our spacetime include superrotations and not just Lorentz rotations/-boosts.
Furthermore, we emphasized in \ref{sec3.1.1} that vector fields that generate superrotations with index $m \in \mathbb{Z}$ have singularities at $0, \infty$ corresponding to the endpoints of a cosmic string. Restricting to the one-sided $\mathfrak{B}_{4 +}$, which is necessary for several deformed quantum groups to have well-defined q-analog, implies that the BMS generators only have singularities for $z, \bar z \rightarrow \infty$. In combination with the previous argument the evidence for cosmic strings would then exclude the possibility of said quantum groups. 
Conversely, the existence of general superrotations ($m \in \mathbb{Z}$) and the closure of the BMS algebra necessitates also general supertranslations with poles \cite{Compere:2018aar}. Some authors argued that these lead to several technical problems, which can, however be remedied (see \cite{Compere:2018aar} and references therein).

\chapter{Conclusion}\label{chap6}

Before summarizing our results let us quickly recap the main ideas and theories underlying this thesis. On the one hand, this is a version of non-commutative geometry implemented by Hopf algebras. We reviewed its mathematical foundations in chapter \ref{chap2} and noted that a distinctive feature is the appearance of an observer independent scale. In this DSR interpretation of the $\kappa$-Poincar\'e and other quantum groups the symmetries realized by the algebra of the momenta are the fundamental objects and the spacetime takes a back seat; its properties as a smooth manifold are merely emergent. A key function of the coalgebra structure defined on the momentum space is to induce covariant representations on the multiparticle states requiring the coassociativity of the coproduct, which in turn implies the (modified) Yang Baxter equation for the associated classical r-matrix. A further important constraint arises when we want to interpret the deformation parameter of the quantum group as an observable quantity so we have to be able to assign a real value to it. 
On the other hand, there is the exciting field of asymptotic symmetries. One of its most striking features is the existence of distinguishable, degenerate vacua, intimately related to sub-algebras (embeddings) isomorphic to the Poincar\'e (or an analogous bulk symmetry algebra) in the infinite dimensional BMS algebras. This fact opens a connection to the theory of quantum groups and motivated our quest to study the deformations on the asymptotic symmetry algebras. In particular, we were interested in consistently implementing those quantum groups that are studied and applied in the Poincar\'e algebra.

After reviewing the relevant asymptotic symmetry algebras in three and four dimensions, the strategy in chapter \ref{chap4} to achieve this goal was to start from the Lie bialgebras which, by the results of Etingof and Kazhdan cited in chapter \ref{chap2}, exhausts all possible QUE. Since the cobracket of a LBA has to satisfy a 1-cocycle condition, a major step in its classification is the calculation of the corresponding cohomology group. For the asymptotically (A)dS spacetimes in three dimensions we first proved that $H^1\left( \mathfrak{W} \oplus \mathfrak{W} , \bigwedge^2 \left( \mathfrak{W} \oplus \mathfrak{W} \right) \right)$ vanishes and, together with the simple observation that there are no ad-invariant elements in (the exterior product of) the $\mathfrak{W} \oplus \mathfrak{W}$ algebra, we thus reduced the search for LBA to the classification of triangular r-matrices up to automorphism. Motivated by the fact that in the interior of the spacetime only the rigid symmetries (Poincar\'e or (A)dS) survive, we imposed the constraint that the Hopf algebra resulting from the deformation should form a sub-Hopf algebra when restricted to the embedding corresponding to the bulk-symmetry. With such a constraint the full r-matrix classification up to $\text{Aut}'\left( \mathfrak{W} \oplus \mathfrak{W} \right)$ was achieved starting from earlier results on the embedded $\mathfrak{o}(4, \mathbb{C})$ algebra. 

For the subsequent quantization of the LBA by a Drinfeld twist we again used results known from the finite dimensional algebras and occasionally automorphisms from the quotient $\text{Aut}(\mathfrak{o}(4))/ \text{Aut}'(\mathfrak{W} \oplus \mathfrak{W})$. By performing the explicit twisting procedure of the coalgebra sector for selected LBA, we noticed that except for a specific abelian twist all twists lead to coproducts containing both infinite power series in the deformation parameter as well as infinitely many different generators. This peculiarity of the infinite dimensional algebras makes a specialization of the deformation parameter to a complex/real value for physical interpretation ill-defined, a conclusion that could only be averted if we restricted to one-sided Witt algebras.

As reviewed in section \ref{sec3.1.2}, the charges associated with the asymptotic symmetries satisfy the Lie algebra of the asymptotic Killing vectors up to potential central extensions. The inclusion of the central charges, explicitly calculated in the seminal work by Brown and Henneaux, does not change the fact that all LBA are triangular and coboundary on $\mathfrak{W} \oplus \mathfrak{W}$. However, the classification of the r-matrices is affected in several ways. For once, there are a variety of new r-matrices containing the central element and we classified which combinations with the original $\mathfrak{W} \oplus \mathfrak{W}$ r-matrices are allowed. We furthermore noticed that the Schouten bracket of some r-matrices which are triangular in the algebra without central extension receive extra terms and also that the resulting coalgebra structures corresponding to r-matrices without central elements can nonetheless contain contributions from the central extension.

 While there are some important conceptual differences between AdS and dS in the study of asymptotic symmetries, in the algebraic description we simply view them as different real forms of one complex algebra. The corresponding reality conditions for the r-matrices turned out to be more restrictive for the asymptotically dS spacetime, excluding r-matrices that have important contraction limits in the $\mathfrak{B}_3$ algebra. 
But the deformation of asymptotically (A)dS spacetime symmetries is relevant in its own right for several reasons. 
For once, a positive cosmological constant, which after all seems to be the relevant case in our universe, was also the starting point of \cite{Cianfrani:2016ogm}, where a connection between a model of quantum gravity (Loop Quantum Gravity) and the time-like $\kappa$-Poincar\'e variant was established in a non-perturbative way.
Furthermore, by the AdS/CFT correspondence the asymptotic symmetries in the bulk become the global symmetries of the boundary theory and thus a ``dual'' deformation of the conformal symmetries can be expected. Such quantum groups have been discussed for example in \cite{Falcet:1991xt}.

In the algebraic description the contraction of the (asymptotically) (A)dS to the (asymptotically) flat symmetry algebra is quite straightforward. We thus set out to investigate the properties of the contraction limit on the level of Lie bialgebras and Hopf algebras in three dimensions. Note that the three dimensional structures including the Hopf algebras are contained in the four dimensional case except for the centrally extended algebras. Beyond this trivial sense we did not explicitly study the deformations of the asymptotically (A)dS symmetries and its contractions in four dimensions due to the lack of a notion of quantum groups on Lie algebroids. 

For the contraction of the r-matrices of (real forms of) $\mathfrak{W} \oplus \mathfrak{W}$ we found a remarkably rich behavior. In particular, depending on the parameter choices and their rescaling, there are different possibilities of contracting an r-matrix and this process is also not injective. One could have hoped that if the contraction of triangular r-matrices is surjective, we can use our knowledge of the twists on $\mathfrak{W} \oplus \mathfrak{W}$ to be able to construct all Hopf algebras in $\mathfrak{B}_3$. Even though the surjectivity can be established on the finite dimensional embeddings by including sums of contractions and contractions along different axis, a result consistent with the findings of \cite{Kowalski-Glikman:2019ttm}, this hope does not quite materialize as the contraction on the level of twists can not necessarily be performed in the same way as those of the corresponding r-matrices. The reason behind this is that the triangular classical r-matrices are in an anti-symmetric form which does not need to hold for the twists and thus some terms might not cancel during contraction. Note that the lack of a straightforward contraction limit of the twists is merely a practical inconvenience; we know that they have to exist even though we did not explicitly showed how to construct them except for twists related to the Lie algebra type non-commutativity. 
Without central charges we showed that triangular coboundary LBA are all there is on $\mathfrak{B}_3$ by calculating the relevant cohomology group. When taking into account the central extension this picture is changed considerably; the first cohomology group does not vanish but consists of a one-parameter family of cocycles which consequently define a Lie bialgebra that can not be quantized by a twisting procedure. We nonetheless find the corresponding Hopf algebra by simply guessing its form. It is interesting that this poses a clear example of how the asymptotically flat case is more than just the contraction of asymptotically (A)dS symmetries and their deformations. 

A further example of this distinction became apparent when we compared the process of specialization for Hopf algebras on $\mathfrak{B}_3$ with those in $\mathfrak{W} \oplus \mathfrak{W}$. Even though the contraction limit could be performed on the level of the LBA, the same limit applied to the twist ($\mathcal{F}_{2'}$) was divergent, consistent with the fact that the twist deformed Hopf algebra in $\mathfrak{B}_3$ admits a specialization and the Hopf algebra deformed with $\mathcal{F}_{2'}$ does not. Thus, in this example the conclusion that the quantum groups on the asymptotically flat symmetry algebras behave not just as contraction limits can not be seen already on the infinitesimal Lie bialgebra structure in contrast to the non-coboundary LBA mentioned above. 

We already noted that apart from the central extension the asymptotic symmetry algebras and their deformations in a three dimensional spacetime are sub (Hopf) algebras of the algebras in the realistic four dimensional spacetime. In the asymptotically flat case ($\mathfrak{B}_4$) we were able to show that all LBA are coboundary and triangular just as in $\mathfrak{B}_3$. Besides much more convoluted calculations also a few key differences stood out. First, it is noteworthy that all the deformations corresponding to a Lie algebra type non-commutativity single out a Poincar\'e embedding in $\mathfrak{B}_4$ because there is no longer a supertranslation generator that is present in all of the embeddings.
Furthermore, in the four dimensional case it is possible to find r-matrices and twists containing generators from multiple embeddings and lead to Hopf algebras with q-analog. In the algebraic sector the possible central extension appears in a different form but nonetheless calculating the cohomology group of $\mathfrak{B}_{4c}$ with values in the exterior product of the adjoint module leads to an almost identical result with just a two parameter LBA that is not triangular. The different form of the central extension does, however, have an influence on the classification of triangular r-matrices which we illustrated on a few examples.

With the classifications of Lie bialgebras and Hopf algebras on the three and four dimensional BMS at hand in chapter \ref{chap5} we first investigated the fate of one of our initial motivations to study quantum groups; the non-commutativity of the ``dual'' spacetime. It turned out that the requirement that the dual sector of an embedding closes a subalgebra, interpreted as a deformed Minkwoski spacetime, is a non-trivial constraint. In particular, the commutation relations dual to the LBA of the extended Jordanian twist resemble those of the well-known light-cone $\kappa$-Poincar\'e but also contain extra terms so that the Minkowski sector is indeed no subalgebra, posing another technical obstruction for this particular quantum group. Similarly, other deformations that did not admit a q-analog on the whole BMS algebra also do not clear this hurdle but the LBA corresponding to the abelian and Jordanian twist with its dual do and thereby define a non-commutative Minkowski spacetime of Lie algebra type. This also implies the introduction of a length scale we again call $1/\kappa$ and in the limit $\kappa \rightarrow \infty$ the classical commuting spacetime is recovered.

Our hopes of finding phenomenological effects detectable at mesoscopic scales were mainly resting on the deformed in-vacuo dispersion relations also (but not only) reported in the context of deformed Hopf algebras, albeit derived in various ways. In this thesis we employed a construction of wave operators of a given embedding in terms of a special basis (generating the so-called Quantum Lie algebra endowed with a star product). As all quantum groups we studied, with the sole exception of a quantization of the non-coboundary LBA from sections \ref{sec4.2.5} and \ref{sec4.3.3}, are defined by a twisting procedure the quantum R-matrix and ``deformed generators'' can be readily derived in order to calculate the potentially deformed wave operator describing the propagation through the non-commutative spacetime. By construction it is clear that a deformation constructed from elements of the same ($n=1$) embedding as the wave operator does not lead to deformed dispersion relations so in order to obtain an observable effect we would need to loosen our restriction and allow for deformations including (soft charge) generators from different embeddings. But even then the requirement that a q-analog can be defined prohibits any resulting deformation of the in-vacuo dispersion relation. 
So far, we thus return empty-handed from our search for new mesoscopic phenomena implemented by Hopf algebra structures on the asymptotic symmetry algebras that correspond to non-commutative geometry of Lie algebra type with a physical scale $1/\kappa$. 

Another phenomenon that was investigated is the deformed Leibniz rule resulting from the covariant representations of the symmetry algebra on multiparticle states induced by the coproduct. While this effect occurs in one form or another in all deformations listed in this thesis, potential experiments to detect it lack a way to counter-balance the smallness of the Planck scale. A further obvious possibility that could enable the detection of signatures of a deformed Leibniz rule would be that $1/\kappa$ is significantly larger than $l_{\text{Pl}}$ which, however, has the drawback that it is no longer backed by most of the motivation from quantum gravity like the thought experiment leading to a maximal resolution due to black hole creation.

Irrespective of the status of experimental evidence for a $\kappa$-deformed Leibniz rule in the BMS algebras, a new proposal for its application in the black hole information loss paradox was brought forward in this thesis. The main ideas can be summarized as follows. Strominger, Hawking and Perry suggested that information of the matter falling into the black hole might be stored in the form of ``soft hair'', i.e. supertransformation surface charges defined on the horizon. Bousso, Poratti and others then critically remarked that the Poincar\'e and soft charges decouple and thus concluded that the soft hair can not retain any ``hard'' information. If the momentum conservation in the relevant scattering processes is described using the deformed Leibniz rule, the soft and hard modes might cease to decouple. In particular, we found several twist deformations that in this way generate dependencies of the Poincar\'e momenta on supertranslation charges but, due to the restriction from the specialization requirement, not for superrotation charges. Since the latter act as the soft hair, according to Strominger et.al., we conclude that on the full $\mathcal{B}_4$ the decoupling can not be averted in a way that rescues the original soft hair proposal.

Finally, we briefly discussed how the exact form of the asymptotic symmetry algebras might be inferred from experiments and how this affects the rest of our results. Experimental verification of the existence of ordinary supertranslations as asymptotic symmetry transformations can be reasonably expected to be established in the near future which is enough for some of the most restrictive  constraints on consistent quantum groups to persist. I.e. in the absence of central charges all Hopf algebra structures constructed on said algebras are quasitriangular and derived from a twist. 
Whether supertransformations parametrized by functions of the angular coordinates without poles have to be included marks an important distinction because if this is not the case and we can use the one-sided BMS algebras for describing asymptotic symmetries the specialization of the deformation parameter to real values is possible for a much wider class of deformations. We reviewed that this distinction has been linked to the (snapping of) cosmic strings and experimental evidence consequently constrains the Hopf algebras with potentially physical interpretation in this sense.

\textit{Looking back to what we set out to achieve we made progress in the classification and analysis of quantum groups (QUE) on a variety of asymptotic symmetry algebras. Both technical and physical constraints led to the exclusion of consistent generalizations of quantum groups studied in the Poincar\'e algebra. We found interesting behavior in particular by studying the so-called contraction limit of spacetimes with cosmological constant and the inclusion of central charges on the level of Lie bialgebras and Hopf algebras.
Somewhat disappointingly, the search for phenomenological signatures of momentum spaces with the remaining deformed Hopf algebra structures did not yield promising results at least in a regime accessible by current experiments.}

Future research may follow up on a variety of points. First, it is conceivable to relax both some of the mathematical and physical requirements, e.g. one may consider the $\kappa$-deformed theory to be only an effective theory valid at the first order on $1/\kappa$ and consequently the specialization can be applied trivially. Another straightforward extension of our results is to take into account further newly found variants of BMS-type algebras e.g. from \cite{Compere:2013bya}, \cite{Afshar:2016wfy}.
To address the problem of identifying phenomenological signatures on mesoscopic scales despite a deformation parameter of the order of $l_{\text{Pl}}$, one could turn to deformed discrete symmetries, an idea recently discussed in the context of $\kappa$-Poincar\'e \cite{Arzano:2016egk} (see also \cite{Arzano:2020rzu}). The mechanisms that could lead to large enough effect sizes range from higher precision in time measurements to the appearance of the inverse of small mass differences e.g. in neutral Kaon systems.
Last but not least, one could proceed to explore the application of the Lie bialgebras and Hopf algebras discussed in this thesis in related branches of pure and applied mathematics. For example, in the study of integrable systems and in particular in the quantum inverse scattering method (see \cite{Slav89}, \cite{Fad94} and references therein) quantum R-matrices play a vital role which, incidentally, was one of the first motivations to introduce quantum groups.

\newpage

%
\appendix

\chapter{Proof of the Cohomology Theorems}\label{app-a}

\underline{\bf{Proof of Theorem \ref{Wth}}}
\newline
We structure the proof by first noting that the 1-cocycles $\delta$ can be separated by their degree $d \in \mathbb{Z}$. This degree is derived from the grading of $\mathfrak{W}$, i.e.
\begin{align}
\delta (L_m) = L_i \wedge L_j
\end{align}
has degree $d = i+j-m$. The separation by degree follows from the fact that a cocycle which, applied to elements of $\mathfrak{W}$, results in terms with different degree can be split into cocycles of homogeneous degree which have to fulfill the cocycle condition \eqref{cocyclec} independently.
Let us \textbf{first} consider cocycles of \textbf{degree} $\boldsymbol{d \neq 0}$. We will show that all such cocycles $\delta$ are cohomolog to $0$, i.e. that $\delta'(L_m) = \delta(L_m) - (\partial_0 r)(L_m) = 0$ for all $m \in \mathbb{Z}$. 
Let $\delta$ be a cocycle of homogenous degree $d$ such that
\begin{align}
\delta (L_m) = \sum_{i_m, j_m \in I_m} \alpha^m_{i_m j_m} L_{i_m} \wedge L_{j_m}, 
\end{align}
where $\alpha^m_{i_m j_m} \in \mathbb{R}; i_m, j_m \in \mathbb{Z}$ and $I_m$ are finite subsets of $\mathbb{Z}$.
Choose a 0-cochain
\begin{align}
r = -\sum_{i_0, j_0 \in I_0} \frac{\alpha^0_{i_0 j_0}}{i_0 + j_0} L_{i_0} \wedge L_{j_0}.
\end{align}
 Then we have
\begin{align}
\delta'(L_0) &= \sum_{i_0, j_0 \in I_0} \alpha^0_{i_0, j_0} L_{i_0} \wedge L_{j_0} - [L_0 \otimes 1 + 1 \otimes L_0, r] \nonumber \\
&= \sum_{i_0, j_0 \in I_0} \alpha^0_{i_0, j_0} L_{i_0} \wedge L_{j_0} - \sum_{i_0, j_0 \in I_0} \frac{\alpha^0_{i_0 j_0}}{i_0 + j_0} (i_0 + j_0) L_{i_0} \wedge L_{j_0} = 0.
\end{align}
From the cocycle condition
\begin{align}
\delta'([L_0, L_m]) = [L_0 \otimes 1 + 1 \otimes L_0, \delta'(L_m)] - [L_m \otimes 1 + 1 \otimes L_m, \delta'(L_0)],
\end{align}
for $m \neq 0$, we infer
\begin{align}
m \delta(L_m) &= \sum_{i_m, j_m \in I_m} (i_m + j_m) \alpha^m_{i_m j_m} L_{i_m} \wedge L_{j_m} \nonumber \\
&= (d+m) \delta'(L_m) \\
\Rightarrow  \delta'(L_m) &= 0,
\end{align}
which concludes the proof for cocycles of degree $d \neq 0$.

\textbf{Next}, let us consider cocycles of \textbf{degree} $\boldsymbol{d=0}$ which can be written in the form
\begin{align}
\delta(L_m) = \sum_{i_m \in I} \gamma^m_{i_m} L_{m - i_m}\wedge L_{i_m}.
\end{align}
Note that without loss of generality we can restrict the indices $i_m$ to be smaller than $m/2$ due to the following observation. If there were an index $i_m > m/2$, we simply substitute $i'_m = m-i_m$ and ${\gamma'}^m_{i'_m} = \gamma^m_{i_m}-\gamma^m_{im-m}$ to describe the same cocycle. We will make repeated use of this restriction in the rest of the proof. 

The conditions
\begin{align}
\delta([L_0, L_m]) &= [L_0 \otimes 1 + 1 \otimes L_0, \delta(L_m)] - [L_m \otimes 1 + 1 \otimes L_m, \delta(L_0)] \nonumber \\
&=  (-m) \delta(L_m) - [L_m \otimes 1 + 1 \otimes L_m, \delta(L_0)]  \\
\Leftrightarrow 0 &=  [L_m \otimes 1 + 1 \otimes L_m, \delta(L_0)] 
\end{align}
implie that all degree $0$ cocycles vanish on $L_0$ because it has to hold for all $m$ and there is no ad-invariant element in $\bigwedge^2 \left( \mathfrak{W} \oplus \mathfrak{W}\right)$.

As a \textbf{next step} we show that \textbf{all cocycles are cohomolog to} $\boldsymbol{0}$ on $\boldsymbol{L_{\pm 1}}$. Let us assume without loss of generality that the indices of 
\begin{align}
\delta (L_1) = \sum_{i_1 \in I_1} \gamma^1_{i_1} L_{1-i_1} \wedge L_{i_1}
\end{align}
are given by $i_1 \in I_1 = \{-p_1, -p_2, ..., -p_n\vert p_1 > p_2 > ... > p_n > 1, n \in \mathbb{N}\}$. From the cocycle condition we get
\begin{align}
\delta([L_1, L_{-1}]) &= [L_1 \otimes 1 + 1 \otimes L_1, \delta (L_{-1})] - [L_{-1} \otimes 1 + 1 \otimes L_{-1}, \delta (L_{1})]  \\
\Leftrightarrow 0 &= \sum_{i_{-1} \in I_{-1}} (\gamma^{-1}_{i_{-1}}(2 + i_{-1}) \underbrace{L_{-i_{-1}} \wedge L_{i_{-1}} }_{\text{type} I} + \gamma^{-1}_{i_{-1}}(i_{-1}-1) \underbrace{L_{1 + i_{-1}}\wedge L_{-1-i_{-1}} }_{\text{type} II}) \nonumber \\
& \phantom{=}  + \sum_{j=1}^n (\gamma^1_{-p_j} (2+ p_j) L_{p_j} \wedge L_{-p_j} + \gamma^1_{-p_j}(p_j-1) L_{-1- p_j} \wedge L_{1+ p_j}). \label{p}
\end{align}
Lets focus on the first term in the second line of \eqref{p} with $p_1$; it can only be canceled by any of the other $p_j$ terms if $p_2= p_1 -1$ which we discuss below. In the \textbf{case} $\boldsymbol{p_2 \neq p_1-1}$ there are two terms that can contribute, one from the first and the second summand in the first line in \eqref{p} which we will call type $I$ and type $II$ terms respectively \footnote{Here and in the following we use the index restriction. Otherwise also e.g. a type $I$ term with $i_{-1} = p_j$ could be used.}. The type $II$ term would correspond to  $i_{-1} = -1-p$. If it existed with non-zero $\gamma^{-1}_{-1-p}$ it would imply the existence of a type $I$ term of the form $\gamma^{-1}_{-1-p} (1-p_1) L_{1+p_1} \wedge L_{-1-p_1}$ which in turn can only be canceled by a type $II$ term with $i_{-1} = -2-p_1$. Since also none of the prefactors $(2+i_{-1})$ and $(i_{-1} -1)$ vanishes if $p_1 \neq 1$ this would go on forever so that we need infinitely many terms in  $\delta(L_{-1})$ which is not possible. Thus $\gamma^{-1}_{-1-p}=0$ and we need a type $I$ term with $i_{-1} = -p_1$ 
\begin{align}
\gamma^{-1}_{1-p_1} (2 + p_1) L_{p_1 } \wedge L_{-p_1}
\end{align}
which implies a type $II$ term with the same $i_{-1}$
\begin{align}
\gamma^{-1}_{-p_1} (-1-p_1) L_{1-p_1} \wedge L_{p_1 -1}.
\end{align}
This term can be canceled only by a type $I$ term with $i_{-1} = 1-p_1$ 
\begin{align}
\gamma^{-1}_{1-p_1} (3 - p_1) L_{1-p_1} \wedge L_{p_1-1}
\end{align}
and the corresponding type $II$ term
\begin{align}\label{II1p}
\gamma^{-1}_{1-p_1} p_1 L_{2-p_1} \wedge L_{p_1}
\end{align}
 requires again a type $I$ term with $i_{-1} = -p_1$
\begin{align}\label{Ip}
\gamma^{-1}_{-p_1} (2-p_1) L_{2-p_1} \wedge L_{p_1}
\end{align}
 ending the sequence. The cancellation of \eqref{II1p} with \eqref{Ip} implies the following ratio of the coefficients
\begin{align}
\frac{\gamma^{-1}_{-p_1}}{\gamma^{-1}_{1-p_1} } = \frac{2-p_1}{p_1}.
\end{align}
When considering the 0-cochain
\begin{align}\label{cochain}
r = \gamma_{s} L_{-s} \wedge L_s
\end{align}
with $s = 1-p_1$, implying
\begin{align}
(\partial_0 r)(L_{-1}) \equiv \delta_r(L_{-1}) = \gamma_{1-p_1}( (-p_1) L_{-2 + p_1} \wedge L_{1-p_1} + (-2 + p_1) L_{p_1-1} \wedge L_{-p_1},
\end{align}
we find the same ratio between the two summands. Thus setting $\gamma_{1-p_1} (p_1)= \gamma^{-1}_{1-p_1}$ in the cocycle
\begin{align}
\delta' = \delta + \delta_r
\end{align}
both coefficients ${\gamma'}^{-1}_{-p_1}, {\gamma'}^{-1}_{1-p_1}$ vanish and therefore also ${\gamma'}^1_{p_1}$ has to be zero. 

Next, we have to consider the \textbf{case} $\boldsymbol{p_2 = p_1 -1}$. In \eqref{p} the term 
\begin{align}\label{p1p2}
\gamma^1_{-p_1}(p_1 -1) L_{1+p_1} \wedge L_{-1-p_1}
\end{align}
can be canceled by a type $I$ term with $i_{-1} = 1 - p_1$ or a type $II$ term with $i_{-1} = -2-p_1$. If the second term does not vanish,  this implies the existence of a type $I$ term with the same $i_{-1}$ which can only be eliminated by a type $II$ term with $i_{-1} = -3-p_1$ and so on, so that infinitely many terms are necessary, ruling out this option. Using the same cochain as above in \eqref{cochain} with the same choice for $s$ and $\gamma_s$ we can eliminate the coefficient ${\gamma'}^{-1}_{1-p_1}$ and thus the possibility to cancel \eqref{p1p2} with a type $I$ term is not possible which means that ${\gamma'}^{1}_{-p_1}$ has to vanish. 

For the \textbf{rest of the} $\boldsymbol{p_j, j > 1}$ we can iteratively use the same argumentation. In particular the arguments with the infinite number of terms in $\delta(L_{-1})$ can be extended to the higher $j$ as the sequence would stop at the terms with $i_{-1} = -p_{j-1}$ which already has to vanish. Furthermore, one has to add coboundaries from the cochains 
\begin{align}
r_j = \gamma_{j} L_{-(1-p_j)} \wedge L_{1-p_j}, \quad j>1
\end{align}
with suitable coefficients $\gamma_{j} (p_j) = {{{\gamma'}^{...}}'}^{-1}_{1-p_j}$ where we define
\begin{align}
{{\delta'}'} = \delta' + \delta_{r_2}, ...
\end{align}
iteratively so that the required terms in ${{\delta'}^{...}}' (L_{-1})$ are eliminated.

Finally, let us explicitely consider the \textbf{case} $\boldsymbol{p_1 = 1}$ that was excluded in the argumentation above. In that case 
\begin{align}
\delta (L_1) = \gamma^1_{-1} L_0 \wedge L_1
\end{align}
and from the cocycle condition we infer that 
\begin{align}
\delta(L_{-1}) = \gamma^1_{-1} L_{-2} \wedge L_{1}.
\end{align}
On $L_{\pm 1}$ $\delta$ then coincides with $\delta_r$, where $r = \gamma^1_{-1}/2 L_1 \wedge L_{-1}$ and thus $\delta' = \delta - \delta_r$  is zero on these elements.
This concludes the proof that $\delta$ is cohomolog to $0$ on $L_{1}$.

In the \textbf{next step} it will be shown that $\boldsymbol{\delta(L_{1}) = 0}$ implies that $\boldsymbol{\delta(L_m)=0}$ for $\boldsymbol{m >1}$. Starting from
\begin{align}\label{prot1}
\delta (L_2) = \sum_{i_2 \in I_2} \gamma^2_{i_2} L_{2-i_2} \wedge L_{i_2},
\end{align}
one explicitly obtains by using \eqref{cocyclec} with $m = 1, n=2$; $m = 1, n=3$ and $m=1, n=4$
\begin{align}
\delta (L_3) =& - \sum_{i_2 \in I_2} \gamma^2_{i_2}((i_2 -1)L_{3-i_2} \wedge L_{i_2} - (1-i_2) L_{2-i_2}\wedge L_{i_2 +1}) \\
\delta (L_4) =& \sum_{i_2 \in I_2} \frac{\gamma^2_{i_2}}{2} (i_2-1) ((i_2 -2)L_{4-i_2} \wedge L_{i_2} + 2 (1-i_2) L_{3 -i_2} \wedge L_{i_2 + 1} - i_2 L_{2-i_2} \wedge L_{i_2+2}) \\
\delta(L_5) =& - \sum_{i_2 \in I_2} \frac{\gamma^2_{i_2}}{6} (i_2-1)\bigg((i_2 -2)(i_2 -3)L_{5-i_2} \wedge L_{i_2} + 3 (1-i_2) (i_2 -2) L_{4-i_2} \wedge L_{i_2 +1} \nonumber \\
&  + 3(1-i_2) i_2 L_{3 -i_2} \wedge L_{i_2 +2} + i_2 (1+i_2) L_{2-i_2} \wedge L_{i_2 +3} \bigg).
\end{align}
Using the same argumentation as above, we can restrict $i_2$ to be bigger than $1$ and we consider the largest index $i'_2$. Then, \eqref{cocyclec} with $m=2, n=3$ yields
\begin{align}
0 = &\sum_{i_2 \in I_2} \bigg( L_{5-i_2} \wedge L_{i_2} \gamma^2_{i_2}\left(\frac{1}{6}(i_2 -1)(i_2-2)(i_2-3) - (i_2 -1)^2 - (i'_2 +1) \right) \nonumber \\
&+ L_{4-i_2} \wedge L_{i_2 +1}\gamma^2_{i_2} \left(\frac{1}{2}(i_2-1)^2(2-i_2 )+(i_2-1) i_2 \right) \nonumber  \\
&+  L_{3-i_2} \wedge L_{i_2 +2}\gamma^2_{i_2} \left( \frac{1}{2}(i_2-1)^2 i_2 + (i_2-1)(i_2-2)  \right) \nonumber \\
&+ L_{2-i_2} \wedge L_{i_2 +3} \left((i_2-1)^2 - (3-i_2) - \frac{(i_2-1)(i_2+1) i_2}{6} \right) \bigg) \label{l2} \\
\Rightarrow \quad 0 &=L_{2-i'_2} \wedge L_{i'_2 +3} \left( (i'_2-1)^2 - (3-i'_2) - \frac{(i'_2-1)(i'_2+1) i'_2}{6} \right)  \label{l3}
\end{align}
and \eqref{l3} implies for $i'_2 > 1, \gamma^2_{i'_2} \neq 0$ the solutions $i'_2 = 3, 4$. $i_2$ can therefore only take the values $i_2 = 2, 3,4$ and one can calculate explicitly that e.g. the term proportional to $L_{1} \wedge L_5$ in \eqref{l2} does not vanish so $\gamma^1_{i_2'} = 0$. Thus $\delta (L_2) = 0$ and iteratively one shows that \eqref{cocyclec} with $m =1$ implies $\delta (L_n) =0$ for $n > 2$. For arbitrary positive $m$ one finds
\begin{align}
\delta([L_{-1}, L_m]) &= - [L_m \otimes 1 + 1 \otimes L_m , \delta(L_{-1})] \\
\Rightarrow 0 &= -\sum_{i_{-1} \in I_{-1}} \gamma^{-1}_{i_{-1}} ((m+1 + i_{-1}) L_{m-1-i_{-1}} \wedge L_{i_{-1}} + (m-i_{-1})L_{-1 - i_{-1}} \wedge L_{i_{-1} +m})
\end{align}
which yields $\gamma^{-1}_{i'_{-1}} =0$ for the largest index $i'_{-1}$ and thus $\delta (L_{-1}) = 0$. 

\textbf{Finally}, one shows explicitly that \eqref{cocyclec} with $m=1, n=-2$ results in $\delta (L_{-2}) = 0$ and, similarly to the case of positive $m$ that can be used to show that $\boldsymbol{\delta (L_m) = 0}$ for all $\boldsymbol{m < -2}$, completing the proof of the first theorem.



\underline{\bf{Proof of Theorem \ref{W+Wth}}}
\newline
Note that a 1-cocycle $\delta$ applied to an element of $\mathfrak{W}$ can be split into three parts $\delta^I, \delta^{II}, \delta^{III}$, mapping to $\mathfrak{W} \wedge \mathfrak{W}$, $\overline{ \mathfrak{W}} \wedge \overline{\mathfrak{W}}$ or $\mathfrak{W} \wedge \overline{\mathfrak{W}}$ respectively, which have to satisfy the cocycle condition separately\footnote{In other words the cocycles decompose into $Z( \mathfrak{W} \oplus \overline{\mathfrak{W}}, \bigwedge^2 \mathfrak{W}) \oplus Z( \mathfrak{W} \oplus \overline{\mathfrak{W}}, \bigwedge^2 \overline{\mathfrak{W}}) \oplus Z( \mathfrak{W} \oplus \overline{\mathfrak{W}}, \mathfrak{W} \wedge \overline{\mathfrak{W}}) $.}. From the previous theorem it follows that $\delta^I$ is cohomolog to zero and from \eqref{cocyclec} one can easily see that $\delta^{II}$ has to vanish. Thus we only need to \textbf{consider the part} $\boldsymbol{\delta^{III}}$ which again can be separated by the degree $d$, which we define such that 
\begin{align}
\delta(L_m) = L_i \wedge \bar{L}_j
\end{align}
has $d = i-m$. 
A general cocycle of homogenous \textbf{degree} $\boldsymbol{d \neq 0}$ is given by
\begin{align}
\delta(L_0) = \sum_{j_0 \in I_0} \alpha^0_{j_0} L_{d} \wedge \bar{L}_{j_0}
\end{align}
on $L_0$. Setting 
\begin{align}
r = \sum_{j_0 \in I_0} \frac{\alpha^0_{j_0}}{d} L_{d} \wedge \bar{L}_{j_0}
\end{align}
we then have 
\begin{align}
\delta'(L_0) = \delta(L_0) - \delta_r(L_0) = 0.
\end{align}
Using this in 
\begin{align}
\delta'([L_0, L_m]) = [L_0 \otimes 1 + 1 \otimes L_0, \delta'(L_m)] -  [L_m \otimes 1 + 1 \otimes L_m, \delta'(L_0)]
\end{align}
it follows that 
\begin{align}
-m \delta'(L_m) = -(d+m) \delta'(L_m) \Rightarrow \delta'(L_m) = 0
\end{align}
concluding the proof for $d\neq 0$.

A general \textbf{degree} $\boldsymbol{0}$ cocycle has the form
\begin{align}
\delta (L_m) = \sum_{i_m \in I_m} \gamma^m_{i_m} L_m \wedge \bar{L}_{i_m}
\end{align}
and by choosing
\begin{align}
r = \sum_{i_1 \in I_1} \gamma_{i_1} L_0 \wedge \bar{L}_{i_1}
\end{align}
it follows that
\begin{align}
\delta'(L_1) = \delta(L_1) - \delta_r(L_1) = 0.
\end{align}
Then for $m\neq 1$
\begin{align}\label{lml1}
\delta'([L_m, L_1]) &= -  [L_1 \otimes 1 + 1 \otimes L_1, \delta'(L_m)] \\
\Rightarrow  \sum_{i_{m+1} \in I_{m+1}} (m-1) \gamma^{m+1}_{i_{m+1}}  L_{m+1} \wedge \bar{L}_{i_{m+1}} &= \sum_{i_m \in I_m} (m-1) \gamma^m_{i_m} L_{m+1} \wedge \bar{L}_{i_m}
\end{align}
and it follows that 
\begin{align}\label{m1}
\gamma^{m+1}_{i_{m+1}} = \gamma^m_{i_m}.
\end{align}
 If $m = 0$ in \eqref{lml1} we conclude 
\begin{align}
0 = - \sum_{i_0 \in I_0} \gamma^0_{i_0} L_1 \wedge \bar{L}_{i_0}
\end{align}
and thus $\gamma^0_{i_0}=0$. Because of \eqref{m1} $\gamma^m_{i_m} = \gamma^0_{i_0}$ for $m<0$ and for $m >0$ all coefficients are given by $\gamma^m_{i_m} = \gamma^2_{i_2}$. However, from \eqref{cocyclec} with $m= 2, n=3$ we find
\begin{align}
- \sum_{i_2 \in I_2} \gamma^2_{i_2} L_5 \wedge \bar{L}_{i_2}= \sum_{i_2 \in I_2} \gamma^2_{i_2} (-2) L_5 \wedge \bar{L}_{i_2} 
\end{align}
and thus $\gamma^2_{i_2} = 0$, concluding the proof.
\newline

\underline{\bf{Proof of Theorem \ref{W++W+th}}}
\newline

As a \textbf{first step} we will now show that $\boldsymbol{H^1 \left(\mathfrak{W}_+, \bigwedge^2 \mathfrak{W}_+\right) = \{0\}}$. Before starting the actual proof of the theorem let us note that we can not simply reuse the proof of theorem \ref{Wth} because there we made use of the index restriction which is not possible in the one-sided Witt algebra. However, for cocoycles of degree $d \neq 0$ this does not apply and thus only the $\boldsymbol{d=0}$ \textbf{part} is left. Vanishing degree of a cocycle $\delta$ on $\mathfrak{W}_+$ implies that 
\begin{align}
\delta(L_0) & =  \gamma^0_1 L_-1 \wedge L_1, \quad \delta(L_{-1}) = \gamma^{-1}_1 L_0 \wedge L_{-1}, \\
\delta(L_1) & = \gamma^1_{2} L_{-1} \wedge L_2 + \gamma^1_1 L_0 \wedge L_1, \quad \delta(L_2) = \gamma^2_{3} L_{-1} \wedge L_3 + \gamma^2_2 L_0 \wedge L_2 , \\
\delta(L_3) & = \gamma^3_4 L_{-1} \wedge L_4 + \gamma^3_3 L_0 \wedge L_3 + \gamma^3_2 L_1 \wedge L_2,
\end{align}
for the first few generators. The only possible 0-cochain that induces a coboundary of degree $0$ is of the form $r = \alpha L_1 \wedge L_{-1}$. Setting $\alpha = - \gamma^1_1/2$ the new cocycle $\delta - \delta_r$ (which is cohomolog to $\delta$) has $\gamma^1_1=0$. By imposing the cocycle condition
\begin{align}
\delta([L_1, L_0]) & = [ L_1 \otimes 1 + 1 \otimes L_1, \delta(L_0)] -  [ L_0 \otimes 1 + 1 \otimes L_0, \delta(L_1)] , \\
\Rightarrow  \delta(L_1) & = 2 \gamma^0_1 L_0 \wedge L_1 + \delta (L_1),
\end{align}
we infer $\delta (L_0) =0$. 
Furthermore, the cocycle conditions with $\delta([L_1, L_{-1}]), \delta ([L_2, L_{-1}]),$ $\delta([L_1, L_2])$ yield the equations
\begin{gather}
\gamma^{-1}_1 + 3 \gamma^1_2 = 0, \\
 3 \gamma^1_2 = -4 \gamma^2_3 - \gamma^2_2 + 2 \gamma^{-1}_1, \quad -3 \gamma^2_2 - 3 \gamma^{-1}_1 = 0, \\
- \gamma^3_4 = - 2 \gamma^2_3 , \quad - \gamma^3_3 = 2 \gamma^2_3 - \gamma^2_2 , \quad - \gamma^3_2 = \gamma^2_2 - 3 \gamma^1_2,
\end{gather}
which do not have a non-trivial solution and thus $\delta (L_{-1, 1, 2, 3}) =0$. From cocycle conditions with $\delta ([L_1, L_m])$ it is then easy to iteratively show that $\delta(L_m) =0, m \geq -1$. This concludes the first part of the proof.

Showing that $H^1 \left(\mathfrak{W}_+ \oplus \mathfrak{W}_+, \bigwedge^2 (\mathfrak{W}_+ \oplus \mathfrak{W}_+) \right) = \{0\}$ with the help of the first part is then completely analogous to the proof of theorem \ref{W+Wth} since there is no use of the index restriction.

\underline{\bf{Proof of Theorem \ref{Virth}}}
\newline
We \textbf{first} prove that $\boldsymbol{H^1 \left( \mathfrak{Vir}, \bigwedge^2 \mathfrak{Vir} \right) = \{0\}}$.
Let $\delta$ be a 1-cocycle on $\mathfrak{Vir}$ with values in $ \bigwedge^2 \mathfrak{Vir} $. It can be decomposed into $\delta = \delta_{\mathfrak{W}} + \delta_c$, where $\delta_{\mathfrak{W}}: \mathfrak{Vir} \rightarrow  \bigwedge^2 \mathfrak{W}$ and $\delta_c : \mathfrak{Vir} \rightarrow  c_L \bigwedge \mathfrak{W}$. Let $ X_c = X + \alpha c_L$ and $Y_c = Y + \alpha' c_L$ be generic elements in $\mathfrak{Vir}$ with $X, Y \in \mathfrak{W}, \alpha, \alpha' \in \mathbb{C}$. The cocycle condition then reads
\begin{align}\label{ccc}
\delta([X_c, Y_c]) = [X_c \otimes 1 + 1 \otimes X_c, \delta (Y_c)] - [Y_c \otimes 1 + 1 \otimes Y_c , \delta(X_c)],
\end{align}
resulting in the separate equations
\begin{align}\label{virco1}
\delta_{\mathfrak{W}} ( [X,Y])  & = [\delta_{\mathfrak{W}} (X_c), Y \otimes 1 +1 \otimes Y]_{\mathfrak{W}} -  [\delta_{\mathfrak{W}} (Y_c), X \otimes 1 +1 \otimes X]_{\mathfrak{W}} , \\
  \delta_c ( [X,Y]) & = [\delta_{\mathfrak{W}} (X_c),  Y \otimes 1 +1 \otimes Y]_c - (X \leftrightarrow Y)   + [\delta_c(X_c), Y\otimes  + 1 \otimes Y]_{\mathfrak{W}} - (X \leftrightarrow Y) ,\label{virco2}
\end{align}
where $[ , ]_{\mathfrak{W}}$ denotes the part of the induced bracket on $ \otimes^2 \left(\mathfrak{Vir} \right)$ that conserves the number of central elements in the input (e.g. mapping from $ \otimes^4 \left(\mathfrak{W}\right)$ to $ \otimes^2 \left(\mathfrak{W}\right)$) and $[ , ]_c$ the rest (e.g. mapping $ \otimes^4 \left(\mathfrak{W} \right)$ to  $\mathbb{C} c_L \otimes \mathfrak{W}$). 

From \eqref{ccc} with the commutator $[c_L, Y] = 0$ on the lhs it is easy to see that $\boldsymbol{\delta(c_L)=0}$ as there are no ad-invariant elements (cf. section \ref{sec4.1.1}). Thus \eqref{virco1} is independent from the central charge and one can apply theorem \ref{Wth} to show that  $\delta_{\mathfrak{W}}$ is a coboundary from a cochain $r \in \bigwedge^2 \left(\mathfrak{W} \right)$. 

Redefining $\delta \rightarrow \delta - \delta_r$ we find that $\boldsymbol{\delta_{\mathfrak{W}}(L_m) =0}$ without affecting $\delta_c(L_{0, \pm 1})$.  
Without loss of generality the remaining part of the cocycle is of the form
\begin{align}
\delta_c ( L_m) = \sum_{i_m \in I_m} \gamma^m_{i_m} c_L \wedge L_{i_m},
\end{align}
which again can be separated by the degree $d = m- i_m$.
We still have the freedom to choose a cochain $r \in c_L \bigwedge \mathfrak{W}$ and \textbf{for} $\boldsymbol{d \neq 0}$ we redefine the cocycle as $\delta \rightarrow \delta - \delta_{r_c}$. Setting $r_c = \sum_{i_0 \in I_0} \frac{- \gamma^0_{i_0}}{i_0} c_L \wedge L_{i_0}$ results in $\delta_c(L_0) = 0$ except if there is one index $i_0 =0$. In that case, consider the cocycle condition \eqref{virco2} with $X = L_1, Y = L_0$, yielding
\begin{align}
\sum_{i_1 \in I_1} \gamma^1_{i_1} c_L \wedge L_{i_1} = \sum_{i_1 \in I_1} \gamma^1_{i_1} c_L \wedge (i_1) L_{i_1} - \gamma^0_0 c_L \wedge (-1) L_1.
\end{align}
The last term can only cancel if there is a $i_1 = 1$ but then $\gamma^0_0 = 0$ follows nonetheless. Thus we can set $\delta_c(L_0) = 0$. 

Next, we set $X = L_m, Y = L_0$ in \eqref{virco2} and because $d = m - i_m \neq 0$ we find $\delta_c(L_m) = 0$.

\textbf{For} $\boldsymbol{d =0}$ we instead use the freedom of choosing a cochain to set $\delta_c(L_1) =0$ which can be achieved with $r_c = \gamma^1_1 c_L \wedge L_0$. Then \eqref{virco2} can be used with $X = L_1, Y = L_m$ and one finds that the resulting equations are equivalent to those in the last part of the proof of theorem \ref{W+Wth}. Thus we find $\delta_c (L_m) =0$ in the $d=0$ case as well. Note that this means that after we set $\delta_{\mathfrak{W}} =0$ the remaining part of the proof is basically equivalent to $H^1 \left( \mathfrak{W}, \mathfrak{W} \right) = \{0\}$.

Now we are ready to prove the \textbf{actual theorem} and set $c_L = c_{\bar L}$\footnote{This is physically motivated. If $c_L \neq c_{\bar L}$ a bit more work is needed because also terms of the form $c_L \wedge c_{\bar L}$ can appear.}. Let $\delta$ be a 1-cocycle on $\mathfrak{Vir} \oplus \mathfrak{Vir}$ with values in $\bigwedge^2 \left(\mathfrak{Vir} \oplus \mathfrak{Vir}\right)$ and it again splits according to $\delta = \delta_{\mathfrak{W} \oplus \mathfrak{W}} + \delta_{c} + \delta_{\bar c}$ where $\delta_c$ maps to $c_L \wedge \mathfrak{W}$ and $\delta_{\bar c}$ to $c_L \wedge \overline{\mathfrak{W}} $. The cocycle condition results in separate equations and the analogue of \eqref{virco1} implies that $\delta_{\mathfrak{W} \oplus \mathfrak{W}}$ is a coboundary due to theorem \ref{W+Wth}. Redefining the cocycle accordingly causes $\delta_{\mathfrak{W} \oplus \mathfrak{W}}$ to vanish and the analogue of \eqref{virco2} used with $X = L_m, Y = L_n$ implies an equation with terms in $c_L \wedge \mathfrak{W}$ that can be used to show that $\delta_c$ is cohomolog to zero as in the first part of the proof. Similarly with $X = \bar L_m, Y = \bar L_n$ we infer $\delta_{\bar c}$ is derived from a cochain concluding the proof. 


\underline{\bf{Proof of Theorem \ref{B3th} }}
\newline
Following \cite{Zak97} we start with some general statements on semi-direct products of the form $\mathfrak{g} = \mathfrak{n} \rtimes \mathfrak{h}$, where $\mathfrak{h}$ is semi-simple and $\mathfrak{n}$ is an ideal. If $\mathfrak{h}$ admits only coboundary cocycles then for each cocycle $\delta \in Z(\mathfrak{g}, \bigwedge^2 \mathfrak{g})$ there exists an $r \in \bigwedge^2 \mathfrak{h}$ such that $\delta_0 = \delta - \partial_0 r$ satisfies ${\delta_0}_{\vert_{\mathfrak{h}}} =0$ and it is a cocycle if and only if ${\delta_0}_{\vert_{\mathfrak{n}}} \in  Z( \mathfrak{n}, \bigwedge^2 \mathfrak{g}) \cap \text{Mor}_{\mathfrak{h}} (\mathfrak{n}, \bigwedge^2 \mathfrak{g}) $, where $\text{Mor}_{\mathfrak{h}} (\mathfrak{n}, \bigwedge^2 \mathfrak{g})$ denotes the morphisms between the $\mathfrak{h}$ modules $\mathfrak{n}, \bigwedge^2 \mathfrak{g}$, i.e.
\begin{align}\label{mor}
\delta ( x \triangleright n) = x \triangleright \delta (n), \quad x \in \mathfrak{h}, n \in \mathfrak{n}.
\end{align}
The latter follows from the fact that the cocycle condition involving $[\mathfrak{h}, \mathfrak{n}]$ reduces to \eqref{mor} if ${\delta_0}_{\vert_{\mathfrak{h}}} =0$.

Here we have $\mathfrak{g} \equiv \mathfrak{B}_3, \mathfrak{h} \equiv \mathfrak{W}, \mathfrak{n} \equiv V$ and the plan for the proof is as follows. \textbf{First}, we show that on $\mathfrak{W}$ all cocycles are indeed coboundary, i.e. $\boldsymbol{Z(\mathfrak{W}, \bigwedge^2 \mathfrak{B}_3) = B(\mathfrak{W}, \bigwedge^2 \mathfrak{B}_3)}$. In the \textbf{second step} it is shown that $\boldsymbol{Z( V, \bigwedge^2 \mathfrak{B}_3) \cap \text{\textbf{Mor}}_{\mathfrak{W}} (V, \bigwedge^2 \mathfrak{B}_3)}$ \textbf{vanishes}.

The cocycles (and similarly the coboundaries) can be decomposed according to $$Z(\mathfrak{W}, \bigwedge^2 \mathfrak{B}_3) = Z(\mathfrak{W}, \bigwedge^2 \mathfrak{W}) \oplus Z(\mathfrak{W}, \bigwedge^2 V) \oplus Z(\mathfrak{W}, V \wedge \mathfrak{W}).$$
From theorem \ref{Wth} it is known that $Z(\mathfrak{W}, \bigwedge^2 \mathfrak{W}) = B(\mathfrak{W}, \bigwedge^2 \mathfrak{W})$ and we can use the same arguments to show $Z(\mathfrak{W}, \bigwedge^2 V) = B(\mathfrak{W}, \bigwedge^2 V)$. Namely, all the formula in the proof of theorem \ref{Wth} with $L_m \rightarrow T_m$ in all terms with a wedge product, including the r-matrices, and $L_m \equiv l_m$ in the rest can be used verbatim to establish the latter statement.

Defining the degree of a cocycle $\delta \in Z(\mathfrak{W}, V \wedge \mathfrak{W}) $ to be $i_m + j_m -m$ for
\begin{align}	
\delta (l_m) = l_{i_m} \wedge T_{j_m}
\end{align}
the statement $Z(\mathfrak{W}, V \wedge \mathfrak{W})  = B(\mathfrak{W}, V \wedge \mathfrak{W})$ follows for cocycles of \textbf{degree} $\boldsymbol{d \neq 0}$ similarly as in the proof of theorem \ref{Wth}. 
For \textbf{degree zero cocycles} we again have $\delta (l_0) =0$ from the cocycle condition with $l_0, l_m$. Considering the cocycle condition
\begin{align}
\delta ([ l_1, l_{-1}]) = &  \sum_{i_{-1} \in I_{-1}} \gamma^{-1}_{i_{-1}} \bigg( (2 + i_{-1}) l_{-i_{-1}} \wedge T_{i_{-1}} + (1- i_{-1}) l_{-1-i_{-1}} \wedge T_{1 + i_{-1}}\bigg) \nonumber \\
& -\sum_{i_1 \in I_1} \gamma^1_{i_1} \bigg( (-2 + i_1) l_{-i_1} \wedge T_{i_1} + (-1-i_1) l_{1-i_1} \wedge T_{-1+i_1}\bigg)\label{coc4}, \\
 =& 0,
\end{align}
where we used the ansatz
\begin{align}
\delta (l_1) = \sum_{i_1 \in I_1} \gamma^1_{i_1} l_{1-i_1} \wedge T_{i_1}, \quad \delta (l_{-1}) =\sum_{i_{-1} \in I_{-1}} \gamma^{-1}_{i_{-1}} l_{-1-i_{-1}} \wedge T_{i_{-1}}, 
\end{align}
the situation is a bit different than in the proof of theorem \ref{Wth} as we can not make use of the index restriction. Instead, we have to distinguish two different cases for the ordered set of non-zero indices $i_{1} = - p_j, j = 1, ..., n$. If the largest index is bigger than zero we call it $p_1 \geq p_j$. Then the corresponding fourth term (type $IV$ term) of \eqref{coc4} can not be canceled by a type $III$ term. Thus it can be only canceled by a type $I$ or $II$ term and we use a coboundary to eliminate the potential type $I$ term analogously to the proof of theorem \ref{Wth}. The type $II$ term would require $i_{-1} = -2-p_1$ to cancel the first type $IV$ term which implies the existence of a type $I$ term of the form $l_{2+p_1} \wedge T_{-2 - p_1}$ which necessitates again a type $II$ term with $i_{-1} = -3 -p_1$. This would lead to an infinite tower of terms which is not allowed as $I_{-1}$ has to be a finite set. The only caveat arises if $p_1 = 1$ because a factor in the first line of \eqref{coc4} vanishes in that case, which we discuss below. If the largest index is smaller or equal to zero we call the smallest index $p_1$ and use the coboundary to eliminate the type $II$ term. Then a type $I$ term would be necessary to cancel the type $IV$ term with $p_1$. Again this would lead to an infinite tower except for the case $p_1 =0$. Using the same iterative arguments as in the proof of theorem \ref{Wth} yields that $\gamma^1_{-p_j} = 0$ except for the two special cases $p_1 = 1,0$, i.e. 
\begin{align}
\delta (l_1) = l_0 \wedge T_1, \quad \delta(l_1) = l_1 \wedge T_0.
\end{align}
One can use the r-matrices 
\begin{align}
r = l_{-1} \wedge T_1, \quad r = l_1 \wedge T_{-1}
\end{align}
with appropriate scaling to explicitly show that also in these cases $\delta$ is cohomolog to zero on $l_1$.

When proving that $\delta (l_2)$ is cohomolog to zero we can reuse the equations \eqref{prot1}-\eqref{l3} where every right tensor leg is replaced by $T$ and the left with $l$. The only difference is that without the index restriction there is also the solution $i'_2 = -1$ to \eqref{l3}. Plugging it into \eqref{l2} one finds that e.g. the term proportional to $l_{4} \wedge T_{1}$ does not vanish and thus $\delta(l_2) = 0$. The rest of the proof of $Z(\mathfrak{W}, V \wedge \mathfrak{W})  = B(\mathfrak{W}, V \wedge \mathfrak{W})$ is completely analogous to that of theorem \ref{Wth}.

In the \textbf{second step} we start by noting that
$$ Z(V, \bigwedge^2 \mathfrak{B}_3) = Z(V, \bigwedge^2 \mathfrak{W}) \oplus Z(V, \bigwedge^2 V) \oplus Z(V, V \wedge \mathfrak{W}),$$
and similar for the morphisms. For a $\boldsymbol{\delta \in Z(V, \bigwedge^2 \mathfrak{W})}$ we make the ansatz
\begin{align}
\delta (T_m) = \sum_{i_m, j_m} \alpha^m_{i_m, j_m} l_{i_m} \wedge l_{j_m} 
\end{align}
and if the degree of $\delta$, i.e. $d = i_m + j_m -m$,does not vanish then the condition (from \eqref{mor}) yields
\begin{align}\label{mor2}
l_0 \triangleright \delta (T_m) & = \delta (l_0 \triangleright T_m), \\
\sum_{i_m, j_m} (-i_m-j_m)\alpha^m_{i_m, j_m} l_{i_m} \wedge l_{j_m} & = \sum_{i_m, j_m} (-m)\alpha^m_{i_m, j_m} l_{i_m} \wedge l_{j_m}
\end{align}
and $\delta$ has to be zero. For a degree zero cocycle we set 
\begin{align}
\delta (T_m) = \sum_{i_m} \alpha^m_{i_m} l_{m-i_m} \wedge l_{i_m},
\end{align}
with the index restriction $i_m > m/2$. From the cocycle condition
\begin{align}\label{yac}
0 = \delta([T_0, T_1])  = [T_0 \otimes 1 + 1 \otimes T_0, \delta (T_1)] - [T_1 \otimes 1 + 1 \otimes T_1, \delta (T_0)]
\end{align}
we get
\begin{align}
0 = & \sum_{i_0} \alpha^0_{i_0} ((1+i_0) T_{1-i_0} \wedge l_{i_0} +(1- i_0) l_{-i_0} \wedge T_{1+i_0} ) \nonumber \\
& - \sum_{i_1} \alpha^1_{i_1} ((i_1-1) T_{1-i_1} \wedge l_{i_1} - i_1 l_{1-i_1} \wedge T_{i_1} . \label{coV}
\end{align}
Canceling the term proportional to $T_{1-i_0} \wedge l_{i_0}$ with the largest index $i'_0 \geq i_0$ we find that it can be done with the first term in the first line of \eqref{coV} with $i_1 = i'_0$. A cancellation with the second term and $i_1 = 1-i'_{0}$ is not possible because of the index restriction. Thus we have
\begin{align}\label{lis}
\alpha^0_{i'_0} (1+i'_0) = \alpha^1_{i'_0} (1-i'_0).
\end{align}
From \eqref{mor2} with $m=1$ and $m$ and $0$ exchanged on the other hand one has
\begin{align}
\sum_{i_0} \alpha^0_{i_0} ((1+i_0) l_{1-i_0} \wedge l_{i_0} + (1-i_0) l_{-i_0} \wedge l_{i_0+1}) = \sum_{i_1} \alpha^1_{i_1} l_{1-i_1} \wedge l_{i_1}. 
\end{align}
Cancellation of the first term on the lhs with the largest index can only be done by the rhs and we find
\begin{align}
(1+i'_0) \alpha^0_{i'_0} = \alpha^1_{i'_0},
\end{align}
which would imply $i'_0 =0$ together with \eqref{lis} violating the index restriction. Thus, we conclude that for $\delta \in Z(V, \bigwedge^2 \mathfrak{W}) \cap \text{Mor}_{\mathfrak{W}} (V, \bigwedge^2 \mathfrak{W})$ $\delta (T_0) = 0$ and by \eqref{mor2} also $\delta (T_m) =0$ for all $m$. 

\textbf{Next}, consider a cocycle $\boldsymbol{\delta \in Z(V, V \wedge \mathfrak{W}) \cap \text{\textbf{Mor}}_{\mathfrak{W}} (V, V \wedge \mathfrak{W})}$. If its degree is not zero then the first line in \eqref{mor2} results in $\delta(T_m) =0$ similar to the case before. A degree zero cocycle has the form
\begin{align}
\delta (T_m) = \sum_{i_m} \alpha^m_{i_m} l_{m-i_m} \wedge T_{i_m}
\end{align}
and from the cocycle condition \eqref{yac} we get
\begin{align}
0 = \sum_{i_0} \alpha^0_{i_0} (m+i_0) T_{m-i_0} \wedge T_{i_0} - \sum_{i_m} \alpha^m_{i_m} (i_m-m) T_{m-i_m} \wedge T_{i_m}.
\end{align}
If we set $m$ sufficiently large, the term with the smallest index $i'_0$ on the lhs has to be canceled by $i_m = i'_0$ leading to
\begin{align}\label{mref}
(m+i'_0) \alpha^0_{i'_0} = (i'_0 -m) \alpha^m_{i'_0}.
\end{align}
From the condition that $\delta$ is a morphism of modules (\eqref{mor2} with $m$ and $0$ exchanged) we have furthermore
\begin{align}
\sum_{i_m}\alpha^m_{i_1} m l_{m-i_m} \wedge T_{i_m} = \sum_{i_0} \alpha^0_{i_0} ((m+i_0) l_{m-i_0} \wedge T_{i_0} + (m-i_0)l_{-i_0} \wedge T_{i_0 +m})
\end{align}
leading to 
\begin{align}
\alpha^m_{i'_0}m = \alpha^0_{i_0} (m+i'_0),
\end{align}
which can not be satisfied for all (large enough) $m$ together with \eqref{mref}. Thus, $\delta (T_0) =0$ and by \eqref{mor2} the cocycle vanishes also on all other generators of $V$. 

\textbf{Finally}, a cocycle $\boldsymbol{\delta \in Z(V, \bigwedge^2 V) \cap \text{\textbf{Mor}}_{\mathfrak{W}} (V, \bigwedge^2 V)}$ of degree $d \neq 0$ vanishes also by virtue of \eqref{mor2}. If the degree is zero it can be written as
\begin{align}
\delta (T_m) = \sum_{i_m} \alpha^m_{i_m} T_{m-i_m} \wedge T_{i_m}
\end{align}
and from the morphism condition one extracts
\begin{align}
\sum_{i_m} \alpha^m_{i_m} m T_{m-i_m} \wedge T_{i_m} = \sum_{i_0} \alpha^0_{i_0} ((m+i_0)T_{m-i_0} \wedge T_{i_0} + (m-i_0) T_{-i_0} \wedge T_{i_0+m}).
\end{align}
With the index restriction $i_m < m/2$ and $m <0$ the second term on the rhs with a fixed index $i'_0$ can be canceled by the first with some $i_0 = i'_0 +m$ or by the lhs with $i_m =  i'_0$, requiring that $i'_0 < m/2$. Since the set of indices $i_0$ is finite these two conditions can not be met for sufficiently large $|m|$ and thus  all $\delta \in Z(V, \bigwedge^2 V) \cap \text{Mor}_{\mathfrak{W}} (V, \bigwedge^2 V)$ have to vanish concluding the proof.
\newline

\underline{\bf{Proof of Theorem \ref{B3+th}}}
\newline

We will structure the proof in the same way as the proof of theorem \ref{B3th} and as already noted in the proof of theorem \ref{W++W+th} all results that do not use the index restriction carry over directly. For the first part we are then left to show that $Z \left( \mathfrak{W}_+, \bigwedge^2 V_+ \right) = B \left( \mathfrak{W}_+, \bigwedge^2 V_+ \right) $ which follows from the proof of $H^1 \left( \mathfrak{W}_+, \bigwedge^2 \mathfrak{W}_+ \right) = \{0\}$ if in all wedge products we replace $ L_m \rightarrow T_m$ and $L_m \rightarrow l_m$ in the single generators. 

In the second part of the proof we need to show that degree $0$ cocycles in $Z  \left( V_+ ,\bigwedge^2 \mathfrak{W}_+ \right)  \cap \text{Mor}_{\mathfrak{W}_+} \left( V_+, \bigwedge^2 \mathfrak{W}_+ \right)$ and $Z  \left( V_+ ,\bigwedge^2 V_+ \right)  \cap \text{Mor}_{\mathfrak{W}_+} \left( V_+, \bigwedge^2 V_+ \right)$ vanish. The former cocycle has the form 
\begin{align}
\delta (T_0) = \gamma^0_1 l_0 \wedge l_1, \quad \delta(T_1) = \gamma^1_2 l_{-1} \wedge l_2 + \gamma^1_1 l_0 \wedge l_1 
\end{align}
and imposing the cocycle condition yields
\begin{align}
\delta( [T_0, T_1]) & = [T_0 \otimes 1 + 1 \otimes T_0, \delta(T_1)] - [T_1 \otimes 1 + 1 \otimes T_1, \delta(T_0)], \\
\Rightarrow \gamma^1_2 & = 0, \quad \gamma^1_1 = -2 \gamma^0_1.
\end{align}
From the morphism condition on the other hand we have
\begin{align}
l_1 \triangleright \delta(T_0) = \delta(l_1 \triangleright T_0), \quad \Rightarrow 2 \gamma^0_1 = \gamma^1_1,
\end{align}
which is only possible if $\delta(T_0) =0$. Then we can use cocycle conditions 
\begin{align}
0 = \delta ( [T_0, T_m]) = -m \delta(T_m)
\end{align}
for $m \geq -1$ to show that $\delta$ has to vanish on $V_+$. 

Finally, for a degree $0$ cocycle $\delta \in Z  \left( V_+ ,\bigwedge^2 V_+ \right)$ we use the same argumentation as above with $l_m \rightarrow T_m$ in all wedge products to conclude the proof.

\underline{\bf{Proof of Theorem \ref{B3cth}}}
\newline

Similarly to the proof of theorem \ref{Virth} we start by splitting a cocycle $\delta \in Z \left( \mathfrak{B}_{3c}, \bigwedge^2 \mathfrak{B}_{3c} \right)$ into $\delta = \delta_{\mathfrak{B}} + \delta_c$ so that $\delta_{\mathfrak{B}}$ maps to $\bigwedge^2\mathfrak{B}_3$ and $\delta_c$ to $c_L \bigwedge \mathfrak{B}_3$. Then the cocycle condition \eqref{ccc} reads
\begin{align}\label{bco1}
\delta_{\mathfrak{B}} ( [X,Y])  & = [\delta_{\mathfrak{B}} (X), Y \otimes 1 +1 \otimes Y]_{\mathfrak{B}} -  [\delta_{\mathfrak{B}} (Y), X \otimes 1 +1 \otimes X]_{\mathfrak{B}} , \\
  \delta_c ( [X,Y]) & = [\delta_{\mathfrak{B}} (X),  Y \otimes 1 +1 \otimes Y]_c - (X \leftrightarrow Y)   + [\delta_c(X), Y\otimes  + 1 \otimes Y]_{\mathfrak{B}} - (X \leftrightarrow Y) ,\label{bco2}
\end{align}
where the brackets are defined analogously to the proof of theorem \ref{Virth} and we already omitted any parts of $X, Y \in \mathfrak{B}_3$ containing the central element. Equation \eqref{bco1} then implies with theorem \ref{B3th} that $\boldsymbol{\delta_{\mathfrak{B}}}$ \textbf{is a coboundary} derived from a cochain $r \in \bigwedge^2 \mathfrak{B}_3$ and we redefine $\delta \rightarrow \delta  - \delta_r$. Thus the first terms on the rhs of \eqref{bco2} vanish and \textbf{for} $\boldsymbol{\delta_c}$ we proceed similarly as in the proof of theorem \ref{B3th}. This means we first consider cocycles in $Z \left( \mathfrak{W}, c_L \bigwedge \mathfrak{B}_3 \right) = Z \left( \mathfrak{W}, c_L \bigwedge \mathfrak{W} \right) \oplus Z \left( \mathfrak{W}, c_L \bigwedge V_3 \right)$. For the first summand the proposition $ Z \left( \mathfrak{W}, c_L \bigwedge \mathfrak{W} \right)  =  B \left( \mathfrak{W}, c_L \bigwedge \mathfrak{W} \right) $ follows as in the last part of the proof of theorem \ref{W+Wth} by replacing $L_m \rightarrow l_m$ and $\bar L_m \rightarrow c_L$ in all the formulas. The same formulas can also be used with the replacement $L_m \rightarrow T_m, \, \bar L_m \rightarrow c_L$ in all wedge products including the r-matrices and $ L_m \rightarrow l_m, \, \bar L_m \rightarrow c_L$ in the single tensors to show that $Z \left( \mathfrak{W}, c_L \bigwedge V_3 \right) = B \left( \mathfrak{W}, c_L \bigwedge V_3 \right)$. 

After redefining the cocycle so that $\delta_c|_{\mathfrak{W}} =0$ the cocycle condition involving $l_n, T_m$ reduces to the morphism condition. In particular, for $\boldsymbol{\delta_c \in Z \left( V_3, c_L \bigwedge \mathfrak{W} \right)}$ we additionally have to demand 
\begin{align}\label{morc}
\delta_c ( l_n, T_m]) = - [l_n \otimes 1 +1 \otimes l_n, \delta_c(T_m)].
\end{align}
This equation and the cocycle condition 
\begin{align}\label{coct}
\delta_c ([T_m, T_n]) = [T_m \otimes 1 + 1\otimes T_m, \delta_c(T_n)] -  [T_n \otimes 1 + 1\otimes T_n, \delta_c(T_m)]
\end{align}
have to be satisfied separately for cocycles with different degree which we define to be $d = i_m -m$ for
\begin{align}
\delta_c(T_m) = \alpha^m_{i_m} c_L \wedge l_{i_m}.
\end{align}
For $d \neq 0$ the morphism condition \eqref{morc} with $n =0$ implies $- m \delta_c(T_m) = -i_m \delta_c(T_m) \rightarrow \delta_c(T_m) =0$. If $d =0$ the morphism condition with $l_m, T_0$ yields 
\begin{align}
m \alpha^m_m c_L \wedge T_m = m \alpha^0_0 c_L \wedge T_m
\end{align}
which contradicts the cocycle condition with $T_0, T_m$ leading to $\alpha^0_0 = \alpha^m_m$ unless $\delta_c$ vanishes. 

\textbf{Finally}, we have to deal with cocycles in $\boldsymbol{Z \left( V_3, c_L \bigwedge V_3 \right)}$. The degree is defined analogously and for $d \neq 0$ the morphism condition \eqref{morc} with $n =0$ and the ansatz $\delta_c (T_m) = \alpha^m_{i_m} c_L \wedge T_{i_m}$ yields $- m \delta_c (T_m) = -i_m \delta_c(T_m) \rightarrow \delta_c(T_m) =0$. For $d =0$ on the other hand the morphism condition with $l_m, T_0$ implies $\alpha^m_m = \alpha^0_0$ for all $m \in \mathbb{Z}$. As the other cocycle conditions are trivially satisfied we showed that 
\begin{align}
\delta_{\alpha} : l_m \mapsto 0, \; T_m \mapsto \alpha c_L \wedge T_m
\end{align}
is indeed a cocycle in $Z \left( \mathfrak{B}_3 , \bigwedge^2 \mathfrak{B}_3\right)$. Is it easy to see that it can not be derived from a cochain since the only possibility that has the correct grading ($r \sim c_L \wedge l_0$) produces the wrong coefficients. Thus we showed that $H^1  \left( \mathfrak{B}_3 , \bigwedge^2 \mathfrak{B}_3\right) = \delta_{\alpha}$.

\underline{\bf{Proof of Theorem \Ref{B4th}}}
\newline

By the arguments from the proof of theorem \ref{B3th} for the four dimensional BMS algebra $\mathfrak{B}_4 = (\mathfrak{W} \oplus \overline{\mathfrak{W}}) \rtimes V_4$ it suffices to show that 
\begin{align}\label{coh1}
Z((\mathfrak{W} \oplus \overline{\mathfrak{W}}), \bigwedge^2 \mathfrak{B}_4) = B ((\mathfrak{W} \oplus \overline{\mathfrak{W}}), \bigwedge^2 \mathfrak{B}_4)
\end{align}
and 
\begin{align}
Z(V_4, \bigwedge^2 \mathfrak{B}_4) \cap \text{Mor}_{\mathfrak{W} \oplus \overline{\mathfrak{W}}}(V_4, \bigwedge^2 \mathfrak{B}_4) = 0.
\end{align}
In order to show \eqref{coh1} one can decompose
\begin{align}
Z((\mathfrak{W} \oplus \overline{\mathfrak{W}}), \bigwedge^2 \mathfrak{B}_4) = & Z((\mathfrak{W} \oplus \overline{\mathfrak{W}}), \bigwedge^2 \mathfrak{W}) \oplus Z((\mathfrak{W} \oplus \overline{\mathfrak{W}}), \bigwedge^2 \overline{\mathfrak{W}}) \oplus Z((\mathfrak{W} \oplus \overline{\mathfrak{W}}), \bigwedge^2 V_4) \nonumber \\
&\oplus Z((\mathfrak{W} \oplus \overline{\mathfrak{W}}), \mathfrak{W} \wedge \overline{\mathfrak{W}}) \oplus Z((\mathfrak{W} \oplus \overline{\mathfrak{W}}), \mathfrak{W} \wedge V_4)   \oplus Z((\mathfrak{W} \oplus \overline{\mathfrak{W}}), \overline{\mathfrak{W}} \wedge V_4)  \label{decomp}
\end{align}
and from Theorem \ref{Wth},\ref{W+Wth} and \ref{B3th} we know that 
\begin{gather}
Z(\mathfrak{W}, \bigwedge^2 \mathfrak{W}) = B (\mathfrak{W}, \bigwedge^2 \mathfrak{W}) , \quad
Z(\mathfrak{W}, \mathfrak{W}\wedge \overline{\mathfrak{W}}) = B(\mathfrak{W}, \mathfrak{W}\wedge \overline{\mathfrak{W}}) , \\
Z(\mathfrak{W}, \bigwedge^2 \overline{\mathfrak{W}}) = 0
\end{gather}
and similar for $\mathfrak{W} \leftrightarrow \overline{\mathfrak{W}}$. 
Considering a cocycle $\boldsymbol{\delta \in Z (\mathfrak{W},  \mathfrak{W} \wedge V_4))}$ we make the ansatz
\begin{align} \label{ansc}
\delta(l_m) = \sum_{i_m, j_m, k_m}  \alpha^m_{i_m, j_m, k_m} l_{i_m} \wedge T_{j_m, k_m}. 
\end{align}
Choosing the cochain 
\begin{align}
r = \sum_{i_0, j_0, k_0} \alpha^0_{i_0, j_0, k_0} \frac{-1}{i_0 + j_0 - 1/2} l_{i_0} \wedge T_{j_0, k_0}
\end{align}
the cocycle $\delta'  = \delta - \partial_0 r$\footnote{We redefine the ansatz so that \eqref{ansc} holds for $\delta'$ except that $\alpha^0$ vanishes.} vanishes on $l_0$ and thus the cocycle condition
\begin{align}\label{cocx}
\delta'([l_0, l_m]) & = [l_0 \otimes 1 + 1 \otimes l_0, \delta'(l_m) ] - [l_m \otimes 1 + 1 \otimes l_m, \delta'(l_0) ], \\
\rightarrow  (-i_m + 1/2-j_m) \delta'(l_m) & = - m \delta'(l_m)
\end{align}
implies that it is indeed zero on all generators irrespective of the degree as the indices take only values in $\mathbb{Z}$.

Similar, for $\boldsymbol{\delta \in Z (\mathfrak{W},  \overline{\mathfrak{W}} \wedge V_4))}$ with the ansatz 
\begin{align} \label{ansc2}
\delta(l_m) = \sum_{i_m, j_m, k_m}  \alpha^m_{i_m, j_m, k_m} \bar l_{i_m} \wedge T_{j_m, k_m}
\end{align}
and 
\begin{align}
r = \sum_{i_0, j_0, k_0} \alpha^0_{i_0, j_0, k_0} \frac{-1}{ j_0 - 1/2} \bar l_{i_0} \wedge T_{j_0, k_0}
\end{align}
the cocycle $\delta' = \delta - \partial_0 r$ vanishes on $l_0$. The cocycle condition \eqref{cocx} again yields
\begin{align} 
( 1/2-j_m) \delta'(l_m) = - m \delta'(l_m)
\end{align}
proving that $Z (\mathfrak{W},  \overline{\mathfrak{W}} \wedge V_4)) = B (\mathfrak{W},  \overline{\mathfrak{W}} \wedge V_4)) $.

This leaves only $\boldsymbol{Z (\mathfrak{W}, \bigwedge^2 V_4))}$ in the \textbf{first step} of the proof. For a cocycle $\delta$ of the form
\begin{align}
\delta (l_m) = \sum_{i_m, j_m, k_m, s_m} \alpha^m_{i_m, j_m, k_m, s_m} T_{i_m, j_m} \wedge T_{k_m, s_m}
\end{align}
we define the (left) degree $d_L$ to be $d_L = m-i_m -k_m$ and analogously the right degree $d_R$. Then for a cocycle of degree $d_L \neq 1$ (or similarly for $d_R \neq 1$) the cochain 
\begin{align}
r = \sum_{i_0, j_0, k_0, s_0} \alpha^0_{i_0, j_0, k_0, s_0} \frac{-1}{1- i_0-k_0} T_{i_0, j_0} \wedge T_{k_0, s_0}
\end{align}
can be used to show that $\delta' = \delta - \partial_0 r$ vanishes on $l_0$. In that case the cocycle condition \eqref{cocx} yields
\begin{align}
-m \delta'(l_m) = (1-i_m - k_m) \delta'(l_m),
\end{align}
which has to vanish. 

A cocycle of (left) degree one can be written as
\begin{align}
\delta(l_m) = \sum_{i_m, j_m, s_m} \alpha^m_{i_m, j_m, s_m} T_{m-i_m, j_m} \wedge T_{i_m +1, s_m},
\end{align}
with $i_m < m/2$. From \eqref{cocx} one obtains 
\begin{align}
-m \delta(l_m) = -m \delta (l_m) - [l_m \otimes 1 + 1 \otimes l_m, \delta (l_0)],
\end{align}
which has to hold for all $m$ and thus $\delta(l_0) = 0$. 
Then the cocycle condition 
\begin{align}
\delta ([l_1 , l_{-1}]) =  [l_1 \otimes 1 + 1 \otimes l_1, \delta(l_{-1}) ] - [l_{-1} \otimes 1 + 1 \otimes l_{-1}, \delta(l_1) ]
\end{align}
yields the equation
\begin{align}
0 = & \sum_{i_1, j_1, s_1} \alpha^1_{i_1, j_1, s_1} \left((i_1-1) T_{-i_1, j_1} \wedge T_{i_1 +1 s_1} - (i_1 +1) T_{1-i_1, j_1} \wedge T_{i_1, s_1} \right) \nonumber \\
&- \sum_{i_{-1}, j_{-1}, s_{-1}} \alpha^{-1}_{i_{-1}, j_{-1}, s_{-1}} \left( (2-i_{-1}) T_{-i_{-1}, j_1} \wedge T_{i_{-1} +1, s_{-1}} -i_{-1} T_{-1-i_{-1}, j_{-1}} \wedge T_{i_{-1} +2, s_{-1}} \right).\label{pp}
\end{align}
Comparing with \eqref{p} shows that the index structure (neglecting the second index of the $T$) is almost the same, except that the index of the $T$ on the rhs of the wedge product and the prefactors on the second terms in the first and second line of \eqref{pp} are always higher by one. Because of the index restriction, the ordering of the wedge product is fixed (so that a left and a right term never have to coincide) and thus the same argumentation can be used. This is also consistent with the prefactors as any commutators with the right wedge product legs are higher by one compared to the situation in the proof of theorem \ref{Wth}. Therefore $\delta (l_1) =0$ and we use the same strategy to show that $\delta (l_2) =0$, i.e. we iteratively use the cocycle condition to compute $\delta(l_5)$ and then compare with the cocycle condition with $l_2$ and $l_3$. As a result, we obtain in analogy with \eqref{l3}
\begin{align}
0 = T_{2-i'_2, j_2} \wedge T_{i'_2 +4, s_2} \left(-\left( \frac{1}{2}+i'_2\right) i'_2 +(i'_2-1) - \frac{1}{6} i'_2 (1+i'_2) (2+i'_2) \right)
\end{align}
for the largest index $i'_2$ and since this equation has no solution in $\mathbb{Z}$, $\delta(l_2)=0$. Then $\delta(l_m) =0, m >0$ follows easily and 
\begin{align}
0 = \delta([l_{-1}, l_m]) = - [l_m \otimes 1 +1 \otimes l_m, \delta (l_{-1})]
\end{align}
for all positive $m$ yields $\delta(l_{-1}) =0$. Showing that $\delta(l_{-2}) =0$ and subsequently for all negative $m$ is again obtained in an analogous way and thus $Z (\mathfrak{W}, \bigwedge^2 V_4)) = B (\mathfrak{W}, \bigwedge^2 V_4))$.

In the \textbf{second step} it will be shown that $\boldsymbol{Z( V_4, \bigwedge^2 \mathfrak{B}_4) \cap \text{\textbf{Mor}}_{\mathfrak{W} \oplus \overline{\mathfrak{W}}} (V_4, \bigwedge^2 \mathfrak{B}_4) = \{0\}}$. Using a decomposition as in \eqref{decomp} we \textbf{start with} $\boldsymbol{\text{\textbf{Mor}}_{\mathfrak{W} \oplus \overline{\mathfrak{W}}} (V_4, \bigwedge^2 \mathfrak{W})}$.  The general ansatz
\begin{align}
\delta(T_{mk}) = \sum_{i,j} l_i \wedge l_j
\end{align}
and the morphism condition
\begin{align}\label{mormor}
l_0 \triangleright \delta (T_{mk}) = \delta (l_0 \triangleright T_{mk})
\end{align}
yield
\begin{align}
\sum_{i,j} \left( \frac{1}{2}-m \right) l_i \wedge l_j = \sum_{i,j}(-i-j) l_i \wedge l_j 
\end{align}
and thus $\text{Mor}_{\mathfrak{W} \oplus \overline{\mathfrak{W}}} (V_4, \bigwedge^2 \mathfrak{W})= \{0\}$.
Similarly, one has $\boldsymbol{\text{\textbf{Mor}}_{\mathfrak{W} \oplus \overline{\mathfrak{W}}} (V_4, \bigwedge^2 \overline{\mathfrak{W}}) =\{0\}}$ and \linebreak $\boldsymbol{\text{\textbf{Mor}}_{\mathfrak{W} \oplus \overline{\mathfrak{W}}} (V_4,  \mathfrak{W} \wedge \overline{\mathfrak{W}}) =\{0\}}$. 

For a cocycle $\boldsymbol{\delta \in \text{\textbf{Mor}}_{\mathfrak{W} \oplus \overline{\mathfrak{W}}} (V_4, \bigwedge^2 V_4)}$ \eqref{mormor} results in
\begin{gather}
\sum_{i_{mk}, j_{mk}, k_{mk}, s_{mk}} \alpha^{mk}_{i_{mk}, j_{mk}, k_{mk}, s_{mk}} \left( \frac{1}{2} -m \right)  T_{i_{mk}, j_{mk}} \wedge T_{k_{mk}, s_{mk}} \nonumber \\
= \sum_{i_{mk}, j_{mk}, k_{mk}, s_{mk}} \alpha^{mk}_{i_{mk}, j_{mk}, k_{mk}, s_{mk}} (1-i_{mk} - k_{mk}) T_{i_{mk}, j_{mk}} \wedge T_{k_{mk}, s_{mk}} ,
\end{gather}
with the ansatz 
\begin{align}
\delta(T_{mk} ) = \sum_{i_{mk}, j_{mk}, k_{mk}, s_{mk}} \alpha^{mk}_{i_{mk}, j_{mk}, k_{mk}, s_{mk}} T_{i_{mk}, j_{mk}} \wedge T_{k_{mk}, s_{mk}} 
\end{align}
and therefore $\text{Mor}_{\mathfrak{W} \oplus \overline{\mathfrak{W}}} (V_4, \bigwedge^2 V_4) = \{0\}$.

\textbf{Finally}, for a cocycle $\boldsymbol{\delta \in \text{\textbf{Mor}}_{\mathfrak{W} \wedge \overline{\mathfrak{W}}} (V_4,  V_4 \oplus \mathfrak{W})}$ we define the degree such that 
\begin{align}
\delta (T_{mk}) = \sum_{i_{mk}, j_{mk}, k_{mk}} \alpha^{mk}_{i_{mk}, j_{mk}, k_{mk}} l_{i_{mk}} \wedge T_{j_{mk}, k_{mk}}
\end{align}
has degree $d= i_{mk} +j_{mk} -m$. If $d \neq 0$ the equation
\begin{gather}
\left( \frac{1}{2} - m \right) \sum_{i_{mk}, j_{mk}, k_{mk}} \alpha^{mk}_{i_{mk}, j_{mk}, k_{mk}} l_{i_{mk}} \wedge T_{j_{mk}, k_{mk}} \nonumber \\
 = \sum_{i_{mk}, j_{mk}, k_{mk}} \alpha^{mk}_{i_{mk}, j_{mk}, k_{mk}} \left(\frac{1}{2} - i_{mk}- j_{mk} \right) l_{i_{mk}} \wedge T_{j_{mk}, k_{mk}}
\end{gather}
from \eqref{mormor} implies $\delta =0$. 
Otherwise, i.e. for a degree zero cocycle, we set
\begin{align}
\delta (T_{mk}) = \sum_{i_{mk}, s_{mk}} \alpha^{mk}_{i_{mk}, s_{mk}} l_{m-i_{mk}} \wedge T_{i_{mk}, s_{mk}}
\end{align}
and the cocycle condition 
\begin{align}
0 = \delta ( [T_{0 k}, T_{mk'}]) =  [T_{0 k} \otimes 1 + 1 \otimes T_{0 k}, \delta (T_{mk'})] [T_{m k'} \otimes 1 + 1 \otimes T_{m k'}, \delta (T_{0k})]  
\end{align}
results in 
\begin{align}
0 =& \sum_{i_{mk'}, s_{mk'}}  \alpha^{mk'}_{i_{mk'}, s_{mk'}} \left(- \frac{m-i_{mk} +1}{2} T_{m-i_{mk'}} \wedge T_{i_{mk'}, s_{mk'}} \right) \nonumber \\
&- \sum_{i_{0k}, s_{0k}} \alpha^{0k}_{i_{0k}, s_{0k}} \left( \left( - \frac{1-i_{0k}}{2} -m \right) T_{m-i_{0k}, k} \wedge T_{i_{0k}, s_{ok}} \right), \\
\rightarrow \alpha^{mk}_{i'_{0k} , s_{0k}}  \frac{m -i'_{0k} +1}{2} = & \alpha^{0k}_{i'_{0k}, s_{0k}} \left( \frac{1-i'_{0k}}{2} -m \right) \label{usc}
\end{align}
for the smallest index $i'_{0k}$ and sufficiently large $m$. Additionally, the morphism condition yields
\begin{align}
l_m \triangleright \delta (T_{0k}) =&  \delta (l_m \triangleright T_{0k}) , 
\end{align}
\begin{gather}
\sum_{i_{0k}, s_{0k}} \alpha^{0k}_{i_{0k}, s_{0k}} \left( (m+i_{0k}) l_{m-i_{0k}} \wedge T_{i_{0k}, s_{0k}} + \left( \frac{m+1}{2} -i_{0k}\right) l_{-i_{0k}} \wedge T_{i_{0k} +m, s_{0k}} \right) \nonumber \\
 =  \sum_{i_{mk}, s_{mk}} \alpha^{mk}_{i_{mk}, s_{mk}} \left(\frac{m+1}{2} l_{m-i_{mk}} \wedge T_{i_{mk}, s_{mk}} \right), \\
\rightarrow  \alpha^{0k}_{i'_{0k}, s_{0k}}(m + i'_{0k}) =  \alpha^{mk}_{i'_{0k}, s_{0k}} \frac{m+1}{2}\label{usc2}
\end{gather}
under the same conditions. Since the equations \eqref{usc} and \eqref{usc2} can not be satisfied simultaneously for all $m$ we have  $\text{Mor}_{\mathfrak{W} \wedge \overline{\mathfrak{W}}} (V_4,  V_4 \wedge \mathfrak{W}) = \{0\}$ and analogously $ \text{Mor}_{\mathfrak{W} \wedge \overline{\mathfrak{W}}} (V_4,  V_4 \wedge \overline{\mathfrak{W}}) = \{0\}$. This completes the proof.
\newline

\underline{\bf{Proof of Theorem \ref{B4cth}}}
\newline

The proof is structured similarly as for theorem \ref{B3cth}. A generic cocycle in $Z \left( \mathfrak{B}_{4c}, \bigwedge^2 \mathfrak{B}_{4c} \right)$ can be split into the parts $\delta = \delta_{\mathfrak{B}} + \delta_c + \delta_{\bar c} + \delta_{c,\bar c}$ mapping to $\bigwedge^2 \mathfrak{B}_4$, $ c_l \bigwedge \mathfrak{B}_4$, $c_{\bar l} \bigwedge \mathfrak{B}_4$ and $c_l \bigwedge c_{\bar l} \mathbb{C}$ respectively. The cocycle condition with $X, Y \in \mathfrak{B}_4$
\begin{align}
\delta([X, Y]) = [ \delta(X), Y \otimes 1 + 1 \otimes Y] - ( X \leftrightarrow Y)
\end{align}
splits into
\begin{align}\label{4dco1}
\delta_{\mathfrak{B}}([X,Y]_{\mathfrak{B}})  & = [\delta_{\mathfrak{B}}(X), Y \otimes 1 + 1 \otimes Y]_{\mathfrak{B}} - (X \leftrightarrow Y), \\
\delta_{c} ([X,Y]_{\mathfrak{B}} ) & = [ \delta_{\mathfrak{B}} (X), Y\otimes 1 + 1 \otimes Y ]_{c} + [\delta_c (X) , Y \otimes 1 + 1 \otimes Y]_{\mathfrak{B}}  - (X \leftrightarrow Y), \label{4dco2}\\
 \delta_{\bar c} ([X,Y]_{\mathfrak{B}} ) & = [ \delta_{\mathfrak{B}} (X), Y\otimes 1 + 1 \otimes Y ]_{\bar c} + [\delta_{\bar c} (X) , Y \otimes 1 + 1 \otimes Y]_{\mathfrak{B}}  - (X \leftrightarrow Y), \label{4dco3} \\
 \delta_{c, \bar c} ([X,Y]_{\mathfrak{B}}) & = [\delta_c (X), Y \otimes 1 + 1 \otimes Y]_{\bar c} + [\delta_{\bar c} (X) , Y \otimes 1 + 1 \otimes Y]_c - (X \leftrightarrow Y), \label{4dco4}
\end{align}
where $[ , ]_{\mathfrak{B}}$, $[ , ]_c$ and $[ , ]_{\bar c}$ conserves or respectively increases the number of central charges in the input. Equation \eqref{4dco1} implies with theorem \ref{B4th} that $\boldsymbol{\delta_{\mathfrak{B}}}$ \textbf{is a coboundary} with a cochain $r \in \bigwedge^2 \mathfrak{B}_4$. After redefining $\delta \rightarrow \delta - \delta_r$ we next turn to the semi-simple ``factor'' $\mathfrak{W} \oplus \mathfrak{W}$ of the semi-direct product $\mathfrak{B}$ using equations \eqref{4dco2}-\eqref{4dco3} where the first term on the rhs vanishes. We further decompose 
\begin{align}
Z \left( \mathfrak{W} \oplus \mathfrak{W}, c_l \bigwedge \left( \mathfrak{W} \oplus \mathfrak{W} \right) \right) = & Z \left(\mathfrak{W} , c_l \bigwedge \mathfrak{W} \right) \oplus Z \left(\mathfrak{W} , c_l \bigwedge \overline{\mathfrak{W} }\right)  \nonumber \\
& \oplus Z \left(\overline{\mathfrak{W}} , c_l \bigwedge \overline{\mathfrak{W}} \right) \oplus Z \left(\overline{\mathfrak{W}} , c_l \bigwedge \mathfrak{W} \right) 
\end{align}
and \eqref{4dco2} for $\boldsymbol{\delta_c \in Z \left(\mathfrak{W} , c_l \bigwedge \mathfrak{W} \right)}$ can be used as in the last part of the proof of theorem \ref{W+Wth} by replacing $L_m \rightarrow l_m, \bar L_m \rightarrow c_l$ to show that $\delta_c$ has to be a coboundary from a cochain $r \in c_l \bigwedge \mathfrak{W}$. Similarly $\boldsymbol{Z \left(\overline{\mathfrak{W}} , c_l \bigwedge \overline{\mathfrak{W}} \right) = B \left(\overline{\mathfrak{W}} , c_l \bigwedge \overline{\mathfrak{W}} \right)}$ and the remaining parts have to vanish because e.g. the rhs of \eqref{4dco2} is alwas zero for $\delta_c \in Z \left(\mathfrak{W} , c_l \bigwedge \overline{\mathfrak{W} }\right)$. Equation \eqref{4dco3} completely analogously asserts the same statements for $ c_l \rightarrow c_{\bar l}$.

\textbf{Next} we consider cocycles $\boldsymbol{\delta_c \in Z \left( \mathfrak{W} \oplus \mathfrak{W}, c_l \bigwedge V_4 \right)}$. With the general ansatz 
\begin{align}
\delta_c (l_m) = \sum_{p_m, q_m} \alpha^m_{p_m, q_m} c_l \wedge T_{p_m, q_m}
\end{align}
the cochain $r = \sum_{p_0, q_0} \alpha^0_{p_0, q_0} \frac{1}{1/2-p_0} c_l \wedge T_{p_0, q_0}$ can be used so that $\delta_c \rightarrow \delta_c - \delta_r$ vanishes on $l_0$. Then the cocycle condition \eqref{4dco2} with $X = l_m, Y = l_0$ gives
\begin{align}
m \delta_c(l_m) = \left( p- \frac{1}{2} \right) \delta_c (l_m)
\end{align}
and thus $\delta_c(l_m) =0$ and similarly also $\delta_c(\bar l_m) =0 = \delta_{\bar c} (l_m) = \delta_{\bar c} ( \bar l_m)$.

Equation \eqref{4dco2} for $\boldsymbol{\delta_c \in Z \left(V_4 , c_l \bigwedge \mathfrak{W}\right)}$ (and similarly for $\mathfrak{W} \rightarrow \overline{\mathfrak{W}}$) with $X = l_0, Y = T_{pq}$ is just a morphism condition and with the ansatz 
\begin{align}
\delta_c(T_{pq}) = \sum_{i_{pq}}\alpha^{pq}_{i_{pq}} c_l \wedge l_{i_{pq}}
\end{align}
 it yields
 \begin{align}
\left( \frac{1}{2} -p \right) \delta_c(T_{pq}) =  \delta_c (T_{pq})i_{pq}
\end{align}
which thus has to vanish.

This leaves us with $\boldsymbol{\delta_c \in Z \left(V_4, c_l \bigwedge V_4 \right)}$ where the morphism condition with $X = l_0, Y = T_{pq}$ and the ansatz
\begin{align}
\delta_c (T_{pq})  = \sum_{i_{pq}, j_{pq}} \alpha^{pq}_{i_{pq}, j_{pq}} c_l \wedge T_{i_{pq}, j_{pq}}
\end{align}
results in
\begin{align}
\left( p - \frac{1}{2} \right) \delta_c (T_{pq}) = \delta_c (T_{pq}) \left( i_{pq}- \frac{1}{q} \right),
\end{align}
which implies $\delta_c (T_{pq}) = 0$ unless the left degree $d_L \equiv i_{pq}-m$ is zero. Similarly one finds that any cocycle with right degree $d_R \equiv j_{pq}-m \neq 0$ has to vanish, leaving only the possibility
\begin{align}
\delta_{\alpha, \bar \alpha}: l_m \mapsto 0, \quad \bar l_m \mapsto 0, \quad T_{pq} \mapsto (\alpha c_l + \bar \alpha c_{\bar l}) \wedge T_{pq},
\end{align}
which satisfies all the cocyle conditions and can not be written as a coboundary because the only r-matrices with the correct grading ($r \sim c_{l/\bar l} \wedge l_0, \; r  \sim c_{l/\bar l} \wedge \bar l_0$) do not reproduce the coefficients.  

\textbf{Finally}, we have to consider cocycles mapping to $\boldsymbol{c_l \bigwedge c_{\bar l} \mathbb{C}}$. To this end we simply use the cocycle condition \eqref{4dco4} with $X = l_m, Y = l_n$ (respectively $X = \bar l_m, Y = \bar l_n$) and that $\delta_c$ ($\delta_{\bar c}$) is cohomolog to zero to show that $\delta_{c, \bar c}(l_m)$  ($\delta_{c, \bar c} (\bar l_m)$) has to vanish. The same equations with $X = l_m, Y = T_{pq}$ imply $\delta_{c, \bar c}(T_{pq})=0$ if we take into account that on the rhs $\delta_{c} (T_{pq})$ (or $\delta_{\bar c} (T_{pq})$) is cohomolog to at most $\delta_{\alpha, \bar \alpha}$ and thus the brackets $[, ]_{c} , [ , ]_{\bar c}$ vanish.
 
From cocycle conditions with $X = c_l$ and $Y \in \mathfrak{B}_4$ we again conclude that $\delta( c_l)$ (or $\delta (c_{\bar l})$) has to be ad-invariant. Contrary to the proof of theorems \ref{Virth} and \ref{B3cth} this at first sight leaves open the possibility that $\delta (c_l) \sim c_l \wedge c_{\bar l}$. However, from
\begin{align}
\delta( [l_n, l_{-n}]_c) = [\delta(l_n), l_{-n} \otimes 1 + 1 \otimes l_{-n}]_c - ( n \leftrightarrow -n) 
\end{align}
and the previous results that $\delta$ is cohomolog to zero on $\mathfrak{W}$ we see that $\delta(c_l) = \delta(c_{\bar l}) =0$ concluding the proof.

\chapter{Classification of $r$-matrices}\label{app-b}

\section{Classification of Triangular r-matrices in $\mathfrak{o}(4)$}

First, note that since $\mathfrak{o}(4, \mathbb{C}) = \mathfrak{sl}(2) \oplus  \bar {\mathfrak{sl}}(2) $ and $\bigwedge^2 \mathfrak{o}(4, \mathbb{C}) = \mathfrak{sl}(2) \wedge  {\mathfrak{sl}}(2)  \oplus \mathfrak{sl}(2) \wedge  \bar {\mathfrak{sl}}(2) \oplus \bar {\mathfrak{sl}}(2) \wedge  \bar {\mathfrak{sl}}(2)$  each r-matrix can be split according to 
\begin{align}
r = a + \bar a + b, \quad a \in \mathfrak{sl}(2) \wedge {\mathfrak{sl}}(2),\, \bar a \in \bar{ \mathfrak{sl}}(2) \wedge  \bar {\mathfrak{sl}}(2),\, b \in \mathfrak{sl}(2) \wedge  \bar {\mathfrak{sl}}(2).
\end{align}
Starting with a generic 
\begin{align}\label{ans-a}
a = \alpha_+ L_1 \wedge L_0 + \alpha_0 L_1 \wedge L_{-1} + \alpha_- L_{-1} \wedge L_0,
\end{align}
triangularity $[[a, a]] = 0$ enforces 
\begin{align}\label{alin}
\alpha_{0}^2 = - \alpha_+ \alpha_-. 
\end{align}
Using the automorphism \eqref{au1} with $\gamma = \sqrt{- \frac{\alpha_-}{\alpha_+}}, \epsilon = 1$ in the case $\alpha_0 \neq 0$ and with $\epsilon = -1(+1)$ if $\alpha_-=0 (\alpha_+ =0)$ we find that there are two one-parameter r-matrices in $\mathfrak{sl}(2) \wedge  {\mathfrak{sl}}(2)$
\begin{gather}
a_1 = \alpha ( L_1 \wedge L_0 +  L_1 \wedge L_{-1} -L_{-1} \wedge L_0), \\
a_2 = \alpha L_1 \wedge L_0
\end{gather}
and similar for  $\bar{\mathfrak{sl}}(2) \wedge \bar{\mathfrak{sl}}(2)$
\begin{gather}
\bar a_1 = \bar \alpha ( \bar L_1 \wedge \bar L_0 + \bar L_1 \wedge \bar L_{-1} - \bar L_{-1} \wedge \bar L_0), \\
\bar a_2 =   \bar \alpha \bar L_1 \wedge \bar L_0.
\end{gather}
For r-matrices that only contain terms of type $b$ one has to demand $[[b, b]] =0$ and the general result (before applying any automorphisms) as obtained in \cite{Borowiec:2015nlw} reads
\begin{align}
(\beta_+ L_1 + \beta_0 L_0 + \beta_- L_{-1} ) \wedge (\bar \beta_+ \bar L_1 + \bar \beta_0 \bar L_0 + \bar \beta_- \bar L_{-1}).
\end{align}
Taking into account the automorphisms \eqref{au1}, \eqref{au2} one can represent this as eleven r-matrices with up to four parameters 
\begin{align}\label{bb1}
b_1 = & (\beta L_1 + \beta_0 L_0 + \beta L_{-1}) \wedge (\bar \beta \bar L_1 + \bar \beta_0 \bar L_0 + \bar \beta  \bar L_{-1}) ,\\
b_2 = & ( L_1 +L_0 ) \wedge (\bar \beta \bar L_1 + \bar \beta_0 \bar L_0 + \bar \beta  \bar L_{-1}), \\
b_3 = & (L_1 + L_{-1}) \wedge (\bar \beta \bar L_1 + \bar \beta_0 \bar L_0 + \bar \beta  \bar L_{-1}), \\
b_4 = & \beta( L_1 +  L_0 ) \wedge ( \bar L_1 +  \bar L_0 ), \quad b_5 =  \beta( L_1 + L_{-1}) \wedge  (\bar L_1 +  \bar L_{-1} ),  \\
b_6 = & \beta( L_1 + L_{-1}) \wedge  (\bar L_1 +  \bar L_0 ), \quad b_7 = L_1 \wedge  (\bar \beta \bar L_1 + \bar \beta_0 \bar L_0 + \bar \beta  \bar L_{-1}), \\
b_8 = & L_1 \wedge \bar \beta(\bar L_1 + \bar L_0), \quad b_9 = L_1 \wedge \bar \beta(\bar L_1 + \bar L_{-1}), \\
b_{10} = & L_1 \wedge \bar L_1, \quad b_{11} = L_1 \wedge \bar L_0. \label{bb2}
\end{align}
When combining $a$, $\bar a$ and $b$ terms there are two different cases, $[[b, b]] =0$ and $[[b, b]] = -2[[b, a]] - 2[[b, \bar a]] \neq 0$, that will be analyzed separately. 
In the first case (for the moment considering only $a$ terms) one infers $[[b, a]] = 0$ and $[[a, a ]] =0$. With the general ansatz \eqref{ans-a} for $a$ and 
\begin{align}
b = & \beta_1 L_{1} \wedge \bar L_{1} +  \beta_2 L_{1} \wedge \bar L_{0} +  \beta_3 L_{1} \wedge \bar L_{-1} \nonumber \\
& +  \beta_4 L_{0} \wedge \bar L_{1} +  \beta_5 L_{0} \wedge \bar L_{0} +  \beta_6 L_{0} \wedge \bar L_{-1} \nonumber \\
& +  \beta_7 L_{-1} \wedge \bar L_{1} +  \beta_8 L_{-1} \wedge \bar L_{0}   \beta_9 L_{-1} \wedge \bar L_{-1} \label{ans-b}
\end{align}
we extract the equations
\begin{align} \label{claseq1}
-2 \beta_1 \alpha_0 + \beta_4 \alpha_1 & = 0, \quad - \beta_4 \alpha_{-1} -2 \beta_7 \alpha_0 =0, \\
\beta_1 \alpha_{-1} + \beta_7 \alpha_1 & = 0, \quad -2 \beta_2 \alpha_0 + \beta_5 \alpha_1 =0 , \\
- \beta_5 \alpha_{-1} -2 \beta_8 \alpha_0 & =0, \quad \beta_2 \alpha_{-1} + \beta_8 \alpha_1 = 0, \\
-2 \beta_3 \alpha_0 + \beta_6  \alpha_1  & =0, \quad -\beta_6 \alpha_{-1} -2 \beta_9 \alpha_0 =0, \\
\beta_3 \alpha_{-1} + \beta_9 \alpha_1 & =0 \label{claseq2}
\end{align}
from $[[b, a]]=0$. For the coefficients of $a$ triangularity entails \eqref{alin} and for $b$ we additionally use the automorphisms to bring them in the form \eqref{bb1}-\eqref{bb2}. For $b = b_1$, implying 
$$ \beta_1 = \beta_3 = \beta_7 = \beta_9, \quad \beta_2 = \beta_8, \quad \beta_4 = \beta_6, $$
the equations \eqref{claseq1}-\eqref{claseq2} yield
$$ \alpha_{-1} = - \alpha_1, \quad \beta_4 = 2 \beta_1, \quad \beta_5 = 2 \beta_2, $$
resulting in 
\begin{align}
r = (L_1 + L_{-1} + 2 L_0) \wedge ( \beta_1 (\bar L_1 + \bar L_{-1}) + \beta_2 \bar L_0) + a_1.
\end{align}
Similarly, for the other r-matrix components of type $b$ one has
\begin{align}\label{a+b1}
r \equiv b_2 + a =& L_1 \wedge (\beta_1 (\bar L_1 + \bar L_{-1}) + \beta_2 \bar L_0) + a_2 , \\
r \equiv b_3+a = & b_3 + \alpha (L_1 - L_{-1}) \wedge L_0, \label{disc1} \\
r \equiv b_4 + a =& \beta L_1 \wedge (\bar L_1 + \bar L_0) + a_2, \\
r \equiv b_4 + a = & \beta (L_1 + L_0) \wedge \bar L_1  + a_2, \\
r \equiv b_5 + a = & b_5  + \alpha (L_1  - L_{-1}) \wedge L_0, \\
r \equiv b_6 + a = & b_6 + a_1,  \label{a+b2} \\
r \equiv b_7 + a = & L_1 \wedge \bar L_0 + a_2, \\
r \equiv b_8 + a = & L_1 \wedge (\bar L_1 + \bar L_0) + a_2, \\
r \equiv b_9 + a = & L_1 \wedge (\bar L_1 + \bar L_{-1}) + a_2, \\
r \equiv b_{10} + a = & L_1 \wedge \bar L_1 + a_2.
\end{align}
To classify r-matrices of the form $b + \bar a$ one can use \eqref{au2} and that the coefficients of $\bar b$ with \eqref{ans-b} are just the transposed coefficients (if they are represented by a $3 \times 3$ matrix) of $b$ and a global minus sign. In the symmetric cases $b_1, b_4, b_5, b_{10}$ the results are automorphic to \eqref{a+b1}-\eqref{a+b2} with $\varphi'$ and for the rest one has
\begin{align}
r =- \bar b_2 + a = & \beta (L_1 + L_{-1} + 2 L_0) \wedge (\bar L_1 + \bar L_0) + a_1, \\
r = - \bar b_4 + a = & \beta (L_1 + L_{-1} + 2 L_0) \wedge (\bar L_1 + \bar L_{-1}) +a_1, \\
r = - \bar b_6 + a  = & b_6 + \alpha L_1 \wedge( L_0 +2 L_{-1}), \label{disc2} \\
r = - \bar b_7 + a = & L_1 \wedge (\bar L_1 + \bar L_{-1} +2 \bar L_0)+ \bar a_1.
\end{align}
While the $r$ in \eqref{disc1} and \eqref{disc2} are solutions of $[[b, a]] = 0$, the $a$ part is not triangular so they have to be discarded.
Combining the previous results (and explicitly calculating some ''overlaps'' of the form $[[b, \bar a]]$), we find for $b + a + \bar a $ the following possibilities
\begin{align}
r \equiv & b_1 + a + \bar a = (L_1 + L_{-1} + 2 L_0) \wedge ( \bar L_1 + \bar L_{1} + 2 \bar L_0) + a_1 + \bar a_1, \\
r \equiv & b_2 + a + \bar a = \beta L_1 \wedge (\bar L_1 + \bar L_{-1} + 2 \bar L_0)  + a_2 + \bar a_1, \\
r \equiv & b_4 + a + \bar a = L_1 \wedge (\bar L_1 + \bar L_0) + a_2 + \bar a_2, \\
r \equiv & b_{10} + a + \bar a = L_1 \wedge \bar L_{1} + a_2 + \bar a_2.
\end{align}

In the case $[[b, b]] \neq 0$ we again make use of the results found in \cite{Borowiec:2015nlw}. In particular, the general solution for the equation 
\begin{align}\label{abgl}
0 \neq [[b, b]] = - 2[[b, a]] - 2[[b, \bar a]]
\end{align}
 up to $\text{Aut}(\mathfrak{o}(4, \mathbb{C}))$ has the form
\begin{align}\label{absol}
\alpha L_1 \wedge L_{-1} - \alpha \bar L_1 \wedge \bar L_{-1} + b, \quad \alpha  L_1 \wedge L_0 + \alpha \bar L_1 \wedge \bar L_0 + b', 
\end{align}
with specific $b, b'$ that are not of interest for now. The first r-matrix in \eqref{absol} is quasitriangular with ad-invariant (in $\mathfrak{o}(4, \mathbb{C})$) element containing $\Omega = 4 \alpha^2 L_1 \wedge L_0 \wedge L_{-1} + ...$ and since the solutions of \eqref{abgl} up to $\text{Aut}(\mathfrak{W} \oplus \mathfrak{W})$ are in the orbits of $\mathfrak{o}(4, \mathbb{C})$ automorphisms $\varphi$ containing \eqref{absol} we would need 
\begin{align}
\varphi (\Omega) = 4 \alpha^2 \varphi (  L_1 \wedge L_0 \wedge L_{-1} ) + ... = 0
\end{align}
to obtain a triangular solution. This, however, would entail that the matrix of the coefficients of $\varphi$ has determinant zero but then it would not be invertible and thus $\varphi$ no automorphism. Furthermore, there can be no $\mathfrak{o}(4, \mathbb{C})$ automorphism that maps the $a$ terms of the second solution of \eqref{absol} to $a_1$ because $a_1$ can not be written in the form $ (\alpha_1 L_1 + \alpha_2 L_0 + \alpha_3 L_{-1}) \wedge (\alpha'_1 L_1 + \alpha'_2 L_0 + \alpha'_3 L_{-1}) $. We conclude that only r-matrices of the form $b + a_2 + \bar a_2$ have to be considered. To this end we extract the equations 
\begin{align}
\beta_1 \beta_5 - \beta_2 \beta_4 + \beta_4 2 \alpha & = 0, & - \beta_4 \beta_8  + \beta_7 \beta_5 & =0, \\
\beta_1 \beta_8 - \beta_2 \beta_7 + \beta_7 2 \alpha & =0, & 2 \beta_1 \beta_6 -2 \beta_3 \beta_4 + \beta_5 2 \alpha & = 0, \\
-2 \beta_4 \beta_9 + 2 \beta_6 \beta_7 & =0, & 2 \beta_1 \beta_9 -2 \beta_3 \beta_7 + \beta_8 2 \alpha & =0, \\
- \beta_3 \beta_5 + \beta_2 \beta_6 + \beta_6 2 \alpha & =0, & - \beta_5 \beta_9 + \beta_6 \beta_8 & =0, \\
- \beta_3 \beta_8 + \beta_2 \beta_9 + \beta_9 2 \alpha & = 0
\end{align}
from \eqref{abgl}. Additionally, we also get the same equations with $\alpha \rightarrow - \bar \alpha$ and in the terms proportional to $\bar \alpha$ the coefficients of $b$ are transposed. Solving these equations yields only the solution
\begin{align}
\beta_1 L_1 \wedge \bar L_1 + \beta_2( L_1 \wedge \bar L_0 + \bar L_1 \wedge L_0) + \beta_2 L_1 \wedge L_0 + \beta_2 \bar L_1 \wedge \bar L_0, 
\end{align}
 i.e. the same as in \cite{Borowiec:2015nlw}.
 After removing duplicates, all the r-matrices we found can be casted into the classes \eqref{classa}-\eqref{classb}.
 
 \section{Triangular r-matrices in $\mathfrak{P}_4$}

 In table 1 in \cite{Zak97} the classification of r-matrices in the four dimensional Poincar\'e algebra is given. It is not quite complete, however, only certain quasitriangular r-matrices are missing. After calculating the Schouten brackets we obtain the following triangular part of this classification.
 
 \begin{align}\label{appb1}
r_1 & = \bar k_0 \wedge k_0 + \alpha S_{11} \wedge S_{00}, \qquad r_2  = \bar k_1 \wedge k_1 + \beta S_{11} \wedge \bar k_0, \\
r_3 & = \bar k_1 \wedge k_1 + \alpha S_{11} \wedge S_{01}, \qquad r_4  = \alpha_1 S_{11} \wedge S_{01} + \alpha_2 S_{11} \wedge A_{01}, \\
r_5 & = k_0 \wedge k_1 - \bar k_0 \wedge \bar k_1 + \gamma \bar k_1 \wedge k_1, \qquad r_6  = k_0  \wedge k_1, \\
r_7 & =   S_{11} \wedge k_0 + S_{01} \wedge k_1 + A_{01} \wedge \bar k_1  + \beta S_{11} \wedge k_1, \\
r_8 & = S_{11} \wedge (k_0 + k_1)  + \alpha S_{11} \wedge A_{01}, \\
r_{9} & = S_{01} \wedge \bar k_1 + S_{11} \wedge k_1 + \alpha_1 S_{00} \wedge S_{01} + \alpha_2 S_{11} \wedge A_{01}, \\
r_{10} & = A_{01} \wedge k_1 + \alpha_1 S_{11} \wedge S_{01} + \alpha_2 S_{00} \wedge A_{01}, \\
r_{11} & = S_{11} \wedge k_1 + S_{00} \wedge (\alpha S_{11}  + \alpha_1 S_{01} + \alpha_2 A_{01}) + \tilde \alpha S_{11} \wedge A_{01}, \\
r_{12} & = (S_{11} + S_{00}) \wedge \bar k_0 + \alpha_1 S_{11} \wedge S_{00}  + \alpha_2 S_{01} \wedge A_{01}, \\
r_{13} & =  (S_{11} - S_{00}) \wedge \bar k_0 + \alpha_1 S_{11} \wedge S_{00}  + \alpha_2 S_{01} \wedge A_{01}, \\
r_{14} & = S_{11} \wedge \bar k_0 + \alpha_1 S_{11} \wedge S_{00}  + \alpha_2 S_{01} \wedge A_{01}, \\
r_{15} & = S_{01} \wedge k_0 +  \alpha_1 S_{11} \wedge S_{00}  + \alpha_2 S_{01} \wedge A_{01}, \\
r_{16} & = S_{11} \wedge k_0  +  \alpha_1 S_{11} \wedge S_{00}  + \alpha_2 S_{01} \wedge A_{01}, \\
r_{17} & = S_{11} \wedge (k_0 + \beta \bar k_0) + \alpha S_{01} \wedge A_{01}.\label{appb2}
\end{align}


%
%

\begin{singlespace}

\end{singlespace}




\backmatter


\end{document}